\documentclass[3p,times]{elsarticle}

\usepackage{ecrc}
\usepackage{amsmath}
\usepackage{color}
\def\beq{\begin{equation}}
\def\eeq{\end{equation}}

\volume{00}

\firstpage{1}

\journalname{Progress in Particle and Nuclear Physics}

\runauth{Jun Xu}

\jid{}

\jnltitlelogo{Procedia Computer Science}

\CopyrightLine{2018}{Published by Elsevier Ltd.}

\usepackage{amssymb}
\usepackage{graphicx}
\usepackage[figuresright]{rotating}
\begin{document}

\begin{frontmatter}



\dochead{}
%% e.g. \dochead{17th International Conference on Dynamical Processes in Excited States of Solids}

\title{Transport approaches for the Description of Intermediate-Energy Heavy-Ion Collisions}


\author{Jun Xu\footnote{xujun@sinap.ac.cn}}
\address[1]{Shanghai Advanced Research Institute, Chinese Academy of Sciences, Shanghai 201210, China}
\address[2]{Shanghai Institute of Applied Physics, Chinese Academy of Sciences, Shanghai 201800, China}

\begin{abstract}
The transport approach is a useful tool to study dynamics of non-equilibrium systems. For heavy-ion collisions at intermediate energies, where both the smooth nucleon potential and the hard-core nucleon-nucleon collision are important, the dynamics are properly described by two families of transport models, i.e., the Boltzmann-Uehling-Uhlenbeck approach and the quantum molecular dynamics approach. These transport models have been extensively used to extract valuable information of the nuclear equation of state, the nuclear symmetry energy, and microscopic nuclear interactions from intermediate-energy heavy-ion collision experiments. On the other hand, there do exist deviations on the predications and conclusions from different transport models. Efforts on the transport code evaluation project are devoted in order to understand the model dependence of transport simulations and well control the main ingredients, such as the initialization, the mean-field potential, the nucleon-nucleon collision, etc. A new era of accurately extracting nuclear interactions from transport model studies is foreseen.
\end{abstract}

\begin{keyword}
Boltzmann-Uehling-Uhlenbeck \sep quantum molecular dynamics \sep equation of state \sep symmetry energy \sep mean-field potential \sep nucleon-nucleon collision
\PACS 24.10.Lx \sep 25.70.-z \sep 21.30.Fe
%% or \MSC[2008] code \sep code (2000 is the defuault)

\end{keyword}

\end{frontmatter}

% \linenumber

\section{Introduction}

Understanding the properties of strong-interacting matter under extreme conditions is one of the main goals of nuclear physics. To achieve this goal, facilities such as HIAF in China, RIBF/RIKEN in Japan, SPIRAL2/GANIL in France, FAIR/GSI in Germany, SPES/LNL in Italy, RAON in Korea, NICA/JINR in Russian, as well as T-REX/TAMU and FRIB/NSCL in USA, are built or to be built for heavy-ion collision experiments. Since the whole heavy-ion collision takes a very short time, extracting the dynamics and the properties of the matter formed in these collisions from the final products often relies on transport simulations. In relativistic heavy-ion collisions, where the beam energy is much larger than the nucleon rest mass, transport approaches for various phases are needed since the partonic degree of freedom can be important (see, e.g., Ref.~\cite{Lin05}). In low-energy heavy-ion collisions such as fusion or fission reactions, where the beam energy is of only a few MeV, the nucleon-nucleon collisions are mostly blocked due to the Pauli principle so the contribution of the effective collision is much smaller than the smooth nucleon potential, and the dynamics can be well described by the time-dependent Hartree-Fock (TDHF) theory (see, e.g., Ref.~\cite{Sim18} for a recent review). Here we focus on heavy-ion collisions at intermediate energies, i.e., with the beam energy per nucleon from 50 MeV to 2 GeV, where the nucleon degree of freedom dominates the dynamics, and both the smooth nuclear potential and the hard-core nucleon-nucleon collisions become important. The central purpose of intermediate-energy heavy-ion collisions is to extract the equation of state (EOS) of the hot and dense nuclear matter formed during the collisions, while the uncertainties mostly come from the isospin-dependent part, i.e., the nuclear symmetry energy.

In order to describe properly the dynamics of intermediate-energy heavy-ion collisions, transport approaches at this energy regime are mostly developed along two directions: the Boltzmann-Uehling-Uhlenbeck (BUU) approach and quantum molecular dynamics (QMD) approach. The BUU approach basically models the time evolution of the one-body phase-space distribution based on the BUU equation numerically using the test-particle method, and the QMD approach models the time evolution of nucleons under a many-body Hamiltonian with the wave function of each nucleon represented by a Gaussian wave packet. Both approaches can be derived from many-body theories with various approximations. Without nucleon-nucleon collisions, the BUU approach reduces to the Vlasov approach, which is similar to the TDHF approach. On the other hand, it is difficult to solve exactly the extended TDHF approach, which incorporates the collision contribution into the TDHF approach~\cite{Lac99}. Both the BUU approach and the QMD approach include mainly three ingredients: the initialization of projectile and target nuclei, the mean-field potential for each nucleon, and the nucleon-nucleon collision in the dynamical evolution. The mean-field potential can be obtained from effective in-medium nucleon-nucleon interactions based on the Hartree or Hartree-Fock method, and in this way the microscopic nuclear interaction is related to the macroscopic nuclear EOS through the energy-density functional form.

So far both the BUU approach and the QMD approach have made significant contributions to understanding the nuclear EOS, the nuclear symmetry energy, and microscopic nuclear interactions. On the other hand, each transport code is established for different purposes or at different historical times, so the treatment for different parts could be slightly different, leading to different final results thus different predictions of the nuclear matter properties even from the same experimental results, and the discrepancy from different transport codes usually depends on the observables and collision systems. In order to avoid the discrepancies of transport simulation results or reduce them into minimum, the transport community has embarked on the so-called "transport code evaluation project". In such project, transport code practitioners are assigned homework calculations with the same reaction conditions and interaction parameters designed by committee members of this project, and the committee members then analyze the results, try to understand the source of the discrepancy, and eventually reduce the discrepancy. Such project begins in 2004 at the European Center for Theoretical Studies for Nuclear Physics and Related Fields (ECT*) in Trento, Italy, where the particle productions, rapidity distributions, and transverse spectra from various transport codes in the 1 A GeV regime were compared~\cite{Kol05}. Following this step, the International Workshop on Simulations of Low- and Intermediate-Energy Heavy-Ion Collisions was later held in 2014, Shanghai, China~\cite{transport2014}, mainly focused on the stability of nuclei, anisotropic flows, and collision rates. The latest event of this project was the Transport2017 workshop, embedded in the International Collaboration in Nuclear Theory (ICNT) Program supported by the Facility of Rare Isotope Beams (FRIB) at Michigan State University (MSU), held in 2017, East Lensing, USA~\cite{ICNT2017}, and the emphasis was put on testing the nucleon-nucleon collision rate, the nucleon evolution in the mean-field potential, and the production of pion-like particles in a box system with the periodic boundary condition. Further evaluation of the pion production in heavy-ion collisions is in progress, based on the calibrated treatments from the simulations in the box system.

The transport codes that have participated or partially participated in the transport code evaluation project are listed in Table.~\ref{allcodes} in an alphabetical order. There are totally 12 BUU-type codes and 11 QMD-type codes.
The BUU-type codes are: the Boltzmann-Langevin one body (BLOB) code, the BUU code developed by the Budapest/Rossendorf group (BUU-BR), the BUU code developed jointly in collaboration of Variable Energy Cyclotron Centre (VECC) with McGill University (BUU-VM), the BUU code developed by Bratkovskaya, Effenberger, Larionov, Mosel, and collaborators in Giessen (GiBUU), the hadron string dynamics (HSD) model developed in Giessen by Bratkovskaya, Cassing, and collaborators, the isospin-dependent Boltzmann-Langevin (IBL) code, the isospin-dependent Boltzmann-Uehling-Uhlenbeck (IBUU) code developed by Bertsch, Li, Chen, and collaborators, the code developed by Danielewicz (pBUU) in Michigan, the Relativistic Boltzmann-Uehling-Uhlenbeck (RBUU) code developed by the Munich/Catania/T\"ubingen group, the Relativistic Vlasov-Uehling-Uhlenbeck (RVUU) code developed by the Texas A\&M group, the stochastic mean-field (SMF) code, and the simulating many accelerated strongly-interacting hadron (SMASH) code developed in Frankfurt. The QMD-type codes are: the antisymmetrized molecular dynamics (AMD) code, the constrained molecular dynamics (CoMD) code, the improved quantum molecular dynamics code at China Institute of Atomic Energy (ImQMD-CIAE), the original isospin-dependent quantum molecular dynamics (IQMD) code, the isospin-dependent quantum molecular dynamics code at Beijing Normal University (IQMD-BNU), the isospin-dependent quantum molecular dynamics code at Institute of Modern Physics (IQMD-IMP, also known as the Lanzhou QMD (LQMD) in the literature), the isospin-dependent quantum molecular dynamics code at Shanghai Institute of Applied Physics (IQMD-SINAP), the jet AA microscopic (JAM) transport code, the QMD code developed at Japan Atomic Energy Research Institute (JQMD), the T\"ubingen quantum molecular dynamics (TuQMD) code, and the ultra-relativistic quantum molecular dynamics (UrQMD) code.

\begin{table}[h]\small
  \centering
  \caption{The names and representative references of 12 BUU-type and 11 QMD-type codes that have participated in the transport code evaluation project~\cite{Kol05,Xu16,Zhang18}.
}
    \begin{tabular}{|c|c||c|c|}
    \hline
    BUU-type & references  &  QMD-type &  references\\
    \hline
     BLOB    &   \cite{BLOB1,BLOB2} &AMD &  \cite{AMD1,AMD2} \\
     BUU-BR & \cite{Barzwolf1,Barzwolf2} & CoMD & \cite{CoMD1,CoMD2,CoMD3} \\
     BUU-VM & \cite{BUU-VM1,BUU-VM2} & ImQMD-CIAE &  \cite{ImQMD-CIAE1,ImQMD-CIAE2,Zhang08plb} \\
     GiBUU &  \cite{GIBUU1,GIBUU2,GIBUU3,GIBUU4,GIBUU5}  &  IQMD  &  \cite{IQMD1,Aic91,Har98}  \\
     HSD & \cite{HSD1,HSD2,HSD3}   & IQMD-BNU & \cite{IQMD-BNU1,IQMD-BNU2,IQMD-BNU3}   \\
     IBL    &  \cite{IBL1,IBL2,IBL3} & IQMD-IMP & \cite{IQMD-IMP1,IQMD-IMP2} \\
     IBUU   &   \cite{Li97,IBUU2,Li08,IBUU3} & IQMD-SINAP & \cite{IQMD-SINAP1,IQMD-SINAP2} \\
     pBUU   &   \cite{Dan91,pBUU1} & JAM & \cite{JAM1,JAM2} \\
     RBUU   &  \cite{RBUU1,RBUU2,Fer06,RBUU4,RBUU5} & JQMD & \cite{JQMD1,JQMD2}  \\
     RVUU   & \cite{RVUU1,RVUU2,Son15} & TuQMD & \cite{TuQMD1,TuQMD2,TuQMD3,Fuc01,Dan13,TuQMD5}\\
     SMASH    & \cite{SMASH} & UrQMD & \cite{UrQMD1,UrQMD2,UrQMD3,UrQMD4}\\
     SMF & \cite{SMF1,SMF2,SMF3}  &  & \\
    \hline
    \end{tabular}
    \label{allcodes}
\end{table}

The remaining part of this manuscript is organized as follows. Section~\ref{theory} gives a brief description of the transport theory for the BUU approach and the QMD approach. Section~\ref{achievements} briefly reviews the main achievements that the transport approach has so far made in extracting the nuclear equation of state, the nuclear symmetry energy, and the isospin splitting of the nuclear effective mass. Section~\ref{comparison} talks about the present status of the transport model evaluation project. Section~\ref{discussions} discusses the main ingredients and topics of transport models as well as their suggested treatments according to the state-of-art knowledge. Section~\ref{summary} gives a summary and outlook. %For the formulaes $\hbar=c=1$ is used.

\section{Transport theory}
\label{theory}

The theoretical frameworks of the BUU approach and the QMD approach will be briefly reviewed in this section. Here I am not going to rederive the corresponding formulae for the two approaches in a rigorous way. Instead, I will provide briefly definitions of the nucleon phase-space distribution function, the nuclear interaction as well as the mean-field potential, leading to the equations of motion, and the nucleon-nucleon collision part involved in the two approaches, for the ease of further discussions in this manuscript.

\subsection{Boltzmann-Uehling-Uhlenbeck approach}
\label{BUU}

The Boltzmann equation describes the statistical behavior of a thermodynamic system out of equilibrium, with the time evolution of the one-body phase-space distribution $f(\vec{r},\vec{p};t)$ expressed in terms of the free-propagating part, the mean-field potential part, and the collision part, i.e.,
\begin{equation}
\frac{df}{dt} = \left(\frac{\partial f}{\partial t}\right)_{prop} + \left(\frac{\partial f}{\partial t}\right)_{mf} + \left(\frac{\partial f}{\partial t}\right)_{coll}.
\end{equation}
The collision part was modified according to the Fermi-Dirac or the Bose-Einstein quantum statistics by Uehling and Uhlenbeck~\cite{Ueh33}, and the Boltzmann-Uehling-Uhlenbeck equation is written as
\begin{eqnarray}\label{buueq}
\left( \frac{\partial}{\partial t} + \frac{ \vec{p}}{m} \cdot \nabla_r  - \nabla_r U \cdot \nabla_p \right) f(\vec{r},\vec{p};t)
&=& \frac{1}{(2\pi\hbar)^6}\int d^3p_2 d^3p_3 d\Omega v_{rel} \frac{d\sigma_{12}}{d\Omega} (2\pi\hbar)^3 \delta(\vec{p}+\vec{p}_2-\vec{p}_3-\vec{p}_4) \notag \\
&\times&[f_3 f_4 (1-f) (1-f_2) - f f_2 (1-f_3) (1-f_4)].
\end{eqnarray}
On the left-hand side of the above equation, the second term represents the free-propagating term, and the third term represents the momentum-independent mean-field potential contribution. The right-hand side represents the collision integral, with $v_{rel}$ being the relative velocity of the particles before collision, ${d\sigma_{12}}/{d\Omega}$ being the differential cross section, and $+f_3 f_4 (1-f) (1-f_2)$ and $-f f_2 (1-f_3) (1-f_4)$ being respectively the gain and loss term considering the Fermi-Dirac statistics. The momentum conservation condition is expressed by the $\delta$ function.

The above BUU equation can be derived from the real-time Green's-function formalism, see, e.g., Refs.~\cite{Dan84,Bot88,Bot90,Mao94}. In such derivations, the mean-field potential $U$ is related to the real part of the self-energy, while the differential cross section ${d\sigma_{12}}/{d\Omega}$ is related to the imaginary part of the self-energy. A conceptually similar derivation starting from the von-Neumann equation with the $n$-body density matrix can be found in Refs.~\cite{Wan85,Cas90ppnp}, where the real part of the $G$ matrix is defined as the mean-field potential while all the rest correlation terms are attributed to the collision part. With the similar equations for the time evolution of the $n$-body density matrix, it has been shown in Ref.~\cite{Aic91} that different truncation schemes lead to different limits of transport theories. In the dilute limit where the range of the interaction potential is small compared to the mean free path, it reduces to the classical Boltzmann equation without mean-field potential. On the other hand, if the interval time between effective collisions is much larger than the duration of the potential interaction, it reduces to the TDHF theory. The BUU equation is a specific ansatz for the three-body density matrix and describes the time
evolution of the one-body density in a smoothly varying self-consistent mean field with short-range hard-core interactions.

In the following, I illustrate the relation between the TDHF equation and the Boltzmann-Vlasov equation, with the later neglecting the collision term in the right-hand side of Eq.~(\ref{buueq}). The TDHF equation can be expressed as
\begin{equation}\label{tdhf}
i\frac{\partial \phi_i}{\partial t} = h\phi_i,
\end{equation}
where $\phi_i$ is the wave function of the $i$th particle, and the single-particle Hamiltonian $h$ is expressed as
\begin{equation}
h = -\nabla_r \left(\frac{1}{2m} \nabla_r\right) + U(\vec{r}),
\end{equation}
where $m$ is the particle mass and $U(\vec{r})$ is the mean-field potential, which is generally obtained based on an effective in-medium nucleon-nucleon interaction through the Hartree-Fock method. The phase-space distribution function $f(\vec{r},\vec{p};t)$ is related to the particle wave function through the Wigner transformation
\begin{equation}
f(\vec{r},\vec{p};t) = \alpha \sum_i \int \phi_i(\vec{r}-\vec{s};t)\phi^*_i(\vec{r}+\vec{s};t)\exp(2i\vec{s}\cdot\vec{p}/\hbar)d^3s,
\end{equation}
where $\alpha$ is a normalization constant. The above equation together with Eq.~(\ref{tdhf}) leads to
\begin{equation}
i\frac{\partial f(\vec{r},\vec{p};t)}{\partial t} = \alpha \sum_i \int [h\phi_i(\vec{r}-\vec{s};t)\phi^*(\vec{r}+\vec{s};t)- \phi_i(\vec{r}-\vec{s};t)h\phi^*_i(\vec{r}+\vec{s};t)]\exp(2i\vec{s}\cdot\vec{p}/\hbar)d^3s.
\end{equation}
The kinetic contribution and the potential contribution to the left-hand side are respectively
\begin{eqnarray}
i\frac{\partial f^k(\vec{r},\vec{p};t)}{\partial t} &=& -i\frac{\vec{p}}{m} \cdot \nabla_r f(\vec{r},\vec{p};t) \\
i\frac{\partial f^p(\vec{r},\vec{p};t)}{\partial t} &=& i \nabla_r U \cdot \nabla_p f(\vec{r},\vec{p};t) - \frac{i}{24} \nabla^3_r U \cdot \nabla^3_p f(\vec{r},\vec{p};t) + ...
\end{eqnarray}
This leads to the Boltzmann-Vlasov equation
\begin{equation}\label{vlasov}
\frac{\partial f(\vec{r},\vec{p};t)}{\partial t} + \frac{\vec{p}}{m}\cdot \nabla_r f(\vec{r},\vec{p};t) - \frac{2}{\hbar} \sin \left( \frac{\hbar}{2} \nabla_r \cdot \nabla_p\right) U(\vec{r}) f(\vec{r},\vec{p};t) = 0,
\end{equation}
which is exactly the left hand-side part of Eq.~(\ref{buueq}) if only the first-order term of the sine function is considered.

The above Boltzmann-Vlasov equation can be solved by using the test-particle method~\cite{Won82,Ber88}. The Wigner function can be formally written as
\begin{equation}
f(\vec{r},\vec{p};t) = \int \frac{d^3\vec{r}_0d^3\vec{p}_0d^3\vec{s}}{(2\pi\hbar)^3} \exp\{i\vec{s} \cdot [\vec{p} - \vec{P}(\vec{r}_0,\vec{p}_0,\vec{s};t)]/\hbar\} \delta[\vec{r}-\vec{R}(\vec{r}_0,\vec{p}_0,\vec{s};t)] f(\vec{r}_0,\vec{p}_0;t_0).
\end{equation}
The above equation describes how the Wigner function $f(\vec{r}_0,\vec{p}_0;t_0)$ at $t_0$ evolves to that at $t$, with the phase-space coordinates $(\vec{r}_0,\vec{p}_0)$ evolving to $[\vec{R}(\vec{r}_0,\vec{p}_0,\vec{s};t),\vec{P}(\vec{r}_0,\vec{p}_0,\vec{s};t)]$. Substituting the above equation into Eq.~(\ref{vlasov}) gives the equations of motion
\begin{eqnarray}
\frac{\partial \vec{R}}{d t} &=& \frac{\vec{p}}{m}, \\
\vec{s} \cdot \frac{\partial \vec{P}}{\partial t} &=& U(\vec{R}-\vec{s}/2) - U(\vec{R}+\vec{s}/2).
\end{eqnarray}
It is seen that the canonical equation of motion $\partial \vec{P}/\partial t = -\nabla_r U$ is obtained only in the limit of $\vec{s} \rightarrow 0$. It was also commented that the above method solves the evolution of the Wigner function to the first order in the time increment only~\cite{Kri07}. The above discussions are valid for a momentum-independent mean-field potential. For a momentum-dependent mean-field potential, one needs to add another term $\frac{2}{\hbar} \sin \left( \frac{\hbar}{2} \nabla_p \cdot \nabla_r\right) U(\vec{r},\vec{p}) f(\vec{r},\vec{p};t)$ to the left-hand side of Eq.~(\ref{vlasov}), and the equations of motion become
\begin{eqnarray}
\frac{\partial \vec{R}}{\partial t} &=& \frac{\vec{p}}{m} + \nabla_p U(\vec{r},\vec{p}), \\
\frac{\partial \vec{P}}{\partial t} &=& - \nabla_r U(\vec{r},\vec{p}).
\end{eqnarray}
Generally, the mean-field potential is expressed as a function of the phase-space distribution, with the latter calculated numerically by using the test-particle method from averaging parallel events in the same ensemble
\begin{equation}
f(\vec{r},\vec{p};t) \sim \frac{1}{N_{TP}} \sum_i g(\vec{r}_i-\vec{r}) \tilde{g}(\vec{p}_i-\vec{p}),
\end{equation}
where $N_{TP}$ is the test-particle number, and $g$ and $\tilde{g}$ are normalized shape functions giving the particle a finite volume in the coordinate and the momentum space, respectively. They can be simply a $\delta$ function, or a Gaussian function, or a triangular function, and the choice of the shape function affects the calculation of the mean-field potential as well as the treatment of the Pauli Blocking.

The BUU transport approach with the test-particle method can describe rather well one-body observables in most cases. With infinite number of test particles, the BUU transport model is deterministic and in principle solves exactly the BUU equation numerically. Since the phase-space distribution is more accurately described by increasing the test-particle number, the Pauli blocking can also be carried out accurately. On the other hand, the $n$-body correlations are reduced in the price of the mean-field potential calculated from the test-particle method, so the approach can not be compared in detail to the von-Neumann equation in the framework of the BBGKY hierarchy.

Fluctuations in intermediate-energy heavy-ion collisions may not affect significantly one-body observables, but play an important role in determining the cluster formation. The main source of fluctuations is the finite number of nucleons, and the collision term representing the dissipation is always accompanied with fluctuations~\cite{Abe96}. They should be incorporated properly in transport simulations for heavy-ion reactions. In the BUU approach, statistical fluctuations always exist due to the finite number of test particles, but they are different from physical fluctuations, although sometimes they can be adjusted to reproduce the spontaneous clusterization~\cite{Col93}. On the other hand, physical fluctuations should be incorporated to the BUU transport model~\cite{Bau87}. A stochastic term is added on the right-hand side of the Boltzmann equation~\cite{Abe96}, leading to the Boltzmann-Langevin equation. Various approaches for introducing fluctuations into BUU transport models were further developed, from the early Brownian one-body fluctuation (BOB) method~\cite{Cho94}, to the stochastic mean-field (SMF) formulation with density fluctuations~\cite{SMF1,SMF2,SMF3} as well as the isospin-dependent Boltzmann-Langevin (IBL) approach~\cite{IBL1,IBL2,IBL3}, and to the most recent Boltzmann-Langevin one-body (BLOB) approach~\cite{BLOB1,BLOB2}.

\subsection{Quantum molecular dynamics approach}
\label{QMD}

The wave function of the $i$th particle in the QMD approach is explicitly written as
\begin{equation}
\phi_i(\vec{r};t) = \frac{1}{(2\pi \text{L})^{4/3}} \exp\left[ -\frac{(\vec{r}-\vec{r}_i(t))^2}{4\text{L}}+ \frac{i\vec{p}_i(t) \cdot \vec{r}}{\hbar}\right],
\end{equation}
where $\vec{r}_i(t)$ and $\vec{p}_i(t)$ are the centroid coordinate and momentum of the $i$th particle at time $t$, and $\text{L}$ characterizes the width of the wave function in the coordinate space. The total wave function of the system is the direct product of all the wave functions, i.e., $\Phi(\vec{r};t)=\Pi_i \phi(\vec{r},\vec{r}_i,\vec{p}_i;t)$, except for the antisymmetrized molecular dynamics (AMD) model~\cite{AMD1,AMD2} and the Fermionic molecular dynamics (FMD) model~\cite{Fel00} where the wave function is antisymmetrized. The phase-space distribution function is calculated from the Wigner transformation of the wave function
\begin{eqnarray}
f_i(\vec{r},\vec{p}) &=& \frac{1}{(2\pi\hbar)^3} \int \phi^*_i(\vec{r}-\vec{s}/2) \phi_i(\vec{r}+\vec{s}/2) \exp(-i\vec{p}\cdot\vec{s}) d^3s \notag\\
&=& \frac{1}{(\pi\hbar)^3} \exp \left[ -\frac{(\vec{r}-\vec{r}_i)^2}{2\text{L}} - \frac{2\text{L}(\vec{p}-\vec{p}_i)^2 }{\hbar^2}\right],
\end{eqnarray}
and here it is clear that $\sqrt{\text{L}}$ is the width of the Gaussian wave packet. Similar to the size and shape of the test particle in BUU approach, the value of $\sqrt{\text{L}}$ may affect the calculation of the mean-field potential and the Pauli blocking factor.

The generalized Lagrange function of the system can be defined as
\begin{equation}
\mathcal{L} = \int \Phi^* \left(i\frac{\partial }{\partial t} -H\right) \Phi  d^3r,
\end{equation}
where $H$ is the many-body Hamiltonian of the system expressed as
\begin{equation} \label{HQMD}
H = \sum_i T_i + \frac{1}{2}\sum_{i\ne j} V_{ij},
\end{equation}
with $T_i = p^2_i/2m$ being the kinetic energy of the $i$th particle, and $V_{ij}$ being the potential energy between the $i$th and the $j$th particles. For a direct product of the coherent state $\Phi$, the integral of the Lagrange function can be carried out as
\begin{equation}\label{LQMD}
\mathcal{L} = \sum_i \left( \frac{d\vec{r}_i}{dt} \cdot \vec{p}_i - T_i - \frac{1}{2}\sum_{j,j\ne i}\langle V_{ij} \rangle - \frac{3\hbar^2}{8\text{L}m} \right),
\end{equation}
with
\begin{eqnarray}
\langle V_{ij} \rangle \ &=& \int f_i(\vec{r},\vec{p}) f_j(\vec{r}^\prime,\vec{p}^\prime) V_{ij} d^3r d^3r^\prime d^3p d^3p^\prime. \label{U2}
\end{eqnarray}
The traditional QMD approach consists of the Skyrme-type interaction $V^{sky}_{ij}$, the Yukawa interaction $V^{yuk}_{ij}$, the isospin asymmetric interaction $V^{asy}_{ij}$, the momentum-dependent interaction $V^{md}_{ij}$, and the Coulomb interaction $V^{cou}_{ij}$, i.e.,
\begin{eqnarray}
V^{sky}_{ij} &=& t_0\delta(\vec{r}_i-\vec{r}_j) + t_3\rho[(\vec{r}_i+\vec{r}_j)/2]^{\gamma-1}\delta(\vec{r}_i-\vec{r}_j), \\
V^{yuk}_{ij} &=& c_{yuk} \frac{\exp(-\mu|\vec{r}_i-\vec{r}_j|)}{|\vec{r}_i-\vec{r}_j|},\\
V^{asy}_{ij} &=& c_{asy} \tau_i\tau_j \delta(\vec{r}_i-\vec{r}_j),\\
V^{md}_{ij} &=& t_4 \ln [t_5 (\vec{p}_i-\vec{p}_j)^2+1] \delta(\vec{r}_i-\vec{r}_j),\\
V^{cou}_{ij} &=& \frac{Z_i Z_j e^2}{|\vec{r}_i-\vec{r}_j|},
\end{eqnarray}
where $t_0$, $t_3$, $\gamma$, $c_{yuk}$, $\mu$, $c_{asy}$, $t_4$, $t_5$ are parameters, and $\tau_i$ and $Z_i$ are the isospin and charge number of the $i$th particle. The time evolution of the centroid coordinate and momentum of the $i$th particle can be obtained from Eq.~(\ref{LQMD}) according to the Euler-Lagrange equations
\begin{eqnarray}
\frac{d\vec{r}_i}{dt} &=& \frac{\vec{p}_i}{m} + \frac{1}{2}\sum_{j,j\ne i}\frac{\partial \langle V_{ij} \rangle }{\partial \vec{p}_i} = \frac{\partial \langle H \rangle }{\partial \vec{p}_i}, \\
\frac{d\vec{p}_i}{dt} &=& -\frac{1}{2}\sum_{j,j\ne i}\frac{\partial \langle V_{ij} \rangle }{\partial \vec{r}_i} = -\frac{\partial \langle H \rangle }{\partial \vec{r}_i}.
\end{eqnarray}
It is noteworthy that in order to get a ground-state initialization, the following frictional cooling method is sometimes used by incorporating damping terms into the above equations of motion
\begin{eqnarray}
\frac{d\vec{r}_i}{dt} &=& \frac{\partial \langle H \rangle }{\partial \vec{p}_i} - \lambda_r \frac{\partial \langle H \rangle}{\partial \vec{r}_i}, \label{fric1}\\
\frac{d\vec{p}_i}{dt} &=& -\frac{\partial \langle H \rangle }{\partial \vec{r}_i} - \lambda_p \frac{\partial \langle H \rangle}{\partial \vec{p}_i}, \label{fric2}
\end{eqnarray}
where $\lambda_r$ and $\lambda_p$ are damping coefficients with positive values. It is easy to prove that $d\langle H \rangle/dt$ becomes negative with the frictional terms in Eqs.~(\ref{fric1}) and (\ref{fric2}), which damp the system energy, while the shrink of the system can be prevented by further incorporating the Pauli potential $V^{pau}_{ij}$
\begin{equation}
V^{pau}_{ij} = c_{pau} \left(\frac{\hbar}{q_0 p_0}\right)^3 \exp \left[-\frac{(\vec{r}_i-\vec{r}_j)^2}{2q_0^2}-\frac{(\vec{p}_i-\vec{p}_j)^2}{2p_0^2} \right],
\end{equation}
where $c_{pau}$, $q_0$, and $p_0$ are parameters. The above Pauli potential incorporated in some of the QMD-type codes helps to dilute the phase space in not only the initialization but also during the heavy-ion dynamics. Various initialization methods will be further discussed in Sec.~\ref{init}.

Although the treatment of the nucleon-nucleon collisions in the QMD approach is similar to that in the BUU approach in real transport simulations, there are some differences in the source of the collision term~\cite{Aic91}. For the BUU equation in the framework of the BBGKY hierarchy, all two-body potentials have to be accommodated in the collision term. In the QMD approach, even if the two-body potential contributing to the collision term is expressed by an average potential and a residual interaction, the former is not the potential which enters the Schr\"{o}dinger equation. In such case, one generally obtains the collision cross section from a transition matrix or a scattering amplitude from experiments rather than from a potential.

The event-by-event evolution in the QMD transport approach shows more fluctuations than the BUU transport approach, thus giving a better description of the fragmentation in intermediate-energy heavy-ion collisions. Fluctuations in the QMD transport approach are incorporated from the beginning, and are further enhanced from stochastic collisions. The magnitude of fluctuations in the QMD transport approach is related to the width of the Gaussian wave packet rather than from the fundamental theory. A larger width leads to weaker fluctuations in the coordinate space but stronger fluctuations in the momentum space. For a comparison of the cluster formation in heavy-ion collisions from fluctuations in the SMF and AMD codes, I refer the reader to Ref.~\cite{Riz07}. The cost of fluctuations intrinsically built in the QMD transport approach is the bad performance of the Pauli blocking, due to the fluctuation of the phase-space distribution. The Pauli potential or other treatments such as the phase-space constraint in, e.g., the CoMD code, attempts to improve the Pauli blocking in the QMD approach. How to incorporate physical fluctuations while improving the Pauli blocking in the QMD approach requires further efforts.

\section{Achievements of transport model studies}
\label{achievements}

Since the early studies~\cite{Ber84,Kru85prl,Kru85,Aic85,Aic86} on the dynamics of heavy-ion collisions using transport models, great achievements have been made in constraining the properties of hot and dense nuclear matter, which can only be produced in heavy-ion collision experiments in terrestrial laboratories. Studies using transport models help to extract valuable information of the EOS of nuclear matter, which is one of the main topics of nuclear physics research, and is related to various aspects from large systems such as supernovae explosions to small ones such as finite nuclei. The isospin-dependent part of the EOS, i.e., the nuclear symmetry energy, has attracted considerable attentions in the past twenty years, and it is related to the isospin dynamics in heavy-ion collisions, the isospin-dependent excitation of finite nuclei as well as the neutron skin thickness, and the neutron star properties and cooling process. Formulisms of the nuclear matter EOS as well as the mean-field potential will be given in the following paragraphs for the ease of discussions, while the main achievements of transport approaches on the study of the nuclear matter EOS and microscopic nuclear interactions will be presented in the following subsections.

Generally, the binding energy per nucleon $E(\rho,\delta)$ in nuclear matter of the number density $\rho=\rho_n+\rho_p$ and isospin asymmetry $\delta=(\rho_n-\rho_p)/\rho$ can be expressed as
\begin{equation}
E(\rho,\delta) = E_0(\rho) + E_{sym}(\rho)\delta^2 + O(\delta^4),
\end{equation}
where $E_0(\rho)$ is the binding energy of isospin symmetric nuclear matter, and $E_{sym}(\rho)$ is the nuclear symmetry energy. The odd-order $\delta$ term vanishes as a result of isospin symmetry of microscopic nuclear interactions, i.e., the neutron-neutron interaction is considered approximately the same as the proton-proton interaction after neglecting the Coulomb force. In most cases, the above expansion is a good approximation even up to $\delta=1$, as the coefficients of higher-order $\delta$ terms are much smaller compared to the symmetry energy. Since the pressure is expressed as
\begin{equation}
P(\rho,\delta)=\rho^2\frac{\partial E(\rho,\delta)}{\partial \rho},
\end{equation}
either $E(\rho,\delta)$ or $P(\rho,\delta)$ can be taken as the EOS of isospin asymmetric nuclear matter. $E_0(\rho)$ and $E_{sym}(\rho)$ are thus the key knowledge of nuclear matter EOS. Numerically, $E_0(\rho)$ can be expanded around the saturation density $\rho_0$ as
\begin{equation}
E_0(\rho) = E_0(\rho_0) + \frac{K_0}{2!}\chi^2 + \frac{J_0}{3!}\chi^3 + O(\chi^4),
\end{equation}
with $\chi = (\rho-\rho_0)/3\rho_0$. The first-order term in the above $\chi$ expansion vanishes due to the zero pressure at the saturation density. $K_0=9\rho_0^2(d^2 E_0/d \rho^2)_{\rho=\rho_0}$ is the incompressibility of isospin symmetric nuclear matter defined at the saturation density, while the higher-order coefficient $J_0$ is important in determining the EOS at higher densities. Similarly, $E_{sym}(\rho)$ can be expanded around the saturation density $\rho_0$ as
\begin{equation}
E_{sym}(\rho) = E_{sym}(\rho_0) + L\chi + \frac{K_{sym}}{2!}\chi^2 + O(\chi^3),
\end{equation}
where $L=3\rho_0(d E_{sym}/d \rho)_{\rho=\rho_0}$ is the slope parameter, and $K_{sym}=9\rho_0^2(d^2 E_{sym}/d \rho^2)_{\rho=\rho_0}$ is the curvature parameter. Both $L$ and $K_{sym}$ are important coefficients characterizing the density dependence of the symmetry energy. The incompressibility of isospin asymmetric nuclear matter related to the curvature of the corresponding EOS at the saturation density can be expressed as~\cite{Che09}
\begin{equation}
K_{sat}(\delta) = K_0 + K_{sat,2} \delta^2 + O(\delta^4),
\end{equation}
where $K_{sat,2}=K_{sym} - 6L -\frac{J_0}{K_0}L$ is the coefficient characterizing the isospin dependence of the incompressibility. It is noteworthy that the relation $K_{sat,2}\approx K_{asy}=K_{sym} - 6L$ was extensively used in the literature, but actually the third term $-\frac{J_0}{K_0}L$ is generally not negligible.

Various many-body theories are devoted to obtain the EOS of nuclear matter. In transport approaches, phenomenological nuclear interaction models, such as the Skyrme-Hartree-Fock (SHF) model and the relativistic mean-field (RMF) model as well as other effective interaction models, are usually employed. The direct input in the transport simulation is the mean-field potential, which is related to the EOS through the energy-density functional form depending on the effective interaction in the mean-field level. Generally, the mean-field potential can be obtained from the potential contribution of the energy density through the variational principle. The mean-field potential for neutrons and protons in asymmetric nuclear matter of the number density $\rho$ and the isospin asymmetry $\delta$ can be formally expressed as
\begin{equation}
U_{n/p}(\rho,\delta,k) = U_0(\rho,k) \pm U_{sym}(\rho,k)\delta + U_{sym,2}(\rho,k)\delta^2 +O(\delta^3),
\end{equation}
where $U_0$ is the mean-field potential in isospin symmetric nuclear matter, $U_{sym}$ is the symmetry potential, $U_{sym,2}$ is the second-order symmetry potential, and the $"+(-)"$ is for neutrons (protons). For the momentum-independent interaction, the symmetry potential $U_{sym}$ is twice the potential part of the nuclear symmetry energy. In a more general and realistic case, $U_0$, $U_{sym}$, and $U_{sym,2}$ depend not only on the density $\rho$ of the nuclear medium, but also on the momentum $k$ of the nucleon. Through the Hugenholtz-Van Hove theorem, the symmetry energy and its slope parameter are related to the mean-field potential in the following way~\cite{Xuc10,Che12}
\begin{eqnarray}
E_{sym}(\rho) &=& \frac{1}{3} \frac{\hbar^2k_F^2}{2m_0^*} + \frac{1}{2} U_{sym}(\rho,k_F), \\
L(\rho) &=& \frac{2}{3} \left.\frac{\hbar^2k^2}{2m_0^*}\right|_{k_F} - \frac{1}{6}\left(\frac{\hbar^2k^3}{{m_0^*}^2}\frac{\partial m_0^*}{\partial k} \right)_{k_F} + \frac{3}{2} U_{sym}(\rho,k_F) + \left(\frac{\partial U_{sym}}{\partial k}\right)_{k_F} \cdot k_F + 3 U_{sym,2}(\rho,k_F),
\end{eqnarray}
where $k_F=\hbar(3\pi^2\rho/2)^{1/3}$ and $m^*_0=m\left(1+\frac{m}{\hbar^2 k}\frac{\partial U_0}{\partial k}\right)^{-1}$ are respectively the Fermi momentum and the nucleon effective mass in isospin symmetric nuclear matter. Note here the momentum dependence of the nucleon effective mass and the second-order symmetry potential also contribute to the slope parameter of the symmetry energy.

\subsection{EOS of symmetric nuclear matter}
\label{eos}

The accurate knowledge of the EOS of isospin symmetric nuclear matter is obtained mostly around the saturation density $\rho_0 \approx 0.16$ fm$^3$, where the binding energy per nucleon is about $-16$ MeV from the liquid-drop model fit of masses of various nuclei, and the incompressibility is about $K=240 \pm 40$ MeV from isoscalar giant monopole resonance studies~\cite{Bla80,You99,Shl06,Col09,Pie10,Che12a}. Besides nuclear astrophysics observations, our knowledge of nuclear matter EOS at suprasaturation densities relies on transport model studies. In the following I briefly review efforts on constraining the nuclear matter EOS at suprasaturation densities from kaon productions and collective flows in intermediate-energy heavy-ion collisions.

\begin{figure}[h]
\centering
\includegraphics[scale=0.49,clip]{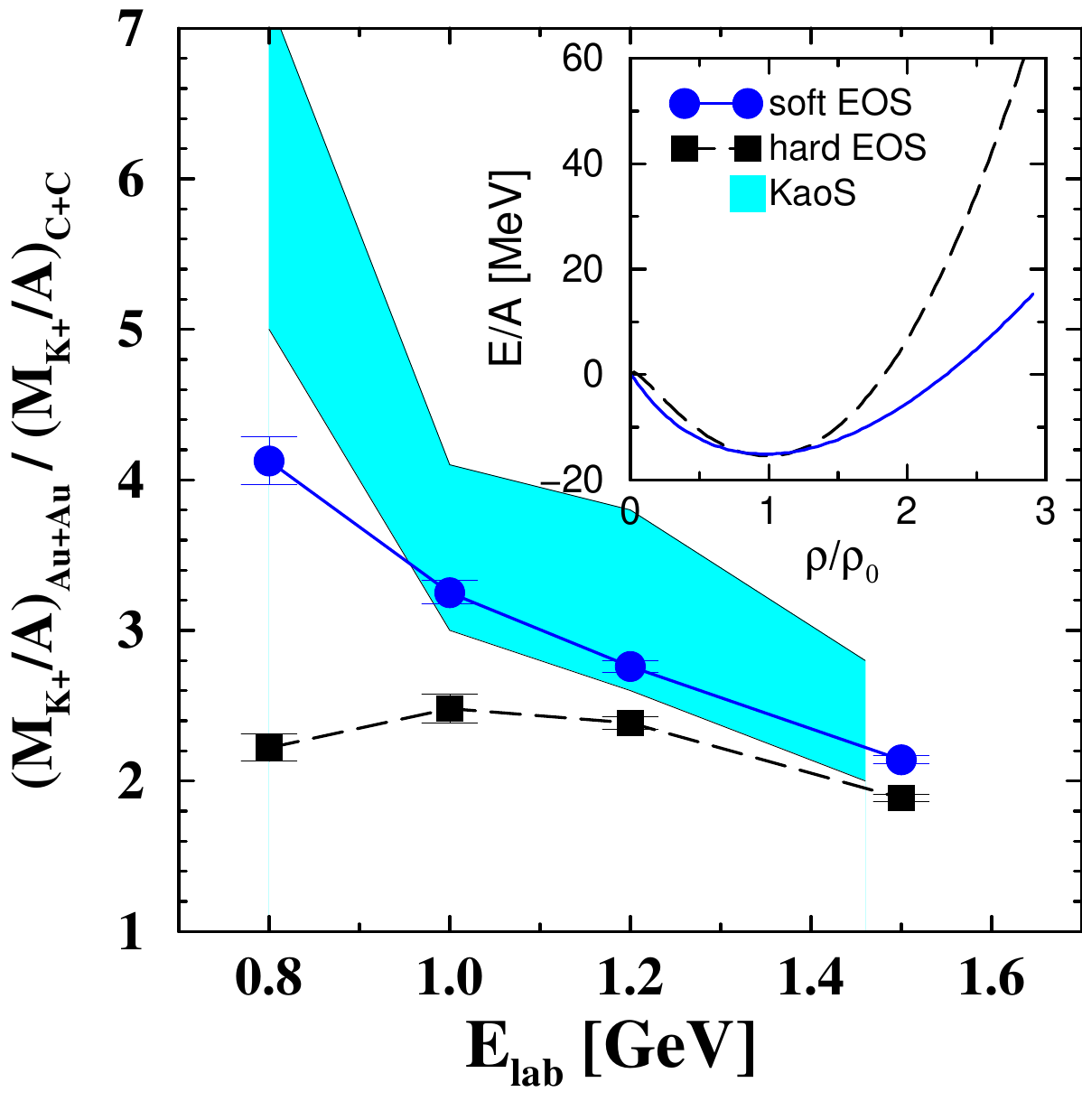}
\includegraphics[scale=0.36,clip]{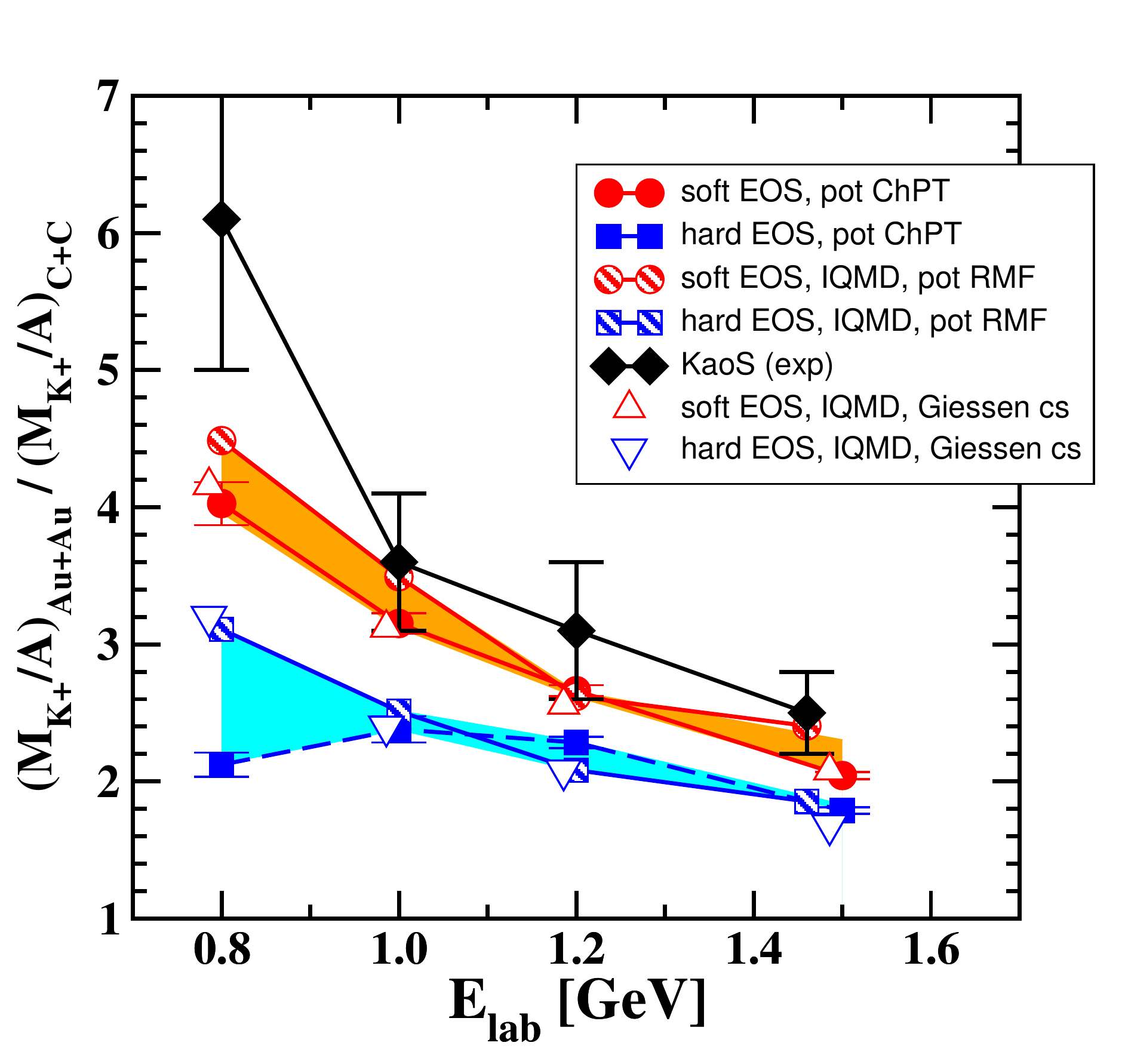}
\includegraphics[scale=0.5,clip]{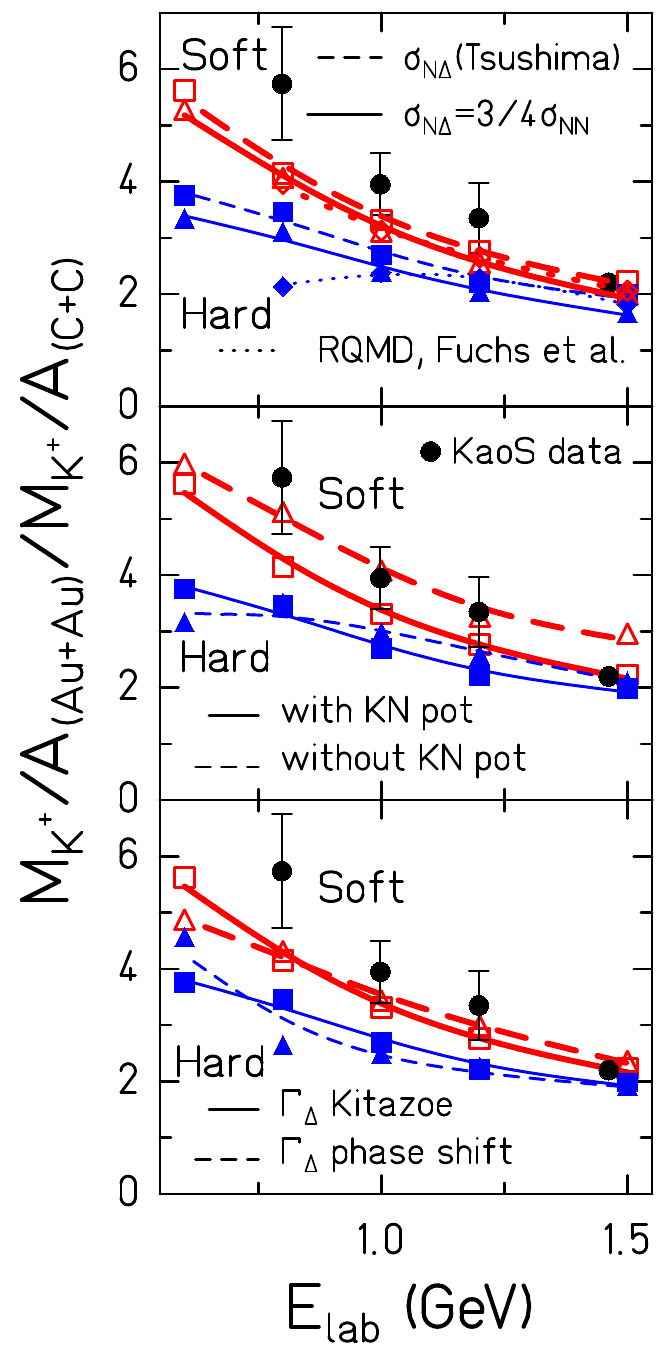}
\caption{(Color online) Comparison of the relative $K^+$ yield per nucleon in Au+Au and C+C collisions as a function of the beam energy from various transport model studies. Left: Taken from Ref.~\cite{Fuc01}; Middle: Taken from Ref.~\cite{Fuc06epja}; Right: Taken from Ref.~\cite{Har06}.}
\label{kaon}
\end{figure}

Since kaons have much weaker interactions with nucleons than pions, they are considered as a better probe of the EOS at suprasaturation densities where they are mostly produced. This is the case especially near the kaon production threshold, where the density and the temperature in the central region of heavy-ion collision system are rather sensitive to the stiffness of the EOS~\cite{Aic85prl}. In heavy-ion collisions with the beam energy of about 1A GeV, the production channels of $K^+$ are mainly $B+B \rightarrow B+Y+K^+$ and $\pi+B \rightarrow Y+K^+$, where $B$ represents baryons and $Y$ represents hyperons, respectively. With a softer (stiffer) EOS, the density of the most compressed stage in heavy-ion collisions is expected to be higher (lower), leading to more (less) violate collisions and stronger (weaker) kaon productions. On the other hand, kaons in nuclear medium are affected by their mean-field potentials based on effective chiral models, and this leads to a repulsive Schr\"odinger equivalent potential for $K^+$, suppressing its yield especially at higher densities and in heavier systems. As shown in the left panel of Fig.~\ref{kaon}, the relative $K^+$ yield per nucleon in Au+Au and C+C collisions is a better probe of the symmetric matter EOS than the yield in a single system, since the yield ratio is rather insensitive to the kaon in-medium interaction. Comparing the transport simulation results with the experimental data by KaoS Collaboration~\cite{Kao01} favors a soft EOS. A later study with more advanced EOSs from the chiral perturbation theory (ChpT) and the relativistic mean-field model (RMF) and using an alternative set of elementary $K^+$ production cross sections (Giessen cs) leads to the similar conclusion, as shown in the middle panel of Fig.~\ref{kaon}. The study in Ref.~\cite{Har06} from a QMD transport model emphasizes the kaon production from the channel of $\Delta+N \rightarrow N+K^+ + \Lambda$. It is shown in the right panel of Fig.~\ref{kaon} that different options of the cross section of $N+\Delta$, the kaon-nucleon potential, as well as the width of $\Delta$ mass do not change much the beam energy dependence of the relative $K^+$ yield per nucleon in Au+Au and C+C collisions, leading to the robust conclusion that the nuclear matter EOS at suprasaturation densities is soft. The kaon production is also compared and discussed in Ref.~\cite{Kol05}, as an important part of the transport model evaluation project. I refer the reader to Refs.~\cite{Ko96,Cas99,Fuc06ppnp} for nice reviews of kaon productions in heavy-ion collisions.

\begin{figure}[h]
\centering
\includegraphics[scale=0.41,clip]{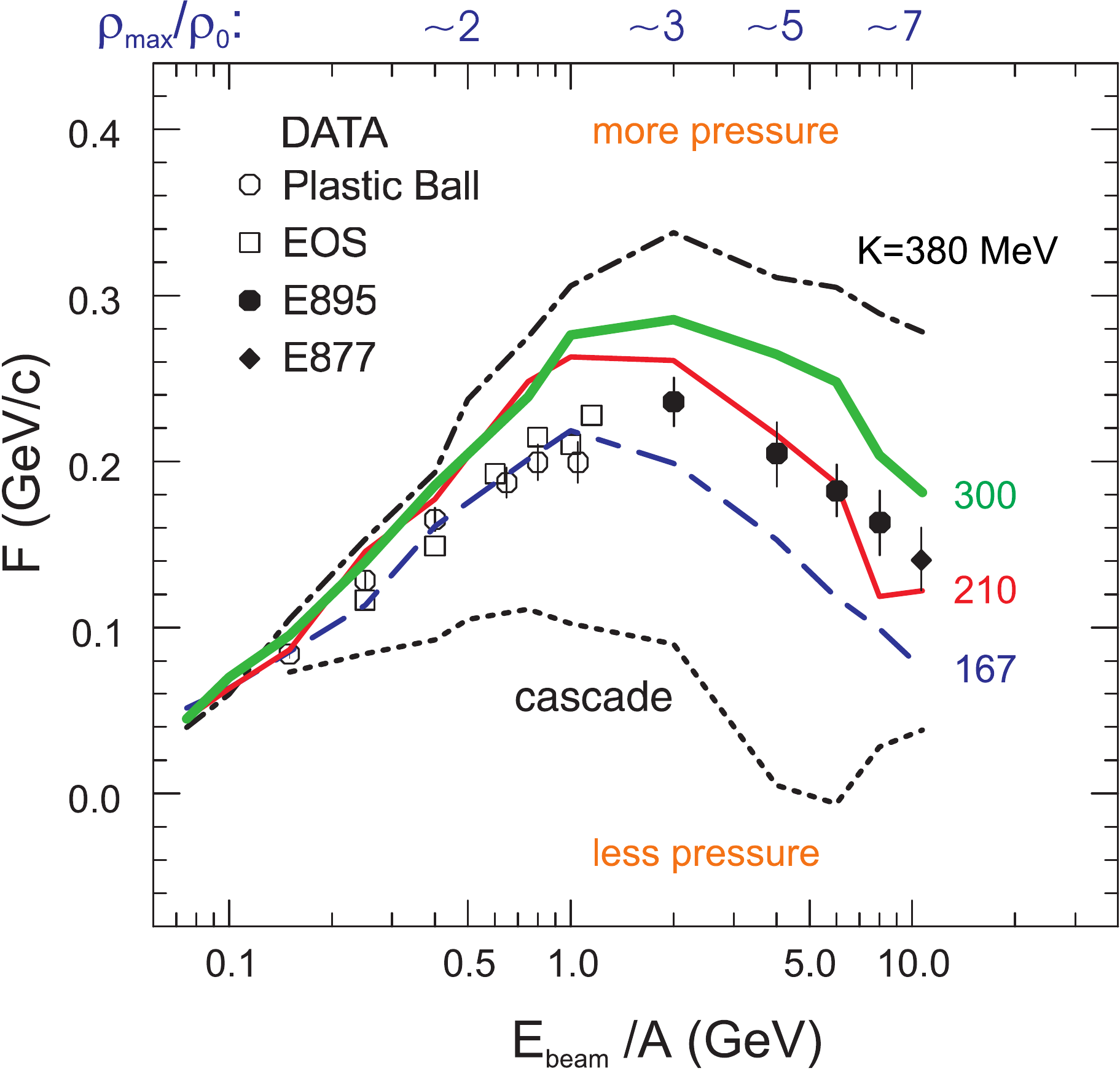}
\includegraphics[scale=0.6,clip]{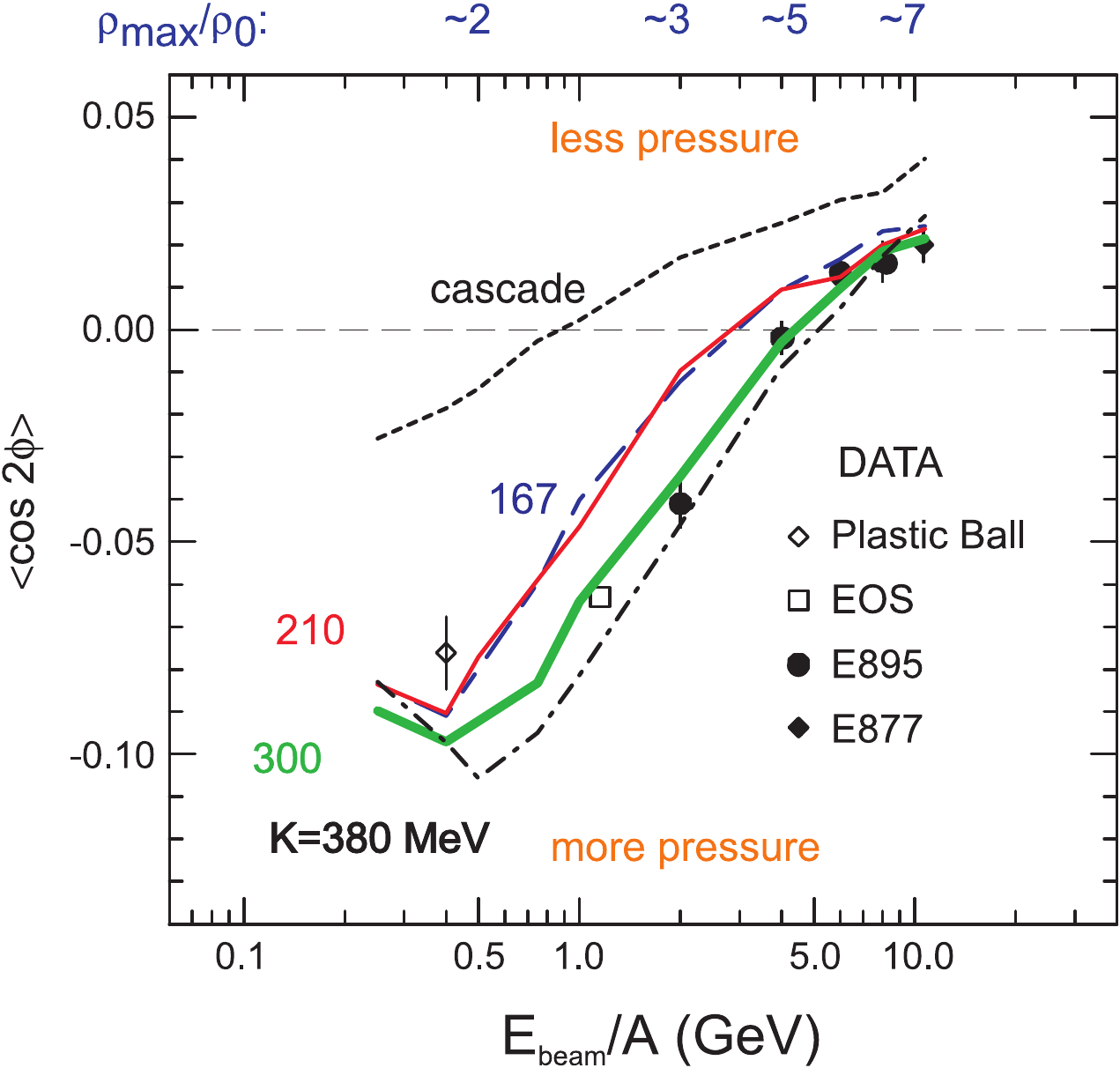}
\caption{(Color online) Studies on collective flows in heavy-ion collisions as probes of symmetric nuclear matter EOS from a BUU transport model~\cite{Dan02}. Left: Transverse flow slope as a function of the beam energy from various EOSs with different values of incompressibility $K$. Right: Elliptic flow as a function of the beam energy from various EOSs with different values of incompressibility $K$. ¡°Plastic Ball,¡±
¡°EOS,¡± ¡°E877,¡± and ¡°E895¡± denote experimental data taken from Refs.~\cite{v1_Ball,v2_Ball}, Refs.~\cite{v1_EOS,v2_EOSE877}, Refs.~\cite{v1_E877,v2_EOSE877}, and Refs.~\cite{v1_E895,v2_E895}, respectively.}
\label{flow_Dan}
\end{figure}

The anisotropic collective flow is another probe of the nuclear matter EOS. The transverse flow is from the deflection of the projectile matter in the positive $x$ direction and the target matter in the $-x$ direction in non-central heavy-ion collisions. The resulting transverse flow is actually from the competition between the repulsive nucleon-nucleon collisions and the mean-field potential, with the latter being attractive at lower densities/beam energies and repulsive at higher densities/beam energies. The slope of the transverse flow $F=[d\langle p_x/A \rangle/d (y/y_{cm})]_{y/y_{cm}=1}$ is a measure of the nuclear interaction, where $y_{cm}$ is the rapidity of particles at rest in
the center of mass, and $A$ is the number of nucleons in the detected particle. As shown in the left panel of Fig.~\ref{flow_Dan}, $F$ is larger with a stiffer EOS or a larger value of incompressibility $K$. The elliptic flow $\langle \cos(2\phi) \rangle$, with $\phi$ the azimuthal angle, is a measure of different numbers of in-plane and out-of plane emitting particles. As seen from the right panel of Fig.~\ref{flow_Dan}, at lower beam energies the expansion of the participant matter is blocked by the spectator matter, which does not pass through that quickly, leading to the squeeze-out of particles perpendicular to the reaction plane and thus a negative elliptic flow. A more repulsive mean-field potential or a stiffer EOS leads to the faster expansion of the participant matter and a stronger squeeze-out effect, resulting in an even more negative elliptic flow. At higher beam energies, the spectator matter passes through more quickly, and the in-plane hydrodynamic flow dominates the dynamics, corresponding to a positive elliptic flow. The comparison between the collective flow and the experimental data leads to the incompressibility $K=200 \sim 300$ MeV. A momentum-dependent mean-field potential with the nucleon effective mass of about $70\%$ of the free nucleon mass and the in-medium nucleon-nucleon collision cross sections are properly incorporated in this study based on a BUU transport approach.

\begin{figure}[h]
\centering
\includegraphics[scale=0.5,clip]{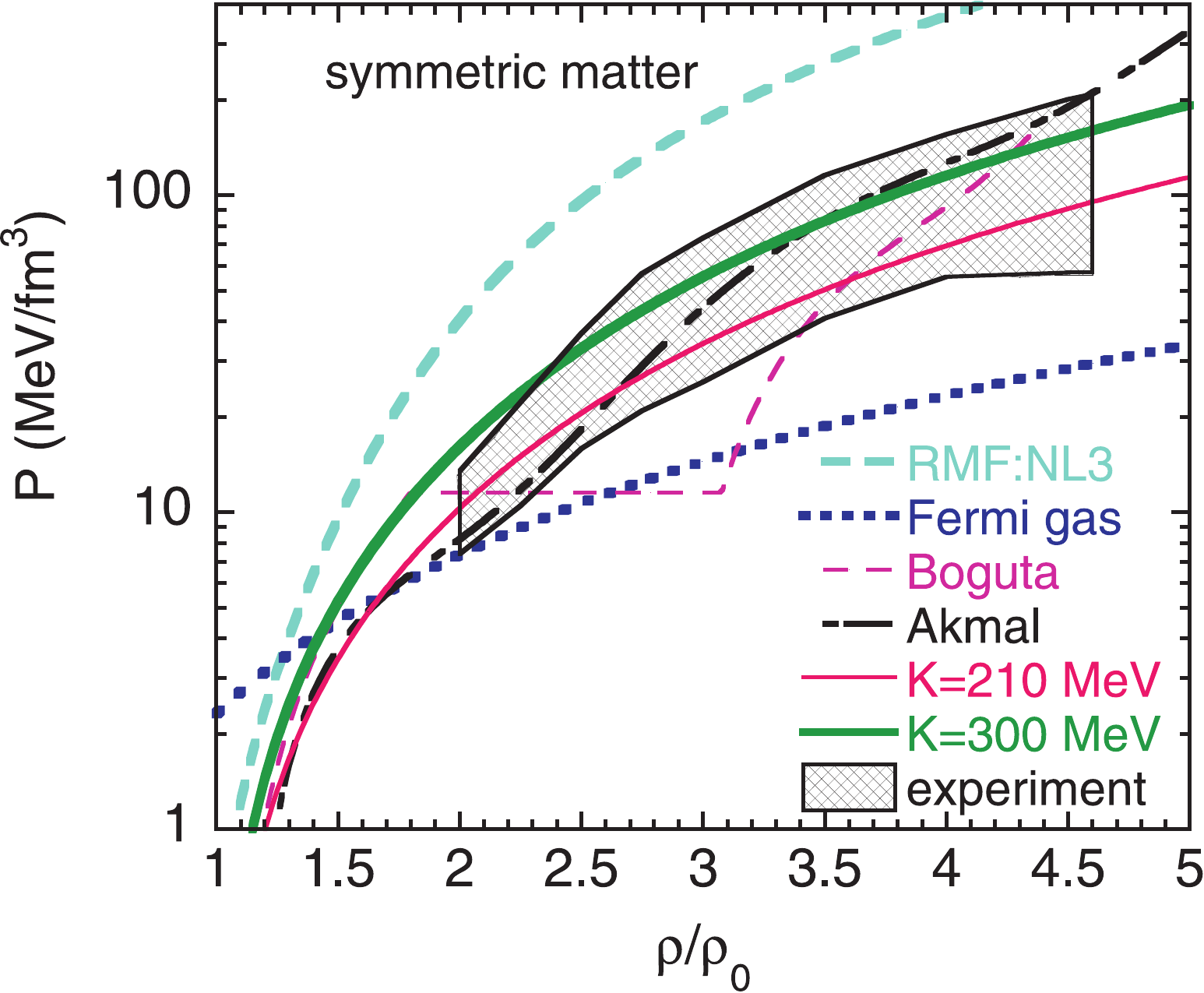}
\includegraphics[scale=0.36,clip]{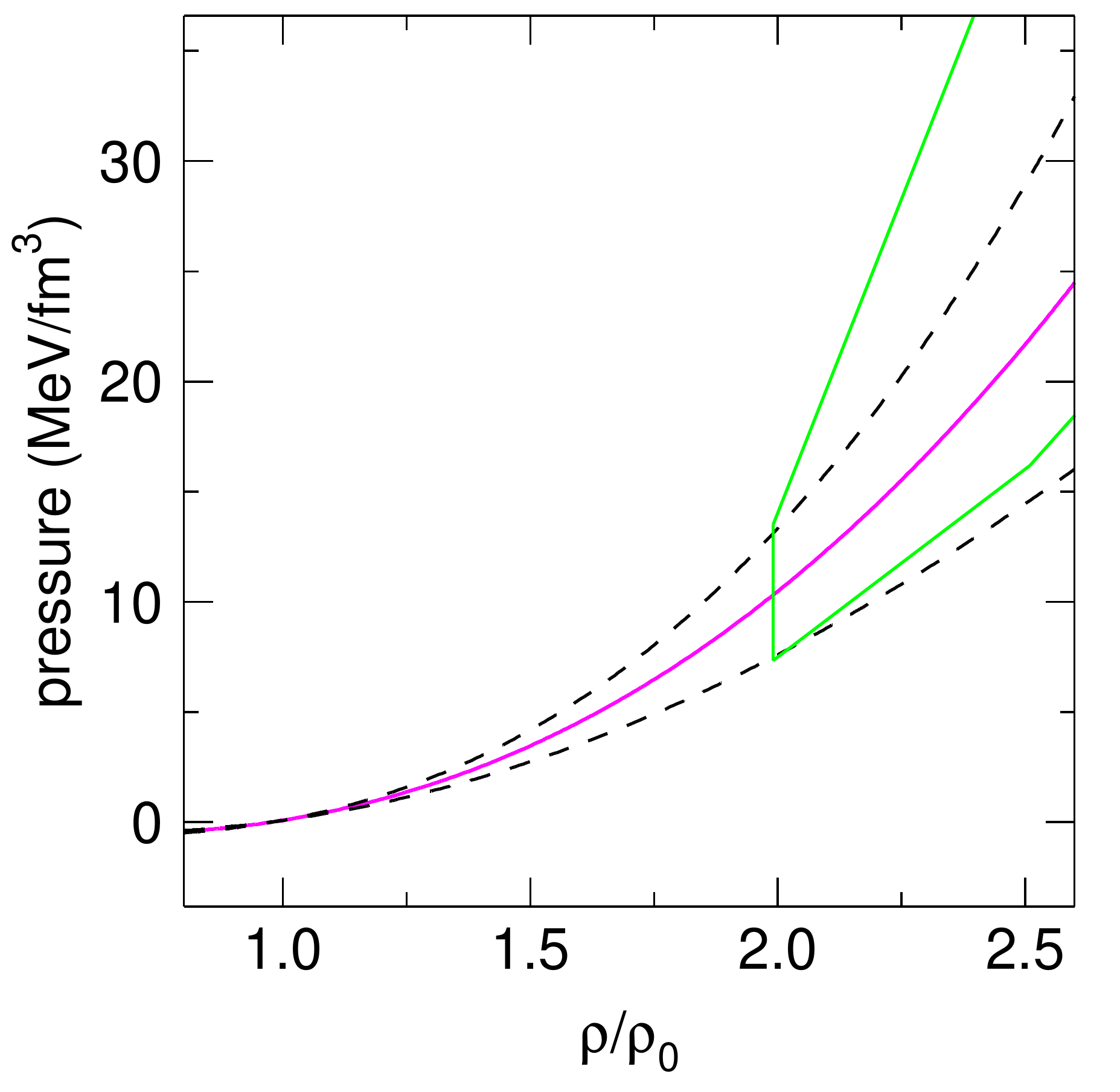}
\caption{(Color online) Constraints on the symmetric nuclear matter EOS from the flow data. Left: Based on a BUU transport approach~\cite{Dan02}; Right: Comparison of the EOS constraint from a more recent study based on a QMD transport approach~\cite{Fev15} (dashed-line region) with that in the left panel (solid-line region).}
\label{EOS_SNM}
\end{figure}

The above flow analysis covers the maximum density range from $2\rho_0$ to $5\rho_0$, where the nuclear matter EOS is stringently constrained. As shown in the left panel of Fig.~\ref{EOS_SNM}, the EOS is constrained within the shaded band, which rules out the very repulsive NL3 parametrization of the relativistic mean-field model. A more recent study was done from a QMD approach based on more extensive analyses of the elliptic flows for protons and light clusters in Au+Au collisions at the beam energies between 0.4 and 1.5 A GeV~\cite{Fev15}. Comparison with the FOPI data~\cite{FOPI} leads to the constraint of the EOS as shown within the dashed-line region in the right panel of Fig.~\ref{EOS_SNM}. It is seen that the EOS within this region is even softer than that within the green solid-line region, with the latter corresponding to the EOS constraint from Ref.~\cite{Dan02}.

\subsection{Nuclear symmetry energy}
\label{esym}

The symmetry energy at saturation density is about 30 MeV from the liquid-drop model fit of the finite nuclei binding energy. The constraints on the density dependence of the nuclear symmetry energy can be classified as those at subsaturation densities and those at suprasaturation densities. At subsaturation densities, information of the symmetry energy can be extracted from finite nuclei analyses as well as low- and intermediate-energy heavy-ion collisions. The former includes measurements of the neutron skin thickness, isovector giant resonances, and so on, while the later includes isospin observables in low- and intermediate-energy heavy-ion collisions, such as the neutron/proton or triton/$^3$He yield ratio, the neutron-proton differential flow, the isoscaling of fragments, the isospin diffusion, and so on. At suprasaturation densities, our knowledge of the symmetry energy relies on nuclear astrophysics observations and intermediate- and high-energy heavy-ion collisions. The former mainly includes neutron star radii as well as the neutrino emission and cooling rate, and the later mainly includes neutron-proton differential collective flows as well as the yield ratios of charged pions and kaons. Transport approach is a useful tool of extracting valuable information of the nuclear symmetry energy at both subsaturation and suprasaturation densities from intermediate-energy heavy-ion collision experiments. For reviews on various topics of the nuclear symmetry energy, I refer the reader to Refs.~\cite{Bar05,Ste05,Lat07,Li08}.

\subsubsection{Probes of nuclear symmetry energy at subsaturation densities from transport model studies}
\label{subsaturation}

\begin{figure}[h]
\centering
\includegraphics[scale=0.4,clip]{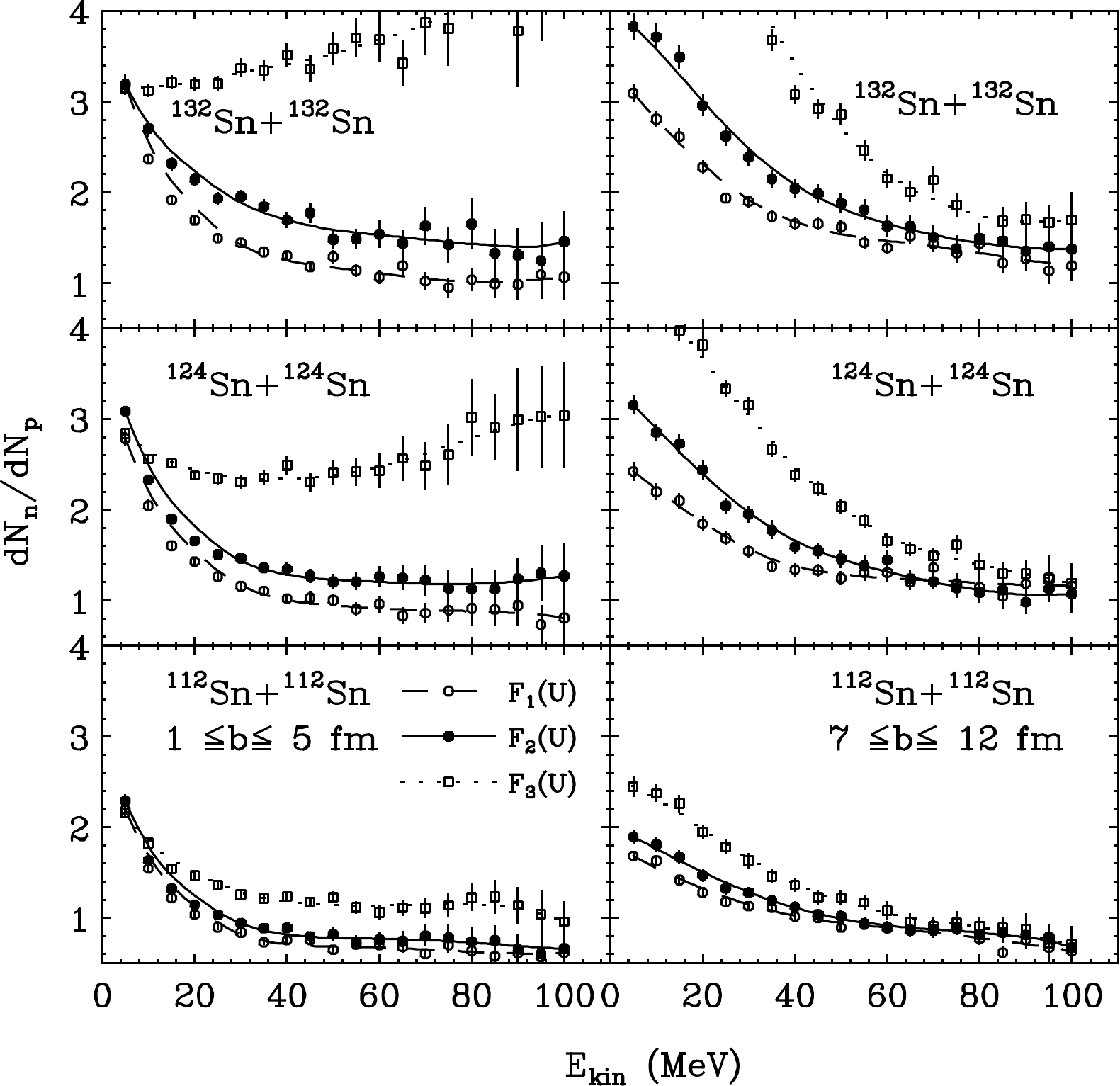}
\includegraphics[scale=0.8,clip]{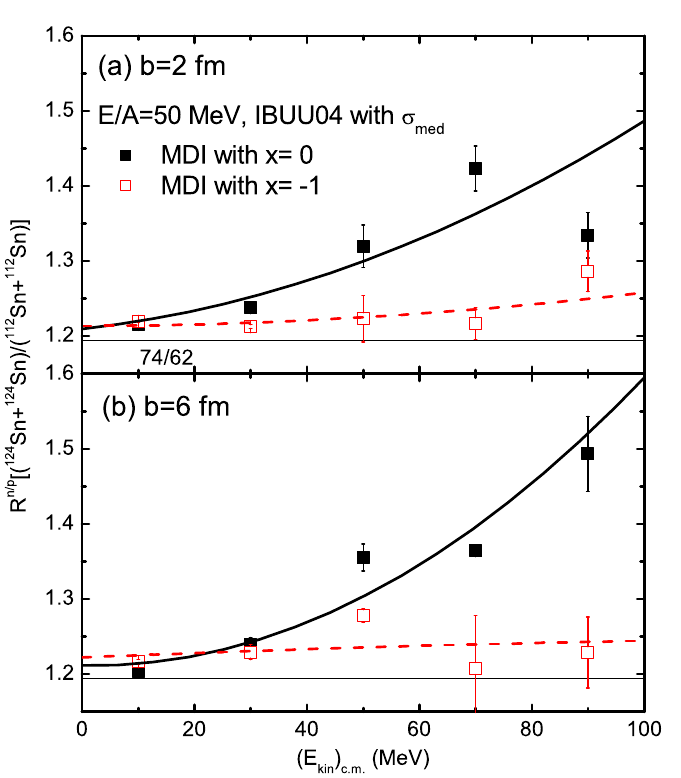}\\
\includegraphics[scale=0.37,clip]{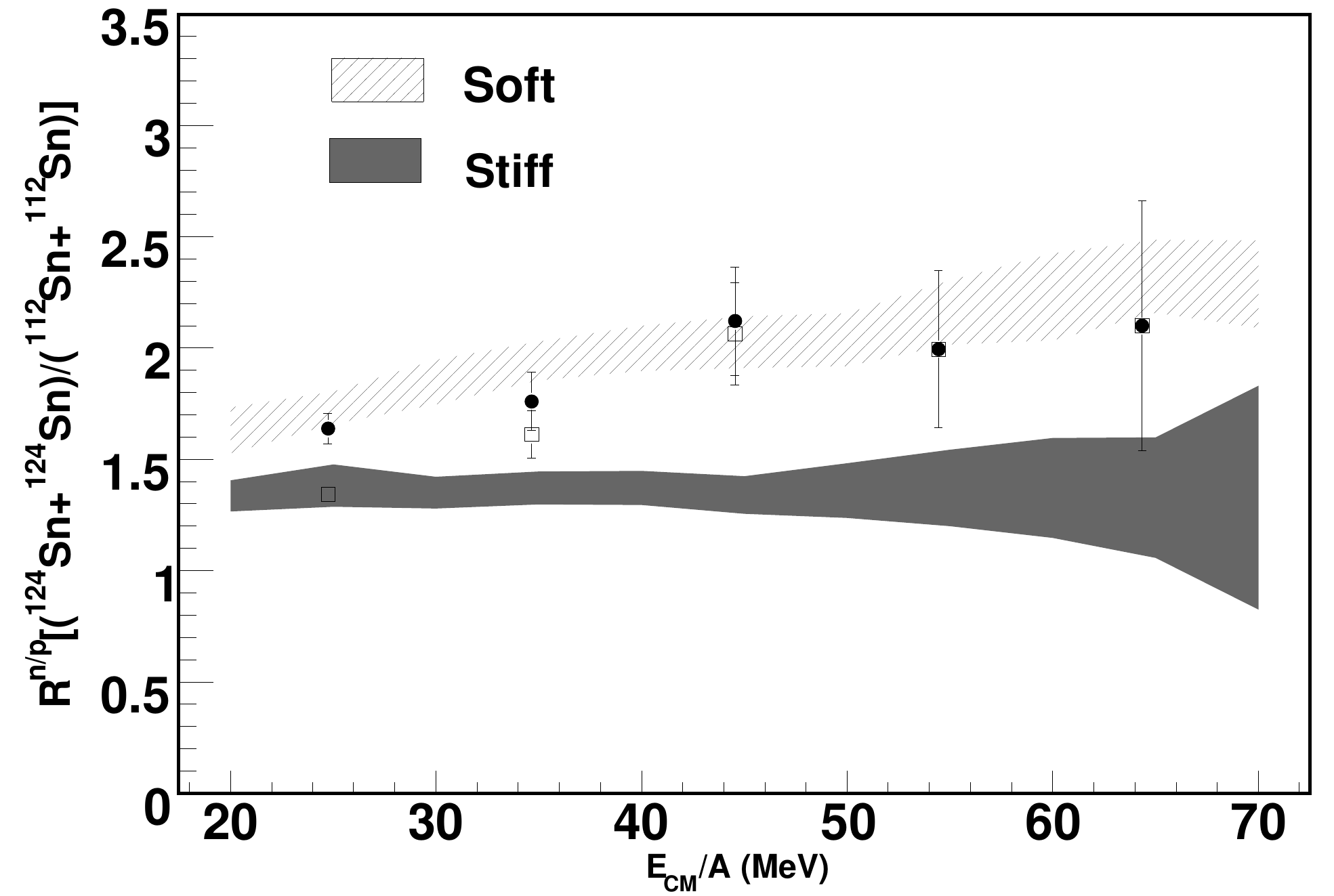}
\includegraphics[scale=0.7,clip]{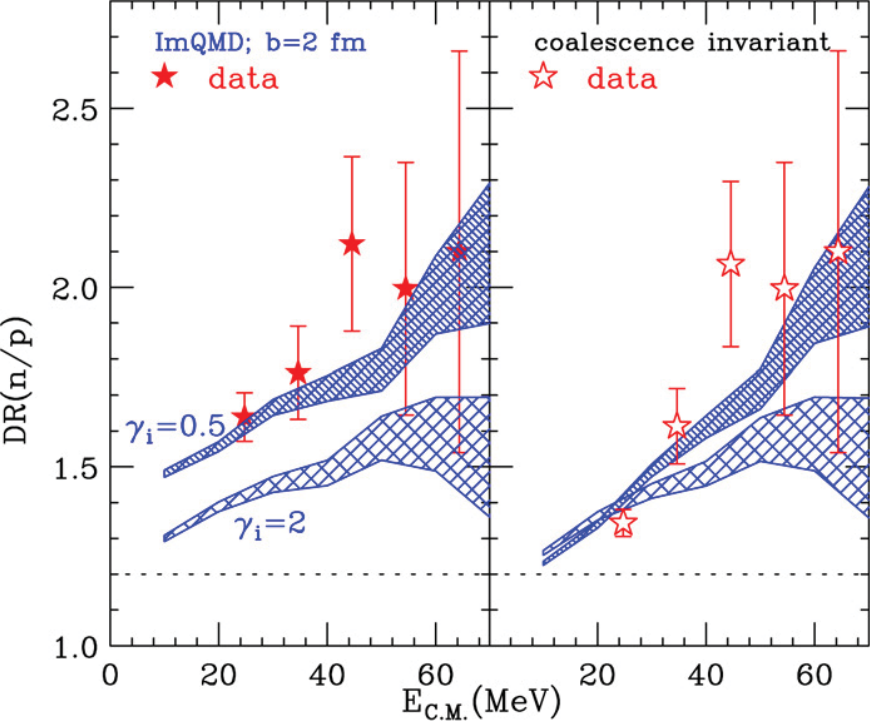}
\caption{(Color online) Upper left: Kinetic energy dependence of the single neutron/proton ratio from the IBUU97 study~\cite{Li97}; Upper right: Kinetic energy dependence of the double neutron/proton ratio from the IBUU04 study~\cite{Li06plb}; Lower left: Kinetic energy dependence of double neutron/proton ratio from the IBUU97 calculation compared with NSCL/MSU experimental data~\cite{Fam06prl}; Lower right: Kinetic energy dependence of double neutron/proton ratio from the ImQMD calculation compared with NSCL/MSU experimental data~\cite{Zhang08plb}. The collision system is Sn+Sn at 40 AMeV for the upper left panel and Sn+Sn at 50 AMeV for the others.}
\label{np}
\end{figure}

The neutron/proton yield ratio is a natural probe of the neutron-proton potential difference as well as the nuclear symmetry energy. A larger symmetry energy generally leads to a more repulsive mean-field potential for neutrons than that for protons in neutron-rich matter. This enhances the emission of neutrons than protons, and makes free neutrons more energetic than protons. This effect is rather insensitive to the isoscalar part of the nuclear EOS as well as the in-medium nucleon-nucleon collision cross sections. The upper left panel in Fig.~\ref{np} compares the kinetic energy dependence of the neutron/proton yield ratio from three different symmetry energies from stiffer to softer, with the potential part $F_1(u)=2u^2/(1+u)$, $F_2(u)=u$, and $F_3(u)=u^{1/2}$, where $u=\rho/\rho_0$ is the reduced density, based on the study using the IBUU transport model of the version IBUU97~\cite{Li97}. The larger neutron/proton yield ratio at lower kinetic energies is due to the Coulomb repulsion, which pushes protons to higher kinetic energies. It is seen that the stronger symmetry energy $F_3$ at subsaturation densities gives a larger neutron/proton yield ratio, especially for a more neutron-rich system. The symmetry energy effect on the neutron/proton yield ratio shows up at higher kinetic energies in midcentral collisions but at lower kinetic energies in peripheral collisions. On the experimental side, the neutron detecting efficiency is generally much lower than that for protons. In order to reduce the influence of the Coulomb interaction and the poor efficiencies of detecting low-energy neutrons, the double neutron/proton yield ratio for $^{124}$Sn+$^{124}$Sn and $^{112}$Sn+$^{112}$Sn was proposed to be a probe of the symmetry energy of practical use~\cite{Li06plb,Fam06prl}. As shown in the lower left panel of Fig.~\ref{np}, the double neutron/proton yield ratio from the NSCL/MSU experiments is compared with the IBUU97 results in Ref.~\cite{Li97}. It is seen that the data favors a potential part of the symmetry energy between $F_3(u)=u^{1/2}$ and $F_2(u)=u$, as shown respectively by the "soft" and "stiff" band in the figure. A more realistic calculation was done using a more advanced version IBUU04 with an isospin- and momentum-dependent mean-field potential for nucleons and in-medium nucleon-nucleon collision cross sections, as shown in the upper right panel of Fig.~\ref{np}. Effects with different symmetry energies from the parameter $x=0$ and $-1$ are qualitatively consistent with those from a momentum-independent mean-field potential, although the sensitivity is lower. A later study from the ImQMD calculation also favors a soft symmetry energy with the potential part of the symmetry energy evolving like $u^{1/2}$, as shown in the lower right panel. In order to be compatible with the IBUU calculations, the so-called coalescence invariant double neutron/proton yield ratio by considering all the neutrons and protons emitted or in clusters was employed, and the conclusion remains the same.

\begin{figure}[h]
\centering
\includegraphics[scale=0.9,clip]{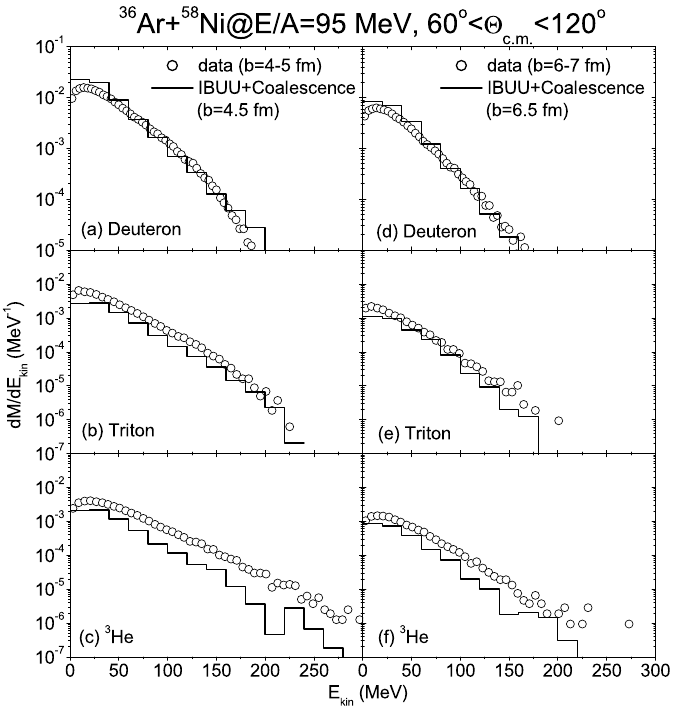}
\includegraphics[scale=0.9,clip]{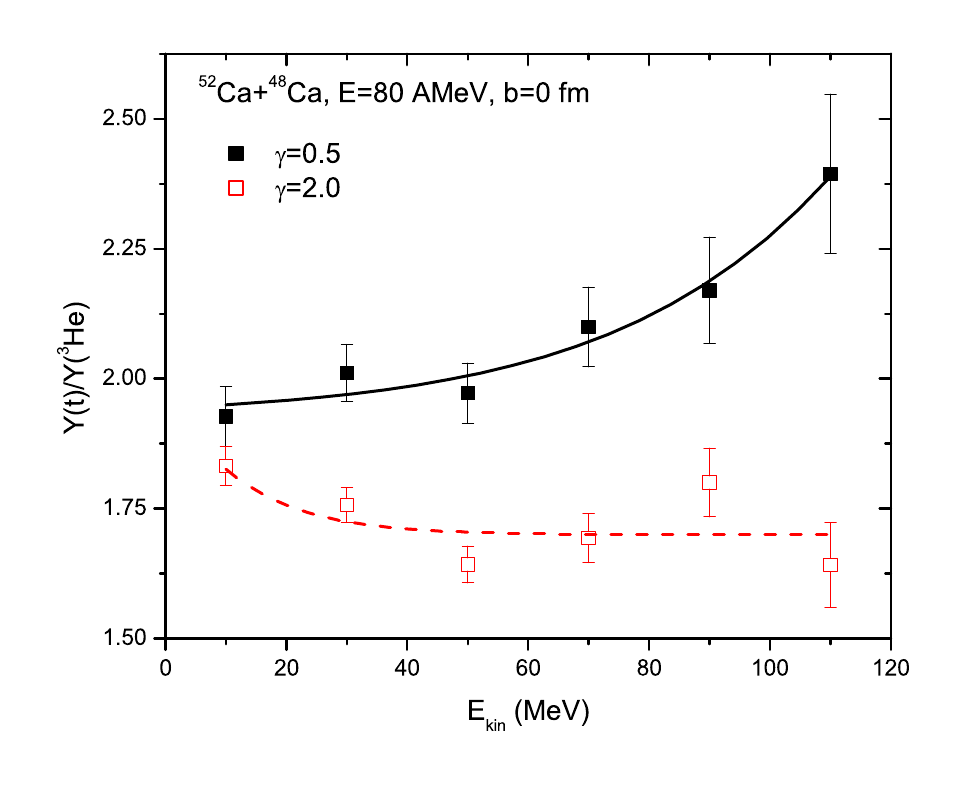}
\caption{(Color online) Left: Kinetic energy spectra of deuterons, tritons, and $^3$He~\cite{Che03npa}; Right: Kinetic energy dependence of the triton/$^3$He yield ratio~\cite{Che03prc}. Results are calculated from a dynamical coalescence model based on the freeze-out phase-space distribution generated by the IBUU transport model.}
\label{clusters}
\end{figure}

The isobaric yield ratio triton/$^3$He is another probe of the nuclear symmetry energy. In the BUU transport approach, the probability for producing a cluster is determined by its Wigner phase-space density and the nucleon phase-space distribution at freeze-out. The freeze-out criterion of a nucleon in BUU simulation is determined in such a way that the local density of a nucleon is lower than a certain small value, e.g., $\rho_0/8$, or its last collision time, and both give similar results. The Wigner functions of triton and $^3$He are obtained from the product of three ground-state wave functions of spherical
harmonic oscillators with parameters adjusted to reproduce the measured root-mean-square radii. As shown in the left panel of Fig.~\ref{clusters}, this method can reproduce reasonably well the kinetic energy spectra of deuterons, tritons, and $^3$He. It is also observed in Refs.~\cite{Che03npa,Che03prc} that a softer symmetry energy, which is stronger at subsaturation densities, leads to an earlier and stronger emission of neutrons than protons, resulting in a larger separation of neutron and proton phase space at freeze-out, thus smaller multiplicities of these light clusters compared with a stiffer symmetry energy. On the other hand, since more neutrons are emitted and accelerated to higher energies than protons from a softer symmetry energy, the isobaric yield ratio triton/$^3$He related to the neutron/proton yield ratio is a good probe of the symmetry energy, especially at higher kinetic energies, as shown in the right panel of Fig.~\ref{clusters}. Again, the result is insensitive to the isoscalar nuclear matter EOS and the in-medium nucleon-nucleon cross sections.

\begin{figure}[h]
\centering
\includegraphics[scale=0.4,clip]{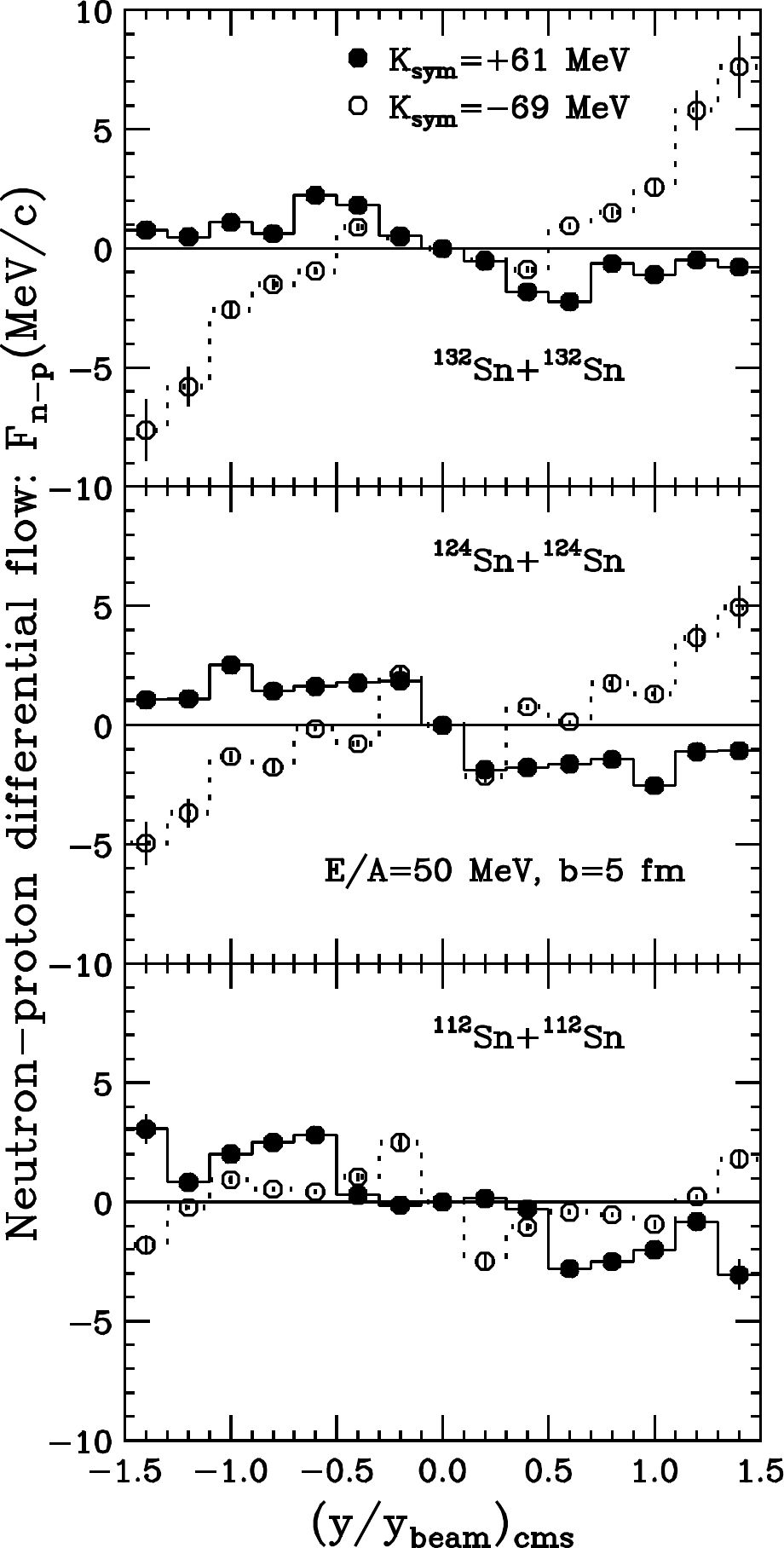}
\includegraphics[scale=0.5,clip]{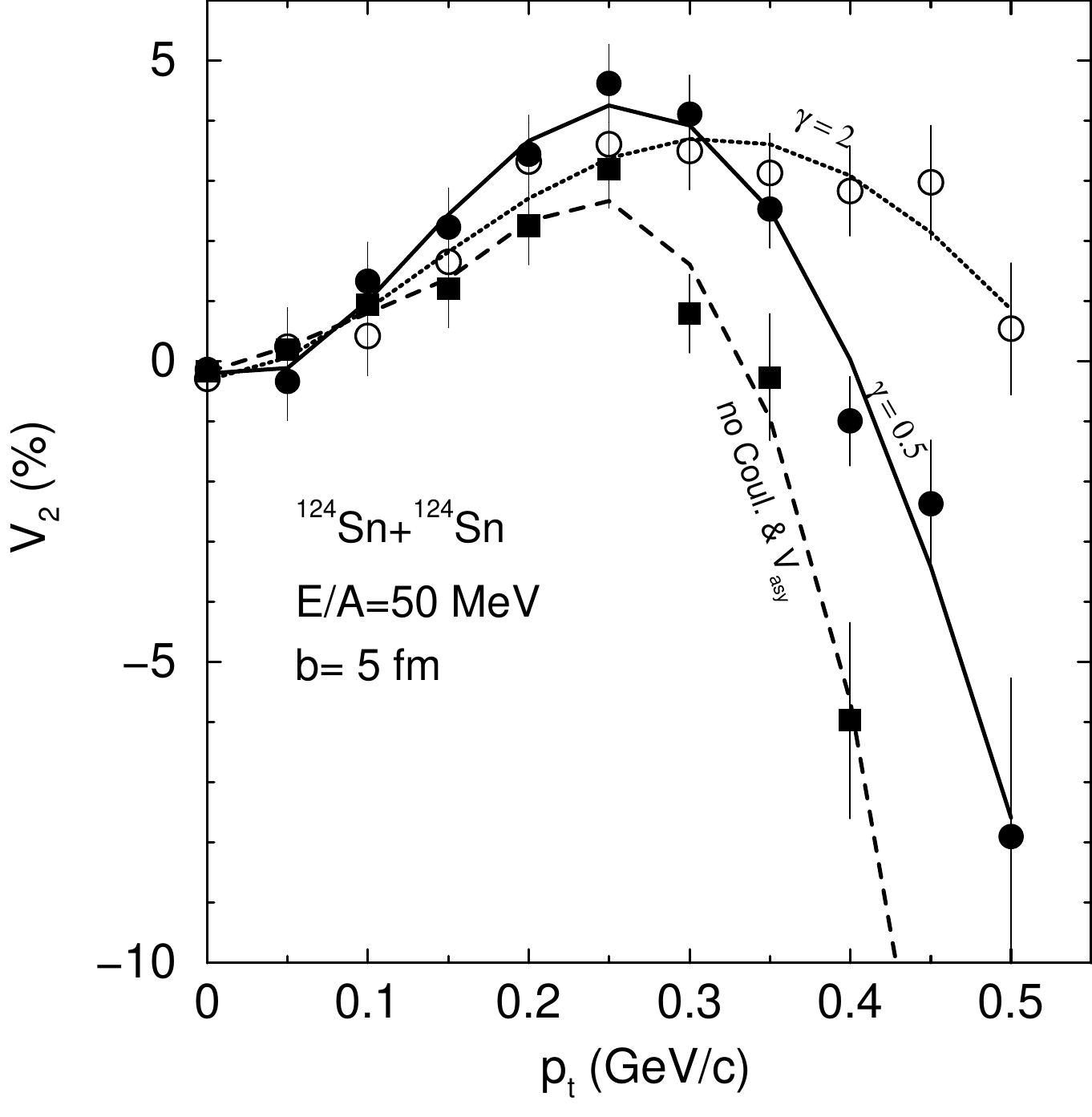}
\caption{Left: Neutron-proton differential transverse flow as a function of the reduced rapidity in mid-central Sn+Sn collisions with different symmetry energies~\cite{Li00}; Right: Proton elliptic flow as a function of the transverse momentum in mid-central Sn+Sn collisions with different symmetry energies~\cite{Li01}.}
\label{flow_Li}
\end{figure}

The neutron and proton collective flows in collisions with nuclei of similar mass along the $\beta$ stability valley and beam energies near the balance energy are proposed as useful probes of the nuclear symmetry energy as well~\cite{Li00,Li01}. Due to the symmetry potential, the neutrons are affected by a more repulsive potential compared to protons in neutron-rich medium, leading to less bound neutrons and larger neutron collective flows. On the other hand, the Coulomb potential for protons, which is well defined, enhances the proton collective flows. The left panel of Fig.~\ref{flow_Li} shows the neutron-proton differential transverse flow defined as
\begin{equation}\label{Fnp}
F_{n-p}(y) = \frac{1}{N(y)} \sum_i (p_x)_i \tau_i,
\end{equation}
with $y$ being the rapidity, $p_x$ being the momentum in the $x$ direction perpendicular to the beam ($z$) direction with x-o-z forming the reaction plane, and $\tau_i=1$ for neutrons and $-1$ for protons. The above definition of the neutron-proton differential transverse flow contains the summation effects of different free neutron and proton numbers as well as their different transverse flows, maximizing the symmetry potential effect while reducing the effects from the isoscalar part of the EOS and in-medium nucleon-nucleon collision cross sections. It is seen from the left panel of Fig.~\ref{flow_Li} that a stiffer symmetry energy with the curvature parameter $K_{sym}=61$ MeV gives a smaller symmetry potential at subsaturation densities, leading to a smaller transverse flow for neutrons than for protons, or a negative slope of $F_{n-p}(y)$, while a softer symmetry energy with the curvature parameter $K_{sym}=-69$ MeV results in a larger transverse flow for neutrons than for protons especially in a more neutron-rich system. Experimentally, it is much easier to measure charged particles like free protons rather than free neutrons. It was further proposed in Ref.~\cite{Li01} that the proton elliptic flow, especially at higher transverse momenta, is more sensitive to the density dependence of the symmetry energy compared to the transverse flow. As shown in the right panel of Fig.~\ref{flow_Li}, the proton elliptic flow at midrapidity $-0.5<y/y_{beam}<0.5$ is calculated based on the IBUU transport model using the density dependence of the symmetry energy $S(\rho) = S(\rho_0)(\rho/\rho_0)^\gamma$ with $\gamma=0.5$ and 2. Particles of higher transverse momenta are emitted at the early stage of the collision and less disturbed by the later dynamics. The pressure gradient increases with increasing $\gamma$, so the in-plane flow dominates the squeeze-out effect, leading to a larger $v_2$ for $\gamma=2$ than $\gamma=0.5$ especially at higher transverse momenta. Without the Coulomb repulsion, the shadowing from the spectator matter lasts longer, enhancing the squeeze-out effect and leading to a more negative $v_2$ especially at higher transverse momenta.

\begin{figure}[h]
\centering
\includegraphics[scale=0.4,clip]{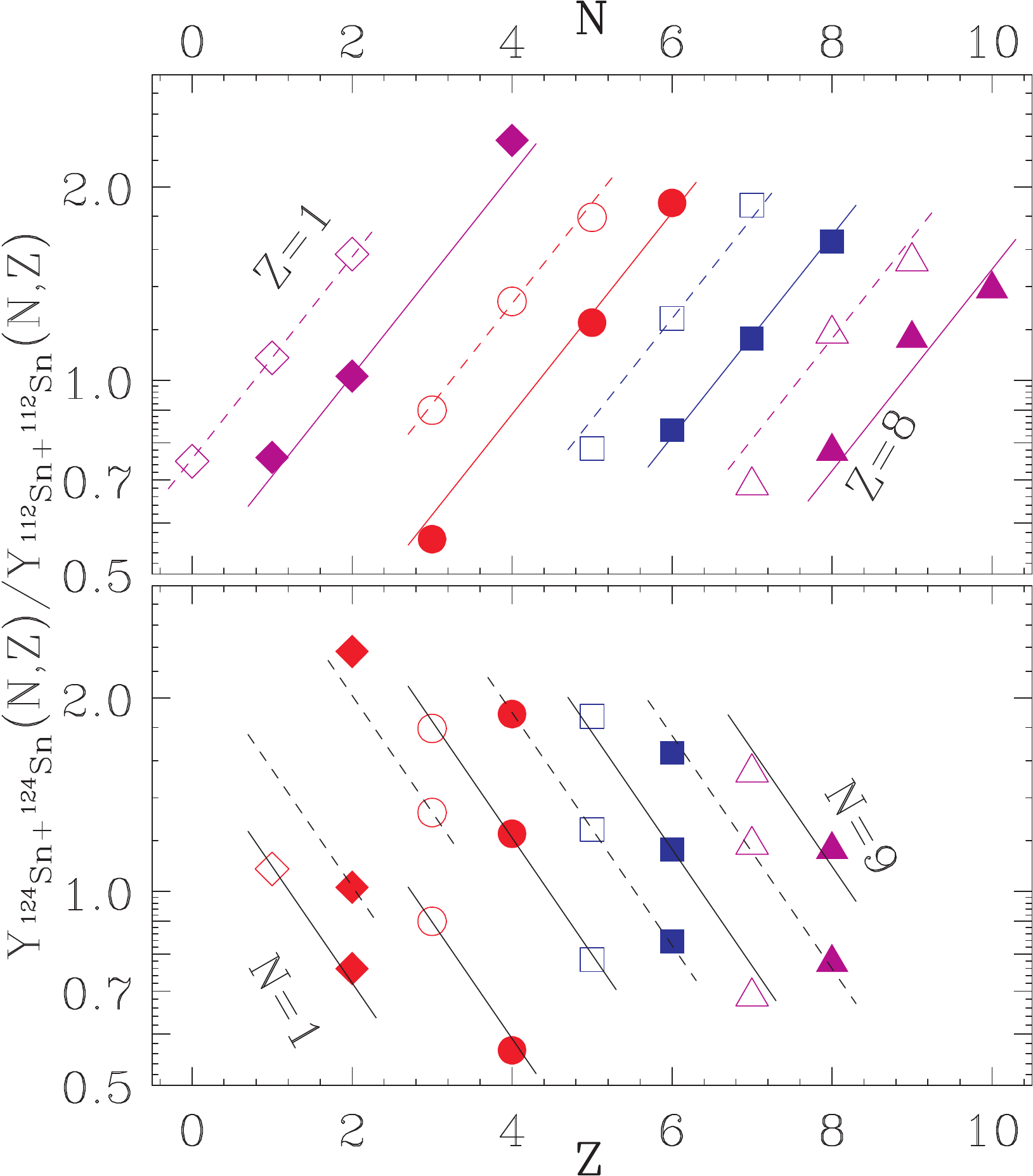}
\includegraphics[scale=0.52,clip]{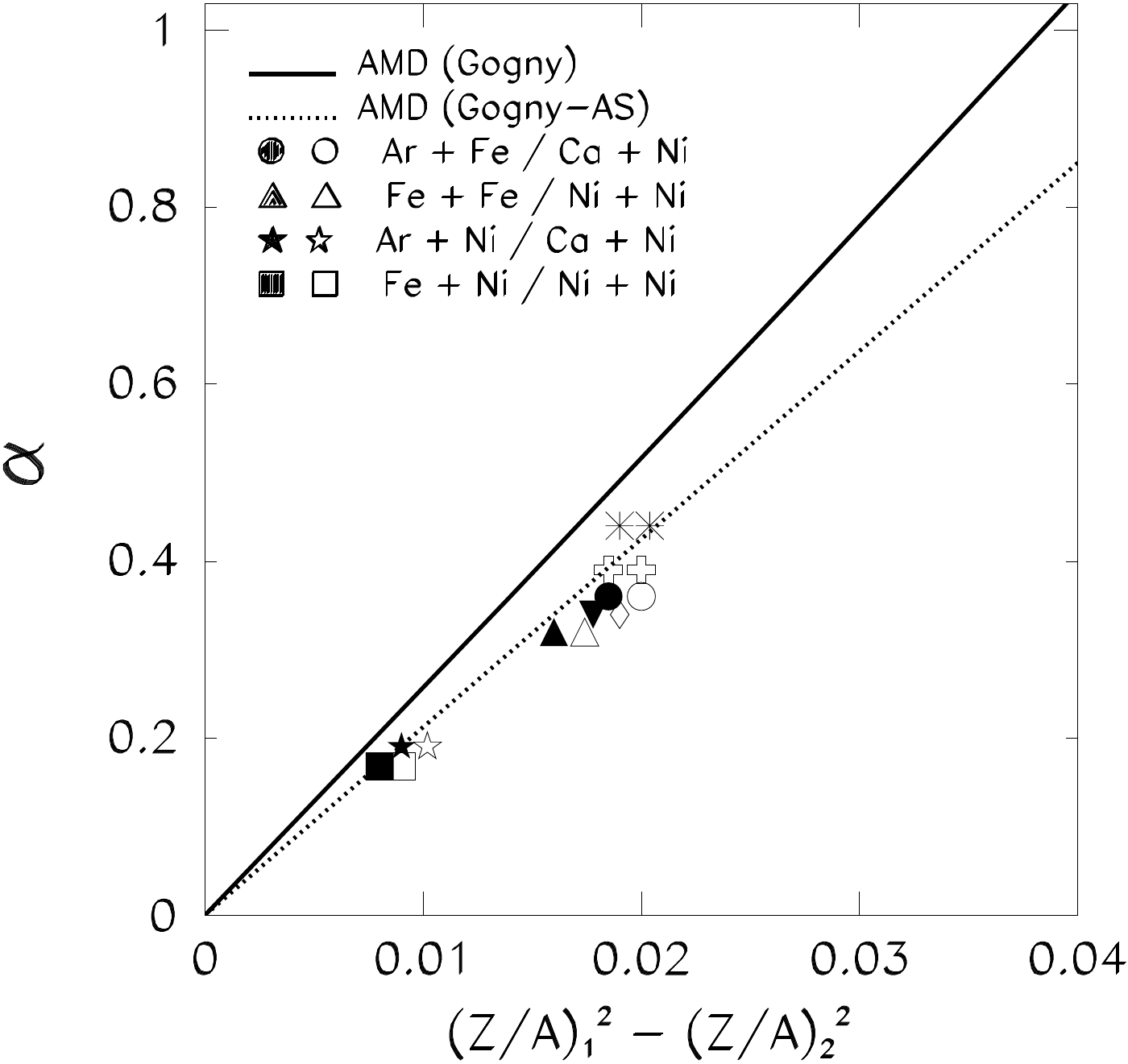}
\caption{Left: Isotope and isotone yield ratios in $^{124}$Sn+$^{124}$Sn and $^{112}$Sn+$^{112}$Sn heavy-ion experiments illustrating the scaling relation for light nuclei production~\cite{Tsa01prl,Tsa01prc}; Right: Isoscaling parameter $\alpha$ as a function of the difference in fragment asymmetry, with lines being the AMD calculation results from Ref.~\cite{Ono03} and scatters being the experimental data from Refs.~\cite{Bot02,Ger04,She07b}.}
\label{isoscaling}
\end{figure}

Based on the theoretical study of the nuclear liquid-gas phase transition, the low-density gas phase is always more neutron-rich than the high-density liquid phase~\cite{Mul95}, and this is called the isospin fractionation. This phenomena was observed from the nuclear multifragmentation analyses, i.e., the gas phase in heavy-ion system represented by free nucleons is more neutron-rich than the liquid phase represented by bound nuclei~\cite{Xu00}. The scaling relation was further found in isotopic and isotone distributions for light nuclei, and the so-called isoscaling law applies for a variety of reaction mechanisms, including the evaporation, the multifragmentation, and deeply inelastic collisions, in heavy-ion reactions~\cite{Tsa01prl}. The left panel of Fig.~\ref{isoscaling} shows the scaling behavior in Sn+Sn collisions from the experimental data~\cite{Xu00,Tsa01prl,Tsa01prc}, i.e.,
\begin{equation}
R_{12}(N,Z) = Y_2(N,Z)/Y_1(N,Z) = C \exp(\alpha N + \beta Z),
\end{equation}
where $Y_2(N,Z)$ and $Y_1(N,Z)$ are the isotope or isotone yields in $^{124}$Sn+$^{124}$Sn (system 2) and $^{112}$Sn+$^{112}$Sn (system 1) collisions, respectively, and $\alpha$ and $\beta$ are isoscaling parameters, approximately the difference of the neutron and proton separation energy between the two systems. For the two systems with comparable mass and collision energies, the volume, surface, and Coulomb contributions to the separation energy largely cancel, so from the statistical model analysis only the symmetry energy term contributes to $\alpha$ through
\begin{equation}
\alpha = 4 C_{sym} [(Z_1/A_1)^2-(Z_2/A_2)^2]/T,
\end{equation}
where $C_{sym}$ is the symmetry energy, $Z_{1(2)}$ and $A_{1(2)}$ are the charge and mass number of the system $1(2)$, and $T$ is the temperature. The problem is then reduced to determining the freeze-out temperature and density with the experimental information in order to extract $C_{sym}$, or equivalently $E_{sym}(\rho,T)$. The right panel of Fig.~\ref{isoscaling}, taken from Ref.~\cite{She07b} as a good summary of isoscaling analyses, compares the isoscaling parameter $\alpha$ from AMD calculations and results from other experimental analyses. The Gogny-AS interaction leads to a stiffer symmetry energy compared with the Gogny interaction, and the slope of $\alpha$ from the Gogny-AS interaction is seen to be smaller compared to that from the Gogny interaction. The isoscaling analysis based on the MSU data leads to the symmetry energy $E_{sym}(\rho)=23.4(\rho/\rho_0)^{0.6}$ (MeV)~\cite{Tsa01prl}, and that based on the TAMU data leads to the symmetry energy $E_{sym}(\rho)=31.6(\rho/\rho_0)^{0.69}$ (MeV)~\cite{She04,She07a}.

\begin{figure}[h]
\centering
\includegraphics[scale=0.3,clip]{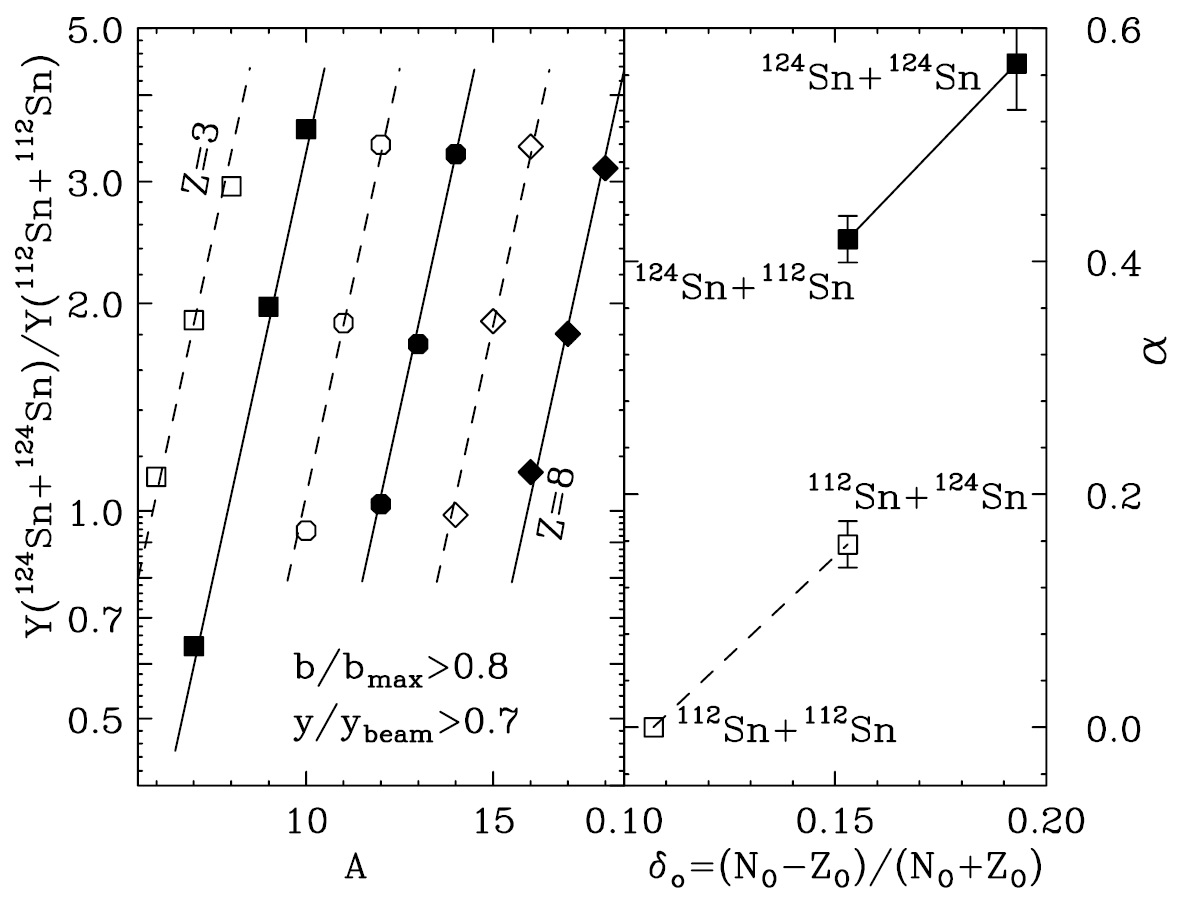}
\includegraphics[scale=0.32,clip]{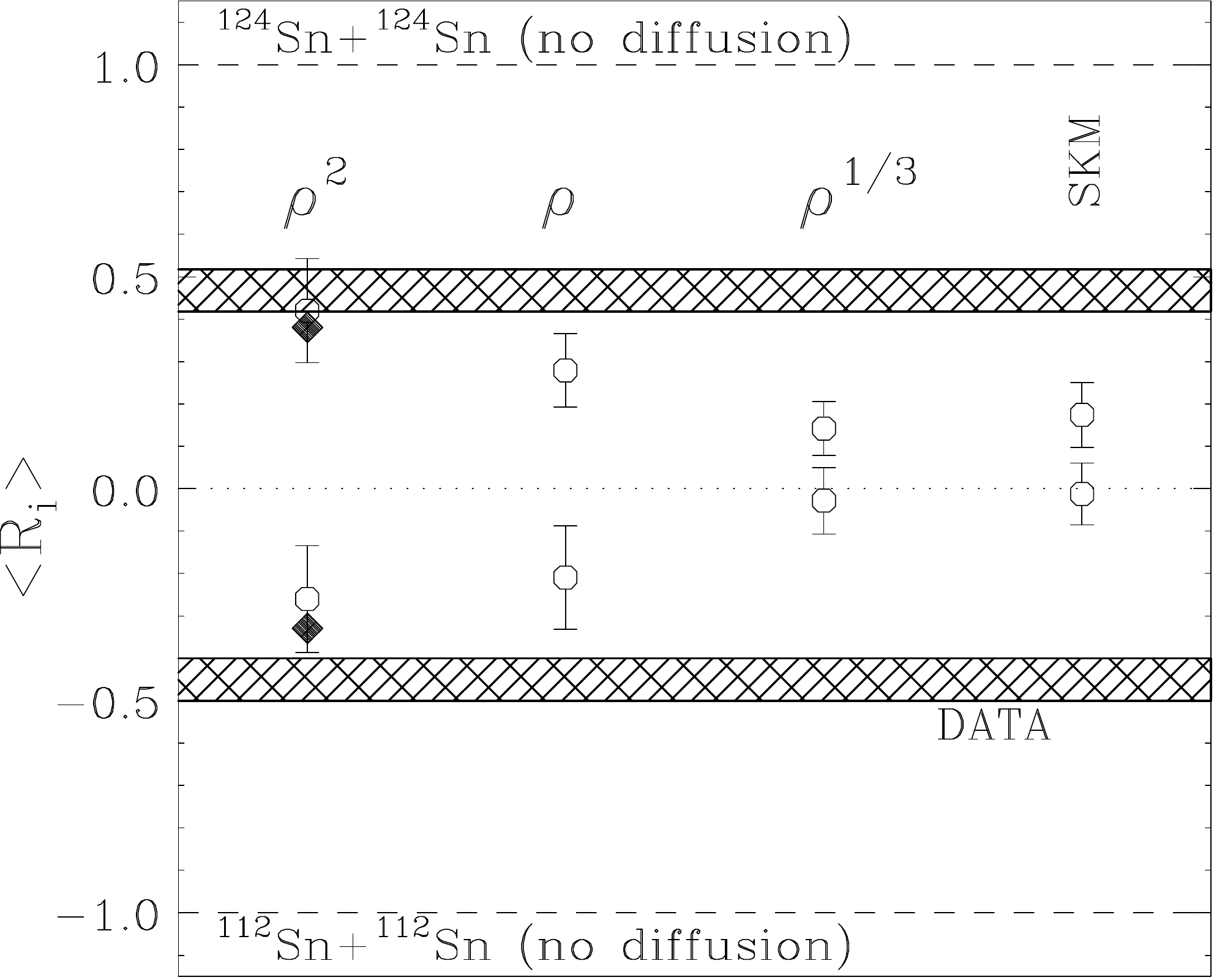}
\includegraphics[scale=0.8,clip]{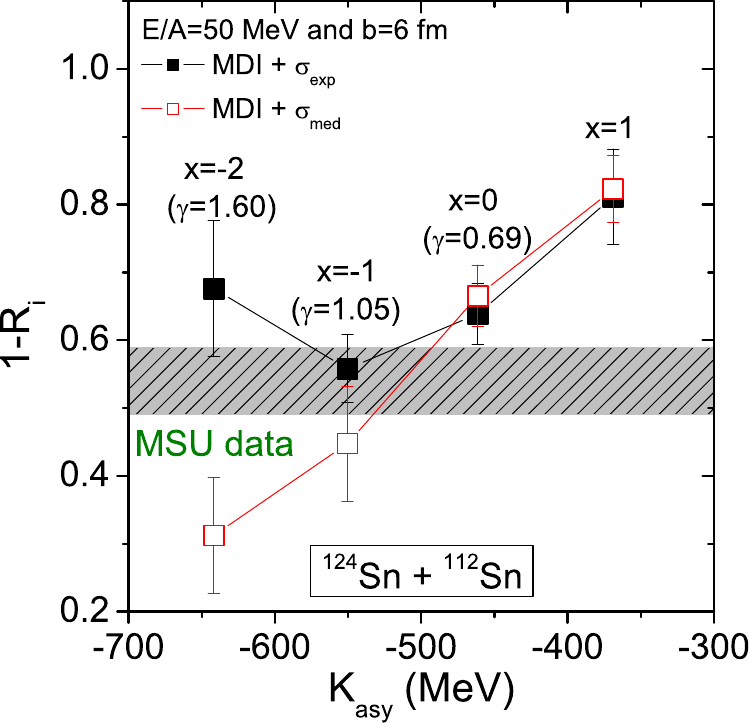}
\includegraphics[scale=0.3,clip]{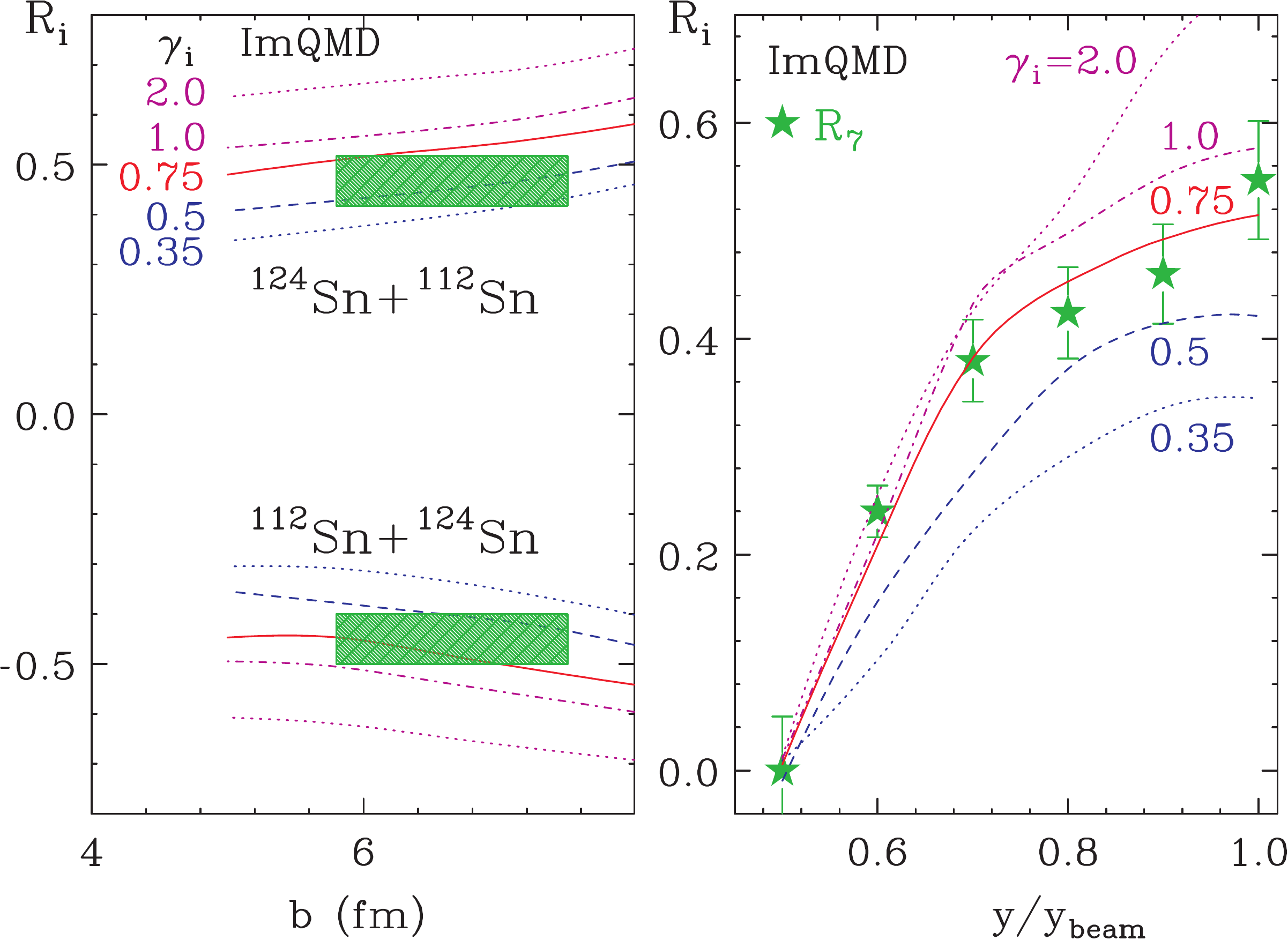}
\caption{(Color online) Upper left: Measured isotope yield ratio around the projectile rapidity with a scaling relation and the isoscaling parameter $\alpha$ in different collision systems~\cite{Tsa04}; Upper right: Measured (bands) and calculated (scatters) isospin transport ratios with increasing softness of the symmetry energy in $^{124}$Sn+$^{112}$Sn and $^{112}$Sn+$^{124}$Sn collisions~\cite{Tsa04}; Lower left: Comparison of the isospin transport ratios from the IBUU04 calculations with the MSU data~\cite{Li05}; Lower right: Comparison of isospin transport ratios at different impact parameters and as a function of the reduced rapidity for $A=7$ isotopes from the MSU data and those from ImQMD calculations~\cite{Tsa09}.}
\label{isodiffusion}
\end{figure}

The isospin diffusion between the projectile and the target nuclei with different isospin asymmetries is a probe of the symmetry energy. I refer the reader to the detailed discussions on the relation between the isospin diffusivity and the nuclear EOS in Refs.~\cite{Shi03,Bar05prc,Riz08}. Generally, a soft symmetry energy leads to a faster isospin transport at subsaturation densities, and the system will be closer to the isospin equilibrium in the final stage. As shown in the upper left panel of Fig.~\ref{isodiffusion}, the isoscaling parameter $\alpha$ can be obtained by taking isotope yield ratios around the projectile rapidity in various collision systems using $^{124}$Sn and $^{112}$Sn nuclei with respect to that in $^{112}$Sn+$^{112}$Sn collisions. For $^{124}$Sn+$^{124}$Sn collisions with the isospin asymmetry of the system $\delta_0=0.193$, the isoscaling parameter is $\alpha=0.57 \pm 0.02$. For $^{112}$Sn+$^{112}$Sn collisions with the isospin asymmetry of the system $\delta_0=0.107$, the isoscaling parameter is $\alpha=0$. However, for $^{124}$Sn+$^{112}$Sn and $^{112}$Sn+$^{124}$Sn collisions with the isospin asymmetry of the system $\delta=0.153$, the isoscaling parameter is not the average value of the above two systems. This indicates that the isospin equilibrium is not reached in $^{124}$Sn+$^{112}$Sn and $^{112}$Sn+$^{124}$Sn collision systems. In order to describe quantitatively the extent of isospin equilibrium, the isospin transport ratio is defined as
\begin{equation}\label{isodiffcoe}
R_i = \frac{2\text{x}-\text{x}_{124+124}-\text{x}_{112+112}}{\text{x}_{124+124}-\text{x}_{112+112}},
\end{equation}
where $\text{x}$ is any isospin sensitive observable. As shown in the upper right panel of Fig.~\ref{isodiffusion}, $R_i=1$ for $^{124}$Sn+$^{124}$Sn collisions and $-1$ for $^{112}$Sn+$^{112}$Sn collisions are obtained by definition. Using the IBUU97 transport model with the potential part of the symmetry energy $12(\rho/\rho_0)^\gamma$ (MeV) with $\gamma=2$, 1, and $1/3$, and $38.5(\rho/\rho_0)-21.0(\rho/\rho_0)^2$ from the SKM interaction, the calculated isospin transport ratio $R_i$ in $^{124}$Sn+$^{112}$Sn and $^{112}$Sn+$^{124}$Sn collisions are compared with the measured one in the upper right panel of Fig.~\ref{isodiffusion}. As expected, the isospin is more equilibrated with a softer symmetry energy. On the other hand, the experimental data favors a stiff symmetry energy with $\gamma=2$. It was later pointed out that the isospin diffusion is sensitive to the momentum dependence of the mean-field potential~\cite{Che05}. By using the isospin asymmetry of the projectile residue as the isospin sensitive observable $\text{x}$ in Eq.~(\ref{isodiffcoe}), the simulation with the IBUU04 transport model using an isospin- and momentum-dependent mean-field potential and nucleon-nucleon cross sections in free space favors the parametrization $x=-1$ compared with the MSU data, as shown in the lower left panel of Fig.~\ref{isodiffusion}. This corresponds to a symmetry energy $E_{sym}=31.6(\rho/\rho_0)^{1.05}$ and the asymmetric part of the isobaric incompressibility $K_{asy} = -550$ MeV at the saturation density. With the in-medium nucleon-nucleon cross sections scaled by the neutron and proton effective masses~\cite{Li05}, the IBUU04 calculations favor a symmetry energy between $x=0$ and $x=-1$, corresponding to the slope parameter of the symmetry energy approximately within $60<L<110$ (MeV). A later study was done by comparing the new MSU isospin diffusion data with the ImQMD calculation~\cite{Tsa09}, as shown in the lower right panel of Fig.~\ref{isodiffusion}. The symmetry energy used in the transport model calculation is $E_{sym}(\rho)=12.5(\rho/\rho_0)^{2/3}+17.6(\rho/\rho_0)^{\gamma_i}$, with $\gamma_i=2$, 1, 0.75, 0.5, and 0.35. Again, the isoscaling coefficient $\alpha$ is used as the isospin tracer $\text{x}$ in Eq.~(\ref{isodiffcoe}) in the experimental analysis, and in the ImQMD calculation it is taken as the asymmetry of the emitted fragments and free nucleons. As confirmed experimentally and theoretically from statistical and dynamical calculations, they become similar for isotopes with mass $A=7$. The symmetry energy from this study was constrained within $0.45\le\gamma_i\le 0.95$ from a $\chi^2$ analysis.

\begin{figure}[h]
\centering
\includegraphics[scale=0.9,clip]{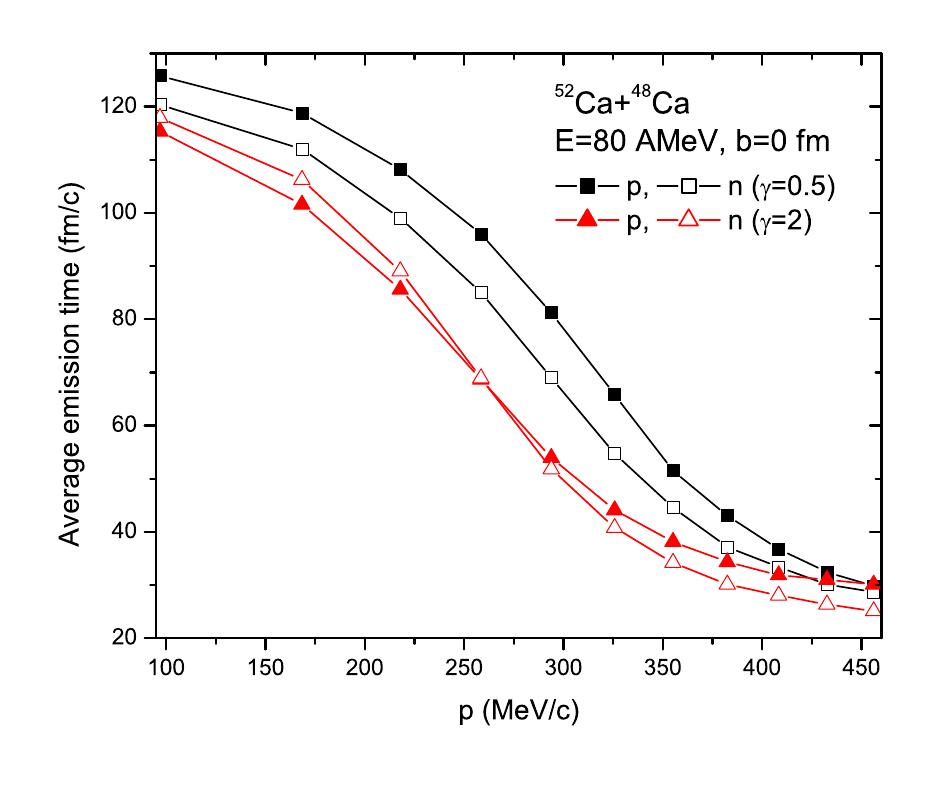}
\includegraphics[scale=0.8,clip]{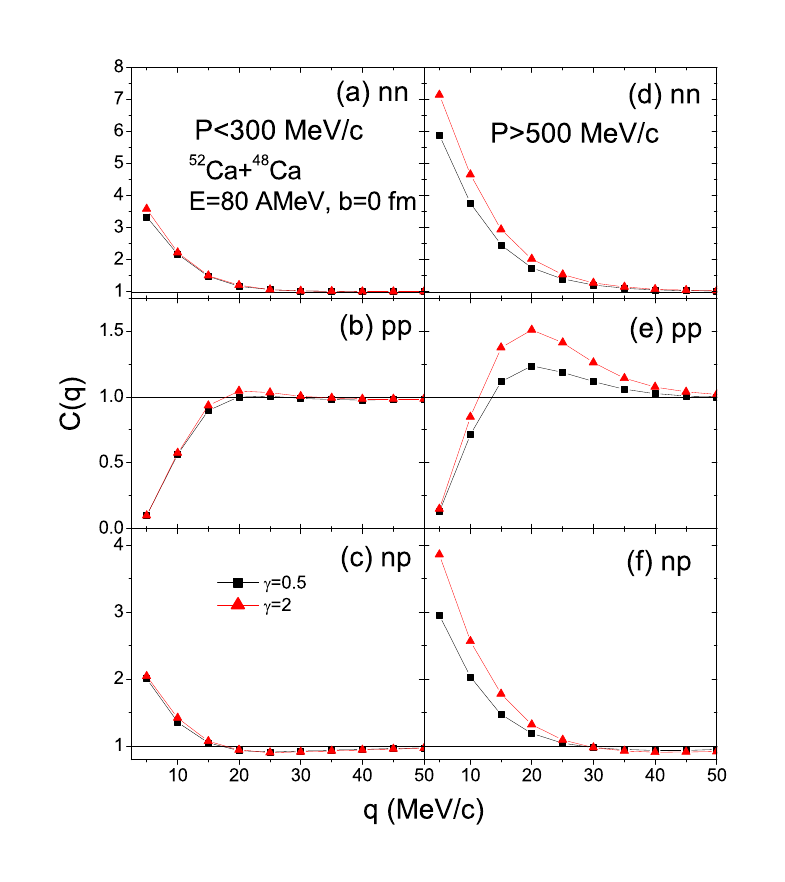}
\caption{(Color online) Left: Average emission times for protons and neutrons from different density dependence of the symmetry energy; Right: Two-particle HBT correlations with total momentum $P<300$ MeV/c and $P>500$ MeV/c from different density dependence of the symmetry energy. Taken from Ref.~\cite{Che03}.}
\label{hbt}
\end{figure}

The two-particle Hanbury-Brown and Twiss (HBT) correlation is a useful tool to study the space-time distribution of the emission source in heavy-ion collisions, and can be helpful in extracting the information of the symmetry energy as well~\cite{Che03}. From the IBUU97 transport model calculations, the average emission times for protons and neutrons in central $^{52}$Ca+$^{48}$Ca collisions are obtained using the symmetry energy $E_{sym}(\rho) = 35 (\rho/\rho_0)^\gamma$ (MeV) with $\gamma=0.5$ and 2, as shown in the left panel of Fig.~\ref{hbt}. Generally, a stiffer symmetry energy with $\gamma=2$ leads to a higher pressure of the excited matter during the collisions, and thus overall smaller average emission times for both neutrons and protons, compared with $\gamma=0.5$. On the other hand, neutrons emit earlier than protons for $\gamma=0.5$ due to the more repulsive potential for neutrons and the more attractive potential for protons at subsaturation densities, while for $\gamma=2$ the emission time difference between neutrons and protons becomes smaller. The information of the emission source can be quantitatively extracted from the two-particle HBT correlation function
\begin{equation}
C({\bf P}, {\bf q}) = \frac{\int d^4 x_1 d^4 x_2 g({\bf P}/2,x_1)g({\bf P}/2,x_2) |\phi({\bf q},{\bf r})|^2}{\int d^4 x_1 g({\bf P}/2,x_1)\int d^4 x_2 g({\bf P}/2,x_2)},
\end{equation}
where ${\bf P} = {\bf p_1} + {\bf p_2}$ and ${\bf q} = ({\bf p_1} - {\bf p_2})/2$ are the total and relative momentum of the two-particle pair, $\phi({\bf q},{\bf r})$ is their relative wave function with ${\bf r}$ being their relative position, and the emission function $g({\bf p}, x)$ describes the probability of emitting a particle with momentum ${\bf p}$ from the space-time point $x=({\bf r}, t)$. Taking the afterburner final-state interaction and the Coulomb interaction into account, the resulting two-particle HBT correlations are shown in the right panel of Fig.~\ref{hbt}. It is seen that the HBT correlation from nucleon pairs with lower total momenta show no sensitivity to the symmetry energy. For nucleon pairs with higher total momenta but lower relative momenta, they are emitted earlier from the high-density phase, and their HBT correlations are good probes of the symmetry energy. For a soft symmetry energy with $\gamma=0.5$, a larger emission time difference between nucleons leads to a weaker space-time correlation. For a stiff symmetry energy with $\gamma=2$, nucleons are emitted at similar times and the nucleon pairs have a stronger space-time correlation. The neutron-proton HBT correlation is proposed to have the largest sensitivity to the symmetry energy, compared with proton-proton and neutron-neutron correlations. The HBT correlations between light clusters such as deuterons, tritons, and $^3$He have also been studied in a later analysis together with the dynamical coalescence model~\cite{Che03npa}. With the increasing mass of clusters, it was found that the symmetry energy effect on their HBT correlations is generally weaker.

\begin{figure}[h]
\centering
\includegraphics[scale=0.3,clip]{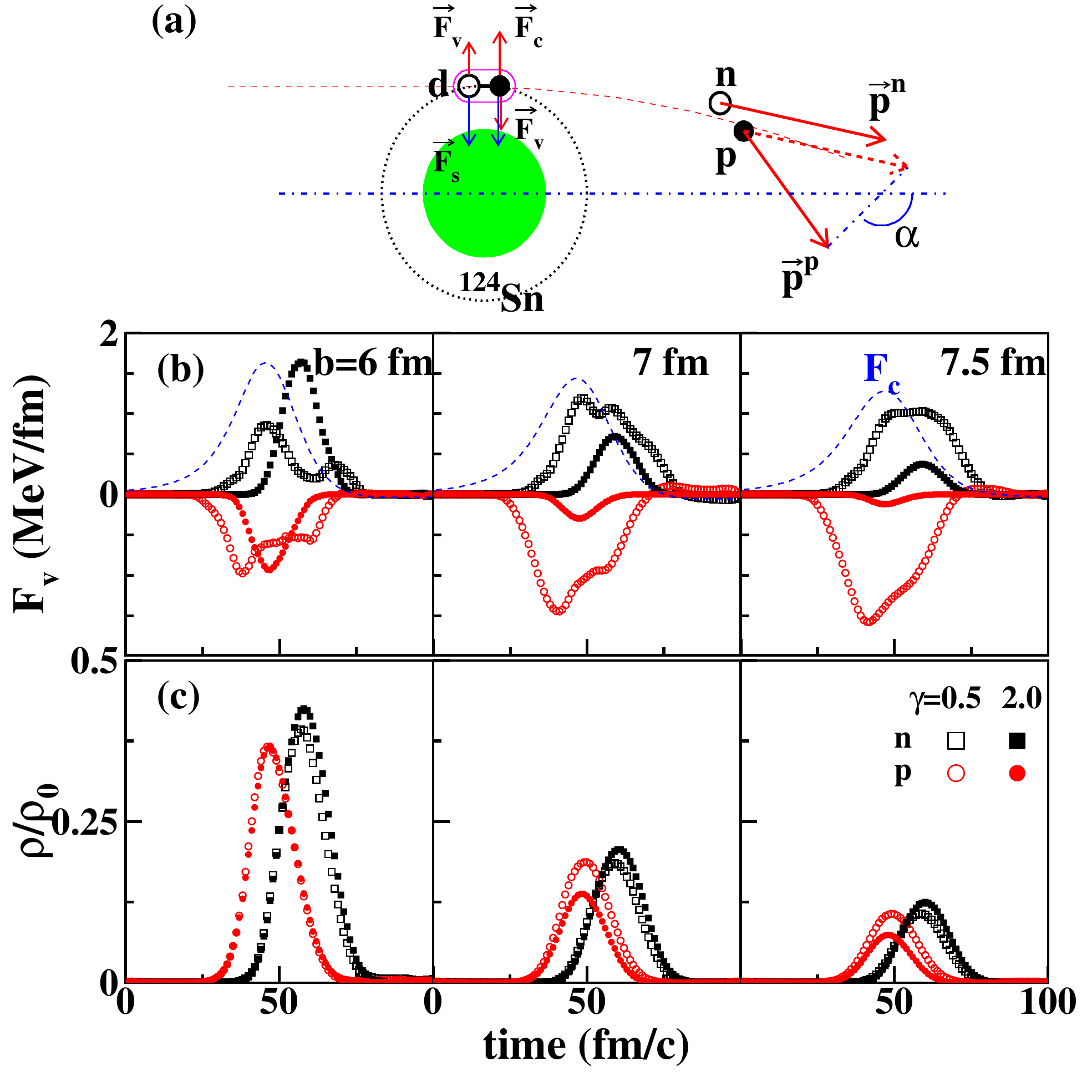}
\includegraphics[scale=0.4,clip]{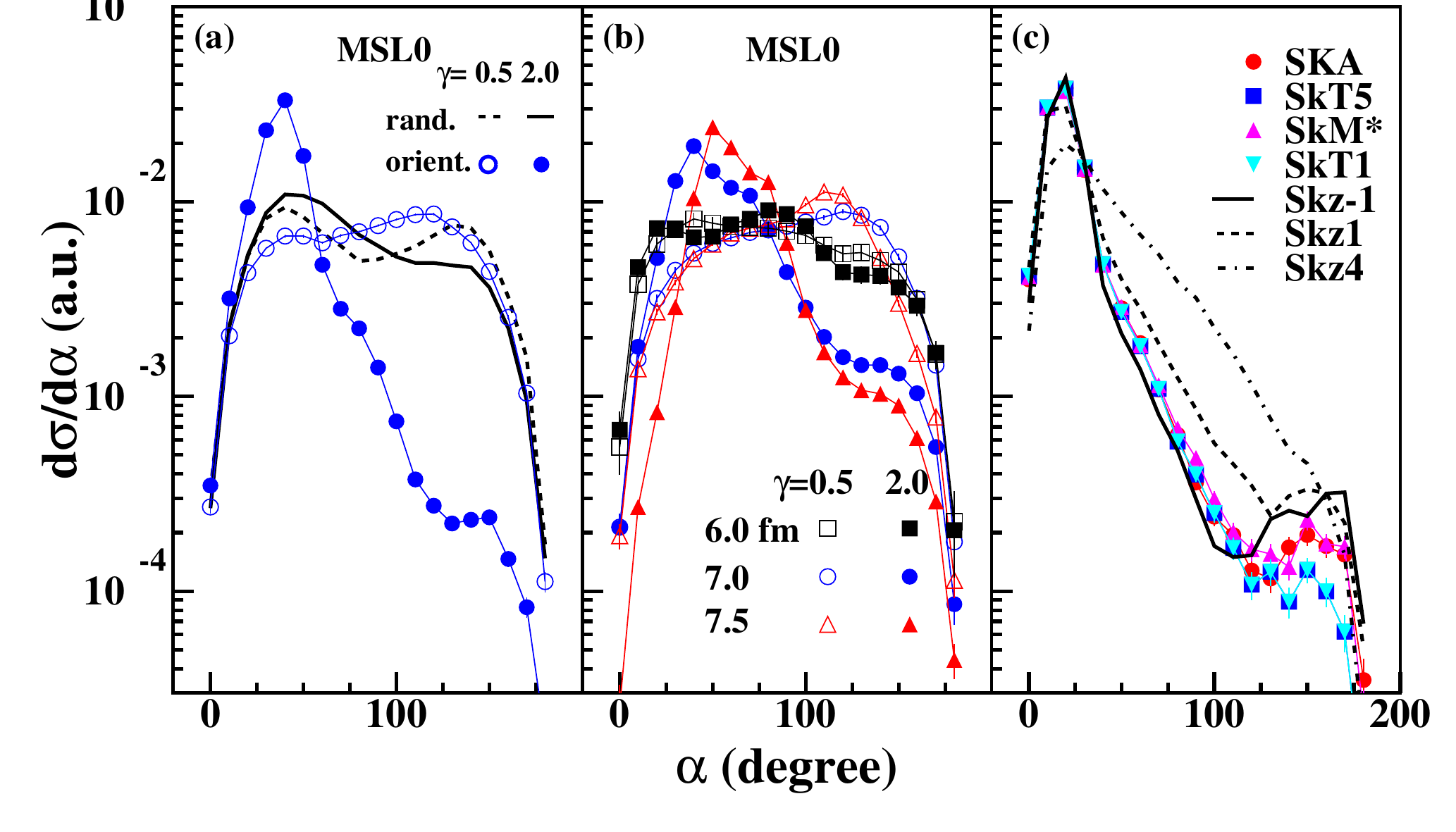}
\caption{(Color online) Left: Schematic view of the deuteron breakup induced by the isovector and Coulomb potential (a) as well as their time evolution (b) and the local neutron and proton density evolution (c) with different density dependence of the symmetry energy; Right: Distribution of the correlation angle between neutrons and protons with randomly oriented or pre-oriented deuteron beams, different density dependence of the symmetry energy, and various Skyrme forces. Taken from Ref.~\cite{Ou15}.}
\label{deuteron}
\end{figure}

Recently, the correlation angular distribution between neutrons and protons from the breakup of deuterons was proposed as a useful probe of the nuclear symmetry energy at subsaturation densities~\cite{Ou15}. As shown in the left panel of Fig.~\ref{deuteron}, a deuteron near a heavy ion is affected by the isoscalar force $F_s$, the isovector force $F_v$, and the Coulomb force $F_c$. The isoscalar force is the same for neutrons and protons and will not affect the deuteron orientation. The isovector force is repulsive for neutrons and attractive for protons, and the Coulomb force is repulsive for protons. They may affect the momenta of neutrons and protons and thus the orientation of deuterons. Based on the ImQMD model, the time evolution of the isovector force, the Coulomb force, and the local neutron and proton densities are all displayed at three impact parameters, with $\gamma=2$ and 0.5 corresponding to a stiff and a soft symmetry energy, respectively. It is seen that the strength of the Coulomb force is comparable to that of the isovector force. The local density is less than $\rho<0.5\rho_0$, and the symmetry energy effect on the local neutron and proton density evolution is more pronounced at larger impact parameters. For a soft symmetry energy with $\gamma=0.5$, the isovector force is stronger at subsaturation densities, and is expected to lead to a larger angular separation for a neutron and a proton after the breakup of a deuteron. To quantitatively illustrate this effect, the correlation angle $\alpha$ is defined as the relative angle between a neutron and a proton with respect to the beam direction. As shown in the right panel of Fig.~\ref{deuteron}, although the symmetry energy effect is largely smeared with a random initial orientation of the incident deuteron beam, significant enhancement at larger $\alpha$ is observed with a soft symmetry energy if the deuteron is pre-oriented parallel to the beam axis, while the Coulomb force dominates the effect and leads to a peak at smaller $\alpha$ with a stiff symmetry energy. A more realistic calculation is done with the relative vector from neutron to proton parallel or antiparallel to the direction of the deuteron wave vector, with each case $50\%$ probability. It is seen that in this case the symmetry energy effect is also pronounced especially at larger impact parameters. Also compared are results from SkA, SkT5, SkT1, SkMD, Skz-1, Skz1, and Skz4 Skyrme forces, with the former four (latter three) sets of parameters having the similar isovector (isoscalar) potential but differ in the isoscalar (isovector) sector. It is seen that the correlation angle distribution between neutrons and protons is sensitive to the isovector potential but rather insensitive to the isoscalar potential.

\subsubsection{Probes of nuclear symmetry energy at suprasaturation densities from transport model studies}
\label{suprasaturation}

\begin{figure}[h]
\centering
\includegraphics[scale=0.6,clip]{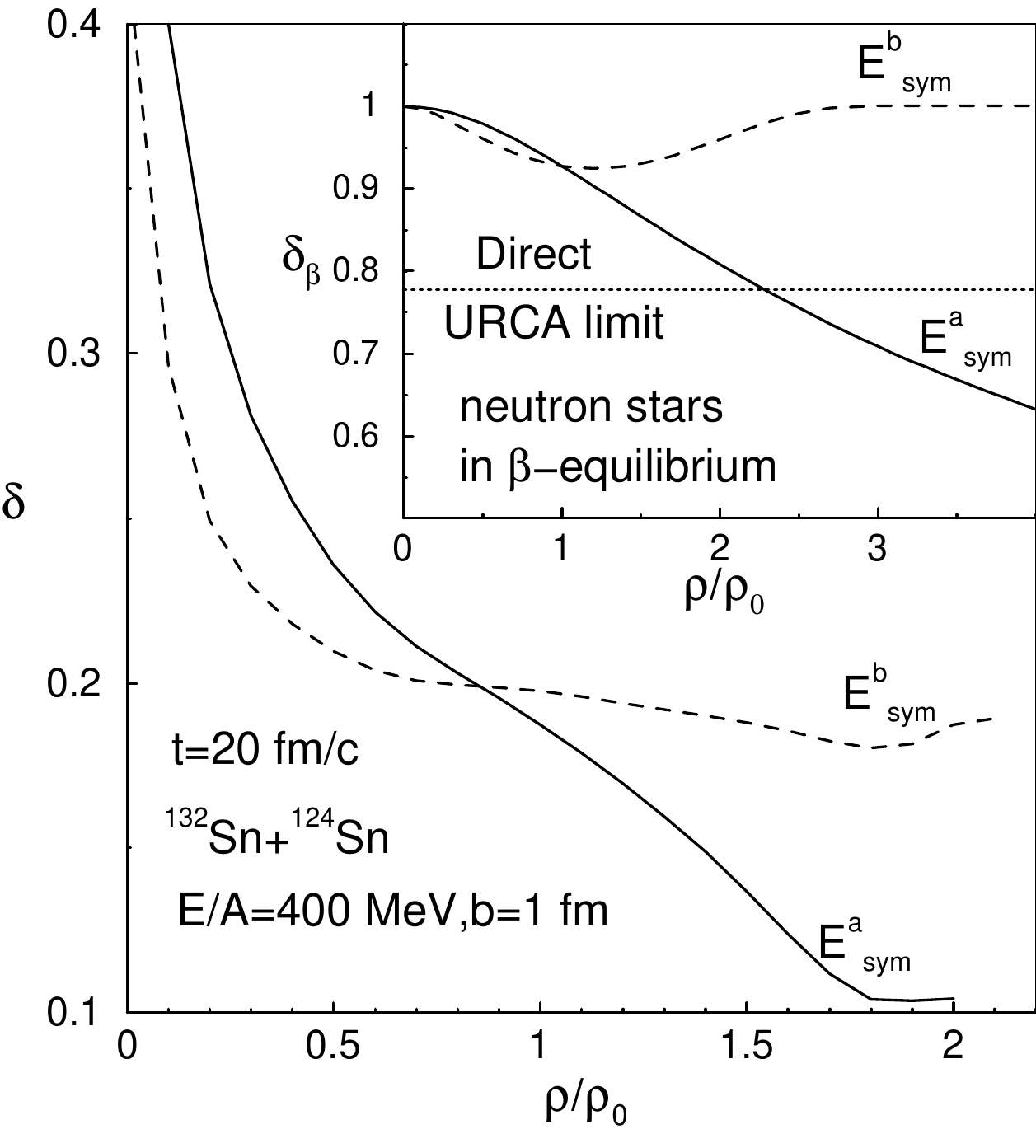}
\includegraphics[scale=0.46,clip]{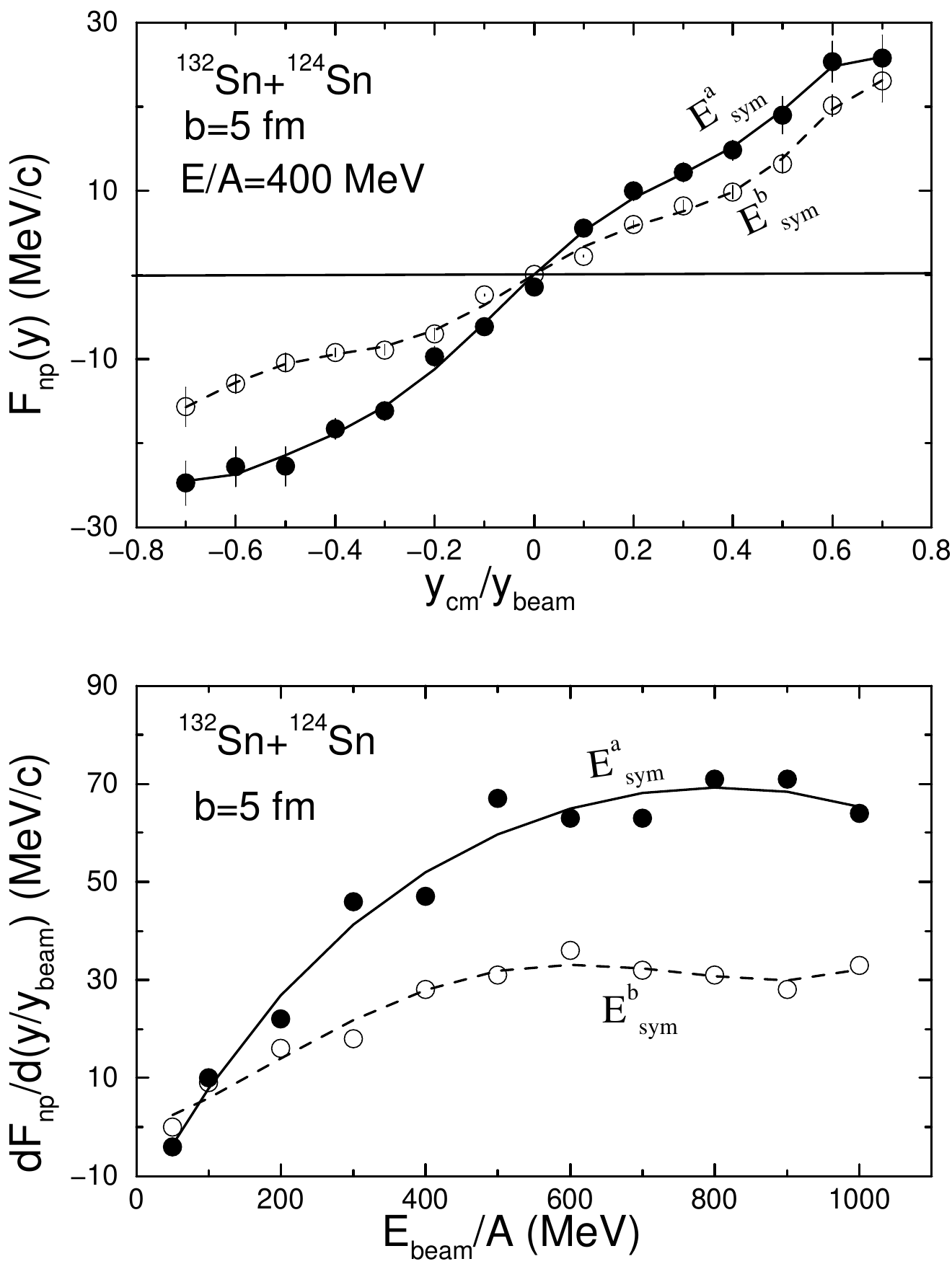}
\caption{Left: Correlation between the isospin asymmetry and the density in $^{132}$Sn+$^{124}$Sn collisions at $t=20$ fm/c and that in neutron star matter from a stiff ($E_{sym}^a$) and a super soft ($E_{sym}^b$) symmetry energy; Right: Neutron-proton differential transverse flow and the beam energy dependence of the excitation function of the neutron-proton differential transverse flow. Taken from Ref.~\cite{Li02prl}.}
\label{Fnp_esym}
\end{figure}

High-energy heavy-ion collisions are the only way in terrestrial laboratories to produce high-density neutron-rich matter, and thus a good way of probing the nuclear symmetry energy at suprasaturation densities. The left panel of Fig.~\ref{Fnp_esym} shows the correlation between the isospin asymmetry and the density in $^{132}$Sn+$^{124}$Sn collisions at $t=20$ fm/c from a stiff ($E_{sym}^a$) and a super soft ($E_{sym}^b$) symmetry energy based on IBUU97 calculations, as well as that in neutron star matter. As mentioned before, the isospin asymmetry is larger in the low-density phase and smaller in the high-density phase. This is especially so for a stiff symmetry energy leading to a stronger isospin fractionation effect, consistent with the stronger symmetry potential at higher densities, leading to stronger emissions for neutrons than for protons. For a super soft symmetry energy which becomes negative above $3\rho_0$, the core of the neutron star becomes a pure neutron matter. The neutron-proton differential transverse flow at the beam energy around 400 AMeV, which is defined in Eq.~(\ref{Fnp}), is a good probe of the symmetry energy at suprasaturation densities. As shown in the right panel of Fig.~\ref{Fnp_esym}, a stiff symmetry energy leads to a larger neutron-proton potential difference and thus a stronger neutron-proton differential transverse flow. With the increasing beam energy, a higher-density region is reached in heavy-ion collisions, and the effect of the symmetry energy on the excitation function of the neutron-proton differential transverse flow becomes even larger. It was proposed in  later studies with an isospin- and momentum-dependent mean-field potential based on the IBUU04 transport model, that the double neutron-proton differential transverse flow~\cite{Yon06}, i.e., the difference of $F_{np}$ between a more neutron-rich and a less neutron-rich system, as well as the different transverse flows between tritons and $^3$He~\cite{Yon09} in high-energy heavy-ion collisions can also be probes of the symmetry energy at suprasaturation densities.

\begin{figure}[h]
\centering
\includegraphics[scale=0.3,clip]{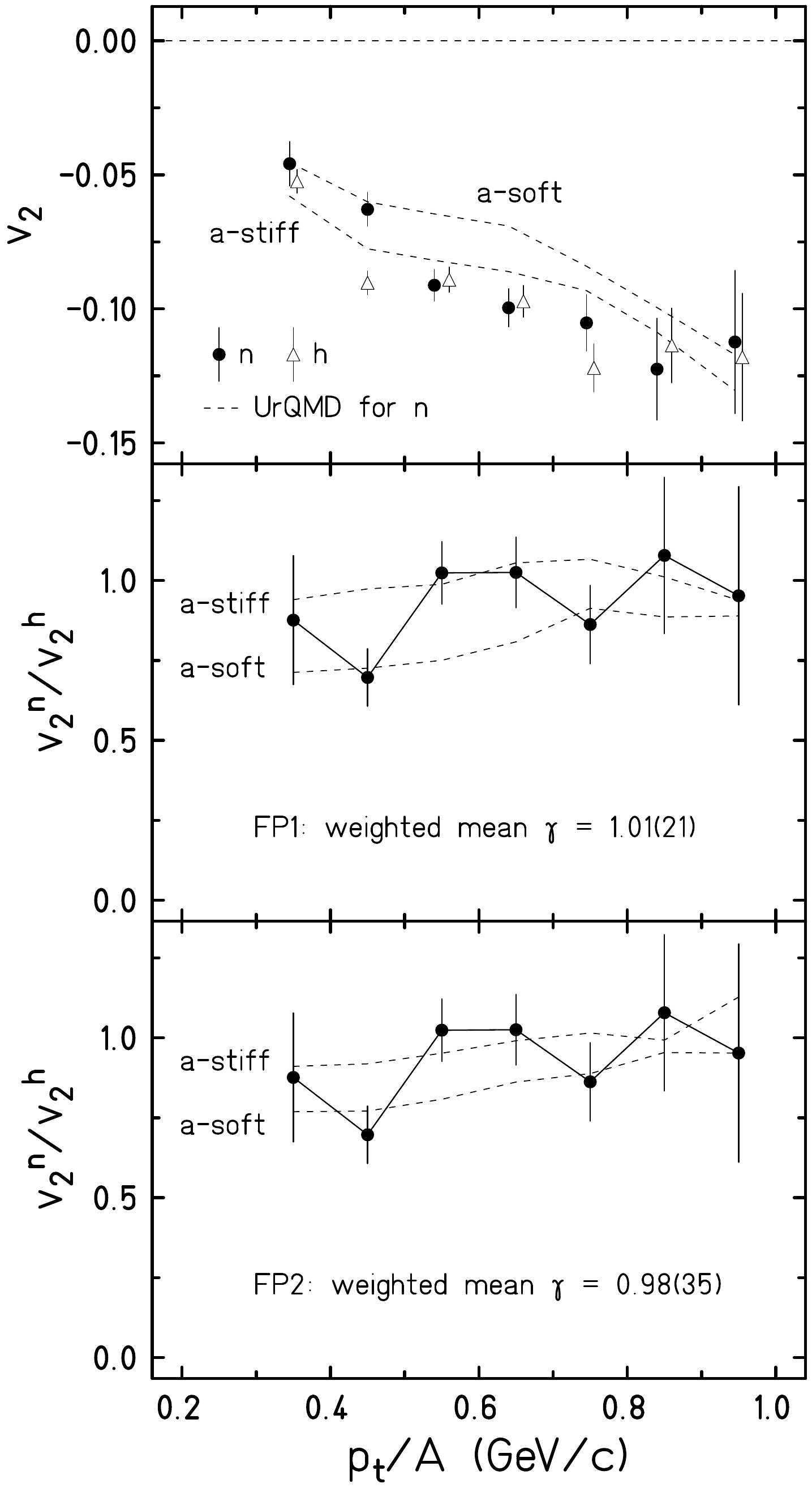}
\includegraphics[scale=0.23,clip]{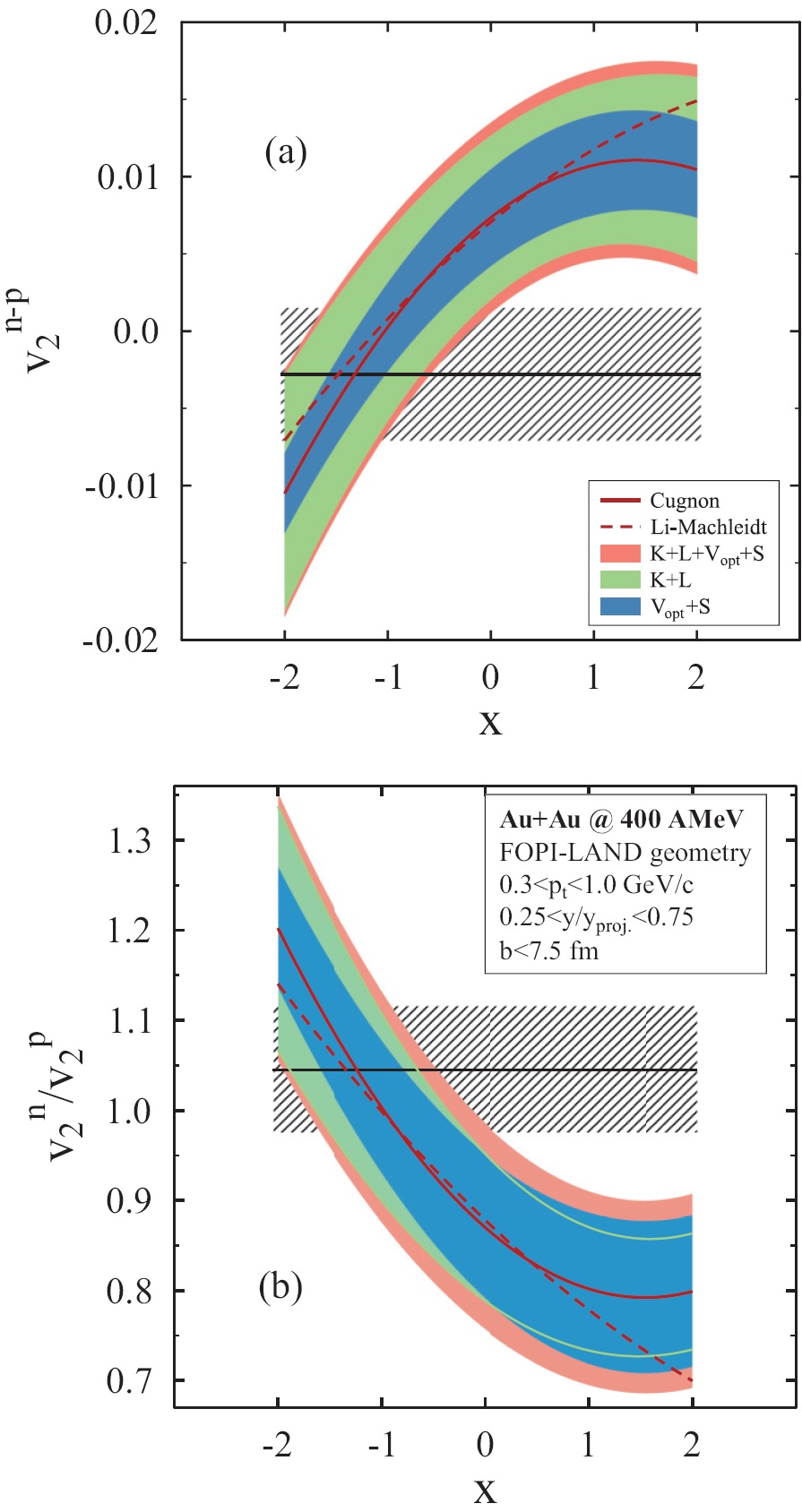}
\includegraphics[scale=0.45,clip]{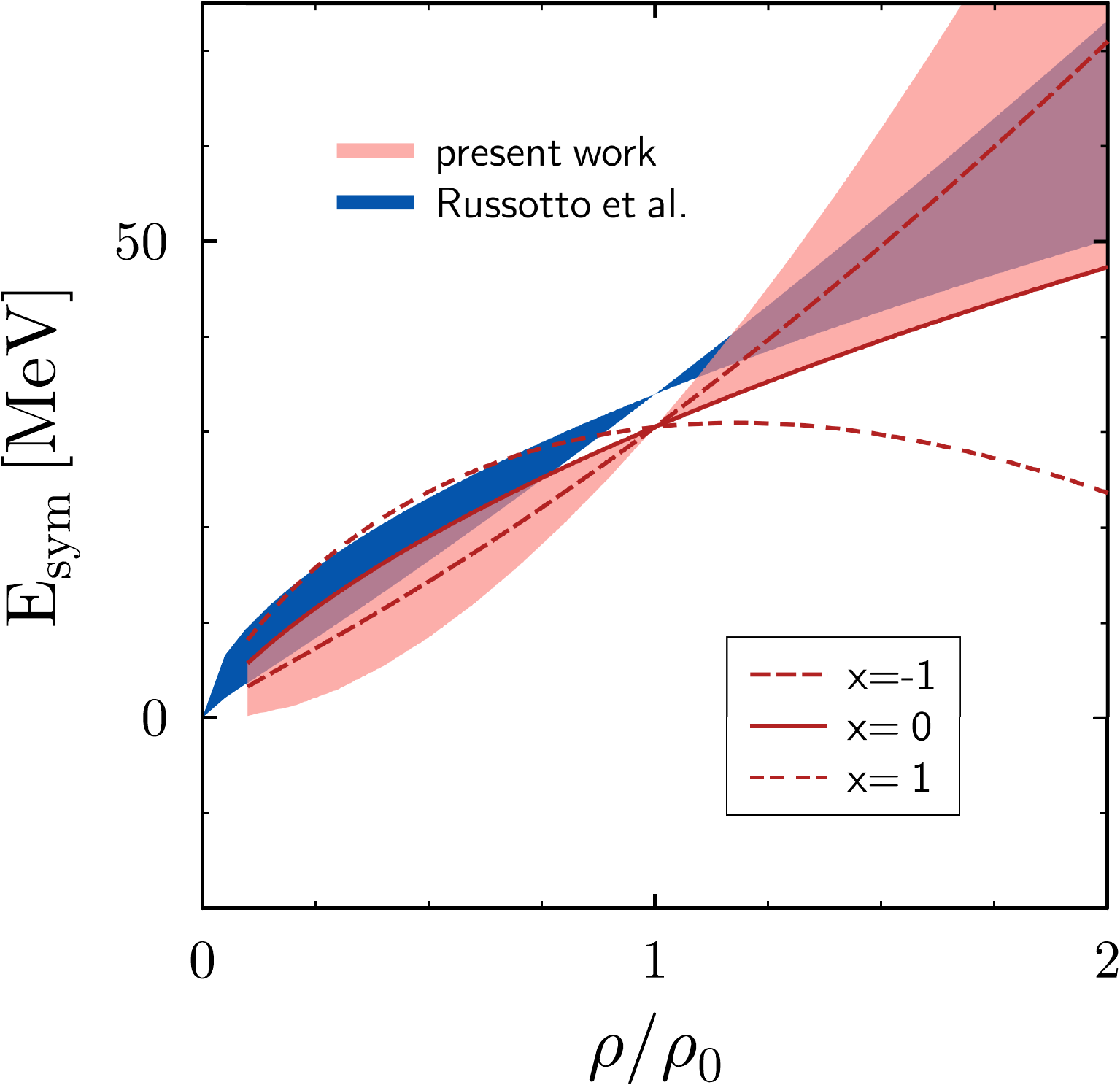}
\caption{(Color online) Left: Elliptic flows of neutrons and hydrogen isotopes in mid-central Au+Au collisions at 400 AMeV from a stiff and a soft symmetry energy based on the UrQMD model, with FP1 and FP2 representing two parameterizations of the in-medium nucleon-nucleon cross sections~\cite{Rus11}; Middle: Neutron-proton elliptic flow difference $v_2^{n-p}$ and ratio $v_2^{n}/v_2^{p}$ integrated over the impact parameter $\text{b}\le 7$ fm from the T\"ubingen QMD model compared with the FOPI data~\cite{Dan13}; Right: A comparison of the density dependence of the symmetry energy extracted from Ref.~\cite{Rus11} and Ref.~\cite{Dan13}.}
\label{v2_esym}
\end{figure}

The elliptic flow in high-energy heavy-ion collisions is another probe of the symmetry energy at suprasaturation densities (see, e.g., Ref.~\cite{Rus14} for a review). Shown in the left panel of Fig.~\ref{v2_esym} is the elliptic flows of neutrons and hydrogen isotopes in mid-central Au+Au collisions at 400 AMeV from a stiff and a soft symmetry energy based on the UrQMD model. FP1 and FP2 are different parameterizations of nucleon-nucleon collision cross sections, representing the uncertainties of their in-medium modifications. The calculation was done with the potential part of the symmetry energy proportional to $(\rho/\rho_0)^\gamma$, with $\gamma=1.5$ corresponding to a stiff symmetry energy (a-stiff) and $\gamma=0.5$ corresponding to a soft one (a-soft). Due to the squeeze-out dynamics, the elliptic flow becomes increasingly negative with increasing transverse momentum $p_t$, especially for a stiff symmetry energy. The ratio of the neutron and hydrogen elliptic flow is closer to 1 for a stiff symmetry energy compared to the smaller ratio from a soft symmetry energy. Comparison with the FOPI data leads to $\gamma=1.01 \pm 0.21$ for the FP1 parametrization and $\gamma=0.98 \pm 0.35$ for the FP2 parametrization. This analysis leads to $\gamma=0.9 \pm 0.4$ and the symmetry energy slope parameter $L=83 \pm 26$ MeV considering the uncertainties of the in-medium nucleon-nucleon collision cross sections. A later study compares the neutron-proton elliptic flow difference $v_2^{n-p}$ and ratio $v_2^{n}/v_2^{p}$ from the T\"ubingen QMD model with the FOPI data as shown in the middle panel of Fig.~\ref{v2_esym}. Here $x$ is the parameter in an effective isospin- and momentum-dependent interaction (MDI)~\cite{Das03,Che05}, which is linearly proportional to the slope parameter $L$ of the symmetry energy, with $x=-1$ corresponding approximately to $L=110$ MeV and $x=0$ corresponding approximately to $L=60$ MeV. A general trend that $v_2^{n-p}$ increases and $v_2^{n}/v_2^{p}$ decreases with the increasing softness of the symmetry energy is observed. Results from nucleon-nucleon collision cross sections by Cugnon and Li-Machleiht are displayed by solid and dashed lines, respectively. The uncertainties due to the incompressibility $K=190-280$ MeV and the width of the nucleon Gaussian wave packet $\text{L}=2.5-7.0$ fm$^2$ are also displayed, and $V_{opt}$ and $S$ represent respectively the different optical potential and symmetry energy parameterizations from different effective interactions used in transport calculations, i.e., the MDI interaction~\cite{Das03,Che05} and the Hartnack optical potential~\cite{Har94}. The comparison leads to the constraints of $x \in (-2.50, 0.25)$ from $v_2^{n-p}$ and $x \in (-2.25, 0)$ from $v_2^{n}/v_2^{p}$. The overall constraints on the symmetry energy is shown in the right panel of Fig.~\ref{v2_esym}, compared with that from Ref.~\cite{Rus11}. The combination of the analyses in Refs.~\cite{Rus11,Dan13} leads to the constraints $L=106 \pm 46$ MeV and $K_{sym}=127 \pm 290$ MeV. The updated results of Ref.~\cite{Rus11} were later reported in Ref.~\cite{Rus16}, which favors a potential contribution of the symmetry energy $22 (\rho/\rho_0)^\gamma$ (MeV) with $\gamma = 0.72 \pm 0.19$, corresponding to $L = 72 \pm 13$ (MeV), from the elliptic flows of neutrons and light charged particles in Au+Au collisions at 400 AMeV based on the UrQMD calculation. In a more recent study~\cite{Dan18} based on the T\"ubingen QMD calculation, an additional term is introduced to the energy density functional in the MDI interaction in order to allow independent variations of $L$ and $K_{sym}$, and the extracted slope parameter of the symmetry energy from the neutron-to-proton elliptic flow ratio becomes $L=84 \pm 30(exp) \pm 19(th)$ MeV.

\begin{figure}[h]
\centering
\includegraphics[scale=0.4,clip]{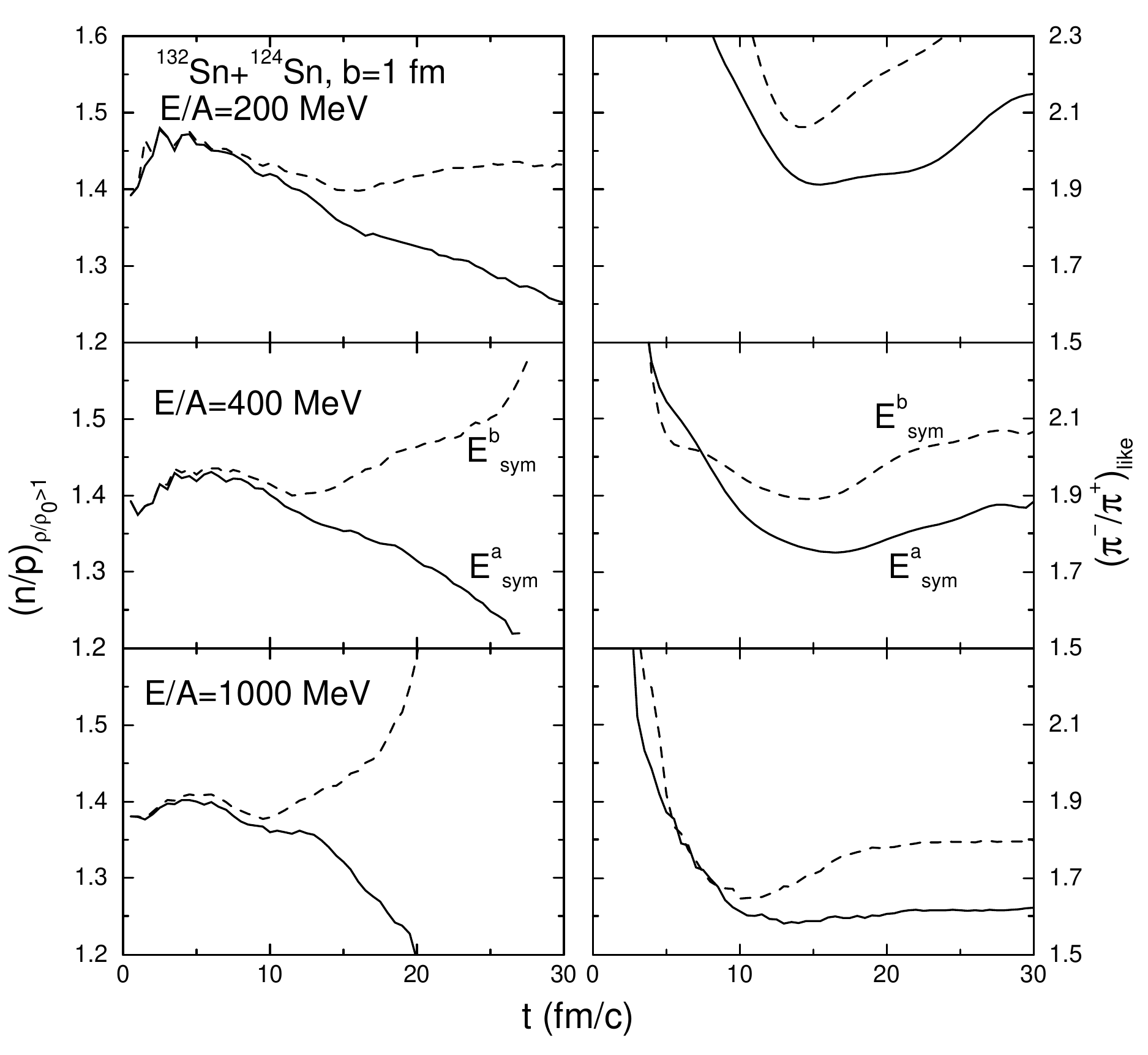}
\caption{Left: Neutron/proton ratios in the high-density phase of heavy-ion reactions with a stiff ($E_{sym}^a$) and a soft ($E_{sym}^b$) symmetry energy; Right: $\pi^-/\pi^+$ ratios including free pions and effective pions in $\Delta$ resonances in the same reactions with a stiff ($E_{sym}^a$) and a soft ($E_{sym}^b$) symmetry energy. Results are from IBUU97 transport model calculations taken from Ref.~\cite{Li02prl}.}
\label{pion_Li}
\end{figure}

The isospin-dependent particle production in heavy-ion collisions is a useful probe of the symmetry energy at suprasaturation densities (see, e.g., Ref.~\cite{Xia14} for a review). The $\pi^-/\pi^+$ yield ratio is a hotly debated probe of the symmetry energy, which was first proposed in Refs.~\cite{Li02prl,Li02npa} as a probe of the isospin asymmetry in the high-density phase in heavy-ion collisions. The left panel of Fig.~\ref{pion_Li} shows the neutron/proton ratios in the high-density phase of $^{132}$Sn+$^{124}$Sn collisions based on IBUU97 transport model calculations, using the same symmetry energies as in Fig.~\ref{Fnp_esym}. As discussed before, a soft symmetry energy ($E_{sym}^b$) leads to a weaker isospin fractionation effect and a more neutron-rich high-density phase. This effect grows as the time evolves especially at a higher beam energy, when the system is highly compressed and a higher density is reached. Pions are mostly produced in the high-density phase of the reactions through $\Delta$ resonances. The dominating isospin-dependent production and decay channels of $\Delta$ resonances are
\begin{eqnarray}
&&n + n \rightarrow \Delta^- + p, ~\Delta^- \rightarrow n + \pi^-;\\
&&p + p \rightarrow \Delta^{++} + n, ~\Delta^{++} \rightarrow p + \pi^+.
\end{eqnarray}
It is thus expected that a more neutron-rich high-density phase from a soft symmetry energy enhances (suppresses) the production of $\Delta^-$ ($\Delta^{++}$) and $\pi^-$ ($\pi^+$), leading to a larger $\pi^-/\pi^+$ yield ratio. This is exactly the case as shown in the right panel of Fig.~\ref{pion_Li}, where the subscript 'like' means total pions including free pions and effective pions in $\Delta$ resonances, with the later eventually decaying into free pions as well. The $\pi^-/\pi^+$ yield ratio is about the square of the neutron/proton ratio in the high-density phase if the system is in chemical equilibrium, and it is larger at lower beam energies with smaller pion multiplicities.

\begin{figure}[h]
\centering
\includegraphics[scale=0.31,clip]{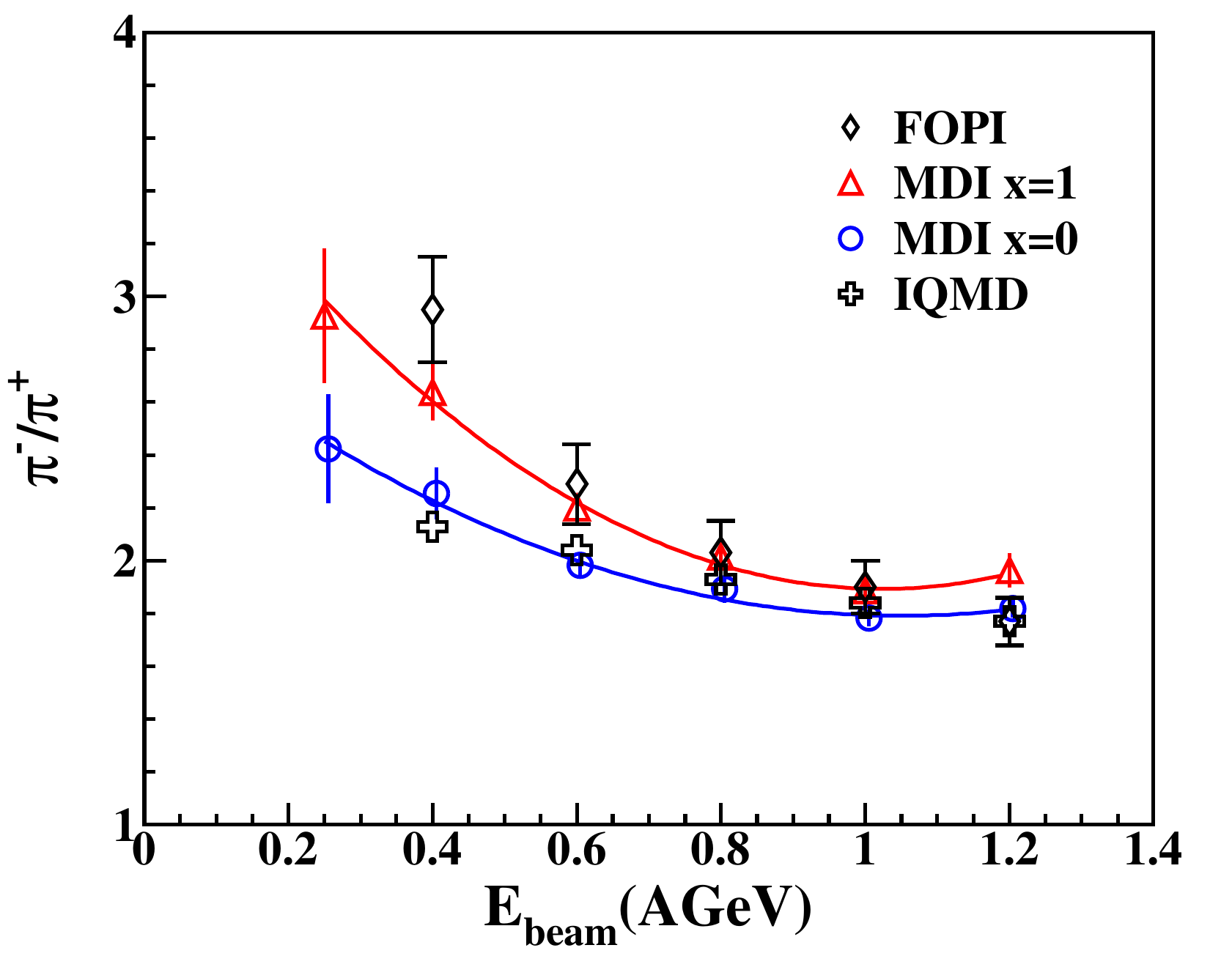}
\includegraphics[scale=0.85,clip]{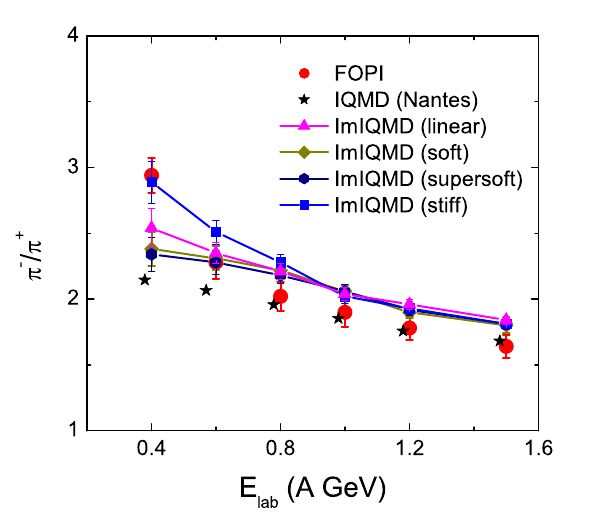}
\includegraphics[scale=0.17,clip]{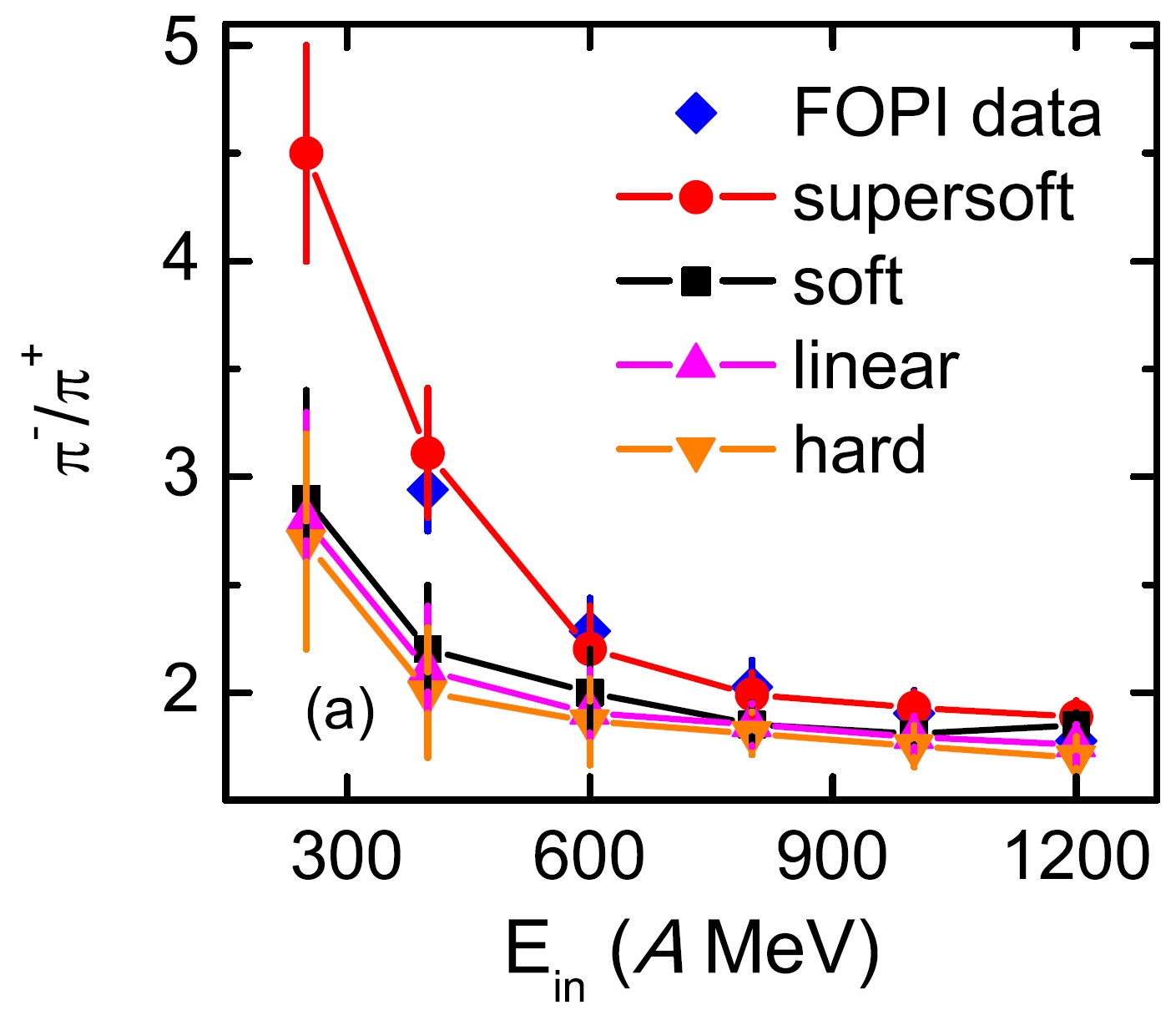}
\caption{(Color online) Beam energy dependence of the $\pi^-/\pi^+$ yield ratio from three different transport model studies. Left: From IBUU04 transport model calculations with the MDI interaction~\cite{Xia09}; Middle: From Lanzhou QMD transport model calculations~\cite{Fen10}; Right: From an isospin-dependent Boltzmann-Langevin approach~\cite{Xie13}. The IQMD results in the left and the middle panels are from Refs.~\cite{Har98,FOPI07}.}
\label{pion_esym1}
\end{figure}

The above study gives a nice picture of measuring simply the $\pi^-/\pi^+$ yield ratio to extract the information of the symmetry energy at suprasaturation densities. However, there are difficulties from the theoretical side in the model dependence and the incomplete physics in transport model studies. The descriptions of particle productions and decays as well as their inverse channels in transport model studies are much more complicated than pure nucleonic dynamics. As shown in the left panel of Fig.~\ref{pion_esym1}, a super soft symmetry energy, which becomes negative above $3\rho_0$ is needed to reproduce the FOPI $\pi^-/\pi^+$ data based on the IBUU04 transport model studies with the MDI interaction. The first evidence of the super soft symmetry energy attracted considerable attentions in the nuclear physics community, because it is not able to support a stable neutron star, unless some exotic mechanism is considered such as the non-Newtonian gravitational potential~\cite{Wen09}. From the Lanzhou QMD transport model calculations, the $\pi^-/\pi^+$ yield ratio is insensitive to the symmetry energy at higher beam energies, but increases with increasing stiffness of the symmetry energy below 1 AGeV, as shown in the middle panel of Fig.~\ref{pion_esym1}. The IQMD results~\cite{Har98,FOPI07} compared in the left and middle panels with an approximately linear density dependence of the symmetry energy also underestimate the FOPI data at lower beam energies. From an isospin-dependent Boltzmann-Langevin transport approach, the $\pi^-/\pi^+$ yield ratio also increases with the increasing softness of the symmetry energy, and a super soft symmetry energy is needed to reproduce the FOPI data, as shown in the right panel of Fig.~\ref{pion_esym1}.

\begin{figure}[h]
\centering
\includegraphics[scale=1.2,clip]{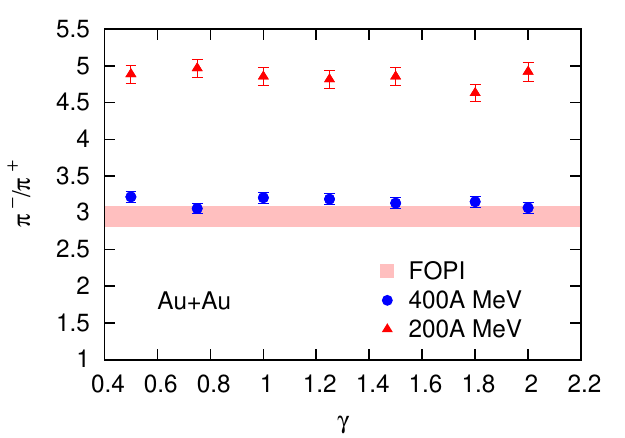}
\includegraphics[scale=0.33,clip]{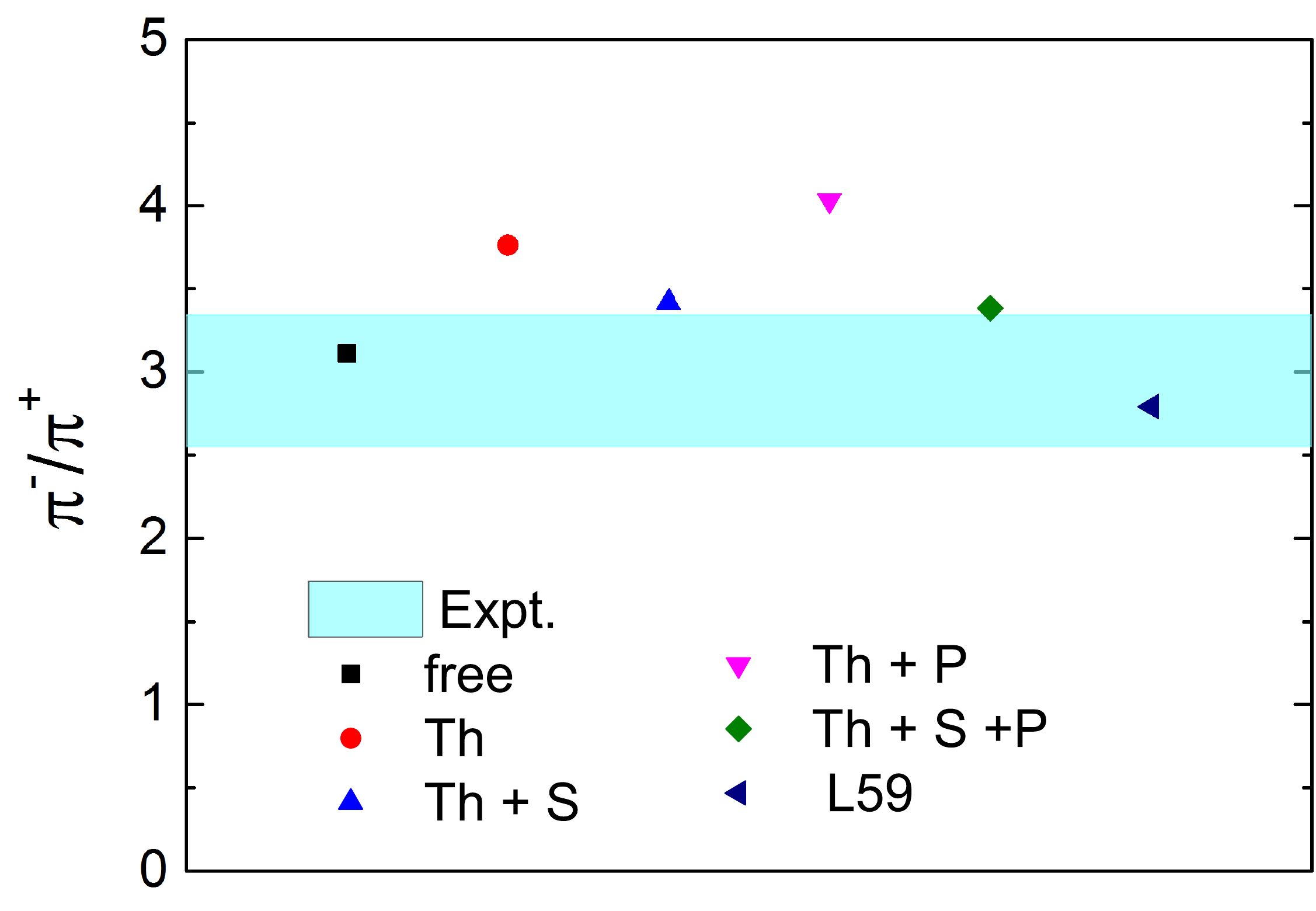}
\caption{(Color online) Left: $\pi^-/\pi^+$ yield ratio in central Au+Au collisions from pBUU transport model calculations~\cite{Hon14} with different stiffness parameters $\gamma$ of the symmetry energy compared with the FOPI data~\cite{FOPI07}; Right: $\pi^-/\pi^+$ yield ratio in Au+Au collisions at the beam energy of 400 AMeV and the impact parameter 1.4 fm based on the RVUU transport model with the NL$\rho$ force~\cite{Zha17} from different scenarios of pion in-medium effects and symmetry energies (see text for details) compared with the FOPI data~\cite{FOPI10}. }
\label{pion_esym2}
\end{figure}

There are also pion in-medium effects that will affect the $\pi^-/\pi^+$ yield ratio. Properties of pions are modified by their $s$-wave interaction with nucleons as well as their $p$-wave interaction through $\Delta$ resonances, and these effects are isospin dependent in neutron-rich nuclear matter. Based on the chiral perturbation theory up to the two-loop order~\cite{Kai01}, the $\pi^-$ mass is enhanced while the $\pi^+$ mass is suppressed in neutron-rich matter. Due to the $p$-wave interaction, pions become softer in nuclear medium, and $\pi^-$ interact strongly with neutrons and become even softer compared with $\pi^+$ in neutron-rich medium~\cite{Xu10}. The pion $s$-wave and $p$-wave interaction have the opposite isospin effect, and from a thermal model study it was found that the $\pi^-/\pi^+$ yield ratio due to the in-medium effect will decrease as a result of the dominating $s$-wave interaction~\cite{Xu10}. Although the $\pi^-/\pi^+$ yield ratio is more sensitive to the symmetry energy near the threshold energy of the pion production, the pion in-medium effects become increasingly important with decreasing beam energy~\cite{Xu13}, as a result of the sharper pion spectral function at lower energies. A study using the pBUU transport model has incorporated the pion-nucleon $s$-wave mean-field potential as $U_{\pi^\pm} = \mp 8 \rho_T S_{int}^0\rho^{\gamma-1}/\rho_0^\gamma$ with $\rho_T$ being the isovector density, and the $s$-wave potential is seen to be related to the interaction part of the symmetry energy $S_{int}(\rho) = S_{int}^0 (\rho/\rho_0)^\gamma$. As shown in the left panel of Fig.~\ref{pion_esym2}, the integrated $\pi^-/\pi^+$ yield ratio is insensitive to the $\gamma$ factor, by changing which the density dependence of the symmetry energy is modified. On the other hand, it was proposed that the kinetic energy dependence of the $\pi^-/\pi^+$ yield ratio, especially for energetic pions, is a better probe of the symmetry energy. The kinetic energy dependence of the $\pi^-/\pi^+$ yield ratio as well as the spectrum of the double $\pi^-/\pi^+$ yield ratio in $^{132}$Sn+$^{124}$Sn collisions with respect to that in $^{108}$Sn+$^{112}$Sn collisions are predicted in a later study~\cite{Tsa17} based on the pBUU transport model. This serves as the predictions for the corresponding experiments, with the experimental data to be available soon. A study with comprehensive pion in-medium effects was done in Ref.~\cite{Zha17} based on the RVUU transport model, where both pion $s$-wave and $p$-wave interactions are incorporated. Since incorporating the off-shell dynamics is still challenging, the pion $p$-wave interaction was incorporated through the pion dispersion relation. $\pi^-/\pi^+$ yield ratios from different combinations of the pion threshold energy effect (Th), the pion $s$-wave interaction effect (S), the pion $p$-wave interaction effect (P), as well as results from free pions (free) and a softer symmetry energy with $L=59$ MeV (L59) are shown in the right panel of Fig.~\ref{pion_esym2}. A parameterized in-medium $\Delta$ production cross section $\sigma_{NN\rightarrow N\Delta}(\rho_N)=\sigma_{NN\rightarrow N\Delta}(0)\exp(-A\sqrt{\rho_N/\rho_0})$, where $\rho_N$ is the nucleon number density, was used in each case in order to reproduce the same pion multiplicity as that in the case with free pions by adjusting the parameter $A$. It is seen that the $\pi^-/\pi^+$ yield ratio from the NL$\rho$ force with the symmetry energy slope parameter $L=84$ MeV without any pion in-medium effects is consistent with the FOPI data, while it is enhanced with the threshold effect, consistent with that observed in Ref.~\cite{Son15}. The isospin-dependent pion $s$-wave and $p$-wave interaction reduces and enhances the $\pi^-/\pi^+$ yield ratio, respectively, with the $s$-wave interaction dominating the results if both are incorporated, consistent with the results from the thermal model study~\cite{Xu10,Xu13}. By adjusting the $\rho$ meson coupling, a softer symmetry energy with $L=59$ MeV with all the pion in-medium effects is able to reproduce the FOPI data.

\begin{figure}[h]
\centering
\includegraphics[scale=0.6,clip]{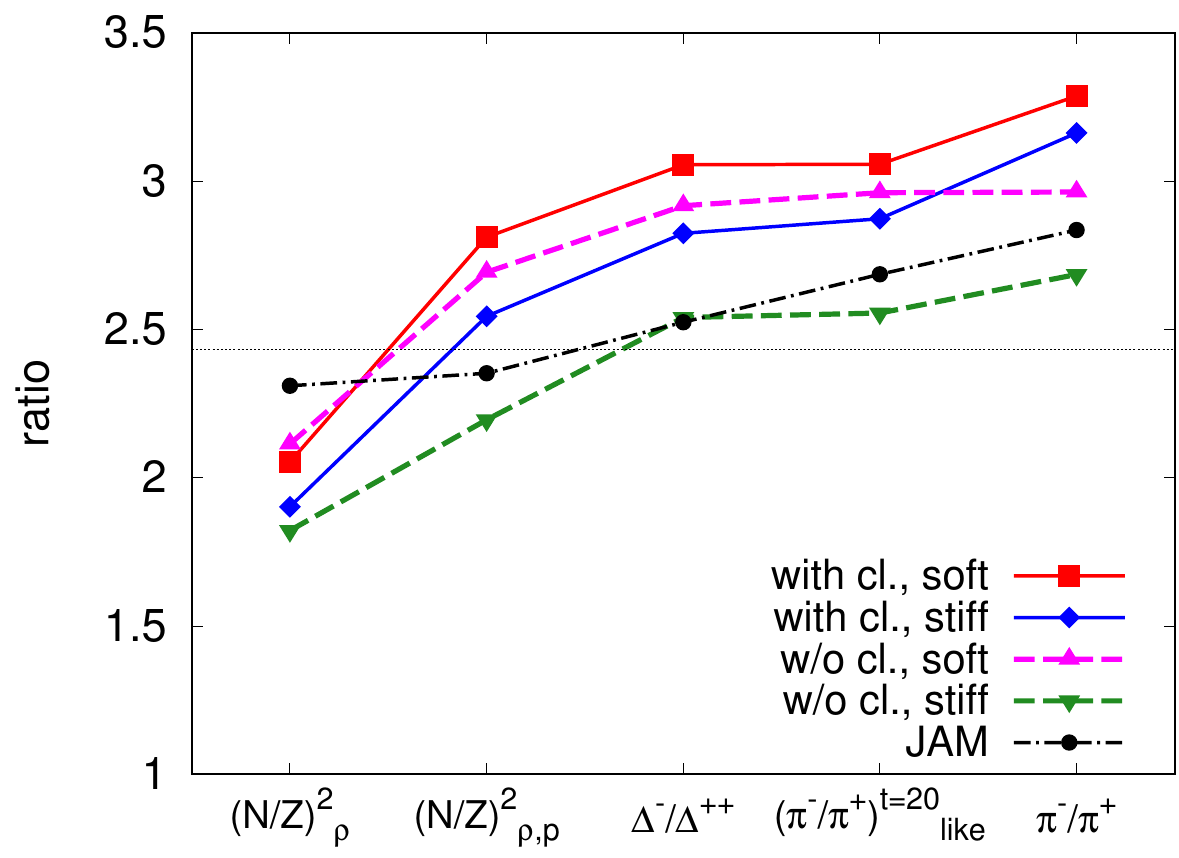}
\includegraphics[scale=0.35,clip]{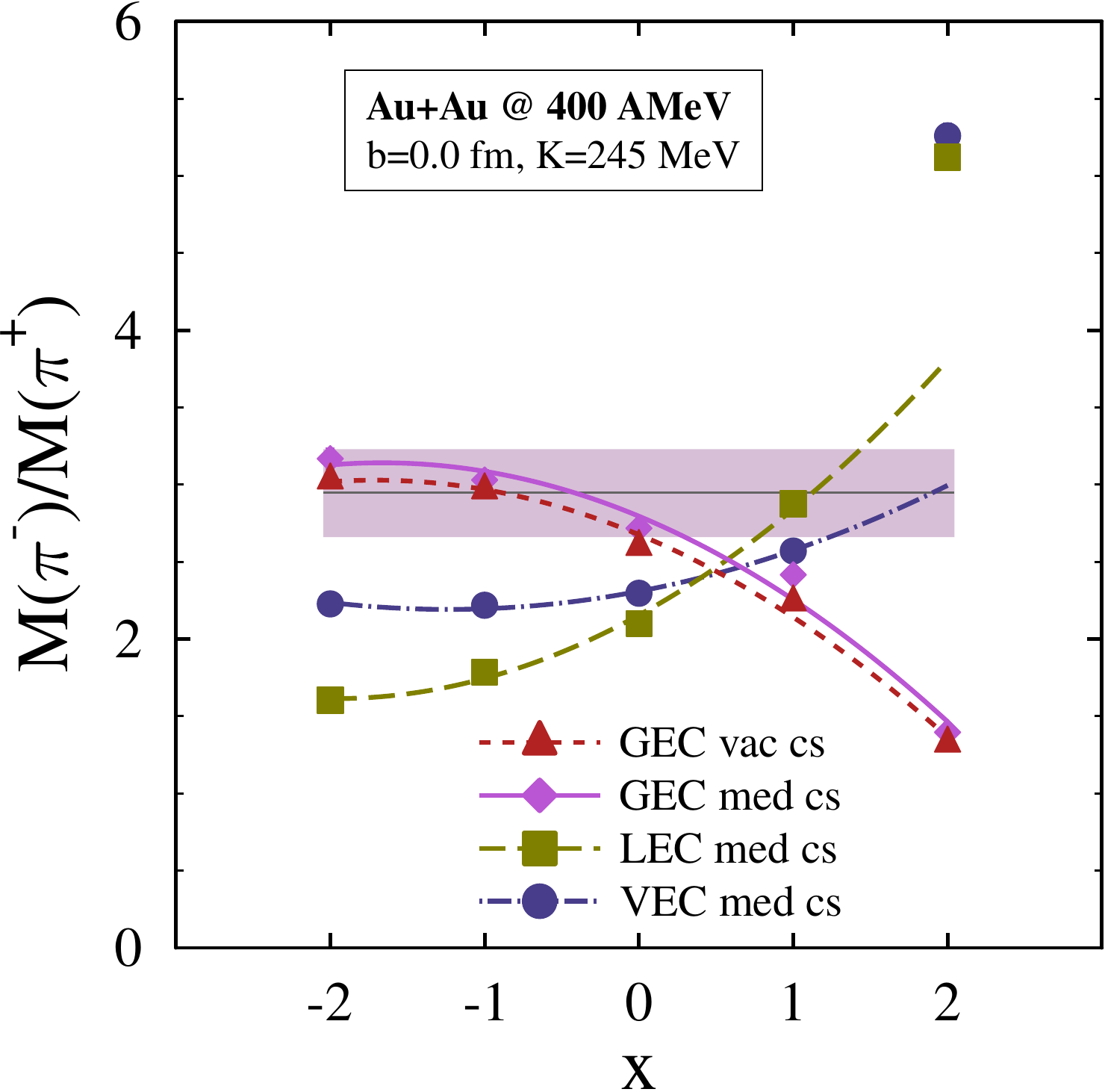}
\caption{(Color online) Left: Various ratios with and without cluster correlations from a stiff and soft symmetry energy in Sn+Sn collisions at 300 AMeV based on a combined transport approach of AMD and JAM~\cite{Ike16}; Right: $\pi^-/\pi^+$ yield ratio as a function of the symmetry energy parameter $x$ from different scenarios of the energy conservation (GEC, LEC, and VEC) and the cross section (vac cs and med cs) in central Au+Au collisions at 400 AMeV based on an improved T\"{u}bingen QMD model~\cite{Coz16} compared with the FOPI data~\cite{FOPI07}.}
\label{pion_esym3}
\end{figure}

There are also other effects that may affect the $\pi^-/\pi^+$ yield ratio. The left panel of Fig.~\ref{pion_esym3} compares various ratios based on a combined transport approach of AMD and JAM, with the former describing properly the cluster correlations, and pion dynamics incorporated in the later. $(N/Z)^2_\rho$ is the neutron/proton multiplicity ratio in the high-density phase ($\rho \ge \rho_0$) integrated over time, while $(N/Z)^2_{\rho,p}$ is that ratio in both high-density and high-momentum phase space integrated over time. It is seen that the $\Delta^-/\Delta^{++}$ yield ratio, the effective $\pi^-/\pi^+$ yield ratio at $t=20$ fm/c, and the overall $\pi^-/\pi^+$ yield ratio are closer to $(N/Z)^2_{\rho,p}$, which is larger compared to $(N/Z)^2_\rho$. This challenges the chemical equilibration in the high-density phase. On the other hand, the effect from the cluster correlation is comparable to the symmetry energy effect. An important finding is that the cluster correlation suppresses the difference of the $\pi^-/\pi^+$ yield ratio from the stiff and the soft symmetry energy. Since the cluster correlation forces some neutrons and protons to move together, the different forces on neutrons and protons are averaged out to some degree. Actually clusters are melted easily in the hot and compressed medium formed in heavy-ion collisions, and their dissociation and formation are expected to happen frequently. It is interesting to see that cluster dynamics can be important in heavy-ion collisions at the beam energy as high as 300 AMeV. A study based on an improved T\"{u}bingen QMD model focused on the energy conservation in transport simulations and the corresponding effect on the $\pi^-/\pi^+$ yield ratio, by improving the energy conservation from the average level to an event-by-event basis. As shown in the right panel of Fig.~\ref{pion_esym3}, the global energy conservation (GEC) scenario guarantees the energy conservation at a global level, while the local energy conservation (LEC) scenario guarantees the energy conservation of colliding particles, with global energy conservation fulfilled also if the mean-field potentials are momentum- and isospin-independent, and the in-vacuum energy conservation (VEC) scenario is the case of the energy conservation without any mean-field potentials in the previous version of the model. It is seen that the modification of the in-medium cross sections from the effective mass scaling only slightly changes the $\pi^-/\pi^+$ yield ratio. The different scenarios of the energy conservation affect the $\pi^-/\pi^+$ yield ratio significantly, and the effect is comparable to the symmetry energy. This shows that one may get different density dependence of the symmetry energy from the same $\pi^-/\pi^+$ yield ratio without treating the energy conservation properly, due to the small pion mass comparable to the possible energy violation.

\begin{figure}[h]
\centering
\includegraphics[scale=0.35,clip]{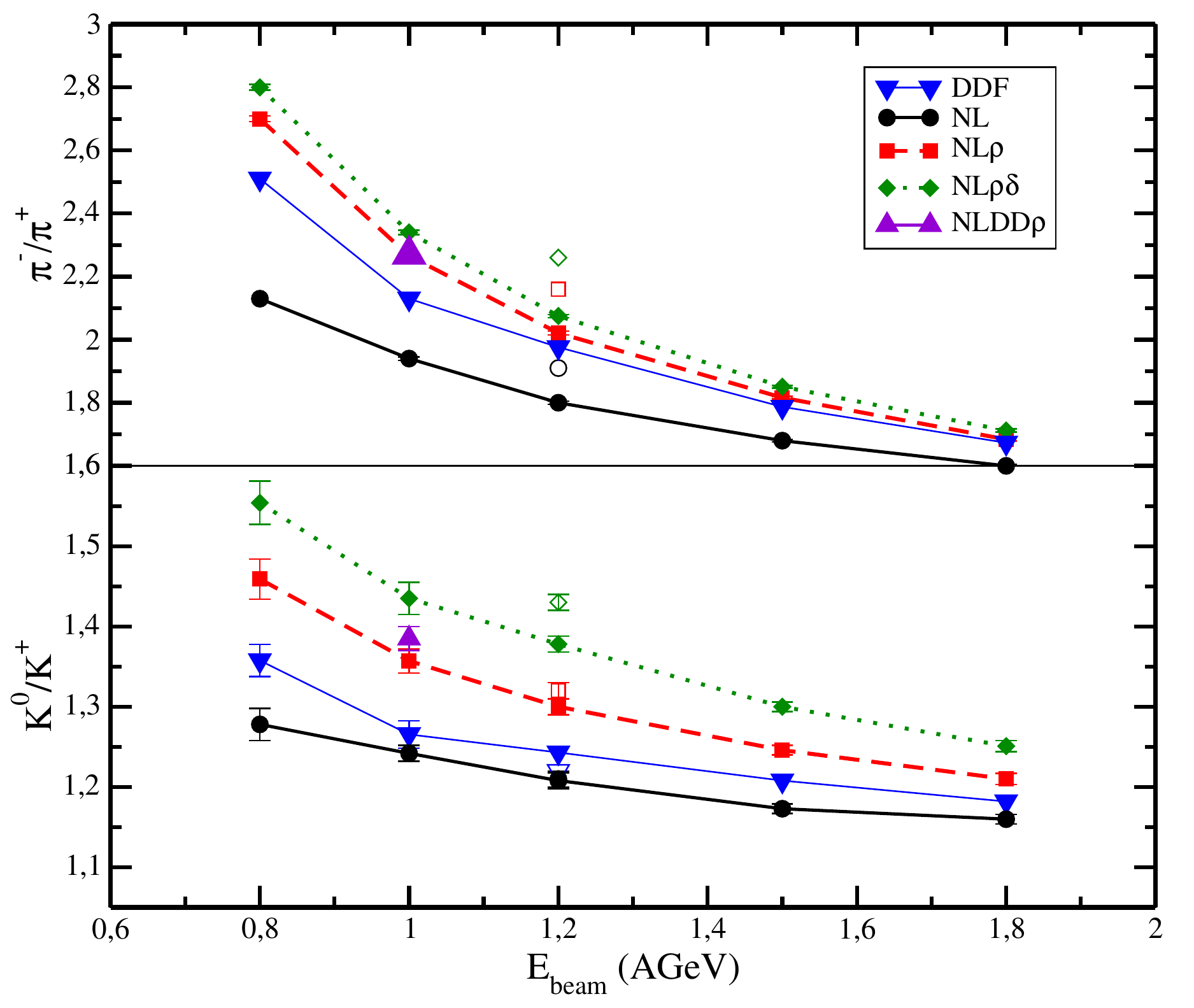}
\includegraphics[scale=0.31,clip]{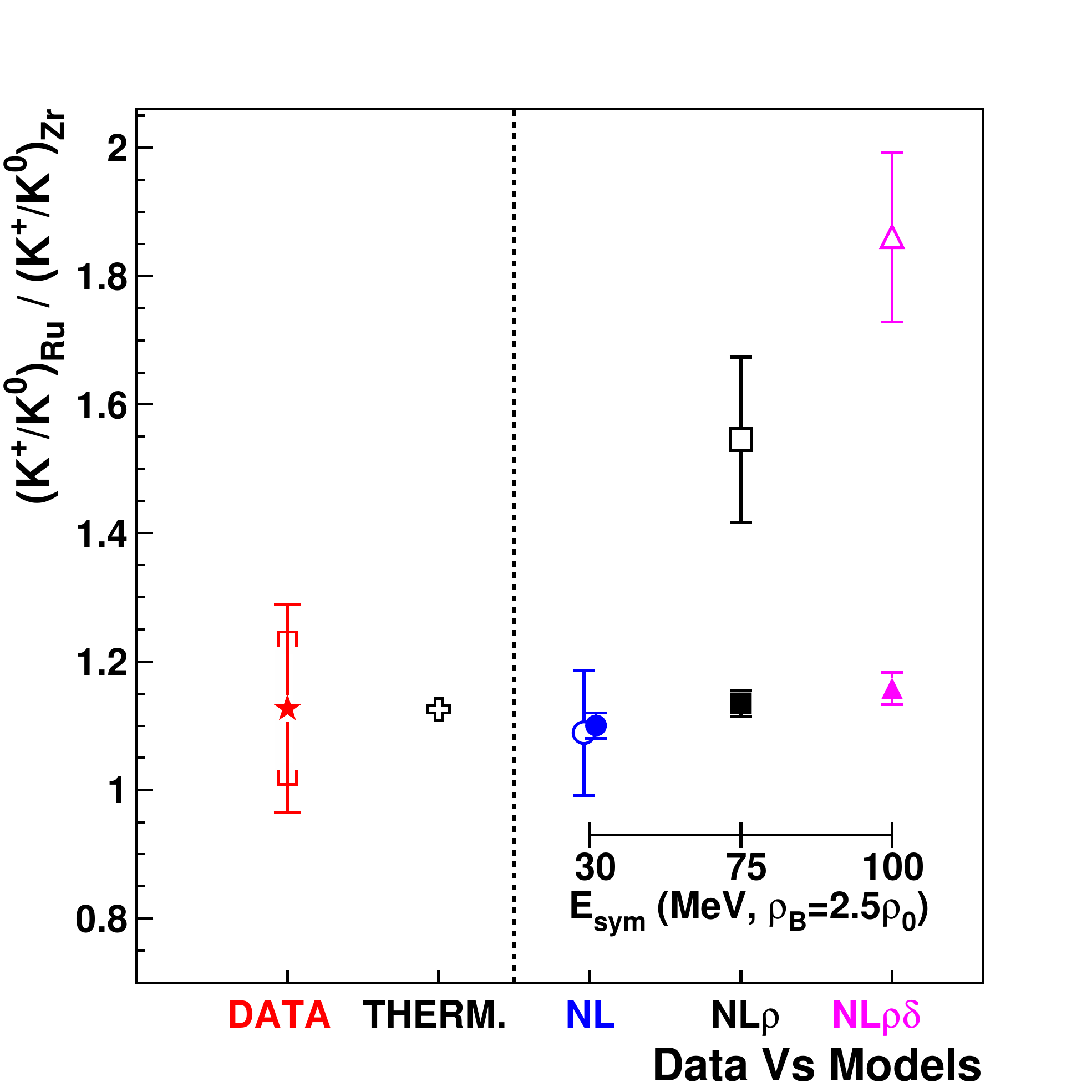}
\caption{(Color online) Left: Beam energy dependence of the $\pi^-/\pi^+$ yield ratio and the $K^0/K^+$ yield ratio in central Au+Au collisions (solid symbols) and $^{132}$Sn+$^{124}$Sn collisions (open symbols) based on the relativistic BUU transport model with different RMF forces~\cite{Fer06}; Right: Experimental relative $K^0/K^+$ yield ratio in isobaric Ru+Ru and Zr+Zr collision systems compared with results from the thermal model and the relativistic BUU transport model with different RMF forces~\cite{Lop07}.}
\label{kaon_esym}
\end{figure}

The $K^0/K^+$ yield ratio could be a cleaner probe of the symmetry energy at suprasaturation densities compared with the $\pi^-/\pi^+$ yield ratio. As is known, kaons are mostly produced at even earlier stage in heavy-ion collisions, while the pion yield is much affected by the reabsorption and isospin exchange processes. Due to the long mean free path for kaons, they are more suitable for the description of on-shell dynamics compared with the situation for pions. Shown in the left panel of Fig.~\ref{kaon_esym} is a comparison of the $\pi^-/\pi^+$ yield ratio and the $K^0/K^+$ yield ratio based on the relativistic BUU transport model with different RMF forces in a box calculation represent an infinite nuclear matter system. The typical isospin-dependent channels for the production of kaons are $n+\pi^- \rightarrow \Sigma^- + K^0$ and $p+\pi^+ \rightarrow \Sigma^+ + K^+$. It is expected that a soft symmetry energy leads to a more neutron-rich high-density phase and enhances the production of $K^0$ than $K^+$. On the other hand, the in-medium mass of $\Sigma^- + K^0$ is larger than that of $\Sigma^+ + K^+$ in neutron-rich matter, leading to an opposite effect on the $K^0/K^+$ yield ratio. With the threshold and the energy conservation effects for both pions and kaons, it is seen that the $\pi^-/\pi^+$ yield ratio and the $K^0/K^+$ yield ratio both increase with the increasing stiffness of the symmetry energy. The variation of the ratio from various symmetry energies is $14-16\%$ for kaons, but it reduces to $8-10\%$ for pions, indicating a higher sensitivity of the $K^0/K^+$ yield ratio to the isovector potential and the behavior of the symmetry energy at suprasaturation densities. The right panel of Fig.~\ref{kaon_esym} compares the relative $K^0/K^+$ yield ratio in $^{96}_{44}$Ru+$^{96}_{44}$Ru and $^{96}_{40}$Zr+$^{96}_{40}$Zr collision systems from FOPI experiments with theoretical calculations. Assuming full stopping and isospin conservation, the thermal model result from the temperature $T=52$ MeV, the baryon chemical potential $\mu_b=756$ MeV, and the isospin chemical potential $\mu_I=-7.1$ MeV for $^{96}_{44}$Ru+$^{96}_{44}$Ru and $-13.5$ MeV for $^{96}_{40}$Zr+$^{96}_{40}$Zr reproduces very well the FOPI experimental data. The relative $K^0/K^+$ yield ratio increases slightly when the RMF force changes from NL, NL$\rho$, to NL$\rho\delta$, i.e., from a soft symmetry energy to a stiff one, in the heavy-ion calculation shown by solid scatters, but dramatically for the box calculation at the temperature 60 MeV and the baryon density $2.5\rho_0$ shown by open scatters. The FOPI data thus favors a soft symmetry energy with $E_{sym}=30$ MeV at $\rho=2.5\rho_0$ from the box calculation, but higher accuracy for the experimental data is needed to pin down the symmetry energy at suprasaturation densities from the heavy-ion calculation.

\subsection{Isospin splitting of nucleon effective mass}
\label{mnp}

The importance of the isospin splitting of the nucleon effective mass on the isospin dynamics in heavy-ion collisions was revisited recently. There are different definitions of the in-medium nucleon effective mass (see, e.g., Ref.~\cite{Li18} for a recent review), and I mostly concentrate on the non-relativistic 'k-mass'
\begin{equation}
m^*_{n/p}=m\left(1+\frac{m}{\hbar^2 k}\frac{\partial U_{n/p}}{\partial k}\right)^{-1}.
\end{equation}
In this case, the isospin splitting of the nucleon effective mass $m_{n-p}^*=(m_n-m_p)/m$ is related to the momentum dependence of the nucleon symmetry potential~\cite{Li04}, and is also related to the nuclear symmetry energy through the Hugenholtz-Van Hove theorem~\cite{Li13plb}. Generally in effective interaction models, it is possible to have different density dependence of the symmetry energy but the same isospin splitting of the nucleon effective mass, or vice versa. So far, most studies favor a larger neutron effective mass than proton in a neutron-rich medium (see Table 2 of Ref.~\cite{Li18}), and this leads to the increasing trend of the symmetry potential with decreasing nucleon momentum~\cite{Li04}. However, the possibility for a larger proton effective mass than neutron in neutron-rich matter can not be completely ruled out yet. The uncertainties of the isospin splitting of the nucleon effective mass complicate the isospin dynamics in intermediate-energy heavy-ion collisions, since all the probes of the nuclear symmetry energy are, to some extent, affected by the isospin splitting of the nucleon effective mass as well. In the following I discuss a few topics related to transport simulations, and I refer the reader to Refs.~\cite{Li15,Li18} for recent reviews.

\begin{figure}[h]
\centering
\includegraphics[scale=0.2,clip]{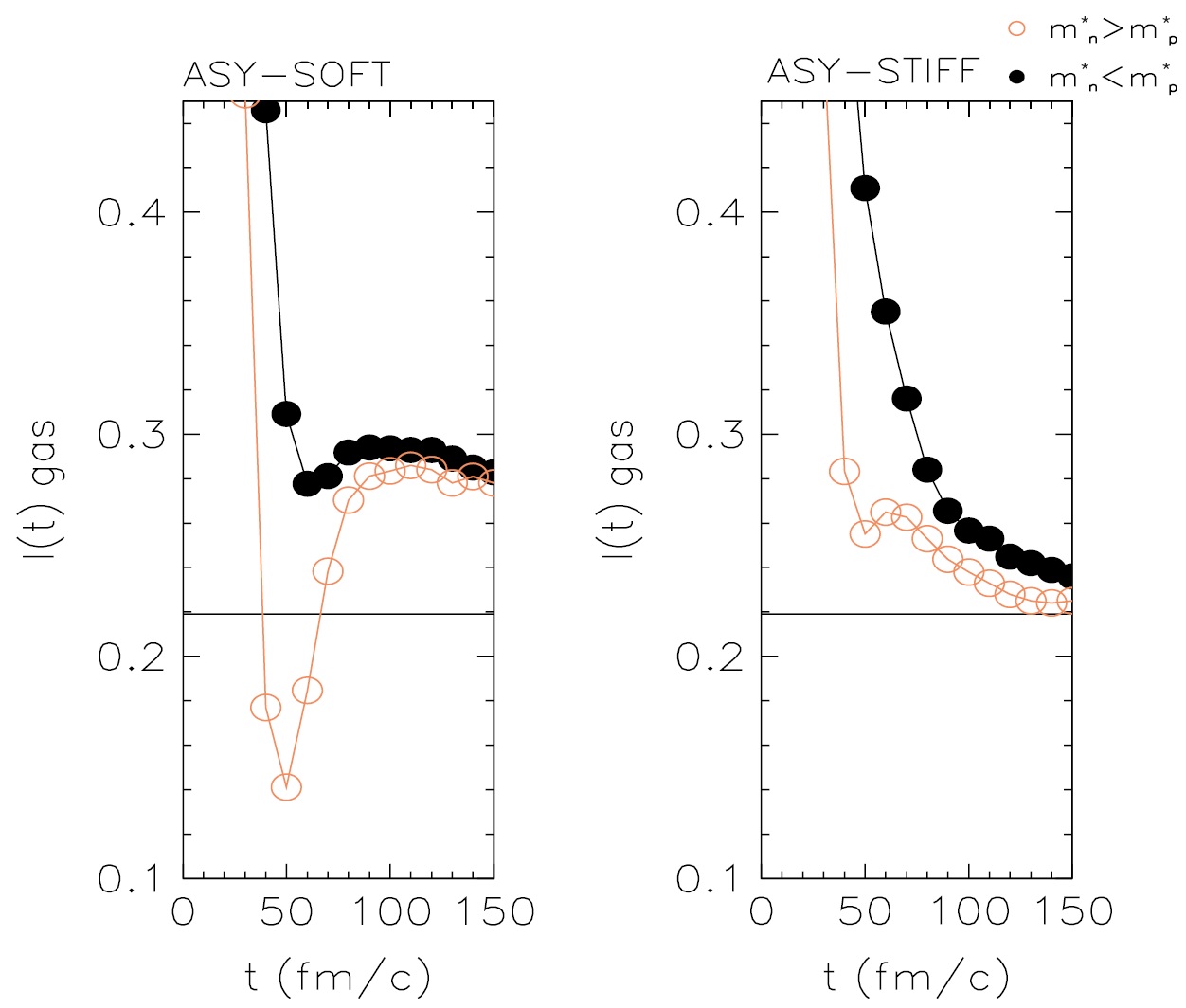}
\includegraphics[scale=0.3,clip]{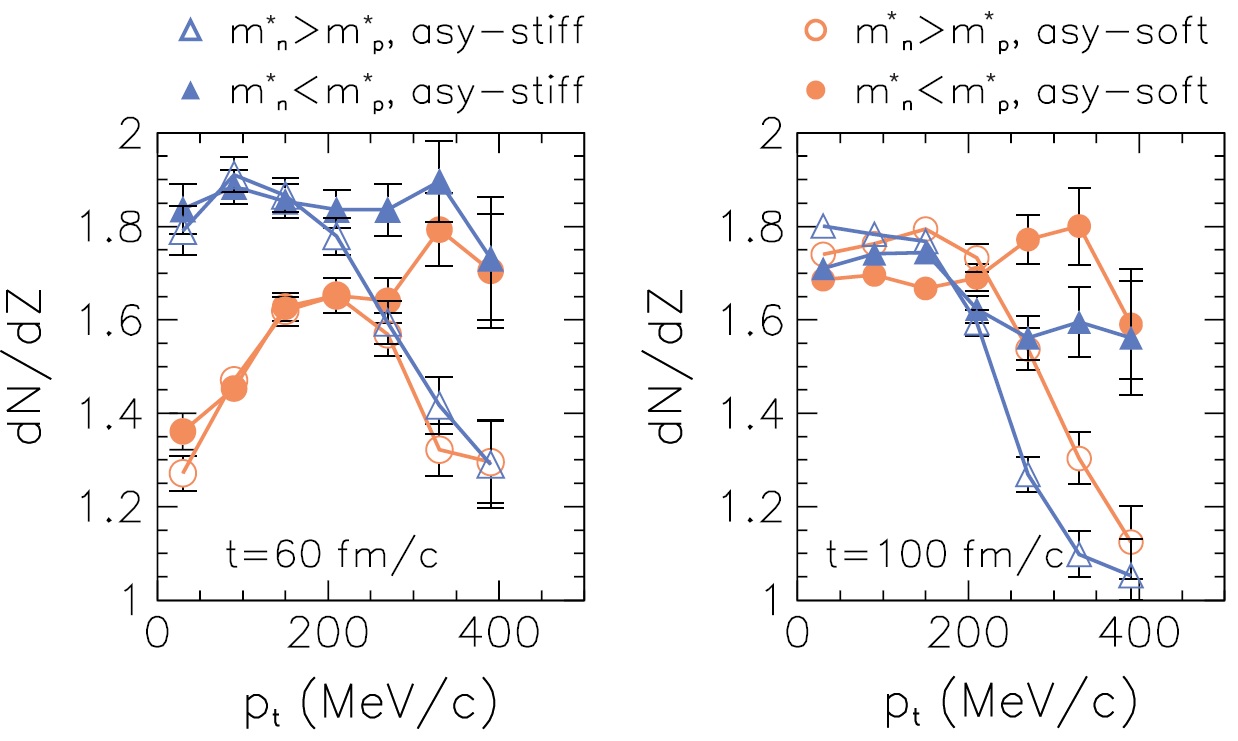}
\caption{(Color online) Left: Isospin asymmetry of the gas phase as a function of time in central $^{132}$Sn+$^{124}$Sn at the beam energy 50 AMeV with two different choices of neutron-proton effective mass splittings, with results in the left (right) panel from a soft (stiff) symmetry energy; Right: Transverse momentum dependence of the neutron/proton ratio in central $^{132}$Sn+$^{124}$Sn at the beam energy of 50 AMeV from two different choices of neutron-proton effective mass splittings and symmetry energies, with results in the left (right) panel at the earlier (later) stage of the collision. Taken from Ref.~\cite{Riz05}.}
\label{np_mnp}
\end{figure}

In the previous subsection, the neutron/proton yield ratio is shown to be a probe of the nuclear symmetry energy. The left panel of Fig.~\ref{np_mnp} compares the isospin asymmetry of the gas phase from different nuclear symmetry energies and neutron-proton effective mass splittings, based on a parameterized Skyrme-like interaction using the Boltzmann-Nordheim-Vlasov transport model. It is seen that a stiff symmetry energy generally leads to a more neutron-rich gas phase at the earlier stage, but a less neutron-rich gas phase at the later stage. The larger isospin asymmetry in the later stage from the soft symmetry energy is due to the more effective isospin distillation, i.e., more neutrons emitted from the low-density phase than protons. On the other hand, at the early preequilibrium stage the isospin-dependent nucleon emission is more sensitive to the neutron-proton effective mass splitting than to the nuclear symmetry energy. A larger neutron effective mass than proton enhances (suppresses) the proton (neutron) emission and leads to a less (more) neutron-rich gas phase at the earlier stage, and this effect is more pronounced for a soft symmetry energy. The isospin asymmetry of the gas phase at the final stage is, however, more sensitive to the symmetry energy and less sensitive to the neutron-proton effective mass splitting. The right panel of Fig.~\ref{np_mnp} helps to further understand this phenomenon. It is clearly seen that at the early stage ($t=60$ fm/c) the neutron/proton ratio at higher transverse momenta is more sensitive to the neutron-proton effective mass splitting, but that at lower transverse momenta is more sensitive to the symmetry energy. At the later stage ($t=100$ fm/c), the sensitivity at lower transverse momenta is largely suppressed while that at higher transverse momenta is still observable.

To get the accurate neutron/proton yield ratio is challenging not only for experimental measurements, but also for transport simulations since there are always difficulties to identify unambiguously free nucleons, especially for BUU-type models. This, together with the complicated isospin dynamics from both the symmetry energy and the isospin splitting of the nucleon effective mass, leads to the so-called 'neutron/proton ratio puzzle'. In order to extract information of the symmetry energy and the neutron-proton effective mass splitting from the MSU data on the kinetic energy dependence of the neutron/proton yield ratio~\cite{Fam06prl,Cou16}, various studies have put considerable efforts based on transport simulations. The feature that a larger neutron effective mass than proton leads to a smaller neutron/proton yield ratio at higher kinetic energies is obtained by most transport model studies. The study based on an isospin-dependent Boltzmann-Langevin approach favors a smaller neutron effective mass than proton in neutron-rich matter~\cite{IBL2}, although better statistics is needed. The IBUU transport model underpredicts the neutron/proton yield ratio, no matter how the symmetry energy and the neutron-proton effective mass splitting are varied~\cite{Kon15}. Once the short-range correlation from the tensor force, which leads to the high-momentum tail in the momentum distribution and enhances the potential contribution of the symmetry energy, is incorporated, it is possible to reproduce the kinetic energy dependence of the neutron/proton yield ratio~\cite{Hen15}, while further efforts on treating more consistently the dynamics are still needed. The study based on the ImQMD model favors a smaller neutron effective mass than proton in neutron-rich matter by comparing with the neutron/proton yield ratio from the MSU experiment~\cite{Cou16}.

\begin{figure}[h]
\centering
\includegraphics[scale=0.28,clip]{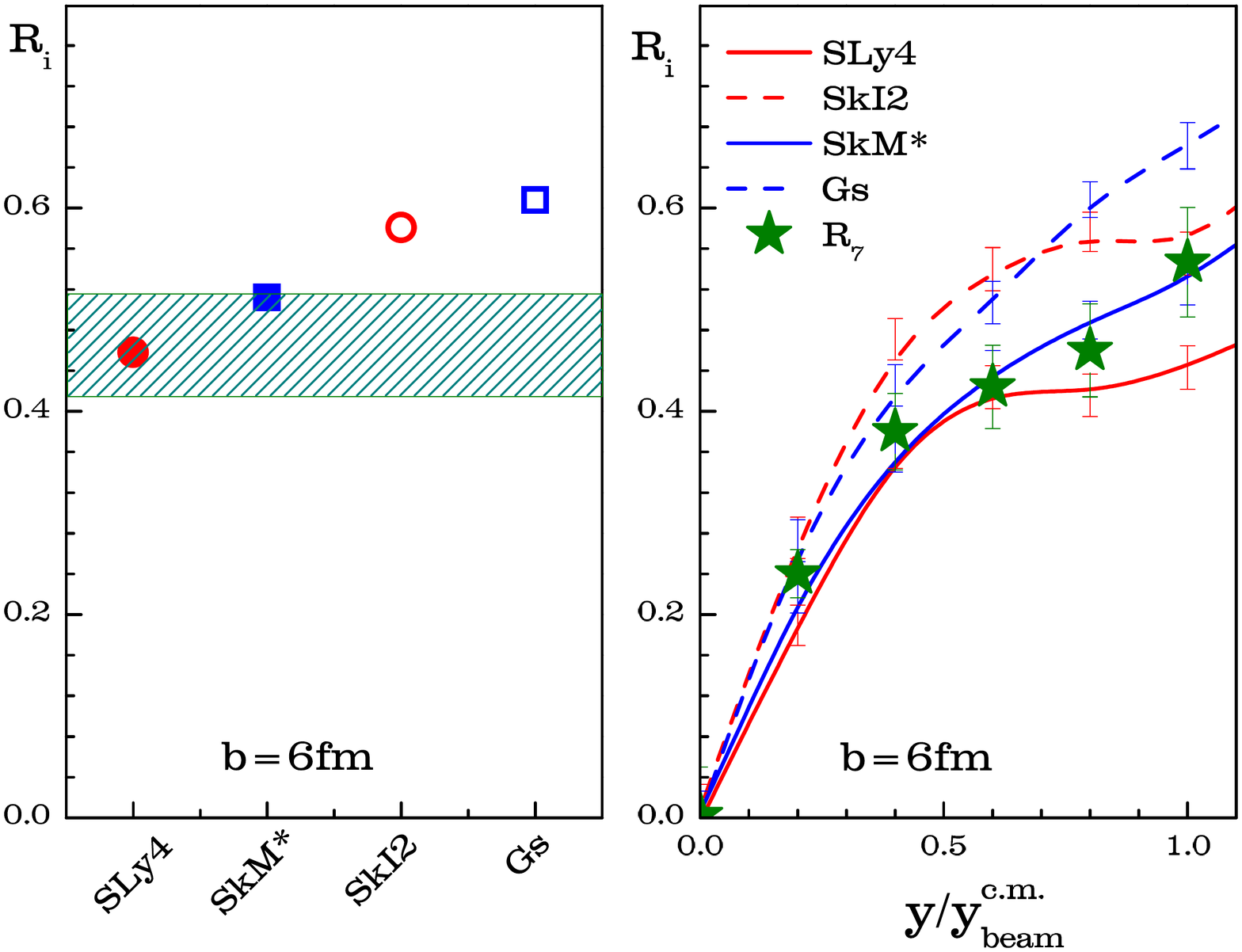}
\includegraphics[scale=0.31,clip]{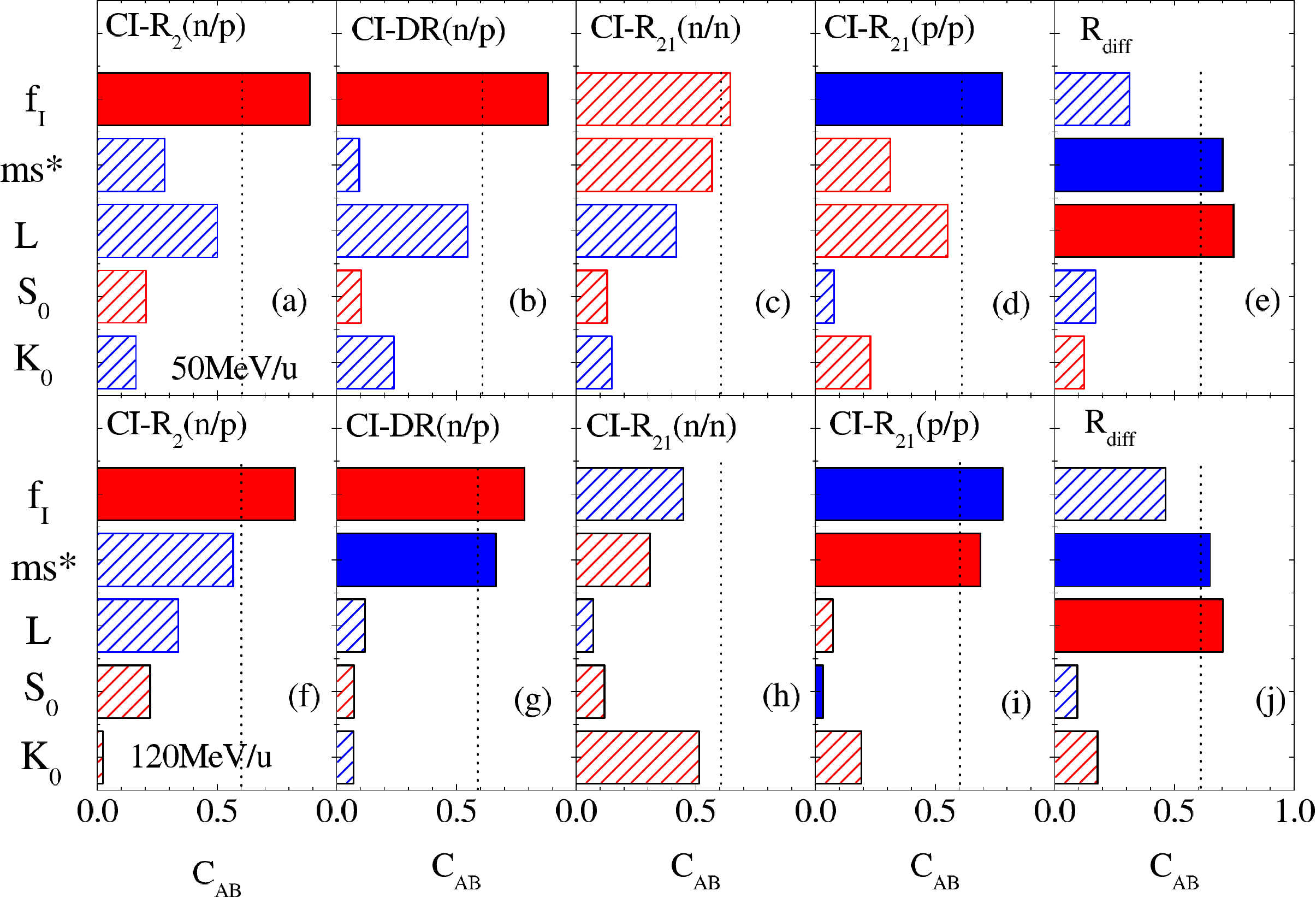}
\caption{(Color online) Left: Comparison of the isospin transport ratios from four Skyrme forces with different symmetry energies and isospin splittings of the nucleon effective mass~\cite{Zha14}; Right: Correlations between five observables and five force parameters in Sn+Sn collisions at 50 (upper) and 120 AMeV (lower)~\cite{Zha15}. }
\label{isodiff_mnp}
\end{figure}

\begin{table}[b]
\vspace{-0.4cm}
\caption{\label{tab:table1}
Corresponding saturation properties of nuclear matter from the SLy4, SkI2, SkM*, and Gs Skyrme forces. All quantities are in MeV, except for $\rho_0$ in fm$^{-3}$ and the dimensionless effective mass ratios for nucleons, neutrons, and protons. The effective mass for neutrons and protons are obtained for the isospin asymmetric nuclear matter with the isospin asymmetry $\delta=0.2$. Taken from Ref.~\cite{Zha14}.}
\centering
\begin{tabular}{lccccccccc}
\hline
\hline
\textrm{Para.}&
\textrm{$\rho_0$}&
\textrm{$E_0$}& \textrm{$K_0$} & \textrm{$S_0$} & \textrm{$L$} & \textrm{$K_{\rm{sym}}$} &
\textrm{$m^*/m$} &$m^*_{\rm{n}}/m$ & $m^*_{\rm{p}}/m$\\
%\multicolumn{1}{c}{\textrm{Decimal}}&
\hline
%\colrule
 SLy4 & 0.160 & -15.97 & 230 &  32 & 46 & -120& 0.69& 0.68 &0.71 \\
 SkI2 & 0.158 & -15.78 & 241 &  33 & 104 & 71 & 0.68& 0.66 &0.71 \\
 SkM* & 0.160 & -15.77 & 217 &  30 & 46 & -156 & 0.79 & 0.82 &0.76 \\
 Gs   & 0.158 & -15.59 & 237 &  31 & 93 & 14 & 0.78 & 0.81 &0.76 \\
\hline
\end{tabular}
\label{tab1}
\end{table}

The isospin splitting of the nucleon effective mass may affect the isospin diffusion as well. The left panel of Fig.~\ref{isodiff_mnp} compares the isospin transport ratios from the Skyrme forces SLy4, SkI2, SkM*, and Gs, with the saturation properties of nuclear matter from these Skyrme forces shown in Table~\ref{tab1}. The calculation is based on the ImQMD model using the momentum-dependent mean-field potential from the Skyrme interaction, and the isospin diffusion results are obtained from reactions with combinations of $^{124}$Sn and $^{112}$Sn as projectile and target nuclei. The bands and the stars are the isospin transport ratio as a measure of the isospin diffusion from the experimental data, by taking respectively the isospin tracer as the isoscaling coefficient $\alpha$ and $\ln[Y(^{7}\text{Li})/Y(^{7}\text{Be})]$. It is seen that by varying the slope parameter $L$ of the symmetry energy from 46 MeV to 104 MeV the isospin transport ratio is changed by about 0.12, while varying the neutron-proton effective mass splitting from $(m^*_n-m^*_p)/m=-0.03$ to 0.06 at $\delta=0.2$ only changes the isospin transport ratio by about 0.06. The rapidity dependence of the isospin transport ratio is also more sensitive to the nuclear symmetry energy than to the isospin splitting of the nucleon effective mass. A later so-called covariance analysis using the same ImQMD model was done in order to extract correlations and sensitivities of five different observables to five force parameters. As shown in the right panel of Fig.~\ref{isodiff_mnp}, R$_2$(n/p) represents the neutron/proton yield ratio in $^{124}$Sn+$^{124}$Sn collisions, DR(n/p), R$_{21}$(n/n), and R$_{21}$(p/p) represent respectively the relative neutron/proton yield ratio, neutron yield, and proton yield in $^{124}$Sn+$^{124}$Sn and $^{112}$Sn+$^{112}$Sn collision systems, and R$_{\rm diff}$ is the isospin transport ratio for the isospin diffusion measurement using $^{124}$Sn and $^{112}$Sn as the projectile and target nuclei. 'CI' is the acronym for the coalescence invariant yield taking not only free nucleons but also those in light clusters into account, and only nucleons with the kinetic energy larger than 40 MeV are considered. The five force parameters are $f_I=\frac{1}{2\delta}(\frac{m}{m_n^*}-\frac{m}{m_p^*})$, the isoscalar nucleon effective mass $m_s^*$, the slope parameter $L$ and the value $S_0$ of the symmetry energy at the saturation density, and the incompressibility $K_0$. The red (blue) bar means the positive (negative) correlation, i.e., the value of the observable increases (decreases) with the increasing force parameter. It is seen that the single and double neutron/proton yield ratios are strongly positively correlated with $f_I$, especially at lower collision energies, with the covariance larger than 0.6 shown by the solid bar. Unlike the double neutron yield, the double proton yield is more sensitive to the isospin splitting of the nucleon effective mass, as a result of the strong Coulomb repulsion. It is interesting to see that the isoscalar effective mass affects significantly the relative neutron/proton yield ratio and the relative proton yield at 120 AMeV. As for the isospin transport ratio, it is positively correlated with $L$ and negatively correlated with $m_s^*$, but rather insensitive to $f_I$, consistent with the finding in the left panel of Fig.~\ref{isodiff_mnp}. This analysis suffers from the uncertainties of the central reference values in the force parameter space, and can be further improved with a Bayesian analysis method.

\begin{figure}[h]
\centering
\includegraphics[scale=0.21,clip]{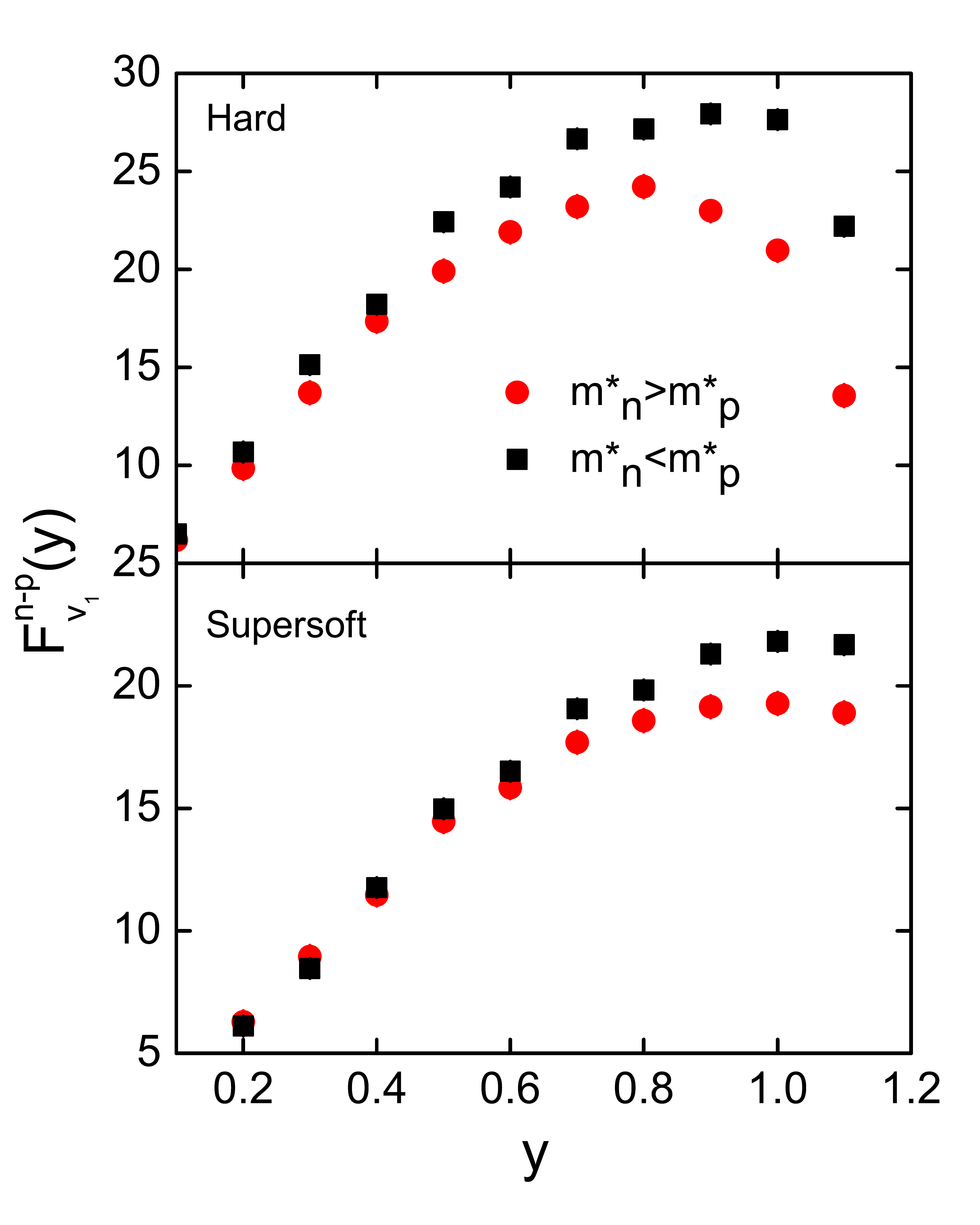}
\includegraphics[scale=0.205,clip]{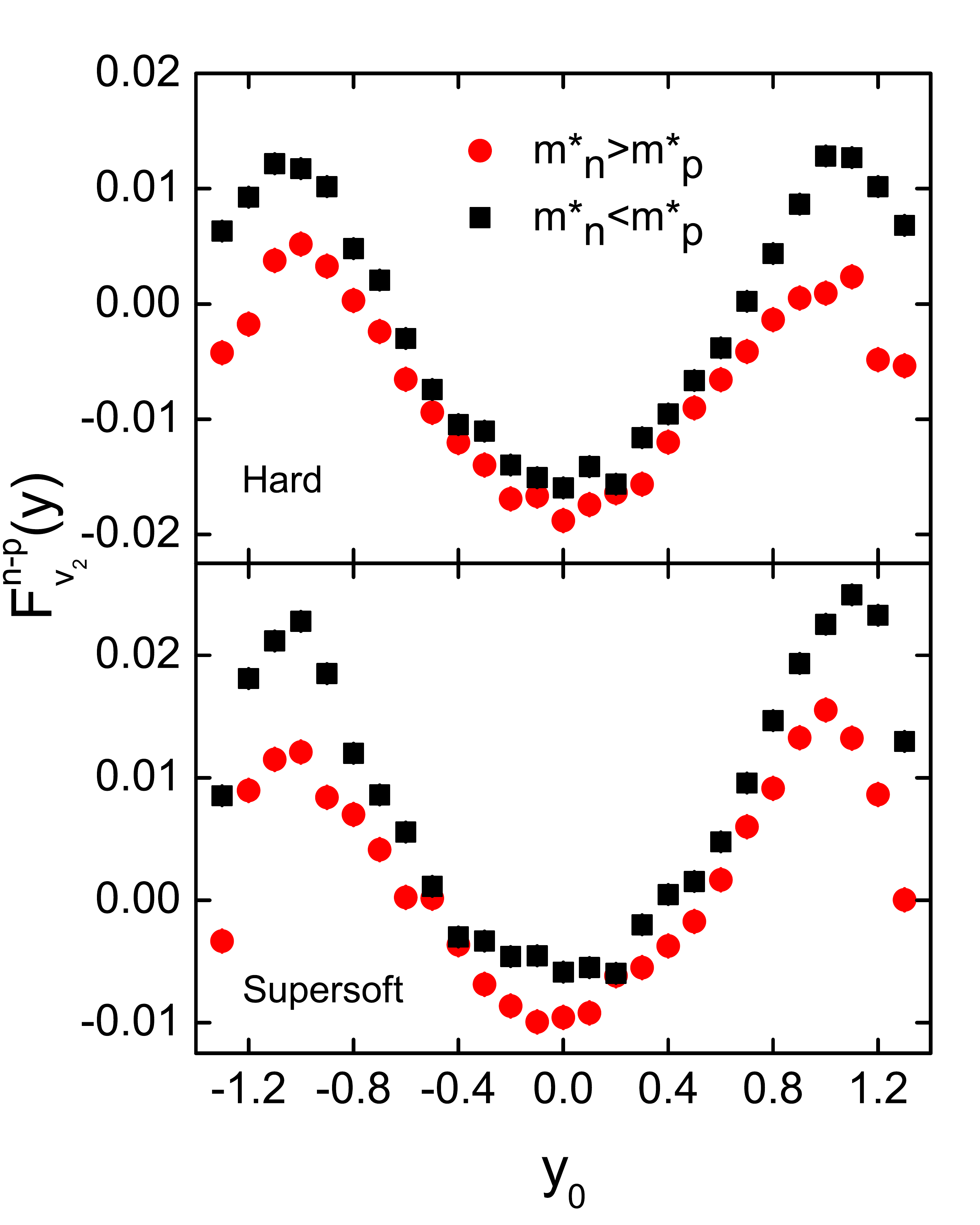}
\caption{(Color online) Rapidity dependence of the neutron-proton differential directed flow (left) and elliptic flow (right) in mid-central Au+Au collisions at 400 AMeV from a hard and a supersoft symmetry energy and different isospin splittings of the nucleon effective mass based on IQMD-BNU calculations~\cite{Xie14}. }
\label{flow_mnp}
\end{figure}

Previously I have shown that the different collective flows for neutrons and protons are probes of the nuclear symmetry energy. Actually they are sensitive to the isospin splitting of the nucleon effective mass as well~\cite{Dit05,Gio10,Fen12,Xie14}. Here I take results from the IQMD-BNU transport model as an example~\cite{Xie14}. In Fig.~\ref{flow_mnp}, the neutron-proton differential collective flows, i.e.,
\begin{equation}
F_{v_i}^{n-p}(y) = \frac{N_n(y)}{N(y)}v_i^n(y) - \frac{N_p(y)}{N(y)}v_i^p(y),
\end{equation}
with $N_n(y)$, $N_p(y)$, and $N(y)$ being the neutron, proton, and nucleon yield at rapidity $y$, and $i=1$ for the directed flow and $2$ for the elliptic flow, are compared with different symmetry energies and isospin splittings of the nucleon effective mass. As shown in the left panel, a hard symmetry energy leads to a larger neutron-proton differential directed flow since neutrons (protons) are affected by a more repulsive (attractive) potential at suprasaturation densities. On the other hand, $m_n^*<m_p^*$ leads to a larger neutron-proton differential directed flow than $m_n^*>m_p^*$, as a result of the lighter (heavier) in-medium neutron (proton) effective mass and thus a stronger neutron (weaker proton) directed flow. As shown in the right panel of Fig.~\ref{flow_mnp}, a hard symmetry energy leads to a more negative neutron elliptic flow than proton at midrapidities, as a result of the more repulsive neutron potential and the stronger squeeze-out effect, and thus a more negative neutron-proton differential elliptic flow. The neutron-proton differential elliptic flow at large rapidities is more sensitive to the isospin splitting of the nucleon effective mass than that at midrapidities, while it seems to be sensitive to the symmetry energy in the whole rapidity range.

\section{Transport code evaluation project}
\label{comparison}

The transport model evaluation project started at European Center for Theoretical Studies in Nuclear Physics and Related Areas (ECT*) in Trento in 2004, when the productions of kaons and pions in heavy-ion collisions at 1 GeV regime were extensively compared~\cite{Kol05}, since they are considered as probes of the nuclear matter EOS at suprasaturation densities. In 2009 there were trials on the comparison of heavy-ion simulations at 400 and 100 AMeV, but unfortunately the uncertainties were not quite understood and the results were not published. In 2014 at Shanghai Jiaotong University~\cite{transport2014} people tried to reemphasize the problems at 400 and 100 AMeV, and the comparisons were mainly about the stability of initial nuclei, stopping and flows, as well as nucleon-nucleon collisions and Pauli blocking rates in heavy-ion simulations~\cite{Xu16}. It was found that in heavy-ion systems different components of transport simulations affect each other, adding to the difficulties of understanding the deviations from various codes. In order to improve the situation, transport simulations in a box with the periodic boundary condition were proposed, where in many cases the limiting results from theoretical derivations are available, and the purpose became to evaluate separately the performance of these transport codes for nucleon-nucleon elastic collisions~\cite{Zhang18}, nucleon evolution in a mean-field potential, as well as production of pion-like particles. The box calculations as well as connections between transport simulations with other nuclear physics research field were extensively discussed at Michigan State University in East Lensing in 2017~\cite{ICNT2017}. It is seen that the comparisons have moved from high-energy heavy-ion collisions to low- and intermediate-energy heavy-ion reactions, and from more realistic heavy-ion simulations to those in an ideal box system representing the nuclear matter. While I was preparing this review, the paper of the box simulation for nucleon-nucleon elastic collisions has been published, while those for the nucleon evolution in the mean-field potential as well as the production for pion-like particles have not been completely finalized. In the following I will review the main results and achievements from the transport code evaluation project in the order of their completion, as well as the current status and the future plan.

\subsection{Comparisons of heavy-ion simulations}
\label{comhic}

The complexity of transport simulations was realized in the comparison in 2004. There are usually several thousand program lines for each transport code. The results of the transport simulations from different codes depend on the input, e.g., elementary cross sections for particle productions. In addition, different parameterizations and treatments for resonances are used in different codes, and they have significant influences on the particle production. The later comparison in 2014 tried to use the initialization as close as possible, and the same mean-field potential as well as the same nucleon-nucleon collision cross sections, i.e., carrying out the comparison in better controlled conditions. Although not completely satisfactory results were obtained in the heavy-ion comparison, useful lessons were learnt and a step was moved towards more robust transport simulation results.

\subsubsection{Particle productions and spectra at 1 AGeV regime}

The information of codes participating the heavy-ion comparison at 1 AGeV regime was listed in Table~\ref{code_Kol05}. They are the BUU-type transport model from the Budapest/Rossendorf group (BUU-BR), the hadron string dynamics (HSD) model developed in Giessen by Bratkovskaya and Cassing, the ultra-relativistic quantum molecular dynamics (UrQMD) model, the isospin-dependent quantum molecular dynamics (IQMD) model, the BUU-type transport model developed by Bratkovskaya, Effenberger, Larionov, and Mosel in Giessen (GiBUU), the Relativistic Vlasov-Uehling-Uhlenbeck (RVUU) transport model developed by the Texas A\&M group, the quantum molecular dynamics model developed by the T\"ubingen group (TuQMD), the Relativistic Boltzmann-Uehling-Uhlenbeck (RBUU) transport model developed by the Munich/Catania/T\"ubingen group, and the BUU-type transport model developed by Danielewicz (pBUU) in Michigan. The code correspondents will be shown in the figures, representing results from the corresponding codes. There are totally 9 transport codes involved in this comparison, including 6 BUU-type codes and 3 QMD-type codes. The comparison was carried out for Au+Au collisions at 0.96 and 1.48 AGeV and Ni+Ni collisions at 1.93 AGeV, at an impact parameter of $\text{b} = 1$ fm. The collision energies are far above the threshold energy ($\sim$ 300 MeV) of pion production, comparable to that ($\sim$ 1500 MeV) of $K^+$ production, and below that ($\sim$ 2500 MeV) of $K^-$ production.

\begin{table}[h]
\vspace{-0.4cm}
\caption{Code correspondents, acronyms of code names, and representative references for the 6 BUU-type and 3 QMD-type transport codes involved in the heavy-ion comparison at 1 AGeV regime~\cite{Kol05}.}
\centering
\begin{tabular}{lccccccccc}
\hline
\hline
\textrm{code correspondents}&
\textrm{acronyms}&
\textrm{representative references}\\
%\multicolumn{1}{c}{\textrm{Decimal}}&
\hline
%\colrule
 Barz/Wolf & BUU-BR & \cite{Barzwolf1,Barzwolf2}  \\
 Cassing & HSD &  \cite{HSD1,HSD2,HSD3} \\
 Reiter & UrQMD &  \cite{UrQMD3,UrQMD4} \\
 Hartnack   & IQMD & \cite{Har98,IQMD1}  \\
 Larionov  & GiBUU &  \cite{GIBUU4,GIBUU5} \\
 Chen  & RVUU &  \cite{RVUU1,RVUU2} \\
 Fuchs  & TuQMD &  \cite{TuQMD3,Fuc01,TuQMD5}  \\
 Gaitanos   & RBUU &  \cite{RBUU4,RBUU5} \\
 Danielewicz   & pBUU & \cite{Dan91}  \\
\hline
\end{tabular}
\label{code_Kol05}
\end{table}

\begin{figure}[h]
\centering
\includegraphics[scale=0.4,clip]{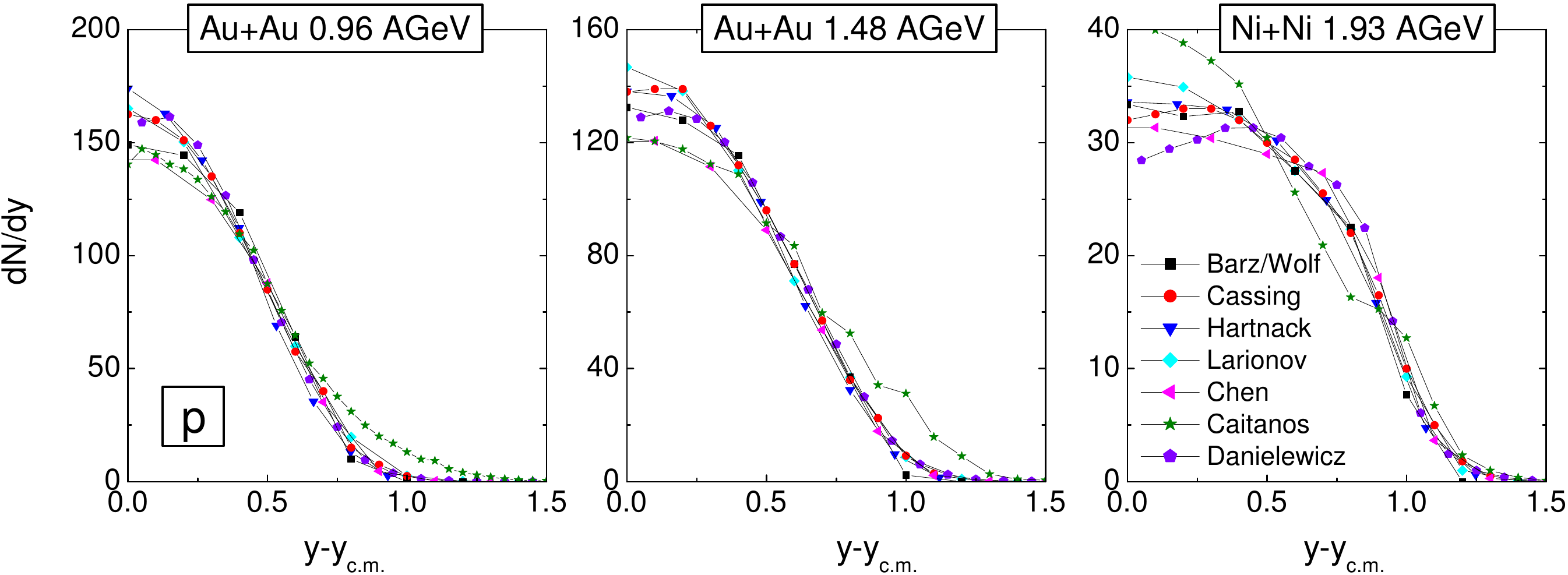}
\includegraphics[scale=0.4,clip]{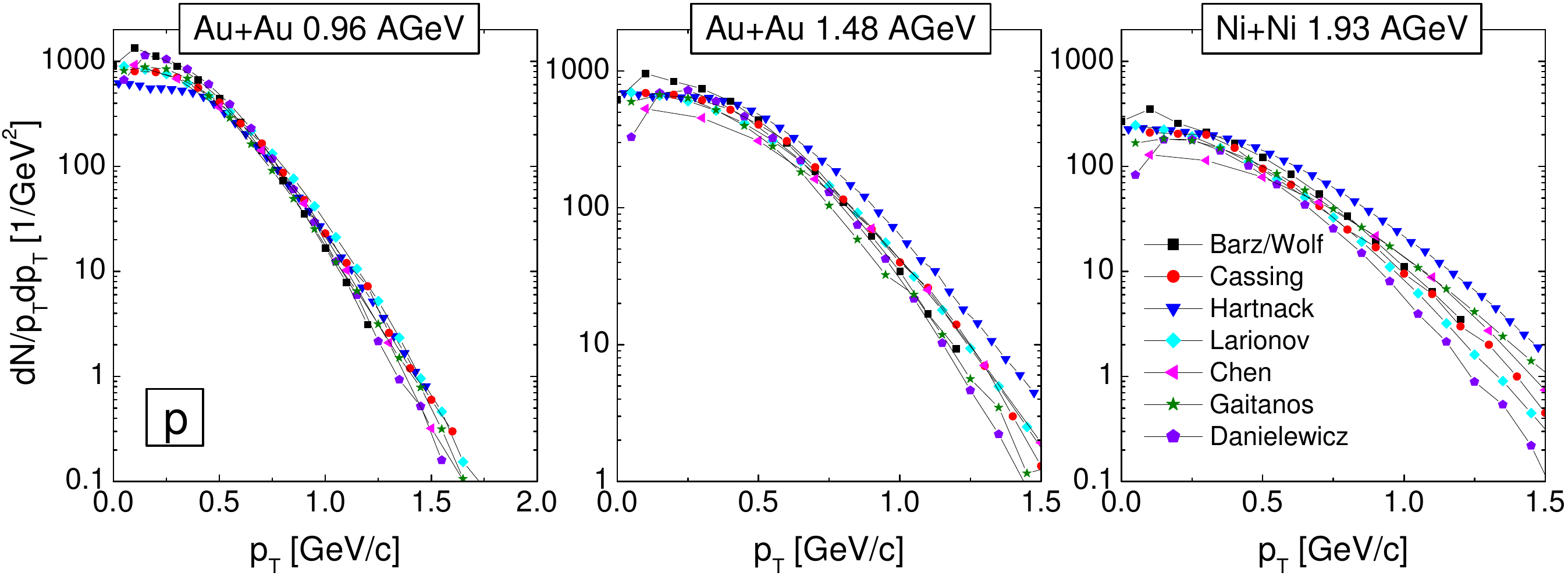}
\caption{(Color online) Rapidity (upper) and transverse momentum (lower) distributions of protons at the final stage of Au+Au collisions at 0.96 (left) and 1.48 (middle) AGeV and Ni+Ni collisions at 1.93 AGeV (right) at an impact parameter of $\text{b} = 1$ fm from various transport codes. Taken from Ref.~\cite{Kol05}. }
\label{proton_Kol05}
\end{figure}

The rapidity distribution of protons reflects the stopping power in heavy-ion collisions. Comparison with the transverse momentum spectra shows how much the longitudinal motion is transferred to transverse energies for particle productions. The variances of the rapidity distribution and the transverse momentum distribution are similar if the system is in thermal equilibrium. At 1 AGeV regime, the stopping and the thermalization of heavy-ion systems from transport simulations depend on nucleon-nucleon collisions, and the Pauli blocking is expected to be less important. At the initial stage, proton distributions are peaked at rapidity $y_{cm}=0.68$, 0.80, and 0.89 for the three different collision systems, with the distribution width representing the Fermi motion. After heavy-ion collisions, both the rapidity and transverse momentum distributions become broader, especially at higher collision energies, as shown in Fig.~\ref{proton_Kol05}. A further detailed comparison between the variance of the rapidity and transverse momentum distribution shows that the stopping power becomes weaker at higher energies, as a result of more forward-peaked nucleon-nucleon collision cross sections. The lighter Ni+Ni collision system is expected to be less thermalized, where the discrepancies among results from different codes are larger, compared to the heavier Au+Au collision system. It is seen that the discrepancies are mostly at smaller transverse momenta, where many collisions are needed to fill this part of the phase space, i.e., a measure of nucleon-nucleon collision numbers in heavy-ion simulations.

\begin{figure}[h]
\centering
\includegraphics[scale=0.4,clip]{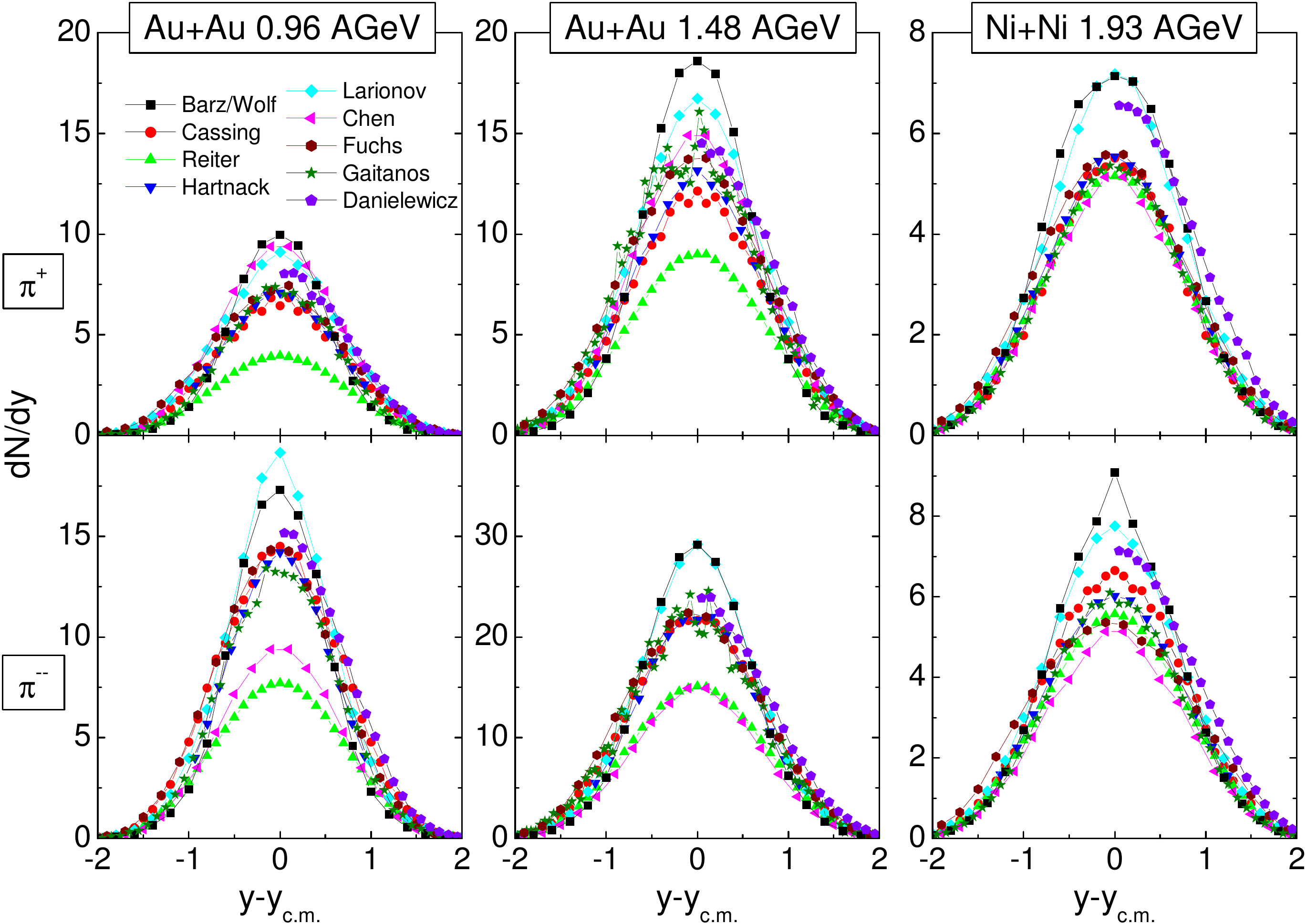}
\includegraphics[scale=0.42,clip]{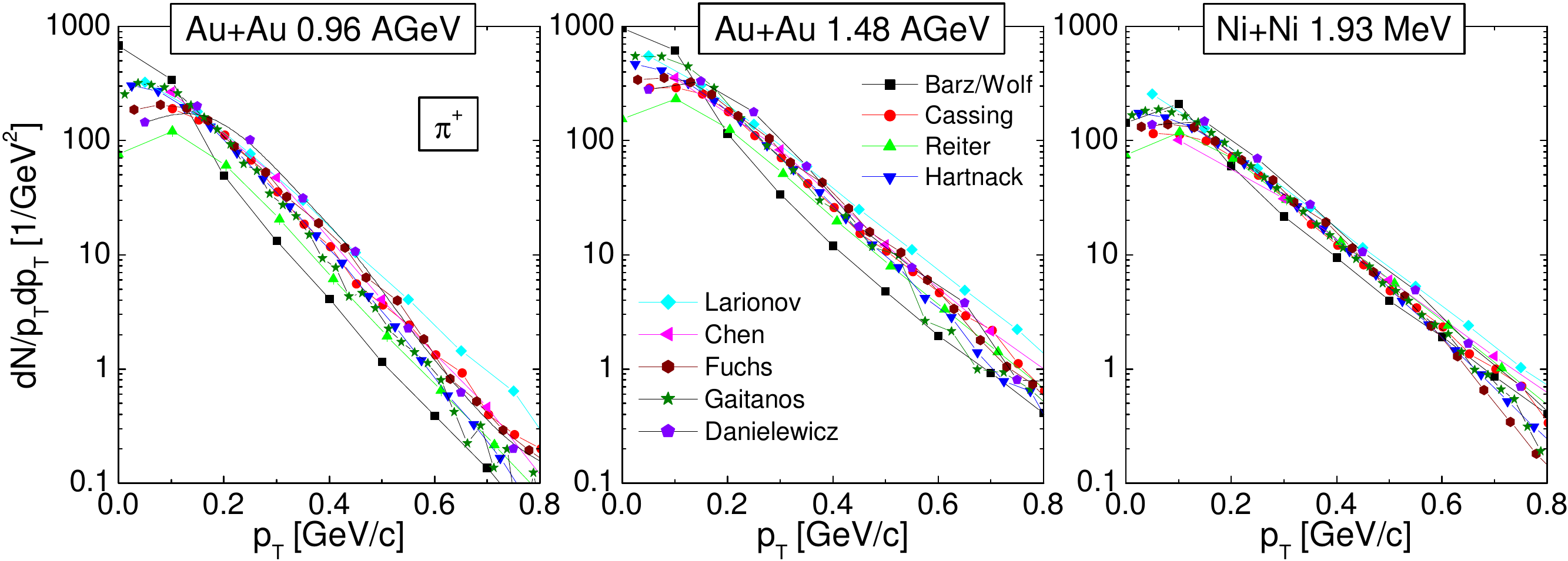}
\caption{(Color online) Rapidity distribution (upper) and transverse momentum distribution at midrapidities (lower) of charged pions at the final stage of Au+Au collisions at 0.96 (left) and 1.48 (middle) AGeV and Ni+Ni collisions at 1.93 AGeV (right) at an impact parameter of $\text{b} = 1$ fm from various transport codes with the same $\Delta$ mass width of 120 MeV. Taken from Ref.~\cite{Kol05}.}
\label{pion_Kol05}
\end{figure}

As shown in the previous sections, multiplicities or energy spectra of charged pions are probes of the nuclear symmetry energy. The direct production of pions is generally less important, and pions are mostly produced through $\Delta$ resonances. One of the discrepancies of pion yields in transport simulations comes from the uncertainties of different $\Delta$ decay widths as well as their energy dependence employed in different transport codes. In order to reduce the discrepancy, all the codes use a constant $\Delta$ mass width of 120 MeV, and the resulting rapidity and transverse momentum distributions are displayed in Fig.~\ref{pion_Kol05}. The UrQMD code used by Reiter is for even higher energies without potentials, and special treatments for particle productions near threshold are employed, so it gives somehow different results from others in Au+Au collisions at 0.96 and 1.48 AGeV. The experimental $\pi^-/\pi^+$ yield ratio of about 1.95~\cite{Sen99} around the collision energy of 1 AGeV is reproduced by most codes, consistent with the thermal model prediction and supporting the argument that pions are mostly produced through $\Delta$ resonances. The $\pi^+$ transverse momentum distributions agree reasonably well among different codes, except for that at small transverse momenta, which is sensitive to the behavior of $\Delta$ resonances of very small masses. It is also seen that most codes overestimate the total pion yield in Au+Au collisions around 1 AGeV, which is about 30 obtained from available experimental data~\cite{Sen99}.

\begin{figure}[h]
\centering
\includegraphics[scale=0.35,clip]{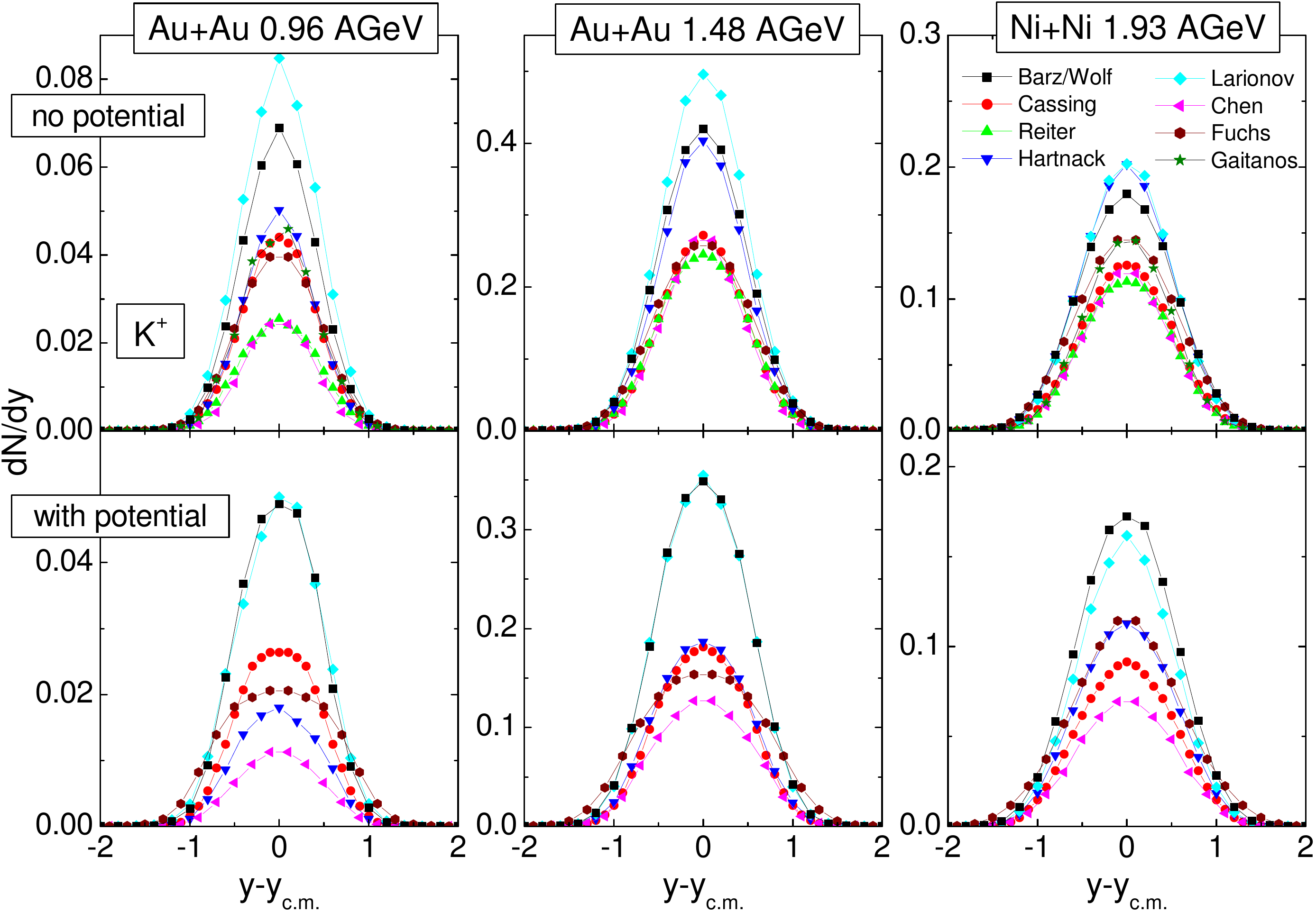}
\includegraphics[scale=0.36,clip]{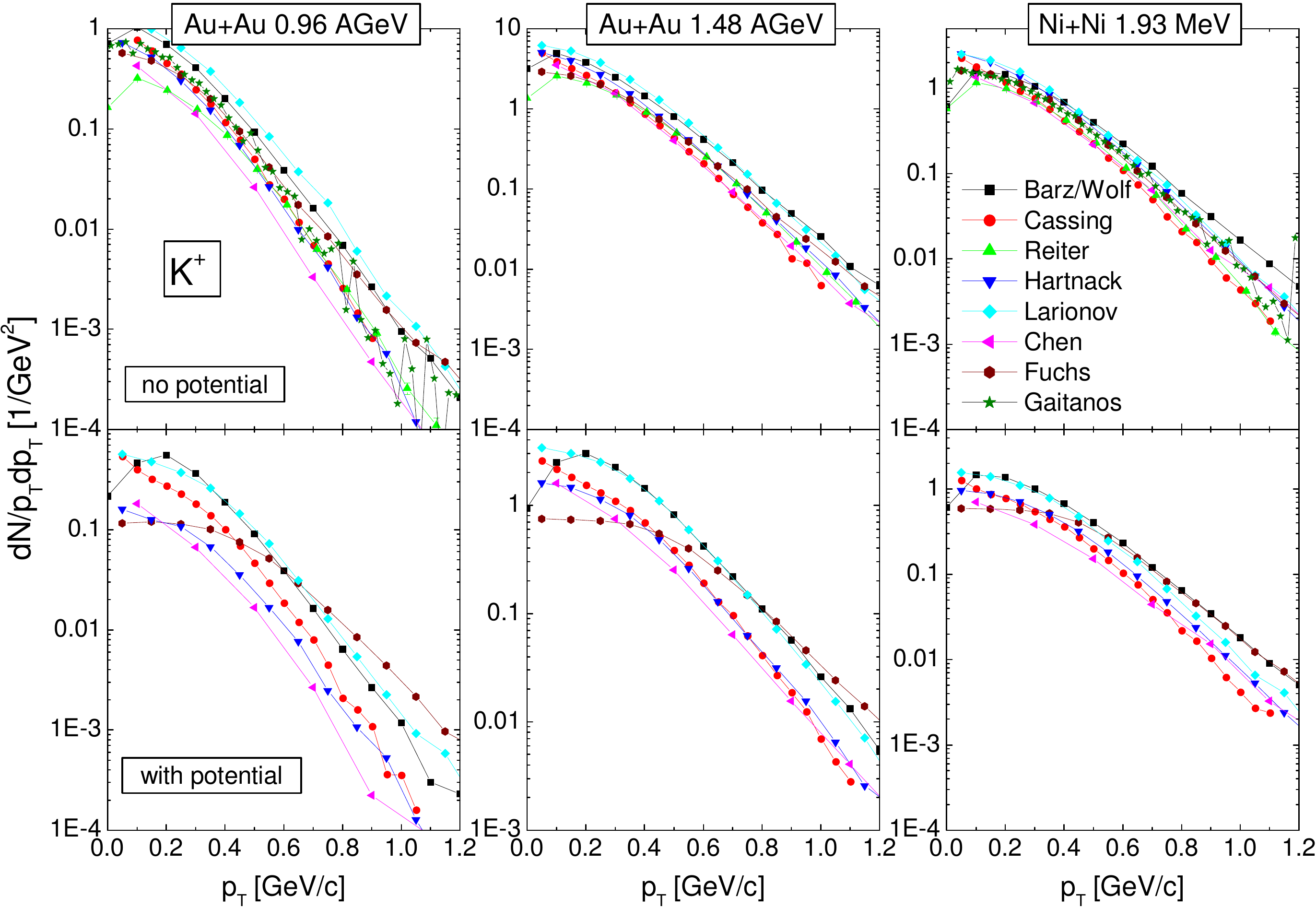}
\caption{(Color online) Rapidity distribution (upper) and transverse momentum distribution at midrapidities (lower) of $K^+$ at the final stage of Au+Au collisions at 0.96 (left) and 1.48 (middle) AGeV and Ni+Ni collisions at 1.93 AGeV (right) at an impact parameter of $\text{b} = 1$ fm from various transport codes with and without the kaon-nucleon potential. Taken from Ref.~\cite{Kol05}.}
\label{kplus_Kol05}
\end{figure}

As a useful probe of the nuclear matter EOS, the $K^+$ production in transport simulations should be compared with care. Using a constant $\Delta$ mass width of 120 MeV, the same $K^+$ production cross sections from $N+\Delta$ and $\Delta+\Delta$ collisions taken from Ref.~\cite{Tsu99}, and the default production cross section from $N+N$ collisions in each transport code, the rapidity and transverse momentum distributions of $K^+$ are compared in Fig.~\ref{kplus_Kol05}. The larger discrepancies for $K^+$ results compared to pions are due to the reason that the collision energy here is around or lower than the threshold energy of $K^+$ production. Since the cross section near the $K^+$ production threshold energy increases exponentially with increasing center-of-mass energy, a small difference in the parameterization of the initial Fermi motion or the baryon potential may affect the $K^+$ production. On the other hand, kaon properties are modified in the nuclear medium, and the energy of kaons with momentum $k$ in a nuclear medium of baryon number density $\rho_B$ is usually parameterized as
\begin{equation}
\omega_K(\rho_B,k)=\sqrt{[m^*_K(\rho_B)]^2+k^2}
\end{equation}
with the kaon effective mass
\begin{equation}\label{km}
m^*_K(\rho_B) = m_K^0\left(1-\alpha \frac{\rho_B}{\rho_0}\right),
\end{equation}
where $m_K^0$ is the kaon mass in vacuum and $\rho_0$ is the saturation density. The coefficient $\alpha$ ranges from $-0.04$ to $-0.075$ for different transport codes involved in this comparison. The rapidity and transverse momentum distributions with the kaon in-medium potential are also compared in Fig.~\ref{kplus_Kol05} for some of the available transport codes. Depending on the value of $\alpha$, a decrease of the total $K^+$ yield of about a factor of 2 is observed with the kaon potential in heavy systems, while the reduction rate is lower in light systems. A covariant form of the kaon potential is used in TuQMD by Fuchs, where both the space and time components are incorporated, making the transverse momentum spectrum of $K^+$ stiffer than others. The comparison shows that incorporating the $K^+$ potential generally adds to the discrepancies of the $K^+$ yield.

\begin{figure}[h]
\centering
\includegraphics[scale=0.35,clip]{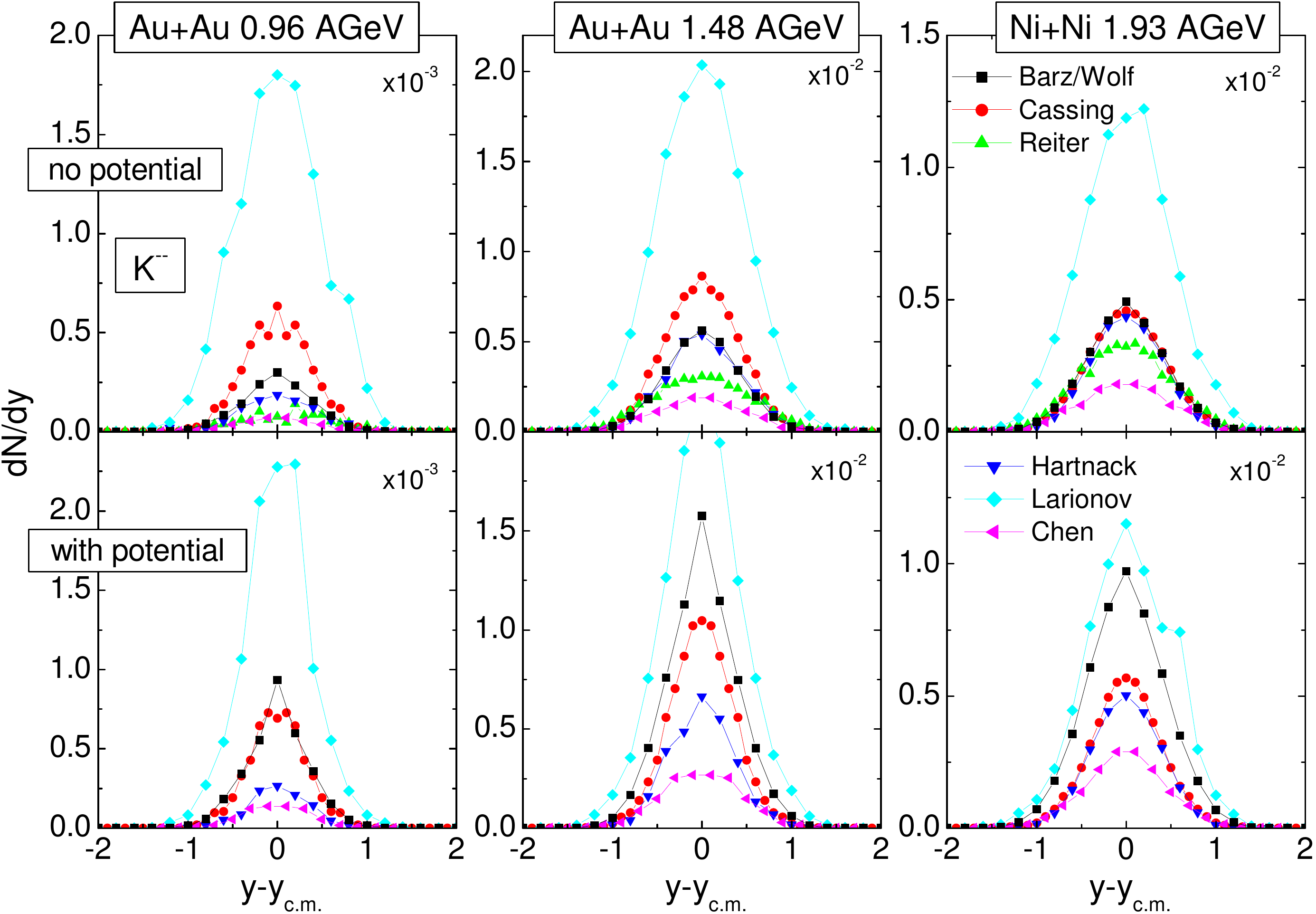}
\includegraphics[scale=0.36,clip]{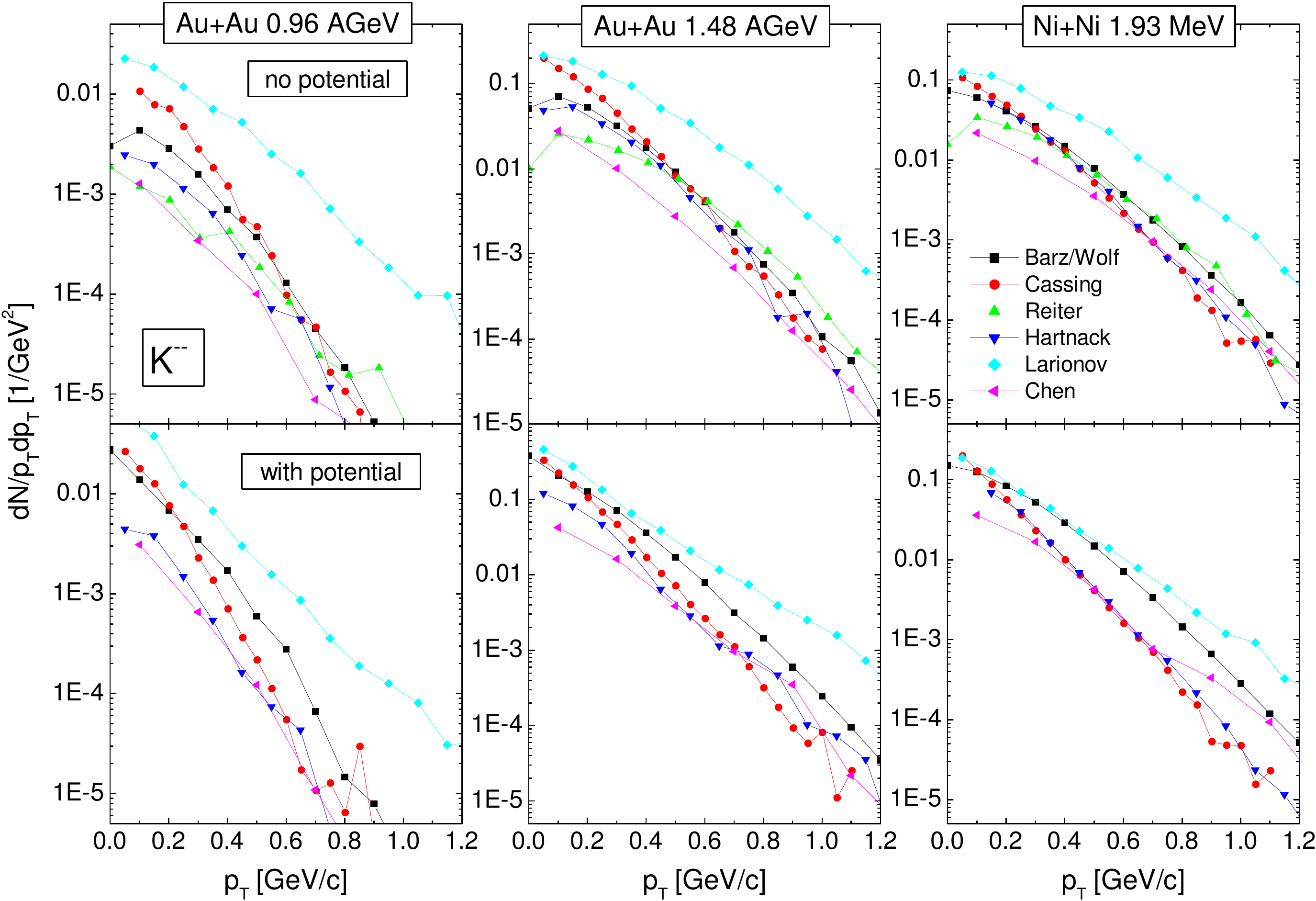}
\caption{(Color online) Same as Fig.~\ref{kplus_Kol05} but for $K^-$. Taken from Ref.~\cite{Kol05}.}
\label{kminus_Kol05}
\end{figure}

The production process of $K^-$ containing an $s$ quark is more complex than $K^+$, since the production of a $K^-$ is generally accompanied with the production of a $K^+$, i.e., $N+N \rightarrow K^- + K^+ + N + N$, while the production of a $K^+$ is generally accompanied with changing a nucleon to a hyperon with a smaller mass difference, i.e., $N+N \rightarrow N+Y+K^+$. For the collision energies much below the threshold production energy of $K^-$ in this comparison, $K^-$ are mostly produced through the channel $\Lambda(\Sigma)+\pi \rightarrow K^- + N$. Since the $\Lambda(\Sigma)$ production is accompanied with the $K^+$ production, the $K^-$ production is coupled with the $K^+$ production. Due to the lower collision energies compared with the threshold energy of the $K^-$ production, $K^-$ are mostly produced at the early stage of the collision from a dense phase, and large discrepancies in their rapidity and transverse momentum distributions are observed in Fig.~\ref{kminus_Kol05}. The in-medium $K^-$ mass can also be parameterized according to Eq.~(\ref{km}), where the coefficient $\alpha$ ranges from 0.10 to 0.22 for various codes involved in the comparison. Since the $K^-$ potential becomes more attractive with increasing density, the $K^-$ yield is much enhanced when its potential is turned on.

\subsubsection{Collisions, stopping, and flow at 100 and 400 AMeV}

There are totally 9 BUU-type and 9 QMD-type transport codes participating in the comparison of Au+Au collisions at 100 and 400 AMeV. The 9 BUU-type codes are the Boltzmann-Langevin one body (BLOB) code, the Giessen Boltzmann-Uehling-Uhlenbeck code with the relativistic mean-field potential (GIBUU-RMF) and the Skyrme-Hatree-Fock potential (GIBUU-Skyrme), the isospin-dependent Boltzmann-Langevin (IBL) code, the isospin-dependent Boltzmann-Uehling-Uhlenbeck (IBUU) code, the Pawel's Boltzmann-Uehling-Uhlenbeck (pBUU) code, the relativistic Boltzmann-Uehling-Uhlenbeck (RBUU) code, the relativistic Vlasov-Uehling-Uhlenbeck (RVUU) code, and the stochastic mean-field (SMF) code. The 9 QMD-type codes are the antisymmetrized molecular dynamics (AMD) code, the constrained molecular dynamics (CoMD) code, the improved quantum molecular dynamics code at China Institute of Atomic Energy (ImQMD-CIAE), the original isospin-dependent quantum molecular dynamics (IQMD) code, the isospin-dependent quantum molecular dynamics code at Beijing Normal University (IQMD-BNU), the isospin-dependent quantum molecular dynamics code at Institute of Modern Physics (IQMD-IMP, also known as the Lanzhou QMD (LQMD) in the literature), the isospin-dependent quantum molecular dynamics code at Shanghai Institute of Applied Physics (IQMD-SINAP), the T\"ubingen quantum molecular dynamics (TuQMD) code, and the ultra-relativistic quantum molecular dynamics (UrQMD) code. The code correspondents as well as the representative references are listed in Table~\ref{code_Xu16}. The improved quantum molecular dynamics code at Guang-Xi Normal University (ImQMD-GXNU) participated in the comparison, but the nucleon-nucleon collision part of this code is completely different from other codes, so the results are not shown in the final published paper~\cite{Xu16}.

\begin{table}[h]\small
  \centering
  \caption{The names, authors and correspondents, and representative references of 9 BUU-type and 9 QMD-type codes participating in the comparison of Au+Au collisions at 100 and 400 AMeV~\cite{Xu16}.
%  The ImQMD-GXNU model listed here is more suitable for fusion energy and the results are not compared.
}
    \begin{tabular}{|c|c|c|c|c|c|c|c|c}
    \hline
    BUU-type & code correspondents & references  &&  QMD-type & code correspondents &  references\\
    \hline
     BLOB    & P.Napolitani,M.Colonna &  \cite{BLOB1,BLOB2} &&AMD & A.Ono & \cite{AMD1,AMD2} \\
     GIBUU-RMF & J.Weil & \cite{GIBUU1,GIBUU2,GIBUU3}  && IQMD-BNU  & J.Su,F.S.Zhang &  \cite{IQMD-BNU1,IQMD-BNU2,IQMD-BNU3} \\
     GIBUU-Skyrme & J.Weil & \cite{GIBUU1,GIBUU2,GIBUU3}   && IQMD & C.Hartnack,J.Aichelin  & \cite{IQMD1,Aic91,Har98}  \\
     IBL    &  W.J.Xie,F.S.Zhang & \cite{IBL1,IBL2,IBL3} && CoMD  & M.Papa & \cite{CoMD1,CoMD2,CoMD3}\\
     IBUU   &  J.Xu,L.W.Chen,B.A.Li & \cite{Li97,IBUU2,Li08,IBUU3} && ImQMD-CIAE & Y.X.Zhang,Z.X.Li & \cite{ImQMD-CIAE1,ImQMD-CIAE2,Zhang08plb} \\
     pBUU   &  P.Danielewicz & \cite{pBUU1,Dan91} && IQMD-IMP & Z.Q.Feng & \cite{IQMD-IMP1,IQMD-IMP2}\\
     RBUU   & K. Kim,Y.Kim,T.Gaitanos& \cite{RBUU1,RBUU2,Fer06} && IQMD-SINAP & G.Q.Zhang & \cite{IQMD-SINAP1,IQMD-SINAP2} \\
     RVUU   & T.Song,G.Q.Li,C.M.Ko & \cite{RVUU1,RVUU2,Son15} && TuQMD & D.Cozma & \cite{TuQMD1,TuQMD2,TuQMD3,Dan13}\\
     SMF    & M.Colonna,P.Napolitani & \cite{SMF1,SMF2,SMF3} && UrQMD & Y.J.Wang,Q.F.Li & \cite{UrQMD1,UrQMD2,UrQMD3}\\
    \hline
    \end{tabular}
  \label{code_Xu16}
\end{table}

The comparison was intended to be carried out in a better controlled condition, which is detailed as the homework description assigned for the code correspondents as follows. The reaction system is Au+Au collisions at the impact parameter $\text{b}=7$ fm and 20 fm, with the former for real heavy-ion collisions, and the latter being a single-nucleus evolution useful for evaluating the stability. In the comparison there are different modes, and only final results for the B-mode and the D-mode are shown, corresponding to those at the beam energies of 100 AMeV and 400 AMeV, respectively. In the present review I will only talk about the B-Full mode and the D-Full mode with both mean-field potentials and nucleon-nucleon collisions, while detailed results with only mean-field potentials (B-Vlasov) or only nucleon-nucleon collisions (B-cascade) can be found in Ref.~\cite{Xu16}. In the simulations, 100 test particles for each run are used for BUU-type codes and totally 10 runs are needed, while 1000 independent events are needed for QMD-type codes. The density distribution of initial nuclei is required to be the same spherical Woods-Saxon distribution for each code, with the central density being the saturation density 0.16 fm$^{-3}$, and the momentum of each nucleon sampled within the Fermi momentum sphere depending on the local density. In real calculations, BUU-type codes mostly give the designed distribution while QMD-type codes generally give different distributions. The difficulties for the later in giving the same distribution are expected, since only a small percent of the initial configurations are selected as stable ones and useful for simulations for QMD-type codes,  while for BUU-type codes with test particles the stability is generally more robust for different initial configurations. Each code is required to use the same constant isotropic nucleon-nucleon elastic collision cross section of 40 mb, and all the inelastic channels are turned off. The non-relativistic codes are required to use a Skyrme-like momentum-independent mean-field potential with the binding energy $-16$ MeV, the incompressibility 240 MeV, and the symmetry energy about $30$ MeV at saturation density 0.16 fm$^{-3}$. The relativistic codes are required to use the mean-field potential from the relativistic mean-field functional with $\sigma\omega\rho$ couplings that gives the same saturation properties. The surface term, sometimes represented by a density gradient or a Yukawa potential, is required to be turned off. Each code is run until $t=140$ fm/c at 100 AMeV and 100 fm/c at 400 AMeV, when the heavy-ion collision is generally finished. The time step can be 0.5 or 1 fm/c decided by the code practitioners.

\begin{figure}[h]
\centering
\includegraphics[scale=0.55,clip]{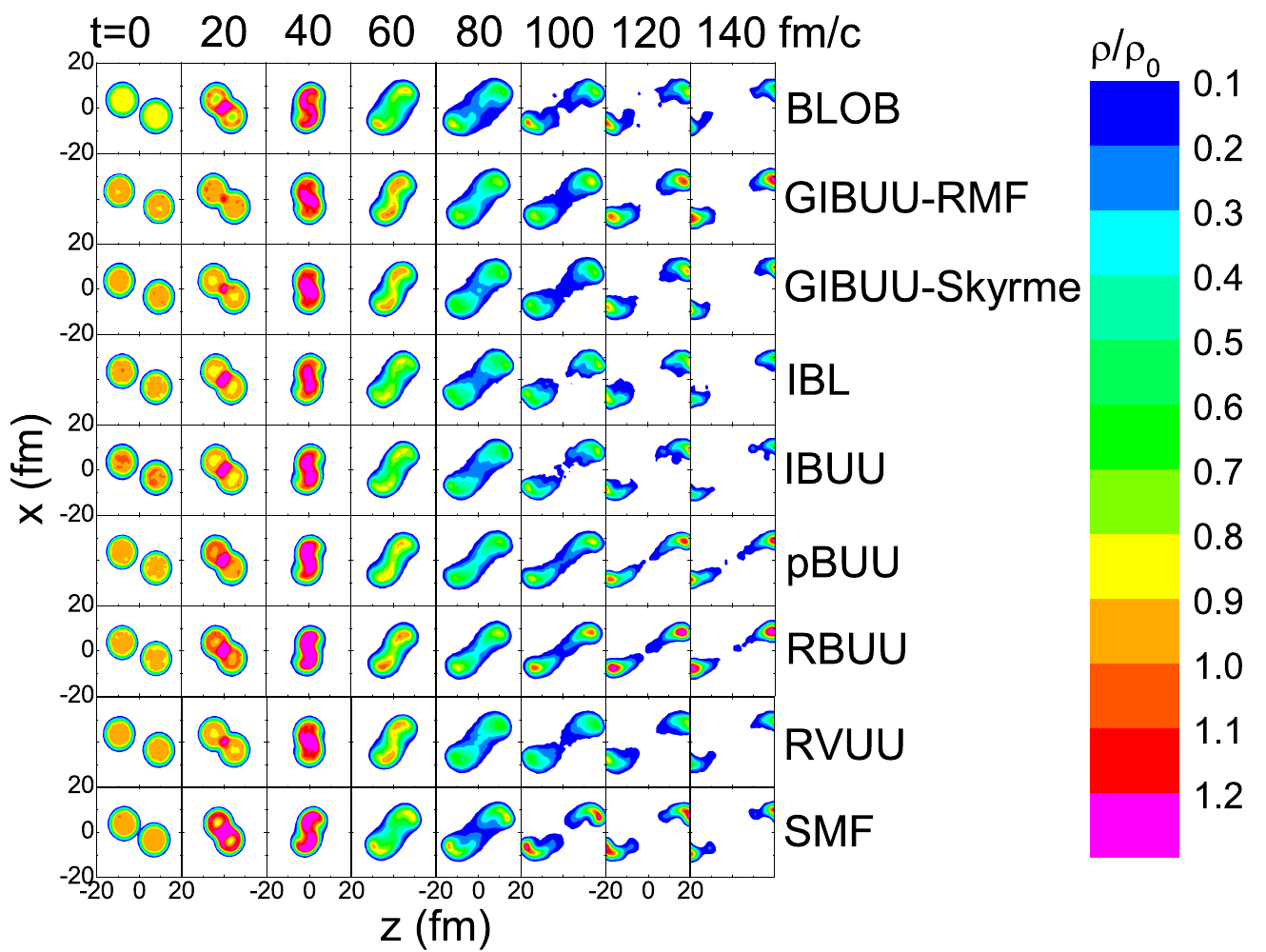}
\includegraphics[scale=0.55,clip]{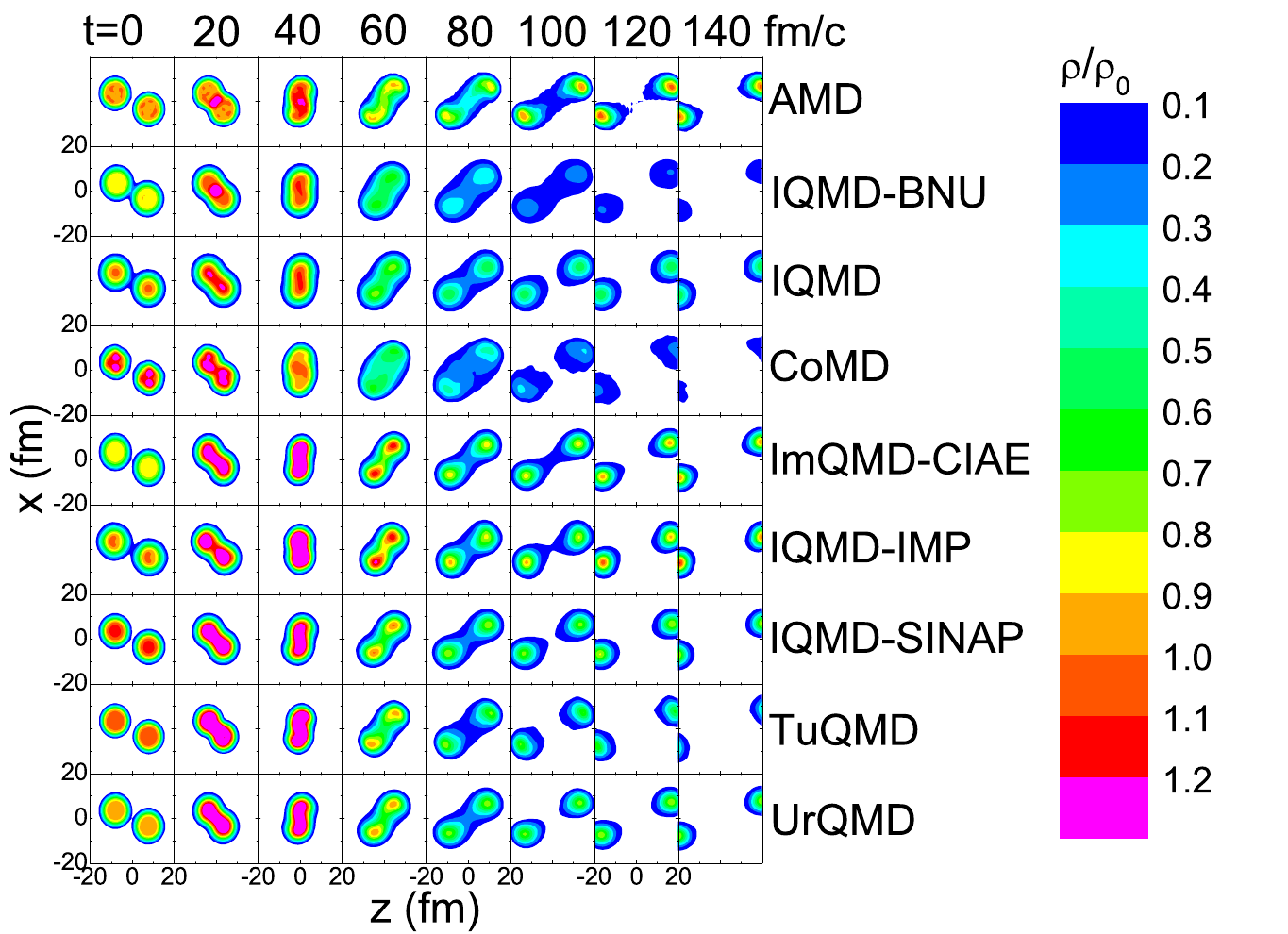}
\caption{(Color online) Density contours in the x-o-z plane every 20 fm/c in Au+Au collisions at the beam energy 100 AMeV and impact parameter $\text{b}=7$ fm from 9 BUU transport codes (left) and 9 QMD transport codes (right). Taken from Ref.~\cite{Xu16}.}
\label{den_Xu16}
\end{figure}

The density contours in the x-o-z plane at every 20 fm/c are displayed in Fig.~\ref{den_Xu16} for 9 BUU codes and 9 QMD codes, after averaging over totally 1000 events. These contours give the impression how the heavy-ion systems of Au+Au collisions evolve with time. The general features are observed for all codes. The time period between $t=20$ and 40 fm/c is important for the stopping and the particle production. The maximum compression happens around 40 fm/c for nearly all codes. The sideward flow is developed between $t=60$ and 80 fm/c. The neck breaks up around 100 fm/c depending on the code. After 100 fm/c, the production and evolution of projectile- and target-like residues are observed, while statistical models are generally needed for the afterburner dynamics after 140 fm/c if one wants to compare the transport simulation results with experimental data, especially for fragments. It is also observed that higher densities are reached in heavy-ion simulations if the initial densities are higher, especially for QMD codes. The density evolution is strongly correlated with the mean-field evolution and attempted nucleon-nucleon collisions to be discussed in the following.

\begin{figure}[h]
\centering
\includegraphics[scale=0.7,clip]{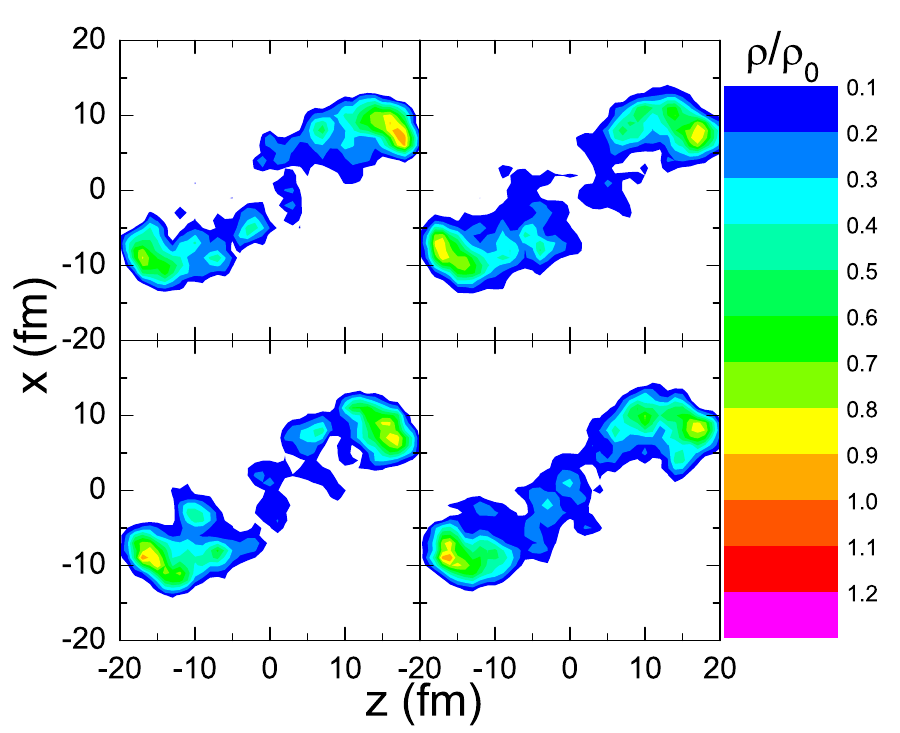}
\includegraphics[scale=0.7,clip]{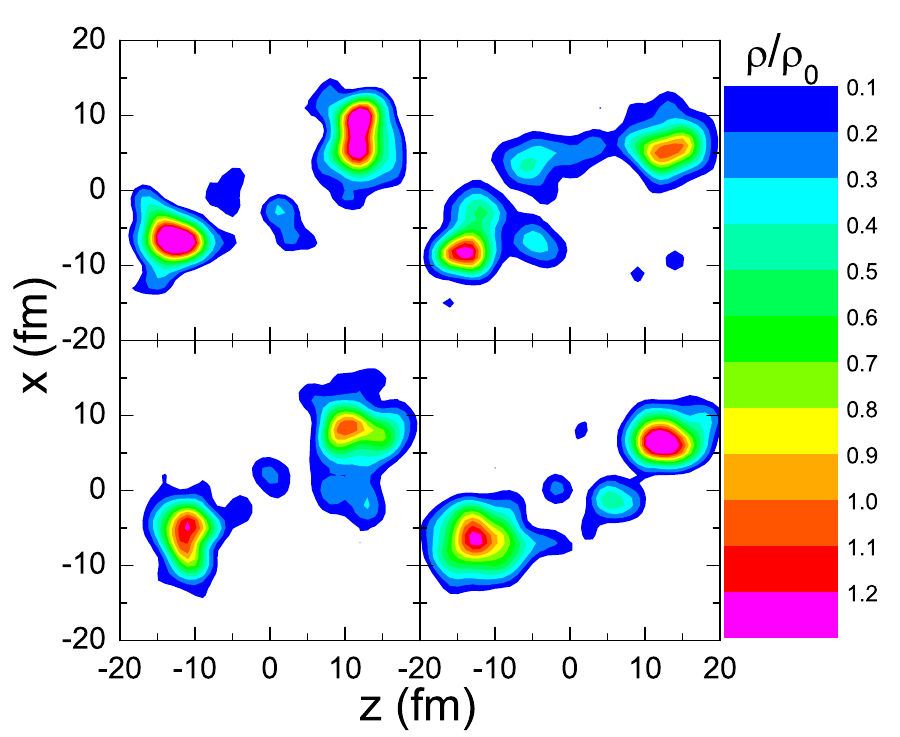}
\caption{(Color online) Four examples of the density contours in the x-o-z plane at $t=100$ fm/c in Au+Au collisions at the beam energy 100 AMeV and impact parameter $\text{b}=7$ fm from the IBUU transport model with 100 test particles (left) and the ImQMD-CIAE transport model for individual events (right). Taken from Ref.~\cite{Xu16}.}
\label{den_1event_Xu16}
\end{figure}

The density distributions for QMD codes look smoother than those for BUU codes in Fig.~\ref{den_Xu16}, while one should keep in mind that these are averaged over 1000 events. For BUU codes with 100 test particles, each event in the 100 parallel collisions shares the same density evolution. For QMD codes, however, the density evolution for each event is independent of the other events. Four examples of the density contours at a fixed time step of $t=100$ fm/c are displayed in Fig.~\ref{den_1event_Xu16}, from the IBUU transport model with 100 test particles as an example of the BUU codes in the left panel, and the ImQMD-CIAE transport model as an representative of the QMD codes in the right panel. One sees that the density evolutions for four different runs are similar for BUU codes, and this reflects the deterministic nature that BUU codes solve the initial-state problem if infinite numbers of test particles are used, unless one introduces the stochastic mechanism in the collision integral, such as in BLOB or SMF. The density evolutions in each run for QMD codes are quite different. One sees various of fragments from ImQMD calculations due to the fluctuation in QMD codes, and the fluctuation depends on the width of the Gaussian wave packet. The breakups of the neck are also different in BUU and QMD codes. In BUU codes the neck is flatter and sketched out farther, the breakup of which may lead to some fine structures. In QMD codes the neck breaks up faster, and the fragments are usually of spherical shape. These difference may not affect the one-body observables in this comparison, but can be important for cluster dynamics.

\begin{figure}[h]
\centering
\includegraphics[scale=0.3,clip]{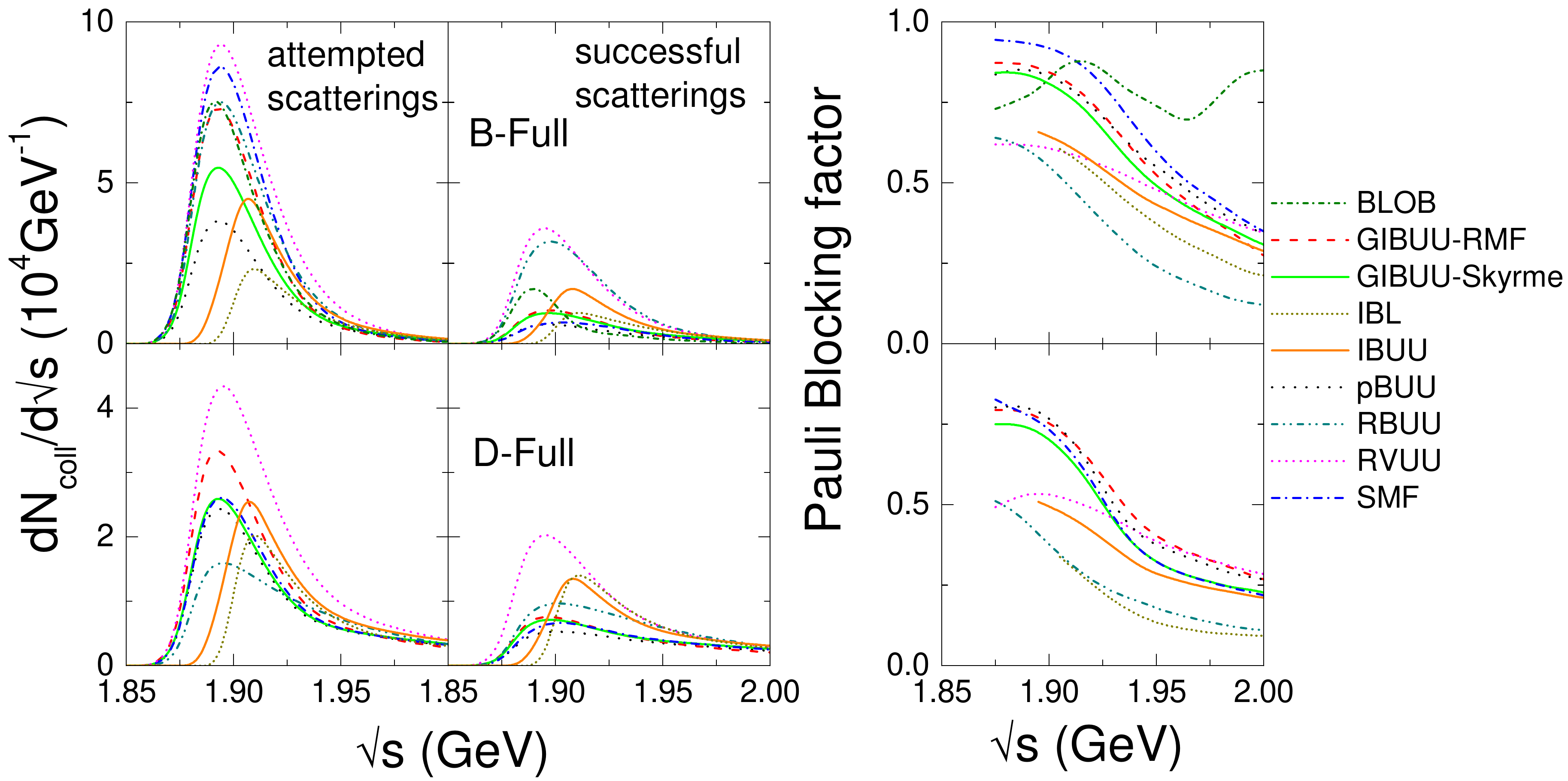}\\
\includegraphics[scale=0.3,clip]{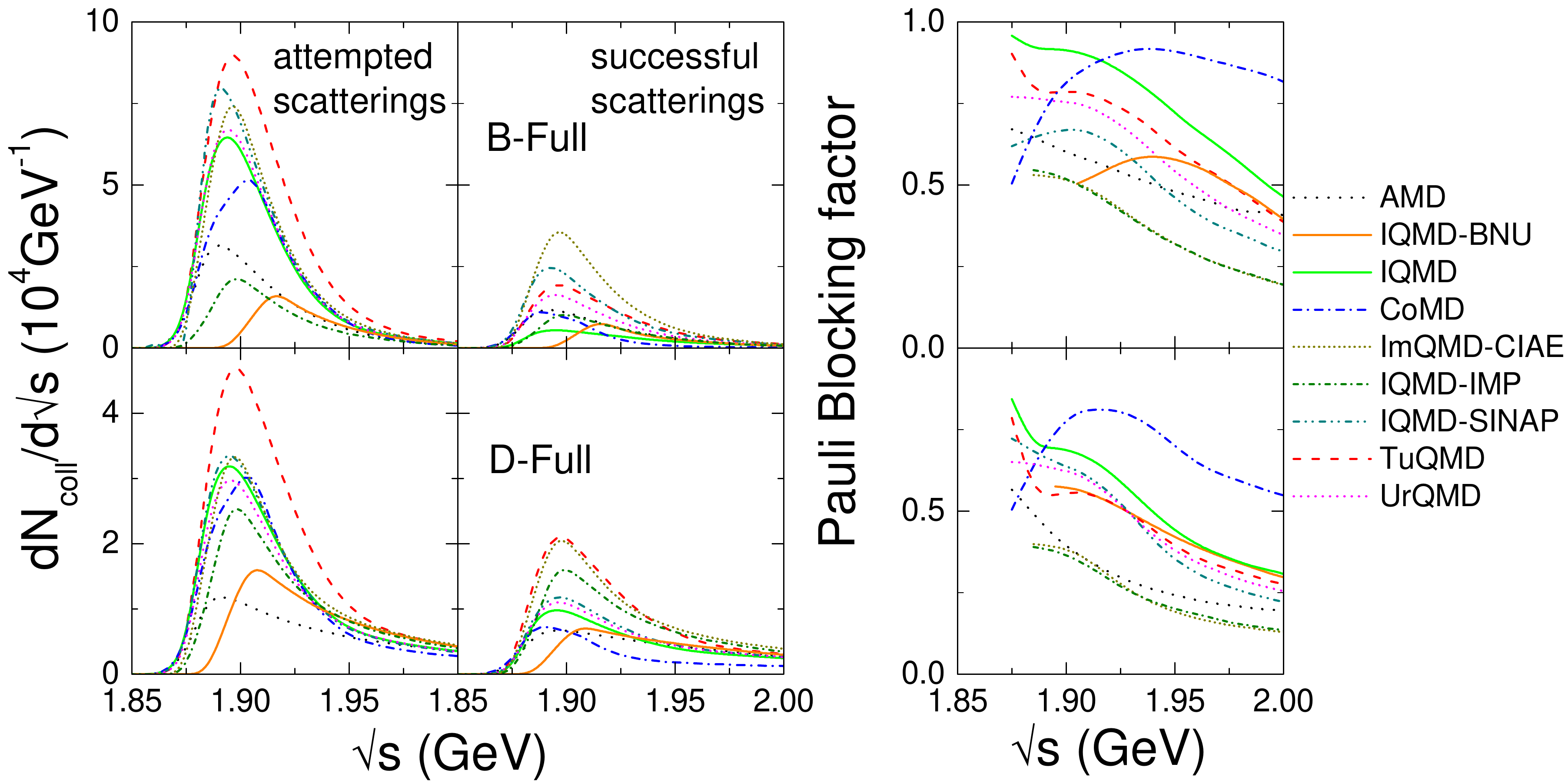}
\caption{(Color online) Center-of-mass energy distribution of the attempted and successful collisions as well as the Pauli blocking factor in Au+Au collisions at the impact parameter $\text{b}=7$ fm and beam energies 100 AMeV (B-Full) and 400 AMeV (D-Full) from 9 BUU transport codes (upper) and 9 QMD transport codes (lower). Modified from figures in Ref.~\cite{Xu16}.}
\label{colls_Xu16}
\end{figure}

Considering the density evolution in Fig.~\ref{den_Xu16}, more attempted nucleon-nucleon collisions happen in the higher-density phase, while the successful collisions, i.e., the attempted collisions that survive the Pauli blocking condition, determine the dynamics of heavy-ion reactions together with the mean-field potential. Most codes in this comparison employ Bertsch's prescription for attempted nucleon-nucleon collisions, as described in the Appendix B of Ref.~\cite{Ber88}, except for AMD, CoMD, BLOB, pBUU, and SMF. The Pauli blocking probability for the collision between particle $i$ and particle $j$ is mostly evaluated according to $1-(1-f_i)(1-f_j)$, and the calculation of the local occupation probability $f_{i/j}$ in the final state of the attempted collision depends on the local phase-space cell selection. Exceptions are AMD and CoMD, where the physical wave packet and the special phase-space constraint are used, respectively. For detailed treatments of nucleon-nucleon collisions and Pauli blockings for various codes in this comparison, I refer the reader to Tables II and III in Ref.~\cite{Xu16}. The center-of-mass energy distribution of both attempted and successful nucleon-nucleon collisions as well as the Pauli blocking factor, i.e., the ratio of unsuccessful collision to the attempted collision numbers, are displayed in Fig.~\ref{colls_Xu16} at both 100 AMeV and 400 AMeV, with results from 9 BUU codes shown in the upper panel, and those from 9 QMD codes shown in the lower panel. Here the bare mass of 0.938 GeV and the momentum of colliding nucleons are used to calculate the center-of-mass energy $\sqrt{s}$ for each nucleon-nucleon collision, and the distribution near the threshold of 1.876 GeV is ought to be a sharp cut-off but smeared out due to the plotting procedure. Note that $\sqrt{s}$ of a free nucleon-nucleon collision is 1.925 and 2.066 GeV for an incident beam at 100 and 400 AMeV, respectively, and 1.894 GeV for nucleons at the Fermi momentum. It is observed from Fig.~\ref{colls_Xu16} that the peak of the distribution is around 1.9 GeV, i.e., most collisions are soft ones between nucleons of small momenta. The distribution has a long tail at larger $\sqrt{s}$ especially at 400 AMeV. The total collision number is generally smaller at 400 AMeV, due to the fast disintegration of the collision system. The shape of the successful collision distribution is similar to that of the attempted collision distribution, while it is not a constant shift, since the Pauli blocking factor is generally larger at smaller $\sqrt{s}$ but smaller at larger $\sqrt{s}$, as a result of larger free phase-space volumes for energetic collisions. For IBL, IBUU, and IQMD-BNU, there are artificial thresholds for nucleon-nucleon collisions, in order to block too soft collisions which are generally spurious. One sees discrepancies for attempted collision distributions from various codes, due to the different density evolutions and treatments of nucleon-nucleon collisions, while even larger discrepancies can be observed for successful collision distributions after Pauli blocking. The Pauli blocking factor converges better at larger $\sqrt{s}$, and CoMD has a different energy dependence of the blocking factor due to its special treatment. I will discuss more about the nucleon-nucleon collisions and the Pauli blocking in the cascade comparison in a box system.

\begin{figure}[h]
\centering
\includegraphics[scale=0.3,clip]{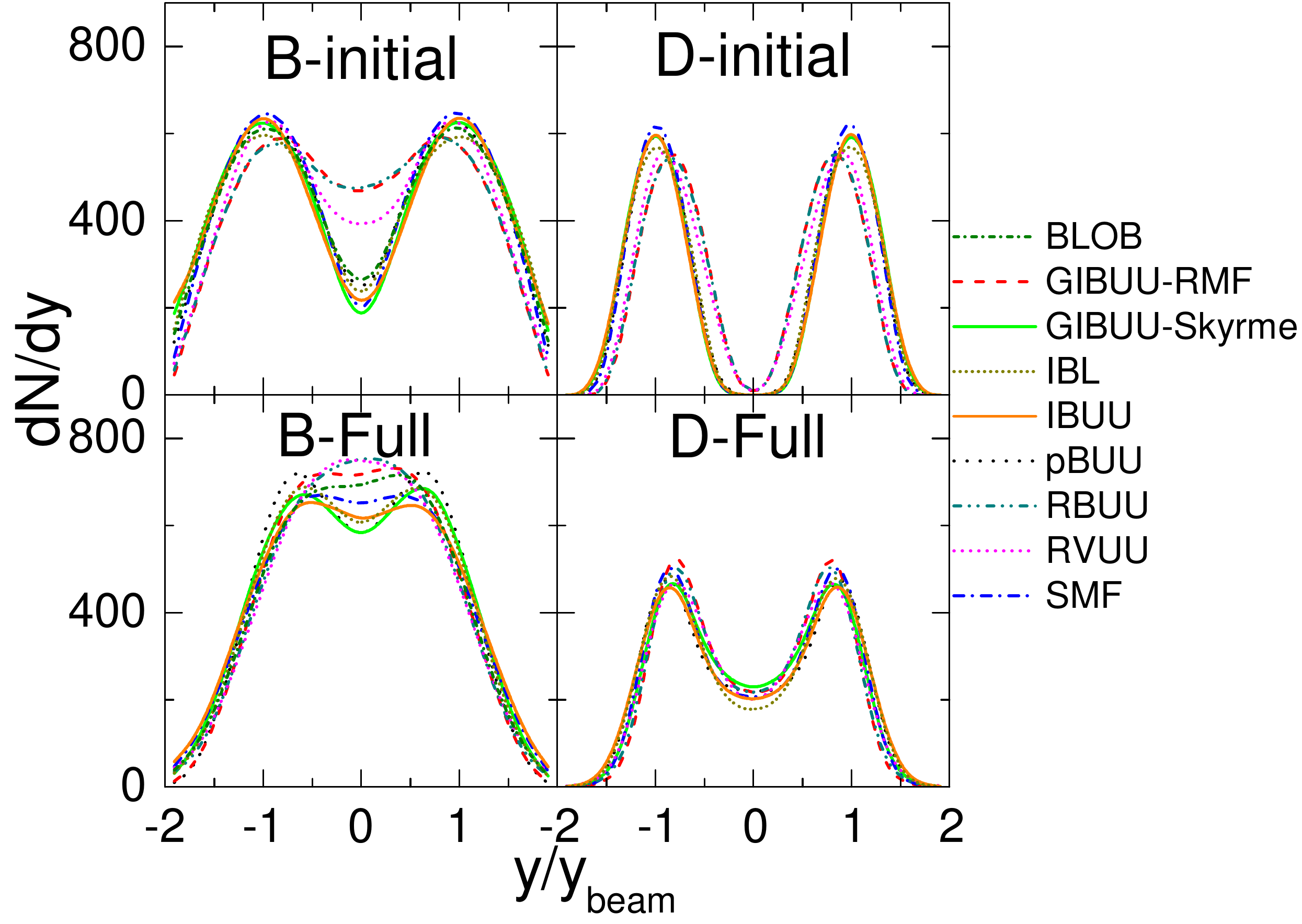}
\includegraphics[scale=0.3,clip]{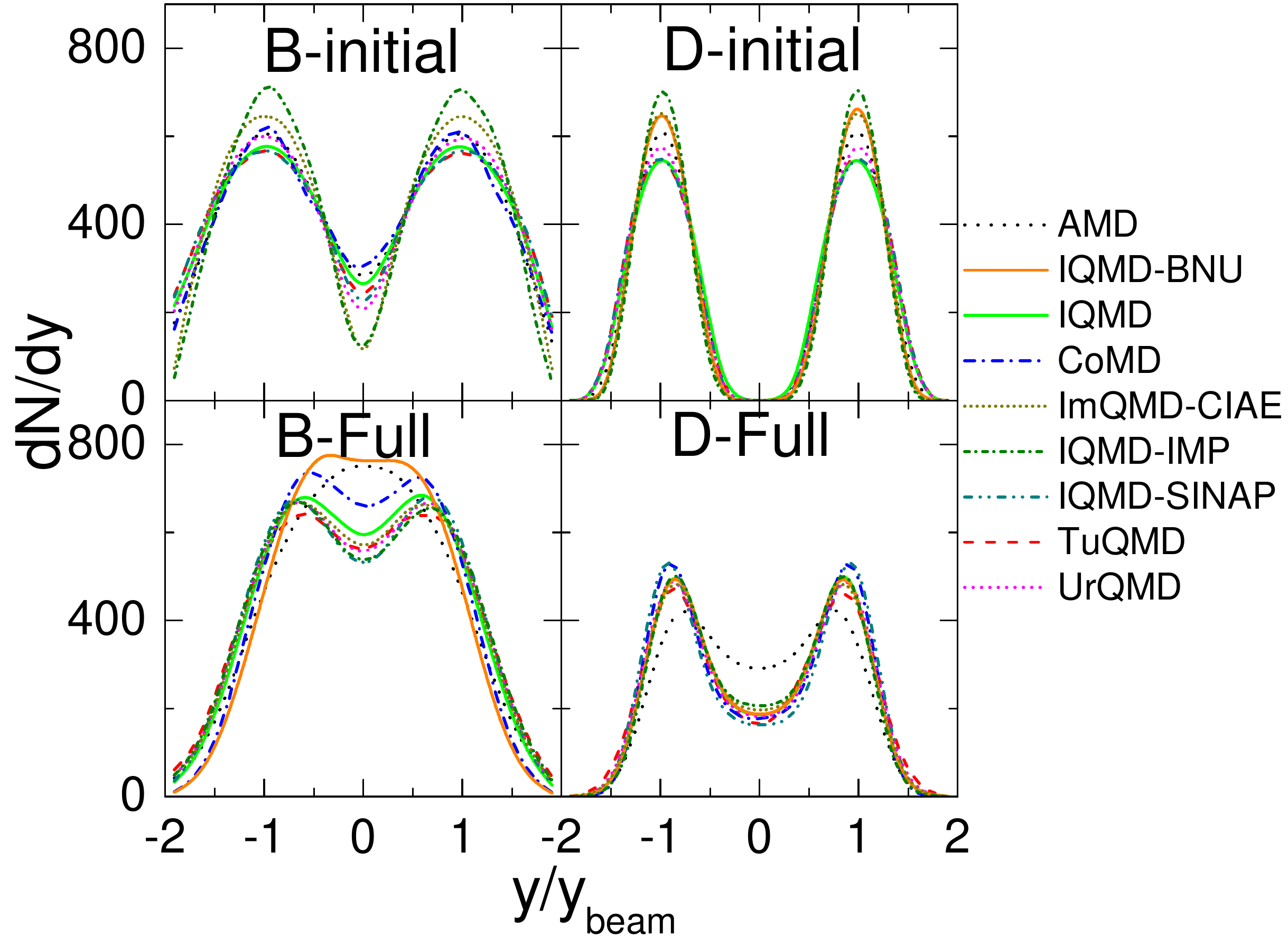}
\caption{(Color online) Initial and final rapidity distributions in Au+Au collisions at the impact parameter $\text{b}=7$ fm and beam energies 100 AMeV (B-Full) and 400 AMeV (D-Full) from 9 BUU transport codes (left) and 9 QMD transport codes (right). Modified from figures in Ref.~\cite{Xu16}.}
\label{rap_Xu16}
\end{figure}

The rapidity distribution characterizes the distribution of particles along the beam axis in a Lorentz invariant manner, and is a measure of the nucleon stopping in heavy-ion collisions. Figure~\ref{rap_Xu16} compares the initial and final rapidity distributions at 100 and 400 AMeV, from 9 BUU codes on the left, and 9 QMD codes on the right. The initial rapidity distribution shows a double-humped structure, with two peaks at $\pm$beam rapidities corresponding to the projectile and target nuclei, with the width corresponding to the Fermi motion of initial nucleons. The peaks are sharper at 400 AMeV than those at 100 AMeV as expected, since the initial Fermi motion becomes more and more negligible at higher beam energies compared with the longitudinal motion. Except for relativistic BUU codes (GIBUU-RMF, RBUU, and RVUU), all BUU codes give similar initial rapidity distributions, consistent with their similar initial density distributions. Due to the different initial density distributions for QMD codes, their initial rapidity distributions are also different. During the heavy-ion collision process, the Coulomb force, the mean-field potential, and the nucleon-nucleon collisions convert the longitudinal motion into random motion. A complete random motion means that the system is fully stopped, otherwise there is some transparency. With an isotropic nucleon-nucleon collision cross section used in this comparison, the stopping is in principle correlated with the successful collision number. A weaker stopping at 400 AMeV corresponds to a smaller successful collision number, as can be seen from Fig.~\ref{colls_Xu16}. For BUU codes, the stronger stopping is observed for relativistic codes at 100 AMeV, probably related to their initial rapidity distributions. For QMD codes, stronger stopping is observed at 100 AMeV for AMD, IQMD-BNU, and CoMD. The stronger stopping for IQMD-BNU seems to be due to the mean-field contribution according to the B-Vlasov results not shown here, while that for CoMD is likely due to the initial rapidity distribution. Although some discrepancies are observed at 100 AMeV, nice agreements are observed for the rapidity distributions at 400 AMeV, except for AMD due to too many nucleons participating in violate collisions as a result of the insufficient precision of the physical-coordinate representation as discussed in Refs.~\cite{AMD1,AMD2}. The nice convergence of the stopping at 400 AMeV is due to the dominating contribution of nucleon-nucleon collisions that are not suffered by the uncertainties of the Pauli blocking at higher beam energies, where the initialization also becomes less important.

\begin{figure}[h]
\centering
\includegraphics[scale=0.7,clip]{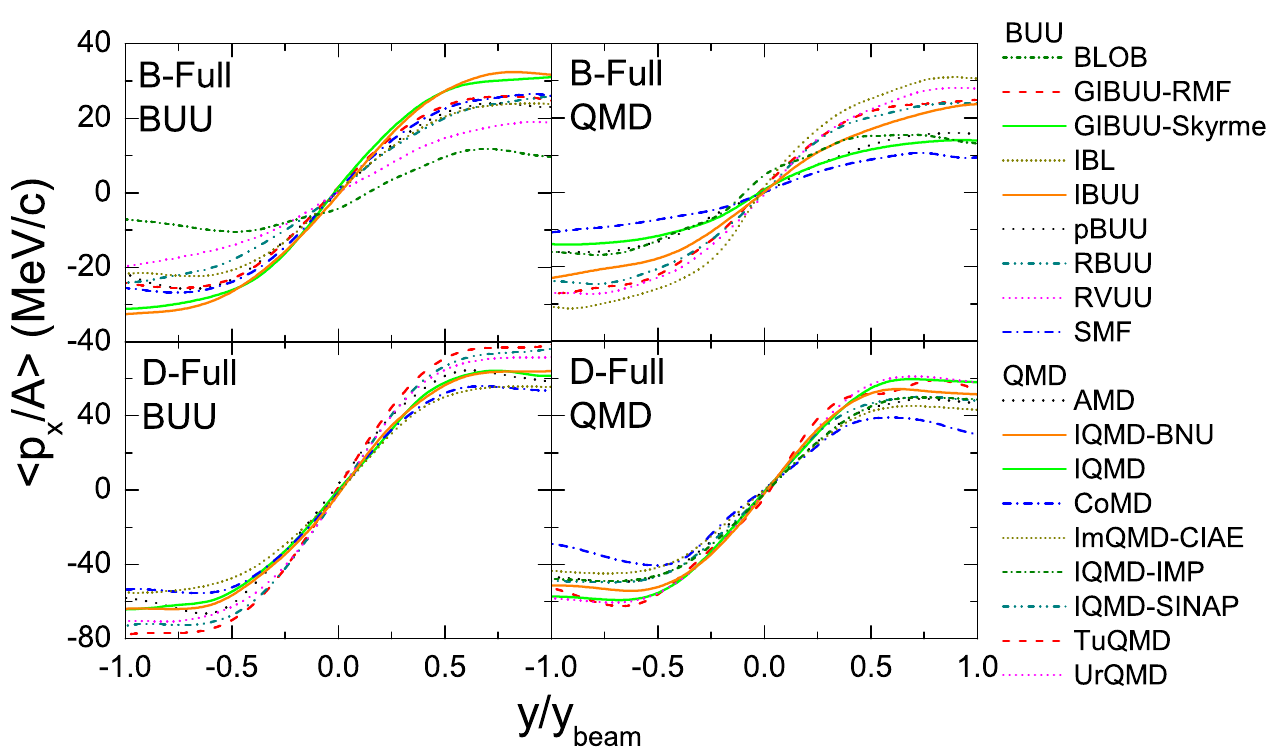}
\includegraphics[scale=0.26,clip]{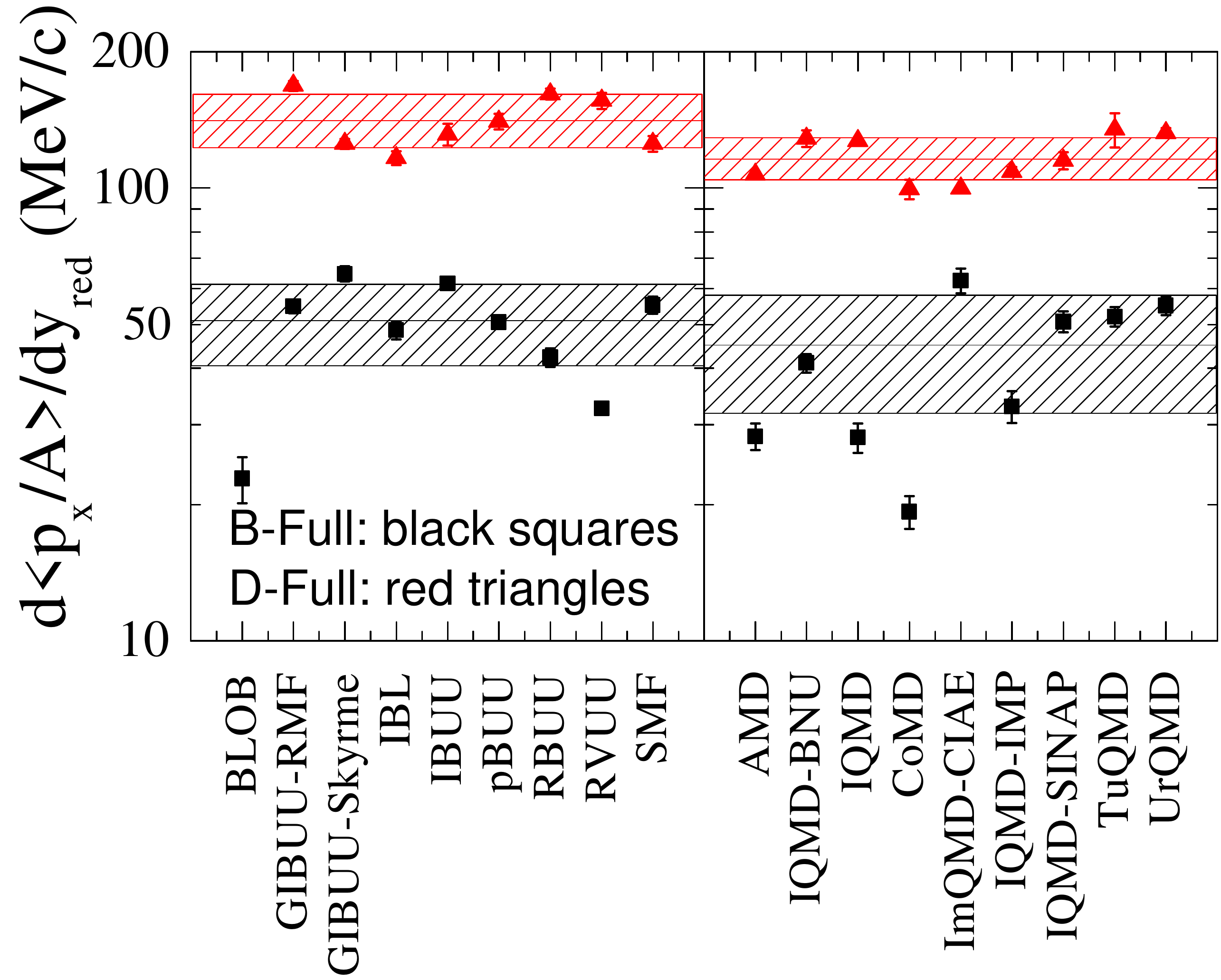}
\caption{(Color online) Transverse flow as a function of the reduced rapidity ($y_{red}=y/y_{beam}$) (left) in Au+Au collisions at the impact parameter $\text{b}=7$ fm and beam energies 100 AMeV (B-Full) and 400 AMeV (D-Full) from 9 BUU transport codes and 9 QMD transport codes as well as its slope at midrapidities (right). Modified from figures in Ref.~\cite{Xu16}.}
\label{v1_Xu16}
\end{figure}

As shown in previous sections, the collective flows, especially the directed flow and the elliptic flow, are mostly developed in the high-density phase, and are thus good probes of the EOS at higher densities. In transport simulations, the collective flows are determined by both mean-field potentials and nucleon-nucleon collisions, with the relative importance depending on the incident energy. Figure~\ref{v1_Xu16} compares the transverse flow, i.e., $\langle p_x/A \rangle$ (slightly different from the directed flow $\langle p_x/p_T \rangle$), as well as its slope at midrapidities from 9 BUU codes and 9 QMD codes. The positive slope of the transverse flow is seen at both 100 and 400 AMeV for all codes, as a result of the dominating repulsive isotropic nucleon-nucleon collisions over the attractive mean-field potential. The transverse flow passes the zero point at rapidity $y=0$, except for BLOB due to the low statistics at midrapidities. The smaller transverse flow for BLOB is due to the stochastic fluctuation in the collision integral, leading to the stronger fragmentation and reducing the flow. Obviously, the agreement among different codes is better at 400 AMeV than at 100 AMeV, since the uncertainties due to the initial configuration and the Pauli blocking are less important at higher energies. At the range of $|y/y_{beam}|<0.38$ the transverse flow shows good linear behavior with respect to the rapidity, and the slope from the linear fit at midrapidities is shown in the right panel of Fig.~\ref{v1_Xu16}. Here a $3\%$ systematic error is assumed, and the mean flow slope is $51\pm11$ MeV/c at 100 AMeV and $143\pm19$ MeV/c at 400 AMeV for BUU codes, and $45\pm13$ MeV/c at 100 AMeV and $116\pm12$ MeV/c at 400 AMeV for QMD codes, respectively, with the error representing the standard deviation. The mean flow slope is systematically larger for BUU codes than for QMD codes although they have overlap within the error. This could be due to the larger transverse flows from relativistic BUU codes (GIBUU-RMF, RBUU, and RVUU), or due to some intrinsic difference in the mean-field potential calculation between BUU and QMD. Although the absolute uncertainty is similar at both incident energies, the relative uncertainty of the transverse flow slope, i.e., the absolute uncertainty divided by the mean slope value, is about $30\%$ at 100 AMeV and $13\%$ at 400 AMeV, reflecting the "theoretical error" of transport simulations. This can be considered as a benchmark of this comparison, and it shows that the flow at higher incident energies is a more robust observable. If one wants to extract valuable information of the nuclear EOS from the experimental flow data at 400 AMeV, the theoretical error of about $13\%$ should be taken into account. The future goal is to reduce the uncertainties of transport simulations to less than $10\%$.

\subsection{Comparisons of box simulations}
\label{combox}

The comparison of transport approaches for heavy-ion simulations is closer to the experimental situations, while different components of transport approaches are coupled together so it is difficult to identify the source of the discrepancies. For example, the different densities in the initialization may affect the density reached in heavy-ion collisions, and thus the mean-field evolution as well as the attempted collisions. The efficiency of the Pauli blocking may affect the number of successful collisions, and thus the density evolution and the mean-field potential.

The box system with the periodic boundary condition is useful to test the reliability of transport approaches. The initialization is no longer a problem, and a common initialization is easily realized for each code. The different components of transport approaches, such as the nucleon-nucleon collision, the mean-field evolution, as well as the particle production, can be tested separately. Most importantly, exact limits of the theoretical answer from the kinetic theory or Landau theory are generally available, which provides the possibility that we can evaluate the performance of the code, while this can not be done in the heavy-ion comparison. In previous studies, the box calculation method has been used to study the kaon production~\cite{Fer05}, the thermalization of gluons~\cite{Xu05}, and the specific shear viscosity of the relativistic hadron gas~\cite{Dem09}. Recently there are also studies using the box calculation method on equilibrium properties of nucleons and pion-like particles with the RVUU transport model~\cite{Zha18} and on nuclear matter thermodynamic and transport properties with the UrQMD transport model~\cite{Mot18}.

The box system with the periodic boundary condition can be constructed with only slight changes of the code for heavy-ion collisions as follows. A dimension of a box is $L_\alpha=20$ fm with $\alpha=x,y,z$, and the center of the box is at $(L_x/2,L_y/2,L_z/2)$. With the periodic boundary condition, a particle that leaves on one side of the box should enter on the other side with the same momentum. The coordinate $r_\alpha$ of the particle should always be within $(0,L_\alpha)$, so it can be expressed as $r_\alpha \rightarrow \text{modulo}(r_\alpha,L_\alpha)$. The separation distance between particle $i$ and $j$, i.e., $\Delta r_{ij,\alpha}=r_{i,\alpha}-r_{j,\alpha}$, should be less than $L_\alpha/2$ in each dimension, so it can be expressed as $\Delta r_{ij,\alpha} \rightarrow \text{modulo}(\Delta r_{ij,\alpha} + L_\alpha/2, L_\alpha) - L_\alpha/2$. The separation distance can be that for the attempted collision judgement, the calculation of the mean-field potential, or that for the phase-space cell or the Gaussian wave packet width in the Pauli blocking treatment. The above description is sufficient to install a box condition into a transport code.

%\subsubsection{Evaluation of nucleon-nucleon collisions}

As a first step, the box calculation is used to evaluate the nucleon-nucleon collision rate as well as the performance of the Pauli blocking for transport approaches. There are totally 15 codes participating in the box comparison of nucleon-nucleon collisions, including 7 BUU-type codes and 8 QMD-type codes. The acronyms, code correspondents, as well as the representative references are listed in Table~\ref{code_Zhang18}. The new codes in additional to those in the heavy-ion comparison at 100 and 400 AMeV~\cite{Xu16} are the BUU code developed jointly in collaboration of Variable Energy Cyclotron Centre (VECC) with McGill
University (BUU-VM), the simulating many accelerated strongly-interacting hadron (SMASH) code, the jet AA microscopic (JAM) transport model, and the QMD code developed at Japan Atomic Energy Research Institute (JQMD). All codes participating in the box comparison of nucleon-nucleon collisions are relativistic ones\footnote{Here the relativistic code means that the nucleon-nucleon collisions are treated relativistically and the particle velocity is $\vec{p}/\sqrt{p^2+m^2}$ instead of $\vec{p}/m$, rather than the relativistic mean-field potential part as in the previous sections.} except for SMF and CoMD.

\begin{table}[h]\small
  \centering
  \caption{The names, authors and correspondents, and representative references of 7 BUU-type codes and 8 QMD-type codes participating in the box comparison of nucleon-nucleon collisions~\cite{Zhang18}.
}
    \begin{tabular}{|c|c|c|c|c|c|c|c|c}
    \hline
    BUU-type & code correspondents & references  &&  QMD-type & code correspondents &  references\\
    \hline
     BUU-VM    & S.Mallik &  \cite{BUU-VM1,BUU-VM2} && CoMD & M.Papa & \cite{CoMD1,CoMD2,CoMD3} \\
     GIBUU & J.Weil & \cite{GIBUU1,GIBUU2,GIBUU3}  &&ImQMD & Y.X.Zhang,Z.X.Li & \cite{ImQMD-CIAE1,ImQMD-CIAE2,Zhang08plb} \\
     IBUU & J.Xu,L.W.Chen,B.A.Li & \cite{Li97,IBUU2,Li08,IBUU3}   && IQMD-BNU & J.Su,F.S.Zhang  & \cite{IQMD-BNU1,IQMD-BNU2,IQMD-BNU3}  \\
     pBUU    &  P.Danielewicz & \cite{pBUU1,Dan91} && IQMD-IMP  & Z.Q.Feng & \cite{IQMD-IMP1,IQMD-IMP2}\\
     RVUU   &  T.Song,Z.Zhang,C.M.Ko & \cite{RVUU1,RVUU2,Son15} && JAM & A.Ono,N.Ikeno,Y.Nara & \cite{JAM1,JAM2} \\
     SMASH   & D.Oliinychenko,H.Petersen & \cite{SMASH} && JQMD & T. Ogawa & \cite{JQMD1,JQMD2}\\
     SMF    & M.Colonna & \cite{SMF1,SMF2,SMF3} && TuQMD & D.Cozma & \cite{TuQMD1,TuQMD2,TuQMD3,Dan13}\\
      &  & && UrQMD & Y.J.Wang,Q.F.Li & \cite{UrQMD1,UrQMD2,UrQMD3}\\
    \hline
    \end{tabular}
  \label{code_Zhang18}
\end{table}

The homework calculations assigned to the code practitioners are as follows. Totally 640 neutrons and 640 protons are uniformly distributed in a cubic box system, corresponding to the nucleon number density of $\rho=0.16$ fm$^{-3}$. The initial momentum distribution of these nucleons follows the Fermi-Dirac distribution, i.e., $f=1/\{\exp[(\epsilon-\mu)/T] + 1 \}$, with the temperature $T=0$ and 5 MeV, the single-particle energy $\epsilon=p^2/2m + m$ for the non-relativistic case and $\epsilon=\sqrt{m^2+p^2}$ for the relativistic case, where $m=938$ MeV is the nucleon mass. At $T=0$ the momenta of nucleons are sampled in a sphere within the Fermi momentum at the saturation density, while at $T=5$ MeV the momentum distribution is smeared out, with the chemical potential $\mu$ determined by $\frac{4}{(2\pi\hbar)^3} \int f d^3 p = \rho$. The system evolves with elastic nucleon-nucleon collisions only, without any inelastic collisions, mean-field potentials, or Coulomb interactions. A constant isotropic cross section of 40 mb is used. The time step can be 0.5 or 1 fm/c decided by the code practitioners. A particle that collides with one particle can collide with another particle within the same time step. All the artificial thresholds of the center-of-mass energy for nucleon-nucleon collisions should be removed, and there is no cut on the energy or distance in order to accelerate the code. Two calculational modes are employed, i.e., cascade without Pauli blocking (C) and cascade with Pauli blocking (CB). For the CB mode, the code practitioners are asked to use their default Pauli blocking treatments for heavy-ion simulations (CBOP1) and the ideal treatment of Pauli blocking by using the blocking probability of $1-(1-f_i)(1-f_j)$ for collisions between particle $i$ and particle $j$ with $f_{i/j}$ the analytical Fermi-Dirac distribution (CBOP2). It was found that all codes give consistent results for CBOP2, and here the main discussions are focused on the 4 modes of the box calculation, i.e., CT0, CT5, CBOP1T0, and CBOP1T5. BUU codes are asked to provide data for 10 runs with each run 100 test particles, and QMD codes are asked to provide data for 200 events. All codes are evolved until $t=140$ fm/c.

\begin{figure}[h]
\centering
\includegraphics[scale=0.31,clip]{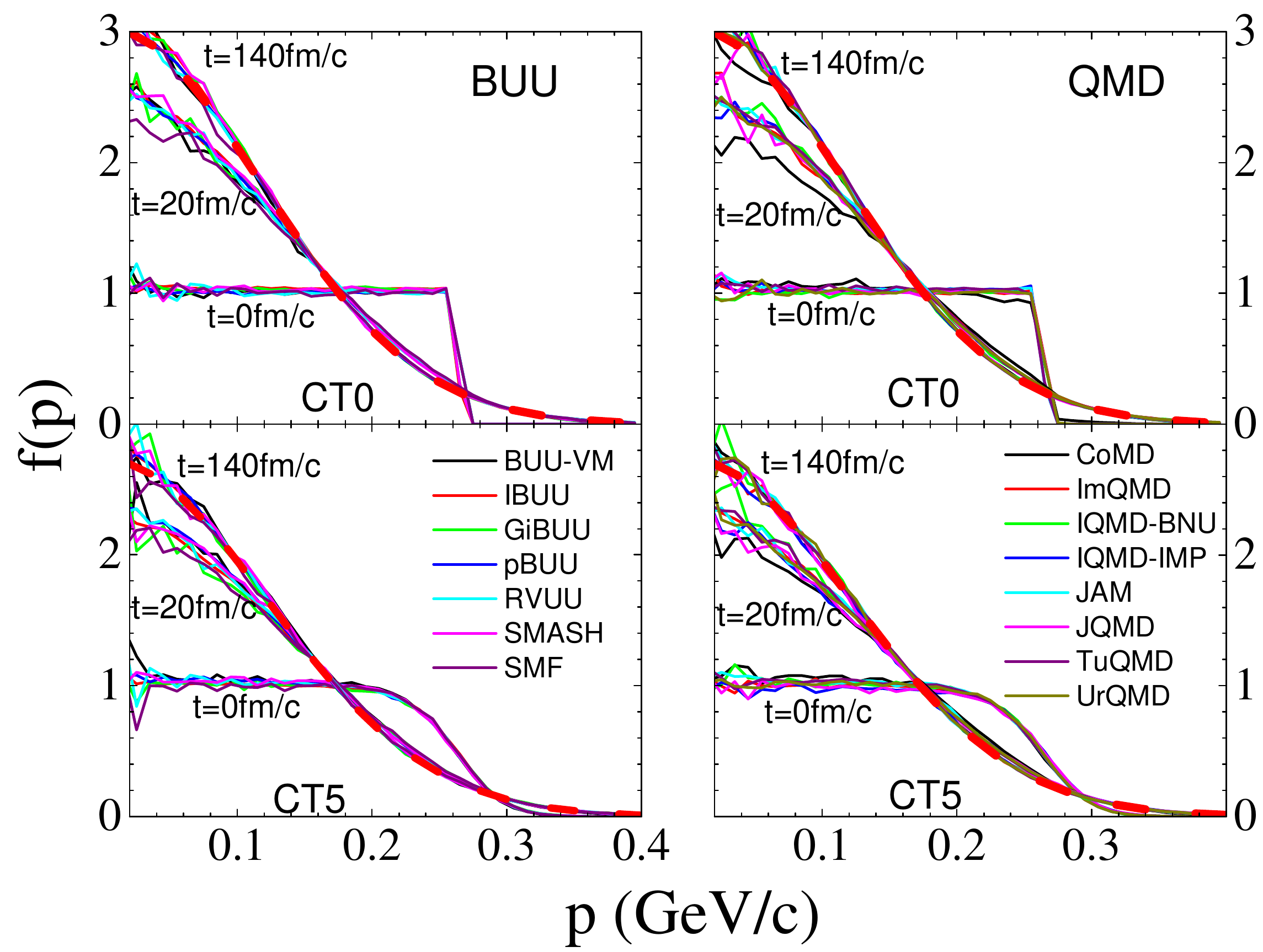}
\includegraphics[scale=0.3,clip]{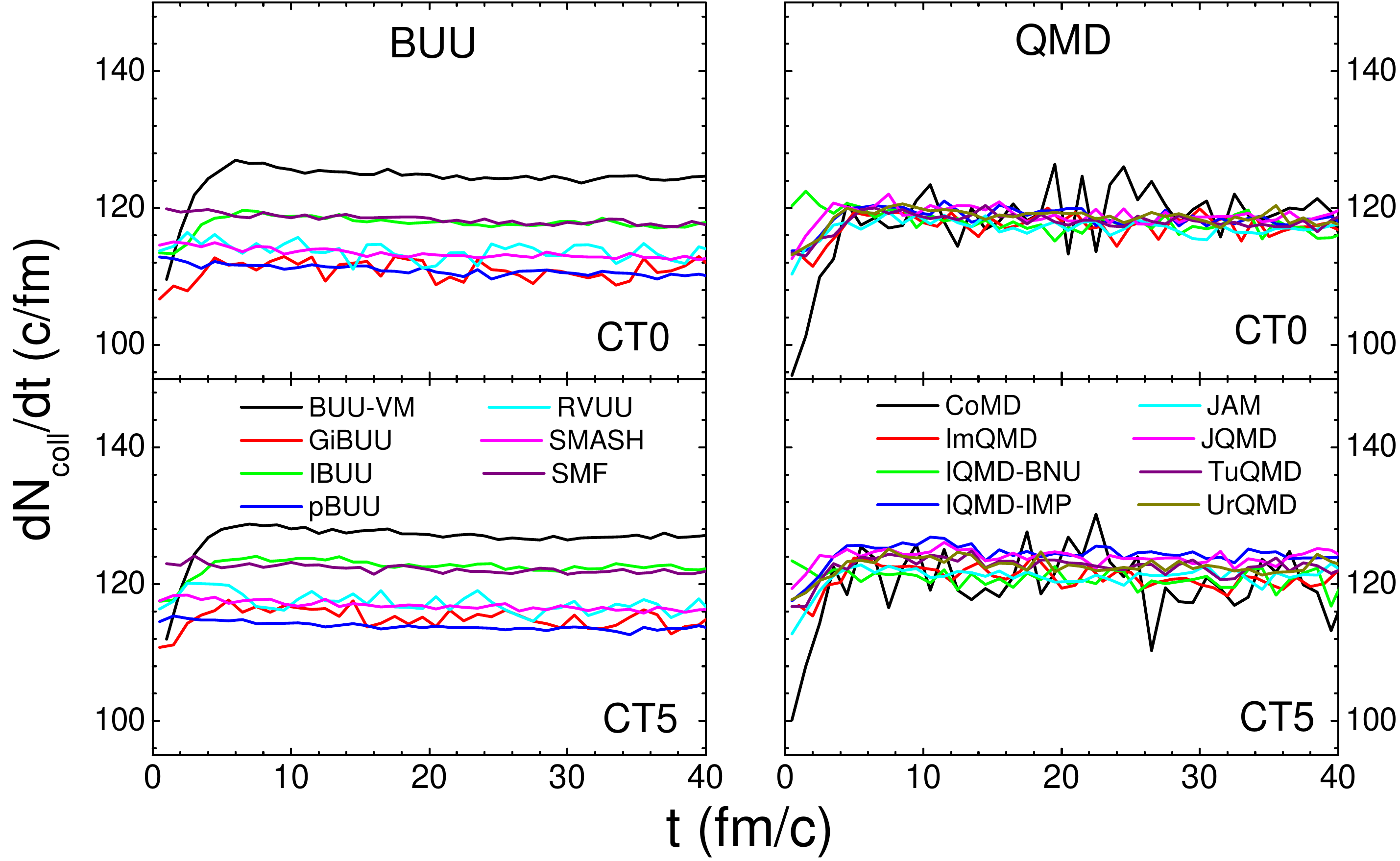}
\caption{(Color online) Left: Momentum distributions at $t=0$, 20, and 140 fm/c from an initial Fermi-Dirac distribution at $T=0$ (CT0) and 5 MeV (CT5) without Pauli blocking, with the thick red dashed lines corresponding to the Boltzmann distribution through the energy conservation condition. Right: Time evolution of the collision rate from an initial Fermi-Dirac distribution at $T=0$ (CT0) and 5 MeV (CT5) without Pauli blocking. Modified from figures in Ref.~\cite{Zhang18}.}
\label{evo_c_Zhang18}
\end{figure}

The time evolution of the momentum distribution without Pauli blocking for the system initialized at $T=0$ and 5 MeV are shown in the left panel of Fig.~\ref{evo_c_Zhang18}. The initial Fermi-Dirac distributions are the same for all codes, satisfying our purpose of using a common initialization for all codes. Since there is no Pauli blocking, the distributions evolve quickly away from the Fermi-Dirac distribution to the Boltzmann distribution shown by the thick red dashed lines, with the temperatures about $T_B=14.8$  and 15.8 MeV from the initialization at $T=0$ and 5 MeV, respectively, theoretically obtained through the energy conservation condition. It is seen that even at $t=20$ fm/c the momentum distributions are already very close to the final Boltzmann distribution.

The expected collision rate for a constant collision cross section can be expressed as
\begin{equation}\label{dncoll}
\frac{dN_{coll}}{dt} = \frac{1}{2}A\rho\sigma \langle v_{rel} \rangle,
\end{equation}
where $1/2$ is from the double counting of collision numbers between identical particles, $A=1280$ is the total nucleon number, and $\langle v_{rel} \rangle$ represents the thermal average of the relative velocity with $v_{rel} = |\vec{p}_1/m-\vec{p}_2/m|$ for the non-relativistic case and $v_{rel} = s/(E_1 E_2)$ for the relativistic case, where $s$ is the square of the invariant mass of the particle pair and $E_{1/2} = \sqrt{m^2+p_{1/2}^2}$ is the particle energy. In the non-relativistic case, the average relative velocity is $\langle v_{rel} \rangle = (36/35)(p_F/m)$ for a Fermi-Dirac distribution at $T=0$, while it becomes $\langle v_{rel} \rangle = (4/\sqrt{5\pi})(p_F/m)$ for a Boltzmann distribution with the same total energy. In the relativistic case, the average relative velocity for a Boltzmann distribution at the temperature $T_B$ is
\begin{equation}\label{vrel}
\langle v_{rel} \rangle = \frac{1}{4m^4 T_B K_2^2(m/T_B)} \int_{2m}^\infty d\sqrt{s} s (s-4m^2) K_1(\sqrt{s}/T_B),
\end{equation}
where $K_n$ is the $n$th-order modified Bessel function. For the final Boltzmann distribution with the temperatures about $T_B=14.8$  and 15.8 MeV from the initializations at $T=0$ and 5 MeV, respectively, the non-relativistic collision rates are 115.9 and 120.1 c/fm, and the relativistic collision rates are 111.4 and 115.4 c/fm, for CT0 and CT5, respectively. It has further been found that the excepted collision rates are generally slightly higher for a Fermi-Dirac distribution compared with the Boltzmann distribution with the same total energy.

The time evolutions of the collision rates in the first 40 fm/c are displayed in the right panel of Fig.~\ref{evo_c_Zhang18}. The relaxation time only takes about 10 fm/c, and after that the collision rates reach their equilibrium values, consistent with the quick equilibration as shown in the left panel of Fig.~\ref{evo_c_Zhang18}. One would expect that the collision rates will slightly drop from the initial Fermi-Dirac distribution into that from the final Boltzmann distribution, while this is not the case especially for some QMD-type codes. The results from CoMD suffer from large fluctuations.

\begin{figure}[h]
\centering
\includegraphics[scale=0.4,clip]{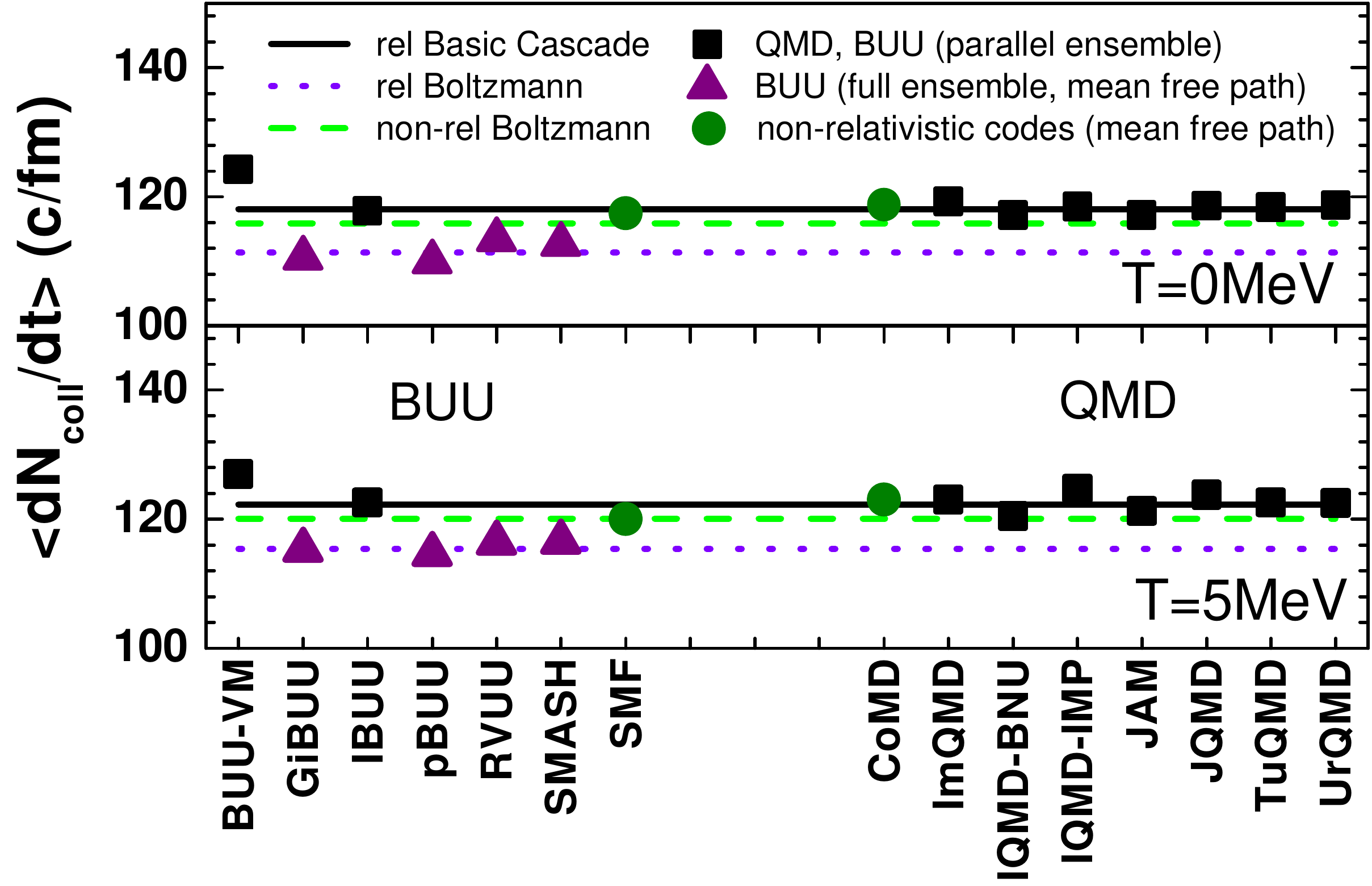}
\caption{(Color online) Collision rates at $t=60-140$ fm/c from various transport codes (scatters) compared with the exact limits (lines). The green dashed lines and the purple dotted lines are exact limits for the non-relativistic and relativistic rates, respectively, and the black solid lines are from a basic relativistic cascade code. Results from each code should be compared with the corresponding limit with the same color. Modified from figures in Ref.~\cite{Zhang18}.}
\label{coll_c_Zhang18}
\end{figure}

The average collision rates in the time interval of $t=60-140$ fm/c from various codes are compared in Fig.~\ref{coll_c_Zhang18}, where the exact limits in the non-relativistic and relativistic case as well as that from a basic relativistic cascade code are also plotted as lines for comparison. Except for SMF, CoMD, and pBUU that use the stochastic approach for the collision treatment, all codes use Bertsch's prescription described in the Appendix B of Ref.~\cite{Ber88}. Here the Bertsch's prescription is slightly modified by removing repeated or spurious collisions, to be detailed in Sec.~\ref{collisions}, otherwise the collision rates can be as high as $150-170$ c/fm. It is seen from Fig.~\ref{coll_c_Zhang18} that the collision rates from various codes shown by black squares are consistent with that from a basic relativistic cascade code, except for BUU-VM. The slight deviation from each other can be due to their slightly different initial conditions, i.e., inaccurate chemical potentials and/or Fermi momenta used for the initialization leading to slightly different total energies. They are, however, still systematically higher than the limit of the relativistic collision rate shown by the purple dotted line obtained from Eq.~(\ref{vrel}). This is due to the higher-order correlations, to be detailed in Sec.~\ref{collisions}. It is suspected that the quick increase of the collision rate in the right panel of Fig.~\ref{evo_c_Zhang18} can be due to the building up of higher-order correlations with time. It is shown in Fig.~\ref{coll_c_Zhang18} that GIBUU, pBUU, RVUU, and SMASH use the full ensemble method for the collision treatment, which suppresses efficiently the higher-order correlations, and give collision rates very close to the exact limits. In this way, these codes give lower collision rates compared to QMD codes and other BUU codes using the parallel ensemble collision treatment. SMF and CoMD are non-relativistic codes, and their collision rates are consistent with the non-relativistic limit. Generally speaking, the collision rates without Pauli blocking are well under control, and the slight deviations and discrepancies are understandable.

\begin{figure}[h]
\centering
\includegraphics[scale=0.32,clip]{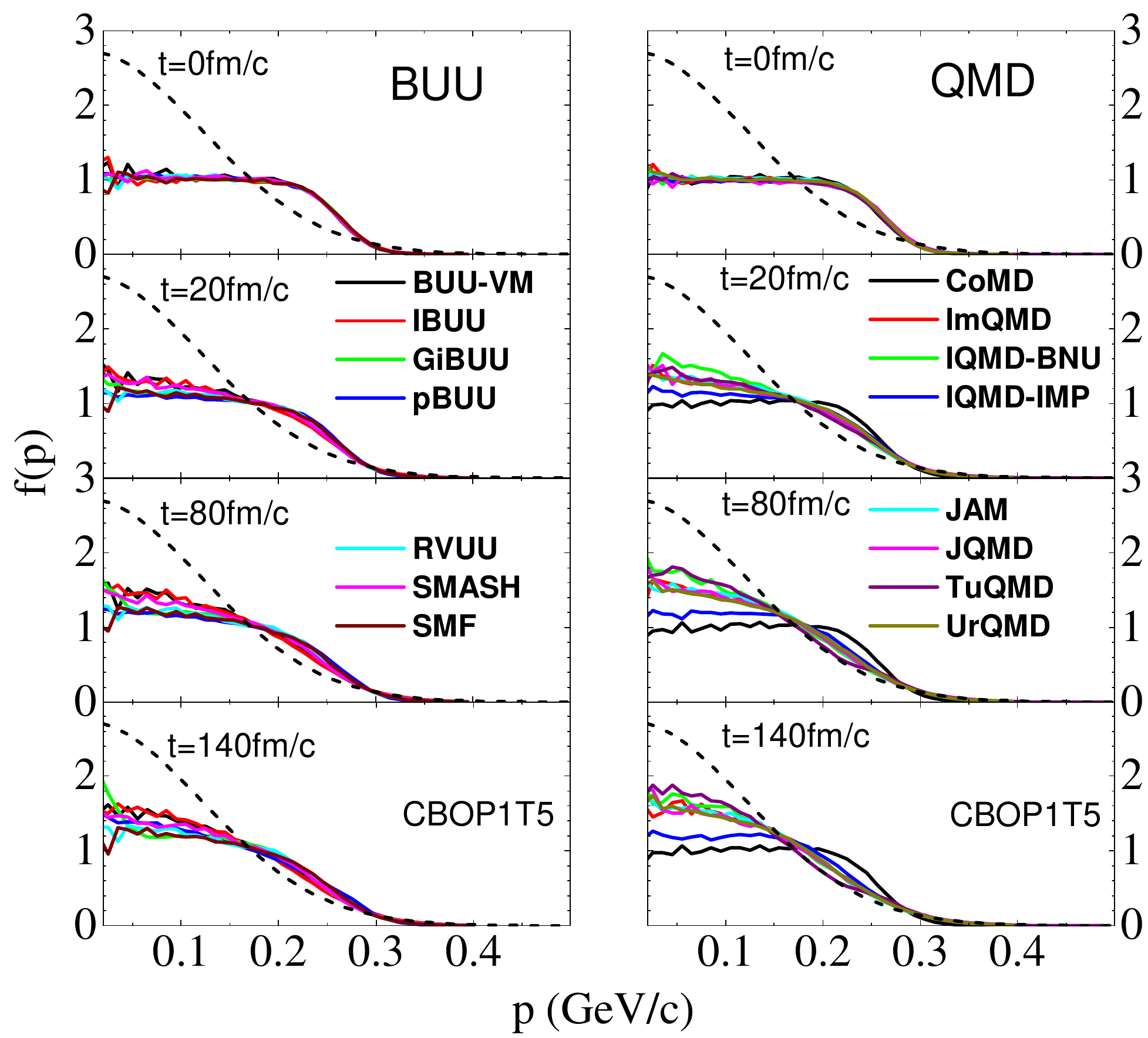}
\includegraphics[scale=0.33,clip]{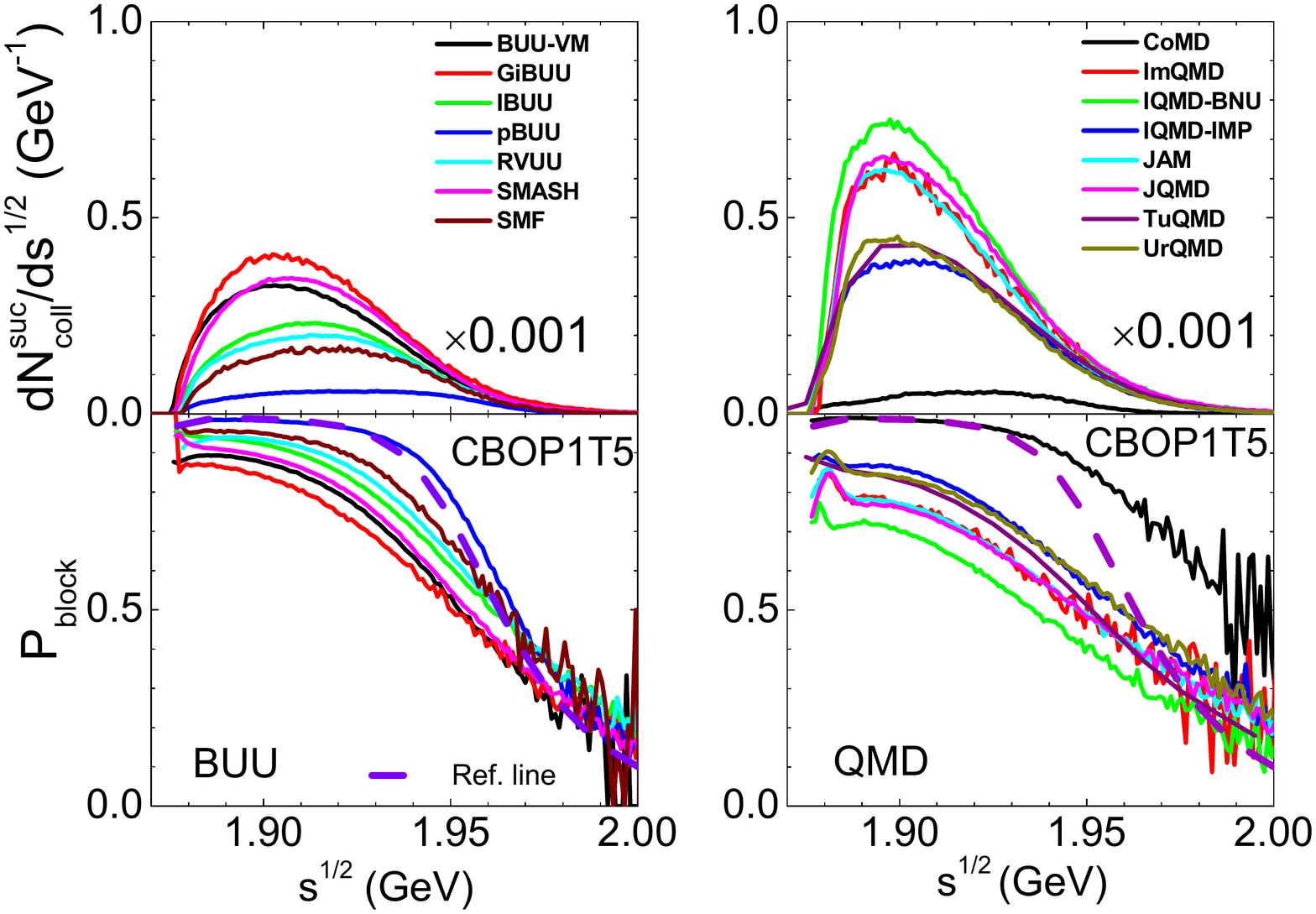}
\caption{(Color online) Left: Momentum distributions at $t=0$, 20, 80, and 140 fm/c from an initial Fermi-Dirac distribution at $T=5$ MeV with Pauli blocking, with the black dashed lines corresponding to the Boltzmann distribution through the energy conservation condition. Modified from figures in Ref.~\cite{Zhang18}. Right: Center-of-mass energy distribution of the successful collision number as well as the Pauli blocking probability from an initial Fermi-Dirac distribution at $T=5$ MeV, with the thick purple dashed lines being the reference Pauli blocking probability from an ideal Fermi-Dirac distribution at $T=5$ MeV. Taken from Ref.~\cite{Zhang18}.}
\label{evo_cb_Zhang18}
\end{figure}

Due to the quantum nature of Fermions, an attempted collision can become a successful one only if the final phase-space is not occupied, otherwise the collision is Pauli blocked. Although the Pauli blocking probability is generally calculated as $1-(1-f_i)(1-f_j)$ in most transport codes, the local phase-space occupation probability $f_{i/j}$ is calculated differently not only in each individual code but also for the two branches of transport codes. In BUU-type codes, $f_{i/j}$ is calculated by averaging over all parallel events depending on the size of the local phase-space cell. In QMD-type codes, $f_{i/j}$ is calculated in an event-by-event manner depending on the width of the Gaussian wave packet. In TuQMD and IQMD-IMP, overlap regions in the coordinate and momentum space are evaluated in order to get the occupation probability. CoMD employs a similar procedure but the collision becomes a successful one only if the occupation probability is less than a chosen small number, e.g., 0.08, i.e., collisions are allowed only into essentially empty phase space cells. pBUU employs an effective temperature approach for the Pauli blocking, i.e., the distribution function in the cell around the final state of the scattered particle is fitted by a weighted sum of two deformed Fermi-Dirac distributions for the Pauli blocking judgement. I refer the reader to Table V of Ref.~\cite{Zhang18} for further details of Pauli blocking treatments in various codes.

The time evolutions of the momentum distributions for the system initialized at $T=5$ MeV from various transport codes with Pauli blocking are displayed in the left panel of Fig.~\ref{evo_cb_Zhang18}. It is seen that although the Pauli blocking is employed, the momentum distributions for most codes evolve away from the initial Fermi-Dirac distribution toward a Boltzmann distribution shown by the dashed line. In addition, BUU codes generally preserve the Fermi-Dirac distribution better than QMD codes except for CoMD, since the two branches use different Pauli blocking treatments as mentioned above. The center-of-mass energy distribution of the successful collision rate as well as the Pauli blocking probability from various codes are compared in the right panel of Fig.~\ref{evo_cb_Zhang18}. For the system initialized at $T=5$ MeV, the average center-of-mass energy for arbitrary nucleon pairs is about 1.92 GeV, while that for two-nucleon pair at the Fermi surface is about 1.95 GeV. Large discrepancies in the successful collision rates due to those in the Pauli blocking probabilities from various codes are seen, and QMD codes systematically give larger successful collision rates and smaller Pauli blocking probabilities. CoMD is an exception, which overestimates the Pauli blocking probability especially at larger center-of-mass energy compared with the reference Pauli blocking probability from an ideal Fermi-Dirac distribution at $T=5$ MeV shown by the thick purple dashed line. On the other hand, pBUU reproduces very well the reference line as a result of the effective temperature approach employed for the Pauli blocking treatment.

\begin{figure}[h]
\centering
\includegraphics[scale=0.4,clip]{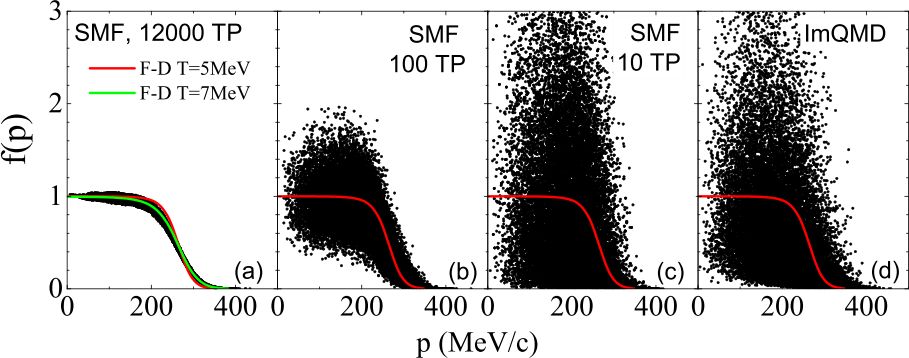}
\includegraphics[scale=0.47,clip]{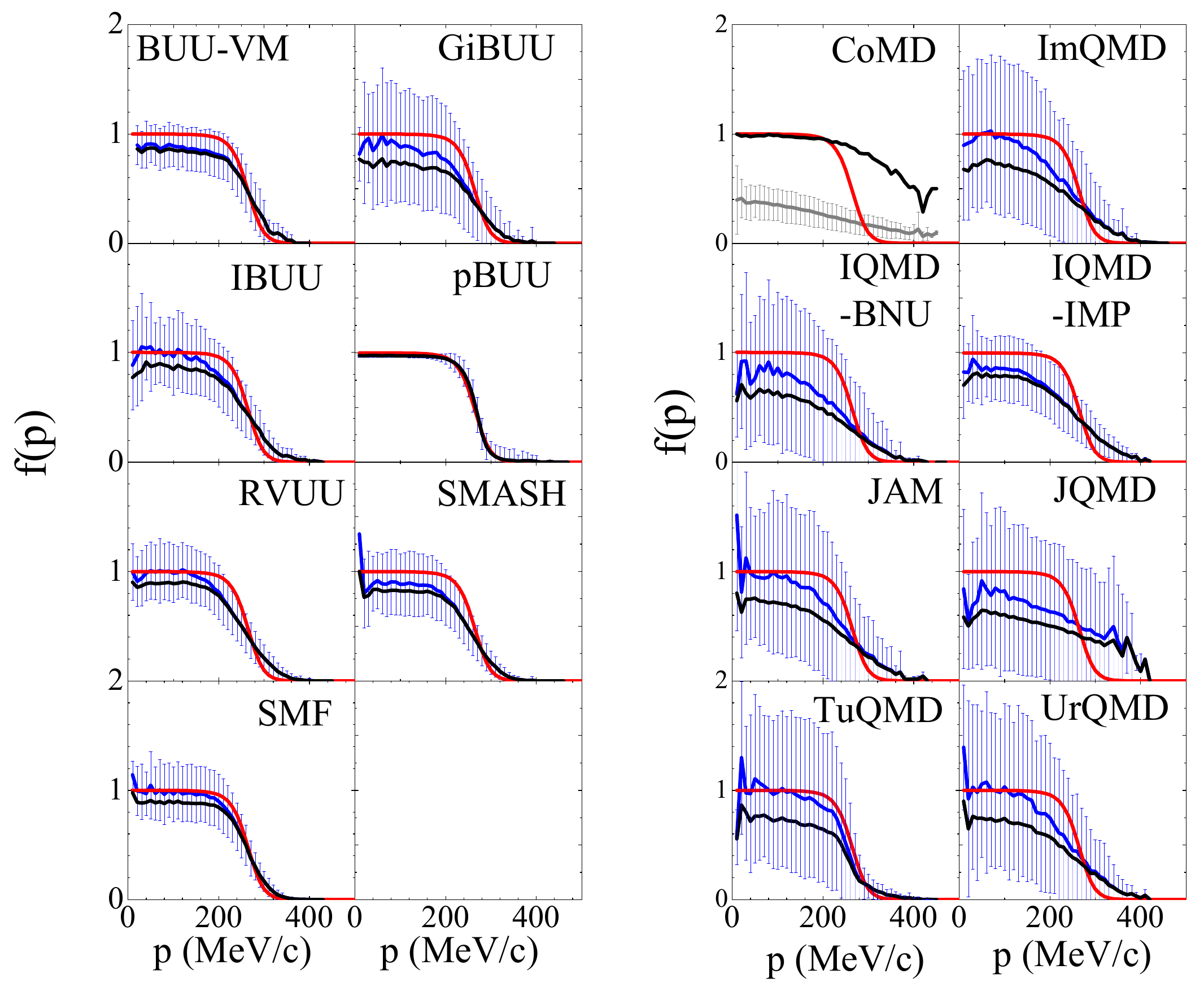}
\caption{(Color online) Upper: Scatter plots in momentum occupational space for the final
states of collisions at the first time step of the box simulation from SMF with different numbers of test particles and ImQMD initialized at $T=5$ MeV. Lower: Distributions of the occupation probability in momentum space for the first time step in the box system initialized at $T=5$ MeV, with the mean value and the variance shown by the blue lines and the blue error bars, and the effective blocking probability shown by the black lines. The ideal Fermi-Dirac distribution is shown by the red lines, while the gray line and the error bar for CoMD needs a separate explanation in the text. Taken from Ref.~\cite{Zhang18}.}
\label{fp1_cb_Zhang18}
\end{figure}

In order to understand the discrepancies in the Pauli blocking probability not only between two branches of BUU and QMD transport codes but also within one branch, the code practitioners are asked to provide the occupation probability in the momentum space for the first time step, when the momentum distribution is still governed by the initial Fermi-Dirac distribution. The scatters in the momentum space in the upper panel of Fig.~\ref{fp1_cb_Zhang18} represent the occupation probabilities used for Pauli blockings at the first time step. For SMF as an example of BUU-type codes, the scatter distribution becomes closer to the ideal Fermi-Dirac distribution at $T=5$ MeV with the increasing number of test particles. The case with the effective test particle number 12000 represents the calculation with 100 test particles but the occupation probability is evaluated as the average over the entire box instead of a local cell. This gives a Fermi-Dirac distribution at a temperature closer to $T=7$ MeV, instead of $T=5$ MeV, due to the Gaussian shape of the test particles in momentum space with a width of 59 MeV/c for SMF. For ImQMD as a representative of QMD-type codes, however, the occupation probability is more scattered. The distribution depends on the width of the Gaussian wave packet, and the fluctuation will not be reduced with the increasing number of events for QMD-type codes, since the occupation probability for each event is independent. The lower panel of Fig.~\ref{fp1_cb_Zhang18} gives an overview of the mean value and the variance of the occupation probability for the first time step in the system initialized at $T=5$ MeV. Generally, the BUU codes with 100 test particles give much smaller fluctuations compared with the QMD codes. For each BUU code, the fluctuation depends on the individual code. pBUU using the effective temperature approach reproduces almost exactly the desired Fermi-Dirac distribution shown by the red line. The fluctuations of the occupation probability can in principle be reduced as small as possible if more test particles are used. For QMD codes, the fluctuations are larger and the mean values shown by blue solid lines are generally smeared out to higher momenta compared to the ideal Fermi-Dirac distribution. If the occupation probability is larger than 1, the collision is Pauli blocked, identical to the case with an occupation probability equal to 1. On the other hand, if the occupation probability is smaller than 1, it is retained. The effective occupation probabilities after taking this effect into account are shown by the black solid lines. So one can see that larger fluctuations of the occupation probability generally underestimate the Pauli blocking, and this is just the case for QMD codes compared to BUU codes. For CoMD, the gray line represents the hard sphere overlap as a function of the momentum, and the error bar represents the fluctuation. The black line is the effective occupation probability, considering that the collision will be Pauli blocked if the overlap is larger than 0.08. This overestimates  the occupation probability at larger momenta, consistent with that observed in the right panel of Fig.~\ref{evo_cb_Zhang18}.

\begin{figure}[h]
\centering
\includegraphics[scale=0.4,clip]{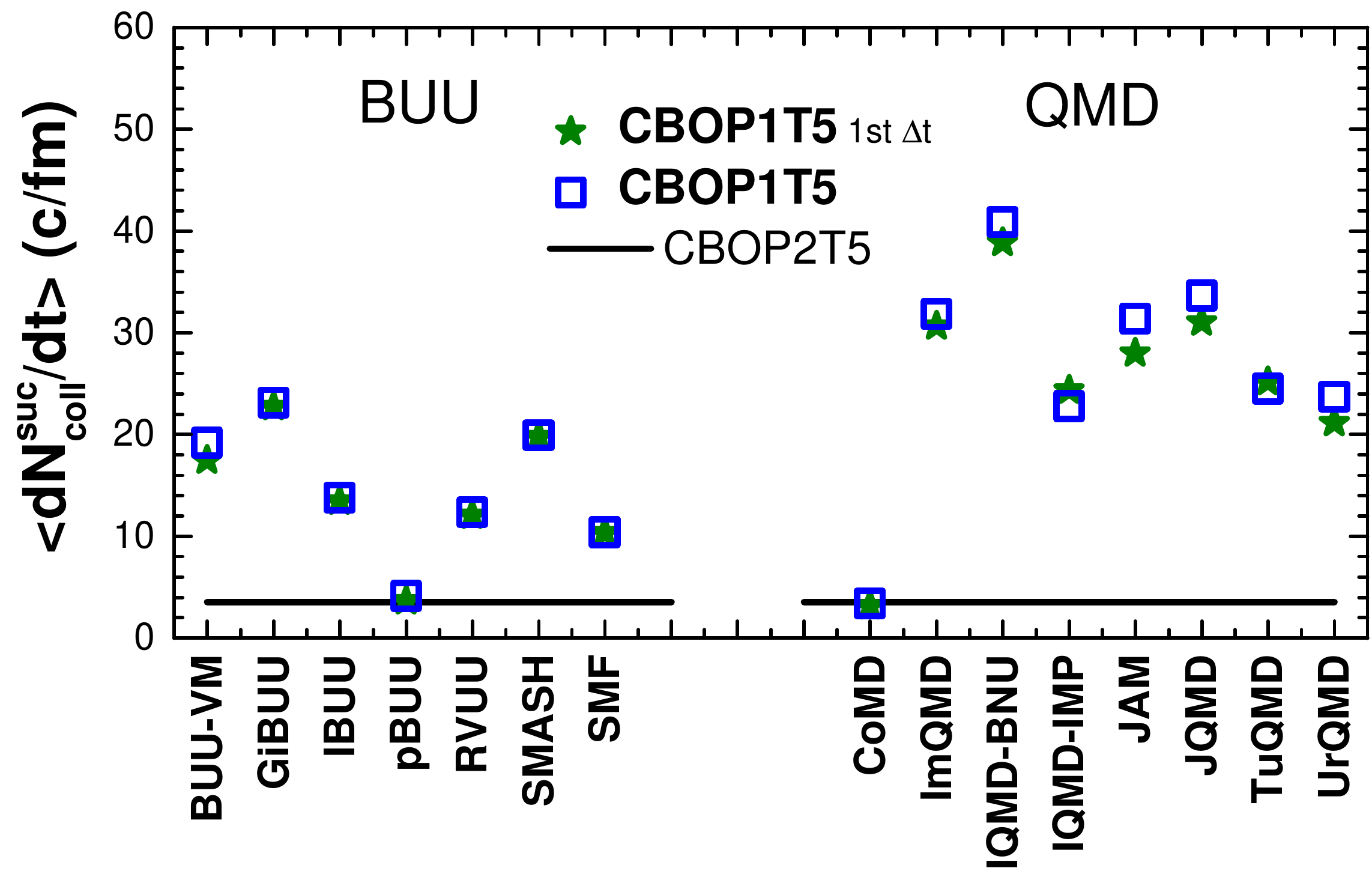}
\caption{(Color online) Successful collision rates with the individual Pauli blocking treatment in each code from an initialization at $T=5$ MeV, with the stars representing those for the first time step, and the open squares for those averaged over $t=60-140$ fm/c. The black line is the limit from a basic cascade code using the ideal Fermi-Dirac distribution at $T=5$ MeV for the Pauli blocking. Modified from figures in Ref.~\cite{Zhang18}.}
\label{coll_cb_Zhang18}
\end{figure}

The successful collision rates with Pauli blocking for the system initialized at $T=5$ MeV for the first time step and the average values in the time interval of $t=60-140$ fm/c from various codes are compared to the exact limit of CBOP2T5 from a basic cascade code in Fig.~\ref{coll_cb_Zhang18}. It should be pointed out that the momentum distribution is still a Fermi-Dirac distribution in the first time step, while it evolves toward a Boltzmann distribution in the later time steps. However, the successful collision rates in the first time step are not much different from those in later stages. This could be due to the energy conservation constraint so that the collision rates are similar for the initial Fermi-Dirac distribution and the Boltzmann distribution. Most QMD-type codes have larger successful collision rates compared to BUU-type codes, and this has been understood due to the larger fluctuation of the occupation probability found in the lower panel of Fig.~\ref{fp1_cb_Zhang18}. The exact limit of the successful collision rate is about 3.5 c/fm at $T=5$ MeV, as shown by the black line. The successful collision rate of SMF with 100 test particles is 10.5 c/fm, while the rate can be dropped to about 6 c/fm with effectively increasing the number of test particles, similar to the exact limit of the successful collision rate at $T=7$ MeV, as illustrated in the upper panel of Fig.~\ref{fp1_cb_Zhang18}. With the effective temperature approach for the Pauli blocking treatment, pBUU reproduces the exact limit of the collision rate rather well. Although the Pauli blocking probability at higher center-of-mass energies is overestimated for CoMD, it reproduces well the exact limit of the successful collision rate since most collisions are soft ones at lower center-of-mass energies. Except for pBUU and CoMD, BUU codes give the successful collision rate in the range of $10-23$ c/fm, while QMD codes give the successful collision rates in the range of $23-40$ c/fm.

%\subsubsection{Evaluation of Vlasov evolution}

%\subsubsection{Evaluation of pion-like particle production}

\subsection{Current status and future plan of the transport code evaluation project}

Currently, the comparisons of box simulations for the mean-field evolution and the production of pion-like particles are still in progress. Meanwhile, since useful knowledge and technique have been learned based on the box simulation, improvements have been made to each individual transport code. In order to have a benchmark calculation before the data from the S$\pi$RIT experiment become available, we go back to the heavy-ion system to see whether we can reduce the transport model uncertainties especially for the $\pi^-/\pi^+$ ratio as a probe of the nuclear symmetry energy at suprasaturation densities mentioned in Sec.~\ref{suprasaturation}. Intended future activities of the transport code evaluation project are also discussed in this subsection.

In the mean-field evolution comparison, the initial density distribution in the box along the $z$ direction is $\rho(z) = \rho_0 + a_\rho \sin(kz)$, where $a_\rho=0.2\rho_0$ is the magnitude of the density fluctuation, and $k=n (2\pi/L_\alpha)$ is the wave number, with $L_\alpha=20$ fm being the size of the cubic box and $n=1$ and $2$ chosen for simulations. The comparison is for isospin symmetric nuclear matter in a box at zero temperature, with only the mean-field potential for nucleons but the Coulomb interaction and nucleon-nucleon collisions turned off. With the same mean-field potential used in the heavy-ion comparison as in Ref.~\cite{Xu16}, it was found that the density fluctuation damps very fast, especially for QMD-type codes. In order to analyze the oscillation of the density fluctuation due to the mean-field evolution for a longer time, simulations with a larger incompressibility were carried out. Through the Fourier transformation of the density distribution, it was found that higher-order oscillation modes appear in addition to $n=1$ or $2$. The main frequency of the density oscillation can be extracted, and it has an analytical solution related to the incompressibility of the nuclear EOS. In this way, it can be evaluated whether the mean-field evolution in the transport simulation is accurately calculated. Besides the main frequency, the damping rate as well as the energy conservation are also the main focuses. Possible comparison with the time-dependent Hartree-Fock calculations is on the schedule.

In the box simulation for the production of pion-like particles, the system is initialized as a uniform nucleonic matter at the temperature $T=60$ MeV. There are several modes in the comparison, such as the one-way $\Delta$ production mode $N+N \rightarrow N+\Delta$, the two-way $\Delta$ production mode $N+N \leftrightarrow N+\Delta$, and the full mode with the two-way $\Delta$ production as well as the $\Delta$ decay and the inverse reaction $\Delta \leftrightarrow N+\pi$. Each mode with a constant $\Delta$ mass width or a more realistic mass-dependent width is calculated with the initial isospin asymmetry $\delta=0$ and $0.2$. In order to maintain the thermalization of the box system, elastic collisions are allowed between baryons with a large constant and isotropic cross section. With the production of $\Delta$ resonances or pions, the temperature of the box system becomes  lower compared to the initial one, and can be obtained through the energy conservation condition. Since the reaction rate for each channel can be calculated at a given temperature, the time evolution of the numbers of nucleons, $\Delta$ resonances, and pions can all be obtained from reaction rate equations, providing a useful reference in the thermalization limit. The above results for the two-way $\Delta$ production mode as well as the full mode at the chemical equilibrium are consistent with the statistical model results, since the detailed balance condition is satisfied for each inverse channel, i.e., $N+N \leftrightarrow N+\Delta$ and $\Delta \leftrightarrow N+\pi$. It was found that the sequence of $N+N \leftrightarrow N+\Delta$, $\Delta \leftrightarrow N+\pi$, and the particle propagation in each individual code may affect significantly the $\Delta$ and pion yield, while using a smaller time step may sometimes help. In addition, the higher-order correlations for inelastic collisions, which are generally larger for larger cross sections, may affect the collision rate and even lead to the violation of the isospin symmetry.

The knowledge learnt from the transport evaluation in the box system should eventually be applied to heavy-ion simulations. In the latest heavy-ion comparison for the production of pion-like particles, the reaction system is non-central $^{132}$Sn+$^{124}$Sn and $^{112}$Sn+$^{108}$Sn collisions at the incident energy of 270 AMeV, same as that in the S$\pi$RIT experiment. The two collision systems with different isospin asymmetries provide the possibility of analyzing the double $\pi^-/\pi^+$ yield ratio. The main parameters for heavy-ion simulations are chosen to be similar to that in Ref.~\cite{Xu16}, while the collision channels for the production of pion-like particles are the same as the full mode in the box simulation for the production of pion-like particles. Improvements from previous efforts on nucleon-nucleon collisions as well as relevant techniques for the $\Delta$ and pion production channels are incorporated in this comparison. From preliminary analysis, the main source of the discrepancy on the pion yield as well as the $\pi^-/\pi^+$ yield ratio is from different density evolutions due to the mean-field potential part in various codes instead of $\Delta$ and pion production channels. In order to remove possible sources of discrepancies, the cascade mode without mean-field potential as well as that without Pauli blocking was further carried out. It was found that the Pauli blocking has negligible effects on the density evolution at this incident energy, but it reduces the pion multiplicities and generally enhances the $\pi^-/\pi^+$ yield ratio. Further analysis with better statistics in order to understand the sensitivity of the $\pi^-/\pi^+$ yield ratio to the nuclear symmetry energy is in progress.

In the foreseeable future, the box simulation for the transport code evaluation project will still be useful. A direct extension of the mean-field evolution is to turn on the isospin asymmetry, in order to evaluate the accuracy of the symmetry potential calculation. It is also useful to do the box simulation at subsaturation densities in a spinodal region, to see how the density fluctuation grows and how clusters are formed. So far the mean-field potential for non-relativistic codes are momentum-independent, and it is useful to do simulation with a more realistic momentum-dependent mean-field potential, so that it is more reasonable to compare results from non-relativistic codes to those from relativistic codes. Further aspects to be checked could be the energy violation due to the momentum-dependent mean-field potential, as well as the violation of the angular momentum conservation for nucleon-nucleon collisions. It is probably useful to develop standard programs, or standard subroutines for the main parts of the transport simulation, e.g., the initialization, the mean-field potential, the nucleon-nucleon collisions, the Pauli blocking, etc., for the BUU-type code and the QMD-type code according to what we have learnt from the transport code evaluation project.

\section{Discussions}
\label{discussions}

In this section I will discuss the main ingredients of transport approaches as well as related topics. This part includes what we have learnt from the transport code evaluation project, and what we suggest for improvements in future studies using transport approaches, especially for the initialization, the mean-field potential calculation, as well as the nucleon-nucleon collision treatment.

\subsection{Initialization}
\label{init}

The initial configurations of projectile and target nuclei are expected to have considerable influences on the heavy-ion observables from transport simulations. For example, the uncertainties of the neutron skin thickness of initial nuclei may affect the neutron/proton yield ratio~\cite{Sun10nskin}, the neutron-proton differential flow~\cite{Yon11}, the triton/$^3$He yield ratio~\cite{Dai14}, the $\pi^-/\pi^+$ yield ratio~\cite{Yon11,Wei14}, and other observables~\cite{Wei15} related to the isospin dynamics in heavy-ion collisions. Generally, BUU-type codes use the initial density distribution from a Woods-Saxon parameterization, from a Skyrme-Hartree-Fock calculation~\cite{SHF}, or from the Thomas-Fermi approximation~\cite{pBUU1,Len89}. QMD-type codes select the stable configurations with the charge radii and binding energies of the initial nuclei close to the experimental data. Usually, additional constraints are used for initialization. For example, a minimum distance between two arbitrary nucleons is required in order to give a more uniform initial density distribution. Also, the frictional cooling method~\cite{Wil77} is sometimes used in order to reach the ground-state configuration, while the nucleon Fermi motion needs to be implemented in a proper way~\cite{Lin16}. On the other hand, the initial density distribution is also related to the shape and size of the test particles in BUU codes and the width of the Gaussian wave packet in QMD codes. For the initial momentum distribution, if nucleons have a finite width in the momentum space, the centroid momenta of nucleons should be sampled with care in order to avoid overestimating the total initial kinetic energy.

\begin{figure}[h]
\centering
\includegraphics[scale=0.35,clip]{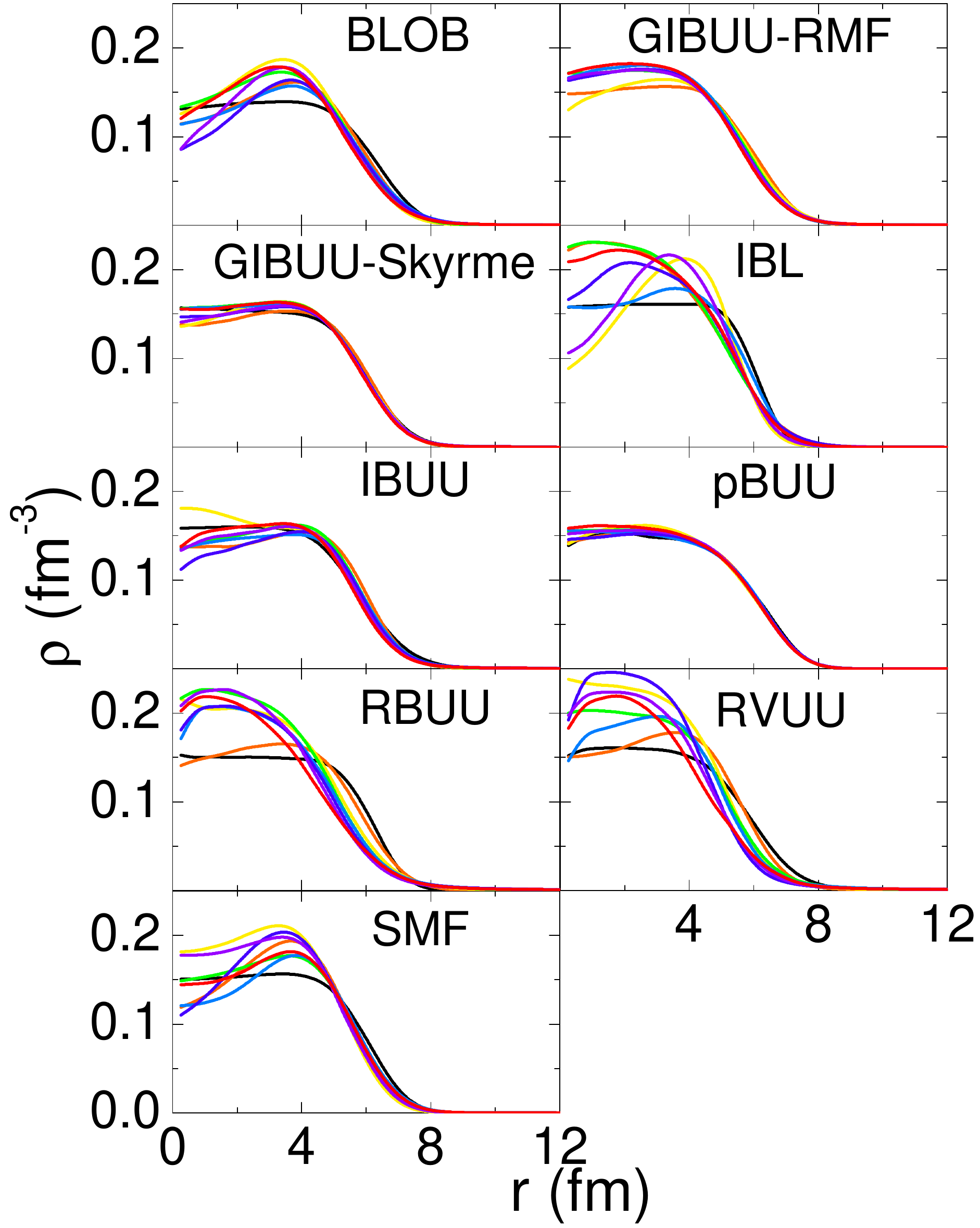}
\includegraphics[scale=0.35,clip]{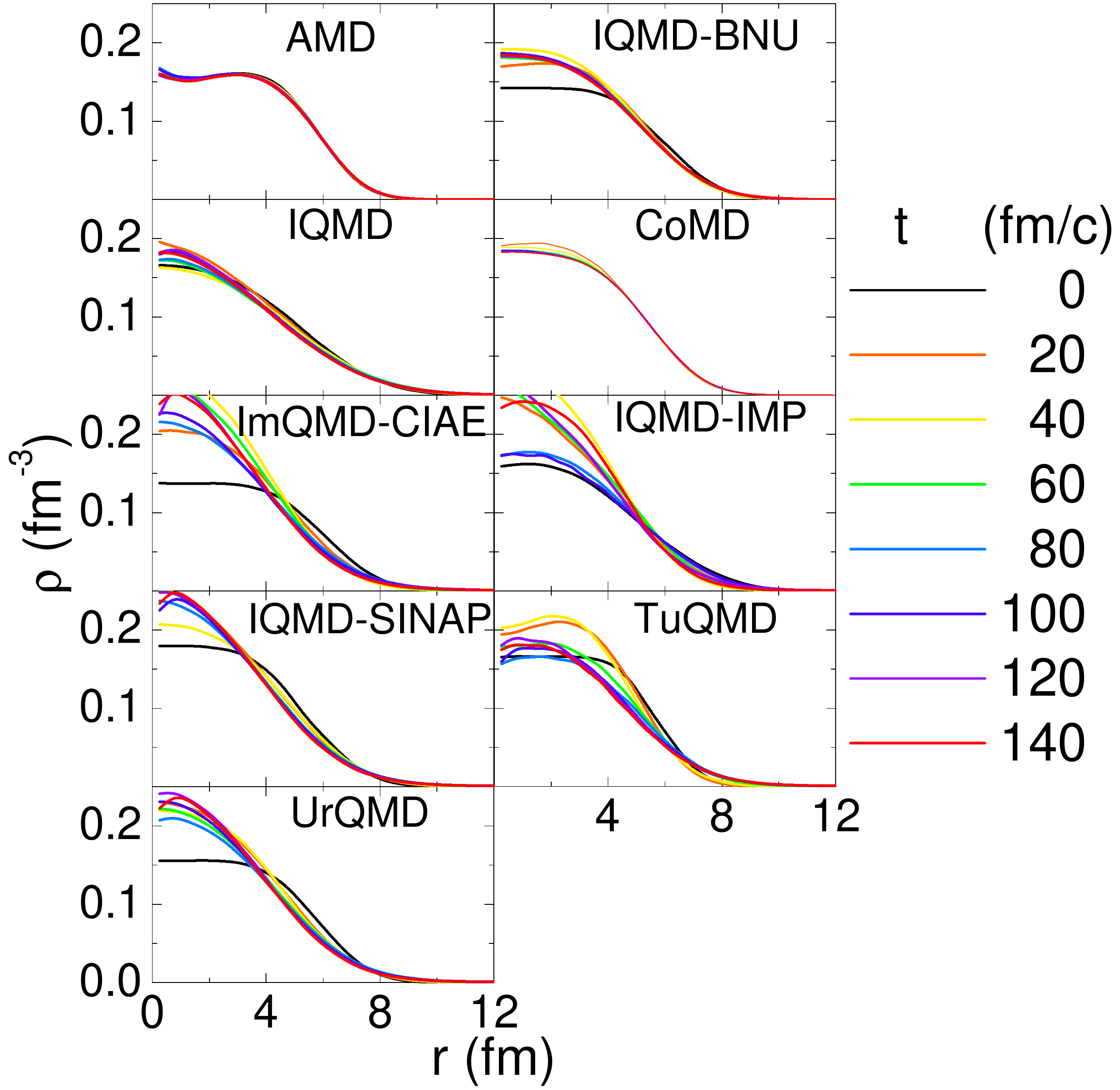}
\caption{(Color online) Time evolution of density profiles for Au nuclei from 9 BUU transport codes (left) and 9 QMD transport codes (right). Modified from figures in Ref.~\cite{Xu16}.}
\label{denevo_Xu16}
\end{figure}

As an example, let us first take a look at the time evolution of density profiles from 9 BUU codes and 9 QMD codes with a given Woods-Saxon distribution in Fig.~\ref{denevo_Xu16}. This is part of the heavy-ion comparison of transport codes in Ref.~\cite{Xu16}, for the purpose of checking the stability of initial nuclei, where the same input mean-field potential is incorporated for each code. If the density distribution is nearly the ground-state configuration and the Pauli blocking is good enough, the density profile should maintain its initial shape, such as that in the GIBUU-Skyrme, pBUU, AMD, and CoMD codes. With the evolution of time, the IBUU, RVUU, SMF, TuQMD, and IQMD-IMP codes show some oscillation behavior like a giant monopole resonance. This is understandable since the Woods-Saxon distribution is not the ground-state configuration for these codes, according to their mean-field potential calculation. On the other hand, the BLOB and IBL codes show some bubble-like structure, while for many other QMD codes the density distribution evolves away from the initial one and stays stable in another one with higher central densities. The behaviors of these codes show that due to the different ways of calculating the same input mean-field potential, the stable or metastable configuration for each code is generally different, and this may be due to their different shapes and/or sizes of the test particles or widths of the Gaussian wave packet, or other different treatments. Also, even if the soft nucleon-nucleon collisions within the same nucleus are not completely Pauli blocked, they sometimes serve as dissipations that damp the density oscillation into a stable state.

\begin{figure}[h]
\centering
\includegraphics[scale=0.4,clip]{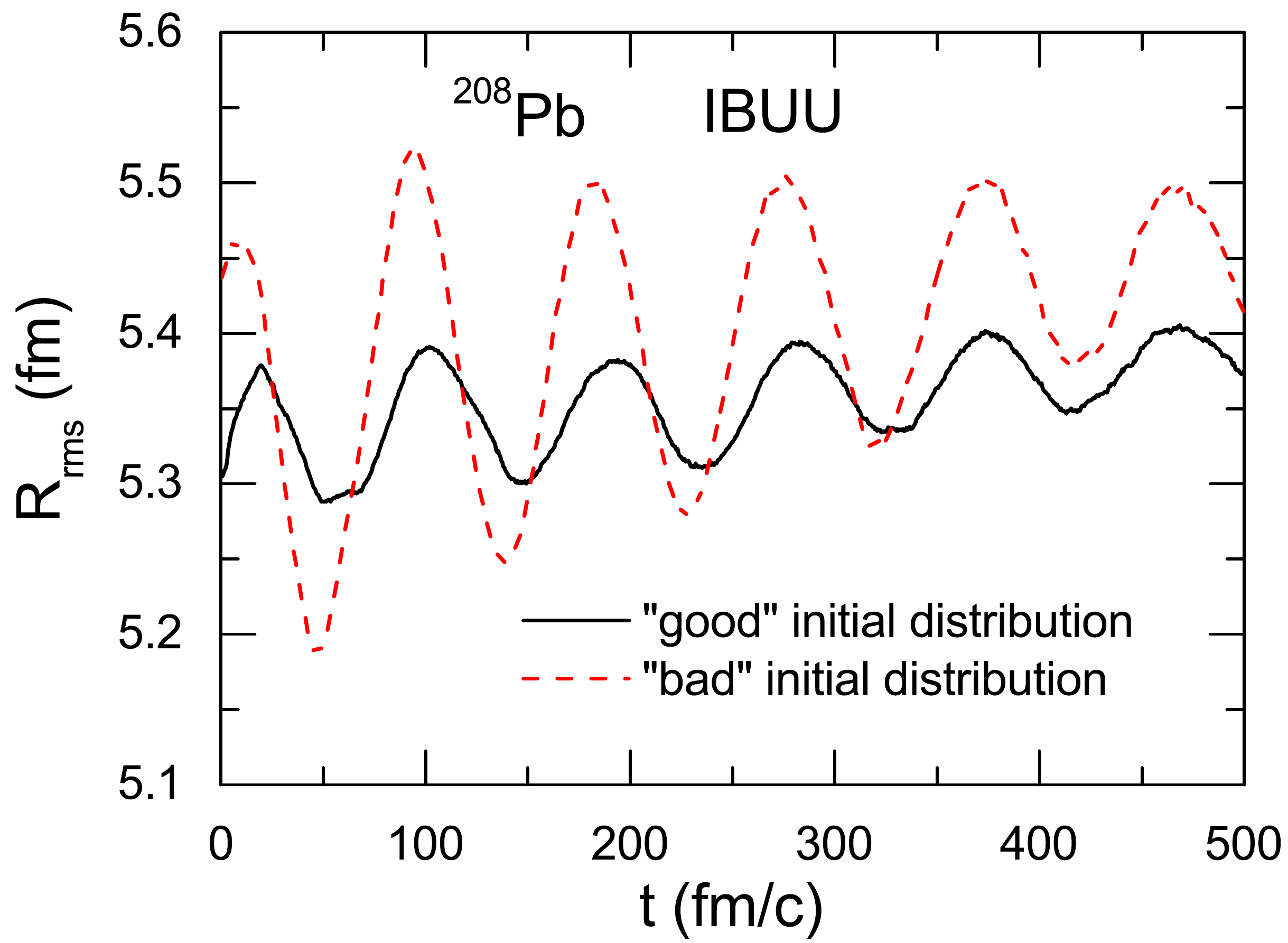}
\caption{(Color online) Time evolution of the root-mean-square radius of $^{208}$Pb nucleus from IBUU calculations, with the initial density distribution generated from the same mean-field potential used in the IBUU calculation ('good') and an arbitrarily parameterized Woods-Saxon initial density distribution ('bad'). }
\label{radius_IBUU}
\end{figure}

Another example shown in Fig.~\ref{radius_IBUU} is the time evolution of the root-mean-square (RMS) radius of a $^{208}$Pb nucleus from the IBUU transport code with different initial density distributions, i.e., one generated from the same mean-field potential using the Hartree-Fock calculation method, and one from an arbitrarily parameterized Woods-Saxon distribution. The Woods-Saxon initialization leads to strong oscillations of the RMS radius. Such oscillations may lead to some spurious dynamics in heavy-ion simulations. The magnitude of the oscillation using the initialization generated from the same mean-field potential used for the transport simulation is much reduced, since the initialization is closer to the ground-state configuration.

In order to reduce the discrepancy due to the initialization in transport simulations, the suggestion for future improvement is that a configuration as closer as possible to the ground state should be used, which can be obtained from the Hartree-Fock calculation~\cite{SHF}, the Thomas-Fermi approximation~\cite{Len89}, or the frictional cooling method~\cite{Wil77}, while the nucleon Fermi motion should also be implemented properly. On the other hand, the mean-field potential should be calculated accurately so that the ground-state configuration suits the mean-field calculation, and that will be discussed in the following subsection.

\subsection{Mean-field potential}
\label{mf}

Extracting the mean-field potential for nucleons is the key purpose of transport simulations for intermediate-energy heavy-ion collisions. However, different methods of calculating the mean-field potential, especially due to the size and shape of the test particles for BUU-type codes or the width of the Gaussian wave packet for QMD-type codes, do introduce errors that lead to different forces from the same mean-field potential we intend to incorporate into the transport code. On the other hand, introducing the finite size of nucleons makes the system more stable as a result of the more diffusive density distribution. I will take the momentum-independent mean-field potential $U=a(\rho/\rho_0) + b(\rho/\rho_0)^\gamma$, corresponding to the potential energy density $\epsilon=\frac{a}{2} \left(\frac{\rho}{\rho_0}\right)^2 + \frac{b}{\gamma+1} \left(\frac{\rho}{\rho_0}\right)^{\gamma+1}$ as an example, and discuss the situation of the mean-field potential calculation from BUU-type codes and QMD-type codes in the following.

For BUU-type codes, the density is calculated by counting test particle numbers in the local cell, i.e.,
\begin{equation}
\rho(\vec{r}) = \frac{1}{N_{TP}} \sum_{i} g(\vec{r}-\vec{r}_i).
\end{equation}
The calculation depends on the choice of the smooth function $g$. If $g$ is a $\delta$ function, i.e., the $i$th particle only contributes to the local cell $\vec{r}$, the calculation of $\rho(\vec{r})$ in the limit of the cell grid is accurate with infinite number of test particles. With a finite numbers of test particles, there are statistical fluctuations that lead to the inaccuracy of the density calculation, and the emission of a few particles will change significantly the whole density distribution, making the system rather unstable. In order to improve the defects of point test particles, one may employ a better smooth function $g$ by using finite-size test particles. Of course, one needs to have the constraint that integrating $\rho(\vec{r})$ over the whole coordinate space leads to the same total number of particles.

A widely used method for BUU-type codes is the lattice Hamiltonian method~\cite{Len89}, where the average density $\rho_L$ at the sites of a three-dimensional cubic lattice is defined as
\begin{eqnarray}
\rho_L(\vec{r}_{\alpha})=\sum_{i}S(\vec{r}_{\alpha}-\vec{r}_i),
\end{eqnarray}
where $\alpha$ is a site index and $\vec{r}_{\alpha}$ is the position of site $\alpha$. $S$ is the shape function describing the contribution of a test particle at $\vec{r}_i$ to the value of the average density $\rho_L(\vec{r}_{\alpha})$ at $\vec{r}_{\alpha}$, and it has the following form
\begin{eqnarray}
S(\vec{r})=\frac{1}{N_{TP}(nl)^6}h(x)h(y)h(z)
\end{eqnarray}
with
\begin{eqnarray}
h(q)=(nl-|q|)\Theta(nl-|q|).
\end{eqnarray}
In the above, $l$ is the lattice spacing, $n$ determines the range of $S$, and $\Theta$ is the Heaviside function. The total potential energy of the system is
\begin{equation}
V = l^3 \sum_\alpha \epsilon_\alpha = l^3 \sum_\alpha \left[\frac{a}{2}\left(\frac{\rho_L(\vec{r}_\alpha)}{\rho_0}\right)^2 + \frac{b}{\gamma+1}\left(\frac{\rho_L(\vec{r}_\alpha)}{\rho_0}\right)^{\gamma+1} \right].
\end{equation}
The force acting on the $i$th particle is
\begin{equation}
\vec{F}_i = -N_{TP} \frac{\partial V}{\partial \vec{r}_i} = -N_{TP}l^3 \sum_\alpha \frac{\partial S(\vec{r}_\alpha-\vec{r}_i)}{\partial \vec{r}_i} \left[ a\left(\frac{\rho_L(\vec{r}_\alpha)}{\rho_0}\right) + b\left(\frac{\rho_L(\vec{r}_\alpha)}{\rho_0}\right)^\gamma\right].
\end{equation}
One sees that the spatial derivative with respect to the density becomes that with respect to the shape function $S$ in the lattice Hamiltonian framework, reducing the possible inaccuracy due to the numerical derivative. This framework is good for the energy conservation of the system evolving in the mean-field potential. On the other hand, the values of $l$ and $n$ should be chosen carefully. Due to the finite size of the test particles, the density becomes more diffusive, which makes the system more stable, but leads to possible inaccuracies of the density or force calculation, especially when the real density gradient is sharper than that of the shape function.

The energy density $\epsilon=\frac{a}{2} \left(\frac{\rho}{\rho_0}\right)^2 + \frac{b}{\gamma+1} \left(\frac{\rho}{\rho_0}\right)^{\gamma+1}$ can be obtained from a Skyrme-type two-body interaction $V_{ij} = t_0 \delta(\vec{r}_i-\vec{r}_j) + t_3 \rho^{\gamma-1}[(\vec{r}_i+\vec{r}_j)/2] \delta(\vec{r}_i-\vec{r}_j)$ through the Hartree calculation. For QMD-type codes, the potential energy from the first term of the above Skyrme-type interaction according to Eq.~(\ref{U2}) can be exactly obtained as
\begin{equation}\label{VQMD}
\langle V_0 \rangle = \frac{a}{2} \sum_i \left(\frac{\langle\rho\rangle_i}{\rho_0}\right),
%+ \frac{b}{\gamma+1} \left(\frac{\langle\rho\rangle}{\rho_0}\right)^{\gamma+1},
\end{equation}
where $\langle\rho\rangle_i = \sum_{j, i \ne j} \rho_{ij}$ is the average density contribution at $\vec{r}_i$, with
\begin{equation}
\rho_{ij} = \frac{1}{(4\pi \text{L})^{3/2}} \exp\left[ -\frac{(\vec{r}_i-\vec{r}_j)^2}{4\text{L}}\right].
\end{equation}
The potential energy contribution from the second term of the Skyrme-type interaction cannot be obtained analytically, and the approximation $\langle\rho^\gamma\rangle_i \approx \langle\rho\rangle_i^\gamma$ is generally used\footnote{This is generally not a good approximation, and it is recommended to carry out the corresponding integral numerically.}. In this way, the corresponding potential energy becomes
\begin{equation}\label{VQMD}
\langle V_3 \rangle =  \frac{b}{\gamma+1} \sum_i\left(\frac{\langle\rho\rangle_i}{\rho_0}\right)^{\gamma},
\end{equation}
and $\langle V \rangle =\langle V_0 \rangle+ \langle V_3 \rangle$ is the total potential energy. The force acting on the $i$th particle becomes
\begin{eqnarray}\label{F1QMD}
\vec{F}_i &=& -\frac{\partial \langle V \rangle}{\partial \vec{r}_i} \notag\\
&=& a \sum_{j, i \ne j} \frac{1}{(4\pi \text{L})^{3/2}} \exp\left[ -\frac{(\vec{r}_i-\vec{r}_j)^2}{4\text{L}}\right] \left( -\frac{\vec{r}_i-\vec{r}_j}{2\text{L}} \right) + \frac{\gamma b}{\gamma+1} \langle \rho \rangle_i^{\gamma-1} \sum_{j, i \ne j} \frac{1}{(4\pi \text{L})^{3/2}} \exp\left[ -\frac{(\vec{r}_i-\vec{r}_j)^2}{4\text{L}}\right] \left( -\frac{\vec{r}_i-\vec{r}_j}{2\text{L}} \right)\notag\\
&+& \frac{\gamma b}{\gamma+1} \sum_{j, i \ne j} \langle \rho \rangle_j^{\gamma-1}  \frac{1}{(4\pi \text{L})^{3/2}} \exp\left[ -\frac{(\vec{r}_i-\vec{r}_j)^2}{4\text{L}}\right] \left( -\frac{\vec{r}_i-\vec{r}_j}{2\text{L}} \right).
\end{eqnarray}

%The different calculations of the forces from BUU-type codes and QMD-type codes are illustrated as follows, by starting directly from the mean-field potential $U=a(\rho/\rho_0) + b(\rho/\rho_0)^\gamma$ with the spatial distribution of each particle a Gaussian wave packet as in QMD, i.e., the density at the position $\vec{r}$ is calculated from the summation of that for all particles as
%\begin{equation}\label{denQMD}
%\rho(\vec{r}) = \sum_j \frac{1}{(2\pi \text{L})^{3/2}} \exp\left[ -\frac{(\vec{r}-\vec{r}_j)^2}{2\text{L}}\right].
%\end{equation}
%The force is thus
%\begin{equation}\label{F2QMD}
%\vec{F}_i = - \left(\frac{\partial U}{\partial \vec{r}}\right)_{\vec{r}=\vec{r}_i} =  a \sum_{j}  \frac{1}{(2\pi \text{L})^{3/2}} \exp\left[ -\frac{(\vec{r}_i-\vec{r}_j)^2}{2\text{L}}\right] \left( -\frac{\vec{r}_i-\vec{r}_j}{\text{L}} \right) + b \gamma \rho^{\gamma-1}(\vec{r}_i) \sum_{j}  \frac{1}{(2\pi \text{L})^{3/2}} \exp\left[ -\frac{(\vec{r}_i-\vec{r}_j)^2}{2\text{L}}\right] \left( -\frac{\vec{r}_i-\vec{r}_j}{\text{L}} \right).
%\end{equation}
%Eq.~(\ref{F2QMD}) is used to illustrate the different force calculations between the BUU-type codes and the QMD-type codes, rather than a solution for the approximation in Eq.~(\ref{VQMD}). By comparing Eq.~(\ref{F1QMD}) and Eq.~(\ref{F2QMD}), one can see that the calculations of the force and the mean-field potential in BUU-type codes and QMD-type codes are different even if the shape of the test particle is of the same Gaussian form.

In BUU-type codes, it is recommended to use the lattice Hamiltonian framework with finite-size test particles, although a large number of point test particles can be qualified for calculating accurately the mean-field potential. The former helps to improve the stability of the system and maintain better the energy conservation, and is able to describe the force reasonably well if the density gradient is not so sharp. For QMD-type codes, the size of the particle is controlled by the width of the Gaussian wave packet. A larger width is expected to give a more diffusive density distribution but a weaker force from the mean-field potential. It is also interesting to investigate the different calculations of the mean-field potential as well as the forces in BUU-type codes and QMD-type codes in future studies. %The calculations of the force from the mean-field potential in BUU-type codes and QMD-type codes are different.

\subsection{Nucleon-nucleon collisions} %elastic, Pauli blocking, inelastic
\label{collisions}

The collision integral on the right-hand side of the BUU equation is generally simulated stochastically in transport approaches. The collision term is responsible for energy dissipations, and crucial for the dynamics in heavy-ion simulations. The hard collision process, with the uncertainties of the in-medium nucleon-nucleon collision cross section, sometimes hampers our understandings of the soft mean-field potential interaction. Although it is mentioned in Secs.~\ref{BUU} and \ref{QMD} that the mean-field potential and the in-medium cross section are related to each other through the fundamental nuclear interaction, they are generally determined separately in real heavy-ion simulations.

There are several methods to realize the collision process. The most widely used one is the Bertsch's prescription~\cite{Ber88}. In the Bertsch's prescription, the minimum distance of two colliding particles in their center-of-mass frame perpendicular to their relative velocity is
\begin{equation}
{d_\perp^\star}^2 = (\vec{r}_1^\star-\vec{r}_2^\star)^2 - \frac{[(\vec{r}_1^\star-\vec{r}_2^\star)\cdot \vec{v}_{12}^\star]^2}{{v_{12}^\star}^2},
\end{equation}
where $\vec{r}_1^\star$ and $\vec{r}_2^\star$ are positions of the two particles in their center-of-mass frame, and $\vec{v}_{12}^\star = \vec{v}_1^\star - \vec{v}_2^\star$ is their relative velocity. Here the asterisk represents the quantity in the center-of-mass frame of the colliding particles. The collision can happen if the condition $\pi {d_\perp^\star}^2 < \sigma$ is satisfied. It is an alternative to use the condition $\pi {d_\perp^\star}^2 < \sigma_{max}$ together with the collision probability $P=\sigma/\sigma_{max}$, where $\sigma_{max}$ is larger than any possible value of the real cross section $\sigma$, with the latter generally a function of the center-of-mass energy of the colliding pair. The above condition satisfies that the collision can happen, while whether it happens for this time step is determined by the condition of the closest approach, i.e., $|(\vec{r}_1^\star-\vec{r}_2^\star)\cdot \vec{v}_{12}^\star/{v_{12}^\star}^2|< \frac{1}{2} \delta t$. The choice of $\delta t = \alpha \Delta t$, where $\Delta t$ is the time step in the heavy-ion reference frame, needs to be taken care. Many codes choose $\alpha=1$, i.e., the same time step in the two reference frames. A more suitable choice is $\alpha = \gamma E_1^\star E_2^\star / E_1 E_2$, where $E_{1/2}^{(\star)}$ is the energy of the particles in different frames, and $\gamma=1/\sqrt{1-\beta^2}$ is the Lorentz factor with $\beta$ being the average velocity of the colliding pair in the heavy-ion reference frame. $\alpha=1/\gamma$ is also a good choice for the time dilation effect, by assuming that the two particles have a common velocity. Slightly different choices in various transport codes can be found in Table IV of Ref.~\cite{Zhang18}.

With the original Bertsch's prescription for collisions, the particle pair that collide once have $50\%$ chance to collide again in the subsequent time step for an isotropic cross section, if the final velocities point toward each other. This effect is contradictory to the assumption of the Boltzmann equation that the collisions are independent of each other and are not repeated. These spurious collisions can be avoided by setting that the two particles, that have collided once, can not collide again unless one of them has collided with a third particle. The collision rate will still be different from the exact limit of the collision rate from Eq.~(\ref{dncoll}) as a result of higher-order correlations, i.e., after the particle pair collide, one of them could collide with a third particle and then the same particle pair collide again. This effect is not accounted for in an exact evaluation of the collision integral in the Boltzmann equation, but shows up especially at higher densities or with a larger cross section. In the situation with a strong Pauli blocking or a forward-peaked cross section, both the repeated collisions and higher-order correlations will be suppressed. For BUU-type codes with the test-particle method, the full ensemble method is often incorporated by allowing nucleon-nucleon collisions from different parallel events, while the cross section is divided by the number of test particles. With the full ensemble method, the higher-order correlations of nucleon-nucleon collisions are suppressed, since the chance of colliding back into the vicinity of a preceding collision partner are much reduced with a smaller cross section. It will be of interest to test in more detail the dynamical effect in addition to the overestimated collision rate caused by repeated collisions and higher-order correlations. In addition, although it is difficult to remove all the higher-order correlations, it is possible to remove the next leading order, i.e., allowing the same pair to collide again only if they have experienced at least two collisions with other particles, etc.

The above Bertsch's prescription can be considered as a geometrical approach. Another approach can be considered as the stochastic approach, where the probability for the collision of a pair of particles with their energies $E_1$ and $E_2$ in the volume $(\Delta x)^3$ and the time interval $\Delta t$ to happen is $v_{rel} \sigma \frac{\Delta t}{(\Delta x)^3}$~\cite{Xu05}, with $v_{rel} = s/(E_1 E_2)$ where $s$ is the square of the invariant mass of the particle pair. This method can be understood as the mean-free-path method. The mean free path $\lambda$ of the particle is $1/(\rho \sigma)$ determined by the number density $\rho$ and the cross section $\sigma$, and the relaxation time for two-particle collisions is $\tau=\lambda/v_{rel}$. The probability for the two particles to collide in the time interval $\Delta t$ is thus $\Delta t/\tau = \rho \sigma v_{rel} \Delta t$. This method guarantees the collision rate satisfying Eq.~(\ref{dncoll}).

The order of checking whether the collision between a particle pair can happen should be treated with care. Many codes use a fixed order following the originally chosen particle indices. This can be easily improved by randomizing the order at each time step. A more realistic treatment is that a physical collision time table is built at the beginning of each time step, and the collisions are processed following the order of this table. The collision order treatments for various transport codes can be found in Table IV of Ref.~\cite{Zhang18}.

The above attempted collisions can happen only if the final state of the scattered particles are not occupied. This is the so-called Pauli blocking effect reflecting the quantum statistics of Fermions in the transport simulation. The Pauli blocking probability is generally expressed as $1-(1-f_3)(1-f_4)$, where $f_3$ and $f_4$ are the occupation probabilities of the final states of the scattered particles. In most BUU-type codes, the occupation probability at the phase-space state $(\vec{r}_i,\vec{p}_i)$ is obtained from counting particle numbers in discrete local phase-space cells representing a certain range of the coordinate and momentum space averaging over all parallel events, i.e.,
\begin{equation}
f_i = \frac{h^3}{2V_rV_pN_{TP}} \sum_{k \in \tau (k \ne i)} \delta(\vec{r}_k-\vec{r}_i) \delta(\vec{p}_k-\vec{p}_i).
\end{equation}
In the above, $N_{TP}$ is the number of test particles, $V_r$ and $V_p$ are the volumes of the phase-space cells in coordinate and momentum space, respectively, and they can be cubic or spherical. For the summation within the phase-space cell, $\tau=n,p$ is the isospin index, showing that the Pauli blocking is isospin dependent, and $k \ne i$ means that the $i$th particle itself does not contribute to the occupation probability. The calculation of $f_i$ in BUU-type codes may depend on the size and the shape of the phase-space cell, and there could be interpolation or smearing treatment by replacing the $\delta$ function in the above expression with a smooth function. The calculation becomes more accurate with the increasing number of test particles to reduce the fluctuation in each phase-space cell in BUU-type codes. A more economic procedure used in pBUU code is the effective temperature method, where the distribution function used for the Pauli blocking in the cell around the final state of the scattered particle is fitted by a weighted sum of two deformed Fermi-Dirac distributions. In most QMD-type codes, the occupation probability is calculated at the centroid phase space $(\vec{r}_i,\vec{p}_i)$ of the scattered wave packet with $\vec{p_i}$ the final momentum, i.e.,
\begin{equation}\label{PBQMD}
f_i = 4 \sum_{k \in \tau (k \ne i)} \exp\left[-\frac{(\vec{r}_i-\vec{r}_k)^2}{2\text{L}}\right] \exp\left[-\frac{2\text{L}}{\hbar^2} (\vec{p}_i-\vec{p}_k)^2\right],
\end{equation}
where the summation is similar to that in BUU-type codes. Since QMD-type codes describe event-by-event evolutions, the fluctuation of the phase-space distribution, depending on the width of the Gaussian wave packet, is much larger compared with that in BUU-type codes, underestimating effectively the Pauli blocking rate. It remains an open question how to preserve fluctuations while improving the Pauli blocking treatment properly in QMD-type codes.

\subsection{Particle productions}
\label{parpro}

With particles other than nucleons, the Boltzmann equation becomes a set of equations for different particle species with the collision term connecting each equation. With the production of pion-like particles, the collisions need to be treated with special care involving inelastic channels. In this case, the total cross section $\sigma_{tot}$ of nucleon-nucleon collisions including both the elastic and inelastic channels should be calculated, and whether an attempted collision can happen depends on $\pi {d_\perp^\star}^2 < \sigma_{tot}$ in the geometrical method or with the probability $\rho \sigma_{tot} v_{rel} \Delta t$ in the stochastic method. When an attempted collision is decided to occur, a channel should then be selected based on the ratio of partial cross sections. Typical examples can be the treatment of elastic nucleon-nucleon collisions $N+N \rightarrow N+N$ and inelastic nucleon-nucleon collisions $N+N \rightarrow N+\Delta$. The collision term in the Boltzmann equation includes not only elastic and inelastic collisions but also decay channels $\Delta \rightarrow N+\pi$, where the decay probability is generally expressed as $1-\exp(-\Gamma \delta t)$, with $\Gamma$ being the width of the $\Delta$ mass distribution and $\delta t$ being the time step in the rest frame of the $\Delta$ resonance. The cross sections of the inverse channels $N+\Delta \rightarrow N+N$ and $\pi+N \rightarrow \Delta$ satisfy the detailed balance condition, and the cross sections and decay widths of different isospin channels are related according to Clebsh-Gordan coefficients.

The particle production and annihilation lead to complexities of transport simulations, and also model dependence, since one needs to handle more degrees of freedom. With time-step free codes, it is less problematic because the sequence of the collision, decay, and propagation can be treated in the order of physical time. In codes with time steps, which is needed for the incorporation of mean-field potentials, it is found that the sequence of the production, decay, and propagation may affect multiplicities of pions and $\Delta$ resonances~\cite{pionbox}. For example, a code with first a decay process $\Delta \rightarrow \pi+N$ and then a production process $\pi+N \rightarrow \Delta$ within a time step will have a larger multiplicity of $\Delta$ resonances and a smaller multiplicity of pions at the time step boundary, compared to a code with first a $\Delta$ production process and then a $\Delta$ decay process. In addition, since the cross sections for different isospin channels are different, this leads to different higher-order correlations and may lead to the isospin symmetry violation~\cite{pionbox}. For instance, the cross section of $p+\Delta^- \rightarrow n+n$ is larger than that of $p+\Delta^0 \rightarrow n+p$, enhancing the annihilation of $\Delta^-$ compared to $\Delta^0$ with higher-order correlations, and this is similar for the case of $\Delta^{++}$ and $\Delta^+$. If the inverse channel of $\Delta$ decay is considered, $n+\pi^- \rightarrow \Delta^{-}$ has a larger cross section than $n+\pi^0 \rightarrow \Delta^0$, enhancing the yield of $\Delta^-$ compared to $\Delta^0$ with higher-order correlations. The two effects compete with each other, while BUU-type codes with the full-ensemble method can be away from suffering the higher-order correlations mentioned above.

\section{Summary and Outlook}
\label{summary}

Transport approaches have made great achievements in understanding the dynamics of intermediate-energy heavy-ion collisions, and extracting valuable information of the nuclear equation of state as well as microscopic nuclear interactions, from various observables such as the collective flow, the particle yield and their ratios, as well as correlations and so on. Since different densities and temperatures are reached at different incident energies, the observables are better probes of the equation of state and the microscopic nuclear interactions at their dominating densities. On the other hand, the collision dynamics are more affected by different ingredients of transport models at different incident energies. At lower incident energies, where there are probes for the equation of state at subsaturation densities, the results are more sensitive to the initialization, the mean-field potential, and the Pauli blocking. At higher incident energies, where there are probes for the equation of state at suprasaturation densities, the results are more sensitive to the nucleon-nucleon collisions as well as the detailed treatment of particle productions.

In order to understand more accurately the isospin-dependent part of the nuclear equation of state as well as the different mean-field potentials for neutrons and protons in neutron-rich medium, further reduction of the model uncertainties for various transport codes as well as those between the main branches, i.e., the BUU transport approach and the QMD transport approach, is very much needed. To achieve this goal, there are at least three main activities, from the ECT* transport workshop in 2004, to the Shanghai transport workshop in 2014, and to the latest MSU transport workshop in 2017. The testing system is from higher-energy heavy-ion collisions to lower-energy heavy-ion collisions, and to the box system with the periodic boundary condition. The observables change from kaons and pions to the stopping and flows as well as collision and Pauli blocking rates. With the help of these comparison, especially the box calculation where the theoretical limits are mostly available, guidance on the optimized treatments for the main ingredients of the transport approach can be learned as follows.

It is recommended to use the ground-state configuration for the initialization, based on the mean-field potential incorporated. The initial phase-space distribution from the Thomas-Fermi or Hartree-Fock method is suitable for BUU transport approaches, and the frictional cooling method could be useful for QMD transport approaches. For the mean-field potential calculation, one should use a large number of test particles for BUU approaches, and the finite-size test-particle method such as the lattice Hamiltonian framework, which maintains better the energy conservation, is recommended. For QMD transport approaches, the width of the Gaussian wave packet should be taken with care in the mean-field potential calculation, while it is recommended to calculate the potential from the density-dependent term exactly with the numerical integral method rather than approximately. The stochastic method for nucleon-nucleon elastic collisions is recommended, which guarantees to reproduce the collision rate properly. If the geometrical method is used, one should remove spurious collisions, treat the time dilation properly, and use the full ensemble method for BUU approaches in order to suppress the higher-order correlations if possible. The sequence of the inelastic collision, decay, and propagation should be treated with care, if particle productions enter the game. For the Pauli blocking in BUU approaches, one should use more test particles, and the effective temperature method as in pBUU is recommended. How to improve the Pauli blocking in QMD approaches needs further investigations, if the wave function is not antisymmetrized as in the AMD code. %For the Pauli blocking in QMD approaches, if the wave function is not antisymmetrized as in AMD code, use parallel events to calculate the phase-space occupation and Pauli blocking rate as in BUU approaches.

So far another two box calculations with only the mean-field potential and for the production of pion-like particles are about to be finished. Meanwhile, the comparison of pion productions in intermediate-energy heavy-ion collisions is in progress. In the future, comparisons for the momentum-dependent mean-field potential as well as cluster formations are needed, in order to obtained their optimized treatments in transport simulations. It could be useful to develop a standard BUU code and a standard QMD code, with the optimized treatments for the main ingredients well settled. Meanwhile, we believe that most transport codes are making improvements according to the optimized treatments learnt from the transport code evaluation project. This hopefully leads to an era for the accurate extraction of the nuclear equation of state and microscopic nuclear interactions from intermediate-energy heavy-ion collisions.

\section{Acknowledgements}
I acknowledge helpful discussions with Lie-Wen Chen, Maria Colonna, Pawel Danielewicz, Akira Ono, Betty Tsang, Yong-Jia Wang, Hermann Wolter, and Ying-Xun Zhang. This work was supported by the Major State Basic
Research Development Program (973 Program) of China under Contract
No. 2015CB856904, and the National Natural Science
Foundation of China under Grant No. 11475243 and No. 11421505.

\clearpage
%% References

% macros used by ADS Database BiBTeX entries:
% see http://adsabs.harvard.edu/abs_doc/aas_macros.sty
\newcommand{\apjl}{Astrophys. J. Lett.\ }
\newcommand{\apj}{Astrophys. J. \ }
\newcommand{\prc}{Phys. Rev. C\ }
\newcommand{\prd}{Phys. Rev. D\ }
\newcommand{\mnras}{Mon. Not. R. Astron. Soc.\ }
\newcommand{\aap}{Astron. Astrophys.\ }
\newcommand{\nphysa}{Nucl. Phys. A\ }
\newcommand{\physrep}{Phys. Rep.\ }
\newcommand{\nat}{Nature\ }
\newcommand{\sci}{Science\ }
\newcommand{\plb}{Phys. Lett. B\ }
\newcommand{\ijmpe}{Int. J. Mod. Phys. E\ }
\newcommand{\prl}{Phys. Rev. Lett.\ }
\newcommand{\jpg}{J. Phys. G\ }
\newcommand{\epja}{Euro. Phys. J. A\ }
\newcommand{\annphys}{Ann. Phys.\ }
\newcommand{\rmp}{Rev. Mod. Phys.\ }
\newcommand{\physrev}{Phys. Rev.\ }
\newcommand{\ppnp}{Prog. Part. Nucl. Phys.\ }
\newcommand{\mpla}{Mod. Phys. Lett. A\ }

\bibliographystyle{elsarticle-num}
\bibliography{transport}

\end{document}